\documentclass{lmcs} %%% last changed 2014-08-20

\keywords{Characteristic formulae, prime formulae, bisimulation, simulation relations, branching-time spectrum, modal logics, complexity theory, satisfiability}

\usepackage{hyperref}

\usepackage{amstext,bm}
\usepackage{amsthm}
\usepackage{amsopn}
\usepackage{tikz}
\usetikzlibrary{trees}
\usetikzlibrary{shapes,arrows}
\usetikzlibrary{intersections}
\usepackage[utf8]{inputenc}
\usepackage{amsmath}
\usepackage{amssymb,amsfonts}
\usepackage[T1]{fontenc}
\usepackage{bm}
\usepackage{graphicx}
\usepackage{caption,newfloat}
\usepackage{hyperref}
\usepackage{wrapfig}
\usepackage{blindtext}
\usepackage{array}
\usepackage{makecell}
\usepackage{centernot}
\usepackage{mathtools}
\usepackage{float}
\usepackage{algpseudocode}
\usepackage[linesnumbered,lined,boxed,commentsnumbered,noend]{algorithm2e}
\SetKwFunction{process}{Process}
\SetKwFunction{gcd}{MLB}
\SetKwRepeat{Do}{do}{while}%
\usepackage{stmaryrd}
\usepackage{thmtools}
\usepackage{thm-restate}
\usepackage{cleveref}
\usepackage{bussproofs}
\usetikzlibrary{external}
\usepackage{colortbl}
\usepackage{hhline}
\usepackage{multirow}
\usepackage{booktabs}

\newcommand{\classize}[1]{{\ensuremath{\mathsf{#1}}}\xspace}
\newcommand{\pspace}{\classize{PSPACE}}
\newcommand{\expc}{\classize{EXP}}
\newcommand{\cP}{\classize{P}}
\newcommand{\NP}{\classize{NP}}
\newcommand{\conp}{\classize{coNP}}
\newcommand{\us}{\classize{US}}

\newcommand{\dpc}{\classize{DP}}
\newcommand{\fpt}{\classize{FPT}}
\newcommand{\ap}{\classize{AP}}

\newcommand{\problemize}[1]{{\textsc{#1}}\xspace}

\newcommand{\SAT}{\problemize{Sat}}
\newcommand{\reachability}{\problemize{Reachability}}

\newcommand{\logicize}[1]{{\ensuremath{\mathbf{#1}}}\xspace}

\newcommand{\hml}{\logicize{HML}}
\newcommand{\logikk}{\logicize{K}}

\newcommand{\mathL}{\ensuremath{\mathcal{L}}\xspace}
\newcommand{\mathS}{\ensuremath{\mathcal{S}}\xspace}
\newcommand{\mathP}{\ensuremath{\mathcal{P}}\xspace}
\newcommand{\true}{\ensuremath{\mathbf{tt}}\xspace}
\newcommand{\ff}{\ensuremath{\mathbf{ff}}\xspace}

\newcommand{\curle}{\lesssim}
\newcommand{\notcurle}{\not\lesssim}

\newcommand{\act}{\ensuremath{\mathtt{Act}}\xspace}
\newcommand{\proc}{\ensuremath{\mathtt{Proc}}\xspace}

\newcommand{\compL}{\ensuremath{\overline{\mathcal{L}}}\xspace}

\newcommand{\sub}{\ensuremath{\mathrm{Sub}}}

\newcommand{\subcs}{\ensuremath{\mathrm{ConSub_{CS}}}\xspace}
\newcommand{\conscs}{\ensuremath{\mathrm{Cons_{CS}}}\xspace}
\newcommand{\subrs}{\ensuremath{\mathrm{ConSub_{RS}}}\xspace}

\newcommand{\algos}{\ensuremath{\mathrm{Prime_S}}\xspace}
\newcommand{\algott}{\ensuremath{\mathrm{Prime^{\diamond}}}\xspace}
\newcommand{\algocs}{\ensuremath{\mathrm{Prime_{CS}}}\xspace}
\newcommand{\algosat}{\ensuremath{\mathrm{Prime^{\mathrm{sat}}}}\xspace}

\newcommand{\algorsu}{\ensuremath{\mathrm{Prime_{RS}^u}}\xspace}
\newcommand{\zero}{\ensuremath{\mathbf{0}}}
\newcommand{\expphi}{\ensuremath{\mathrm{Exp}(\varphi)}\xspace}
\newcommand{\md}{\ensuremath{\mathrm{md}}}

\newcommand{\depth}{\ensuremath{\mathrm{depth}}}
\newcommand{\sat}{\ensuremath{\textsc{Prop}}}

\newcommand{\enc}{\ensuremath{\mathrm{enc}}}
\newcommand{\grcd}{\ensuremath{\mathrm{gcd}}}
\newcommand{\simpl}{\ensuremath{\mathrm{simpl}}}

\newcommand{\reach}{\ensuremath{\mathrm{reach}}}

\newcommand{\trim}{\ensuremath{\mathrm{trim}}}

\newcommand{\reacha}{\ensuremath{\textsc{Reach}_{a}}\xspace}
\newcommand{\inlabeli}{\ensuremath{L_i^\mathtt{in}}\xspace}
\newcommand{\finlabeli}{\ensuremath{L_i^\mathtt{fin}}\xspace}
\newcommand{\inlabelone}{\ensuremath{L_1^\mathtt{in}}\xspace}
\newcommand{\finlabelone}{\ensuremath{L_1^\mathtt{fin}}\xspace}

\newcommand{\inlabelthree}{\ensuremath{L_3^\mathtt{in}}\xspace}
\newcommand{\finlabelthree}{\ensuremath{L_3^\mathtt{fin}}\xspace}
\newcommand{\starlabelthree}{\ensuremath{L_3^{\mathrm{sub}}}\xspace}
\newcommand{\wit}{\ensuremath{\mathrm{wit}_{\not\curle_S}}\xspace}
\newcommand{\witproc}{\ensuremath{\mathrm{witproc}_{\not\curle_S}}\xspace}
\newcommand{\cop}{\ensuremath{\mathrm{copy}}\xspace}
\newcommand{\simequiv}{\ensuremath{\mathtt{char1se}}\xspace}

\newcommand{\nsimeq}{\ensuremath{\mathtt{charnse}}\xspace}
\newcommand{\isimeq}{\ensuremath{\mathtt{charise}}\xspace}
\newcommand{\nsimpre}{\ensuremath{\mathtt{primensp}}\xspace}
\newcommand{\nmosimeq}{\ensuremath{\mathtt{char(n-1)se}}\xspace}

\newcommand{\simab}{\ensuremath{\mathtt{Sim_{A,B}}}\xspace}
\newcommand{\nmosimab}{\ensuremath{\mathtt{Sim^{n-1}_{A,B}}}\xspace}
\newcommand{\nsimab}{\ensuremath{\mathtt{Sim^{n}_{A,B}}}\xspace}
\newcommand{\isimab}{\ensuremath{\mathtt{Sim^{i}_{A,B}}}\xspace}

\newcommand{\ruleff}{\ensuremath{\rightarrow_{\mathrm{ff}}}}
\newcommand{\rulezero}{\ensuremath{\rightarrow_{\mathrm{0}}}}
\newcommand{\rulezerosub}{\ensuremath{\rightarrow_{\mathrm{0}}^{sub}}}
\newcommand{\rulezerostar}{\ensuremath{{\rightarrow_{\mathrm{0}}^{sub}}}^*}
\newcommand{\rulett}{\ensuremath{\rightarrow_{\mathrm{tt}}}}
\newcommand{\rulettsub}{\ensuremath{\rightarrow_{\mathrm{tt}}^{sub}}}
\newcommand{\rulettstar}
{\ensuremath{{\rightarrow_{\mathrm{tt}}^{sub}}}^*}
\newcommand{\rulediam}{\ensuremath{\rightarrow_{\mathrm{\diamond}}}}
\newcommand{\rulediamsub}{\ensuremath{\rightarrow_{\mathrm{\diamond}}^{sub}}}
\newcommand{\rulediamstar}
{\ensuremath{{\rightarrow_{\mathrm{\diamond}}^{sub}}}^*}

\newcommand\myarrowa{\mathrel{\stackrel{\makebox[0pt]{\mbox{\normalfont \scriptsize $a$}}}{\longrightarrow}}}
\newcommand\myarrowb{\mathrel{\stackrel{\makebox[0pt]{\mbox{\normalfont \scriptsize $b$}}}{\longrightarrow}}}
\newcommand\myarrowc{\mathrel{\stackrel{\makebox[0pt]{\mbox{\normalfont \scriptsize $c$}}}{\longrightarrow}}}
\newcommand\myarrowd{\mathrel{\stackrel{\makebox[0pt]{\mbox{\normalfont \scriptsize $d$}}}{\longrightarrow}}}
\newcommand\myarrowe{\mathrel{\stackrel{\makebox[0pt]{\mbox{\normalfont \scriptsize $e$}}}{\longrightarrow}}}
\newcommand\myarrowf{\mathrel{\stackrel{\makebox[0pt]{\mbox{\normalfont \scriptsize $f$}}}{\longrightarrow}}}

\newcommand\myarrowasubi{\mathrel{\stackrel{\makebox[0pt]{\mbox{\normalfont \scriptsize $a_i$}}}{\longrightarrow}}}

\newcommand\myarrowan{\mathrel{\stackrel{\makebox[0pt]{\mbox{\normalfont \scriptsize $a_n$}}}{\longrightarrow}}}
\newcommand\notmyarrowan{\mathrel{\stackrel{\makebox[0pt]{\mbox{\normalfont \scriptsize $a_n$}}}{\not\rightarrow}}}

\newcommand\myarrowtau{\mathrel{\stackrel{\makebox[0pt]{\mbox{\normalfont \scriptsize $t$}}}{\longrightarrow}}}

\newcommand\notmyarrowa{\mathrel{\stackrel{\makebox[0pt]{\mbox{\normalfont \scriptsize $a$}}}{\not\rightarrow}}}
\newcommand\notmyarrowtau{\mathrel{\stackrel{\makebox[0pt]{\mbox{\normalfont \scriptsize $t$}}}{\not\rightarrow}}}
\newcommand\notmyarrowb{\mathrel{\stackrel{\makebox[0pt]{\mbox{\normalfont \scriptsize $b$}}}{\not\rightarrow}}}

\newcommand\myarrowone{\mathrel{\stackrel{\makebox[0pt]{\mbox{\normalfont \scriptsize $1$}}}{\longrightarrow}}}
\newcommand\myarrowk{\mathrel{\stackrel{\makebox[0pt]{\mbox{\normalfont \scriptsize $k$}}}{\longrightarrow}}}
\newcommand\myarrowi{\mathrel{\stackrel{\makebox[0pt]{\mbox{\normalfont \scriptsize $i$}}}{\longrightarrow}}}
\newcommand\myarrowasubone{\mathrel{\stackrel{\makebox[0pt]{\mbox{\normalfont \scriptsize $a_{i_1}$}}}{\longrightarrow}}}
\newcommand\myarrowasubtwo{\mathrel{\stackrel{\makebox[0pt]{\mbox{\normalfont \scriptsize $a_{i_2}$}}}{\longrightarrow}}}
\newcommand\myarrowasubm{\mathrel{\stackrel{\makebox[0pt]{\mbox{\normalfont \scriptsize $a_{i_m}$}}}{\longrightarrow}}}
\newcommand\myarrowasubk{\mathrel{\stackrel{\makebox[0pt]{\mbox{\normalfont \scriptsize $a_{i_k}$}}}{\longrightarrow}}}
\newcommand\notmyarrowasubk{\mathrel{\stackrel{\makebox[0pt]{\mbox{\normalfont \scriptsize $a_{i_k}$}}}{\not\rightarrow}}}
\newcommand\myarrowasubkplusone{\mathrel{\stackrel{\makebox[0pt]{\mbox{\normalfont \scriptsize $a_{i_{(k+1)}}$}}}{\longrightarrow}}}
\newcommand\myarrowasubkminusone{\mathrel{\stackrel{\makebox[0pt]{\mbox{\normalfont \scriptsize $a_{i_{(k-1)}}$}}}{\longrightarrow}}}
\newcommand\myarrowasubj{\mathrel{\stackrel{\makebox[0pt]{\mbox{\normalfont \scriptsize $a_j$}}}{\longrightarrow}}}

\newcommand\vartextvisiblespace[1][.5em]{%
  \makebox[#1]{%
    \kern.07em
    \vrule height.3ex
    \hrulefill
    \vrule height.3ex
    \kern.07em
  }% <-- don't forget this one!
}

\theoremstyle{plain}\newtheorem{satz}[thm]{Satz} %\crefname{satz}{Satz}{S\"atze}
\def\eg{{\em e.g.}}
\def\cf{{\em cf.}}

\begin{document}

% % If the title is longer than 55 characters, then specify a shorter running title as the optional argument to \title. The running title should be roughyl at most 55 characters:
% \title[Instructions]{Instructions for Authors\texorpdfstring{\\}{. }How to prepare papers
%   for LMCS using \texorpdfstring{\MakeLowercase{\texttt{lmcs.cls}}}{lmcs.cls}\rsuper*\\Version of
%   2022-04-01}
% \titlecomment{{\lsuper*}OPTIONAL comment concerning the title, \eg,
%   if a variant or an extended abstract of the paper has appeared elsewhere.}
% \thanks{thanks, optional.}	%optional

\title[Deciding characteristic formulae]
{Deciding characteristic formulae: \\ A journey in the branching-time spectrum}\titlecomment{{\lsuper*}This paper combines and extends the results presented in two conference articles, which appeared at CSL 2025 and GandALF 2025.}\thanks{This work has been funded by the projects `Open Problems in the Equational Logic of Processes (OPEL)' (grant no.~196050), `Mode(l)s of Verification and Monitorability' (MoVeMnt) (grant no.~217987),  `Learning and Applying Probabilistic Systems' (grant no.~206574-051) and `Hyperlogics: Expressiveness, Monitorability and Tools (H.-Lo)' (grant no.~2612260-051) of the Icelandic Research Fund.}
%{Deciding characteristic formulae:\texorpdfstring{\\}{. }When the problem becomes hard?}

% affiliations are numbered automatically with a, b, c (see below)
% use the optional argument to indicate the affiliation(s) of each author
% omit the argument if there is only one author, or only one affiliation
\author[L.~Aceto]{Luca Aceto\lmcsorcid{0000-0001-8554-6907}}[a,b]
\author[A.~Achilleos]{Antonis Achilleos\lmcsorcid{0000-0002-1314-333X}}[a]
\author[A.~Chalki]{Aggeliki Chalki\lmcsorcid{0000-0001-5378-0467}}[a]
\author[A.~Ing\'olfsd\'ottir]{Anna Ing\'olfsd\'ottir\lmcsorcid{0000-0001-8362-3075}}[a]

% affiliation 1 (automatically numbered a)
\address{Department of Computer Science, Reykjavik University, Iceland}	%optional
% write emails for all authors having that affiliation
\email{luca@ru.is, antonios@ru.is, angelikic@ru.is, annai@ru.is}  %optional

% affiliation 2 (automatically numbered b)
\address{Gran Sasso Science Institute, L'Aquila, Italy}	%optional
% \email{name2@email2}  %optional
\email{luca.aceto@gssi.it} 
%% etc.

%% required for running head on odd and even pages, use suitable
%% abbreviations in case of long titles and many authors:

%%%%%%%%%%%%%%%%%%%%%%%%%%%%%%%%%%%%%%%%%%%%%%%%%%%%%%%%%%%%%%%%%%%%%%%%%%%

%% the abstract has to PRECEDE the command \maketitle:
%% be sure not to issue the \maketitle command twice!

\begin{abstract}
  \noindent
    Characteristic formulae give a complete logical description of the behaviour of processes modulo some chosen notion of behavioural semantics. They 
    %have been defined for finite, loop-free %processes with respect to each semantics in van Glabbeek's linear-time/branching-time %spectrum and 
    allow one to reduce equivalence or preorder checking to model checking, and are exactly the formulae in the modal logics characterizing classic behavioural equivalences and preorders for which model checking can be reduced to equivalence or preorder checking.

    This paper studies the complexity of determining whether a formula is characteristic for some 
    %finite, loop-free 
    process in each of the logics providing modal characterizations of the simulation-based semantics in van Glabbeek's branching-time spectrum. Since characteristic formulae in each of those logics are exactly the satisfiable and prime ones, this article presents complexity results for the satisfiability and primality problems, and investigates the boundary between modal logics for which those problems can be solved in polynomial time and those for which they become (co)NP- or PSPACE-complete. % and those for which they become computationally hard.
%The abstract has to precede the maketitle command.  Be
  %sure not to issue the maketitle command twice!  In the abstract,
  %mathematical expressions must be kept to the absolute minimum.
  %Otherwise it should consist of plain ASCII text, without
  %\TeX-commands, including explicit references using the
  %\texttt{\textbackslash cite} command.  Presently we are not able to
  %automatically extract an abstract containing such data and reliably
  %turn it into html code.  If you cannot meet these criteria, it is
  %your responsibility to provide us with an html-version of your
  %abstract.  Please keep the abstract fairy short to prevent it from
  %spilling over to the second page!
\end{abstract}

\maketitle

\section{Introduction}\label{sec:intro}

Several notions of behavioural relations have been proposed in concurrency theory to describe when one process is a suitable implementation of another. Many such relations have been catalogued by van Glabbeek in his seminal linear-time/branching-time spectrum~\cite{Glabbeek01}, together with a variety of alternative ways of describing them including testing scenarios and axiom systems. To our mind, modal characterizations 
of behavioural equivalences and preorders are some of the most classic and pleasing results in concurrency theory---see, for instance,~\cite{HennessyM85} for the seminal Hennessy-Milner theorem and~\cite{BrowneCG88,ERPH13,DeNicolaV95,Glabbeek01,Milner81} for similar results for other relations in van Glabbeek's
spectrum and other settings. By way of example, in their archetypal modal characterization of bisimilarity, Hennessy and Milner have shown in~\cite{HennessyM85} that, under a mild finiteness condition, two processes are bisimilar if, and only if, they satisfy the same formulae in a multi-modal logic that is now often called Hennessy-Milner logic. Apart from its intrinsic theoretical interest, this seminal logical characterization of bisimilarity means that, when two processes are \emph{not} bisimilar, there is always a formula that distinguishes between them. Such a formula describes a reason why the two processes are not bisimilar, provides useful debugging information and can be algorithmically constructed over finite processes---see, for instance,~\cite{BispingJN22,Cleaveland90} and~\cite{MartensG23}, where Martens and Groote show that, in general, computing minimal distinguishing Hennessy-Milner formulae is \NP-hard. 

On the other hand, the Hennessy-Milner theorem seems to be less useful to show that two processes \emph{are} bisimilar, since that would involve verifying that they satisfy the same formulae, and there are infinitely many of those. However, as shown in works such as~\cite{AcetoMFI19,AcetoILS12,BrowneCG88,GrafS86a,SteffenI94}, the logics that underlie classic modal characterization theorems for equivalences and preorders over processes allow one to express \emph{characteristic formulae}. Intuitively, a characteristic formula $\chi(p)$ for a process $p$ gives a complete logical characterization of the behaviour of $p$ modulo the behavioural semantics of interest $\lesssim$, in the sense that any process is related to $p$ with respect to $\lesssim$ if, and only if, it satisfies $\chi(p)$.\footnote{Formulae akin to characteristic ones first occurred in the study of equivalence of structures using first-order formulae up to some quantifier rank. See, for example, the survey paper~\cite{Thomas93} and the textbook~\cite{EbbinghausFT1994}. The existence of formulae in first-order logic with counting that characterize graphs up to isomorphism has significantly contributed to the study of the complexity of the Graph Isomorphism problem---see, for instance, \cite{CaiFI92,KieferSS15}.} Since the formula $\chi(p)$ can be constructed from $p$, characteristic formulae reduce the problem of checking whether a process $q$ is related to $p$ by $\curle$ to a model checking problem, viz.~whether $q$ satisfies $\chi(p)$. See, for instance, the classic reference~\cite{CleavelandS91} for applications of this approach.

Characteristic formulae, thus, allow one to reduce equivalence and preorder checking to model checking. But what model checking problems can be reduced to equivalence/preorder checking ones?  
To the best of our knowledge, that question was first studied by Boudol and Larsen in~\cite{BoudolL92} in the setting of modal refinement over modal transition systems.  
See~\cite{AcetoMFI19,AFEIP11} for other contributions in that line of research. The aforementioned articles showed that characteristic formulae coincide with those that are \emph{satisfiable} and \emph{prime}. (A formula is prime if whenever it entails a disjunction $\varphi_1 \vee \varphi_2$, then it must entail $\varphi_1$ or $\varphi_2$.) Moreover, characteristic formulae with respect to bisimilarity coincide with the formulae that are satisfiable and \emph{complete}~\cite{Achilleos18}. (A modal formula is complete if, for each formula $\varphi$, it entails either $\varphi$ or its negation.)
The aforementioned results give semantic characterizations of the formulae that are characteristic within the logics that correspond to the behavioural semantics in van Glabbeek's spectrum. Those characterizations tell us for what logical specifications model checking can be reduced to equivalence or preorder checking. However, given a specification expressed as a modal formula, can one decide whether that formula is characteristic and therefore can be model checked using algorithms for behavioural equivalences or preorders? And, if so, what is the complexity of checking whether a formula is characteristic? Perhaps surprisingly, those questions were not addressed in the literature until the recent papers~\cite{AcetoAFI20,Achilleos18}, where it is shown that, in the setting of the modal logics that characterize bisimilarity over natural classes of Kripke structures and labelled transition systems, the problem of checking whether a formula is characteristic for some process modulo bisimilarity is computationally hard and, typically, has the same complexity as validity checking, which is \pspace-complete for Hennessy-Milner logic and \expc-complete for its extension with fixed-point operators~\cite{Holmstrom89,Larsen90} and the $\mu$-calculus~\cite{Kozen83}.

The aforementioned hardness results for the logics characterizing bisimilarity tell us that deciding whether a formula is characteristic in bisimulation semantics is computationally hard. But what about the less expressive logics that characterize the coarser semantics in van Glabbeek's spectrum? And for what logics characterizing relations in the spectrum does computational hardness manifest itself? 
%Finally, what is the complexity of computing a characteristic formula for a process? 
\paragraph{Our contributions} %\textcolor{red}{Just one comment. For the satisfiability let's mention that to the best of our knowledge these are the first such results for the logics characterizing preorders in van Glabbeek's spectrum. I have written sth in the beginning of the respective section as well. }
%
%\paragraph{Our Contributions} 
The aim of this paper is to answer the aforementioned questions for %\textcolor{orange}{some of [@Luca: do we cover everything?]} 
the simulation-based semantics in the spectrum. % over finite, loop-free processes. 
In particular, we study the complexity of determining whether a formula is characteristic modulo the simulation~\cite{Milner71}, complete simulation and ready simulation preorders~\cite{BloomIM95,LarsenS91}, as well as the trace simulation and the $n$-nested simulation preorders~\cite{GV92}. Since characteristic formulae are exactly the satisfiable and prime ones for each behavioural relation in van Glabbeek's spectrum~\cite{AcetoMFI19}, the above-mentioned tasks naturally break down into studying the complexity of satisfiability  and primality checking for formulae in the fragments of Hennessy-Milner logic that characterize those preorders. 

\iffalse
Our results are summarized in Tables~\ref{table:sat}--\ref{table:characteristic}.

Our complexity results on satisfiability checking may be found in Section~\ref{section:deciding-satisfiability} and those related to primality checking are in Sections~\ref{section:deciding-primality} and~\ref{Sect:primality-nested-semantics}. Finally, we present our findings on the complexity of deciding whether formulae are characteristic in Section~\ref{section:char-preorders}. 
\fi

By using a reduction to the, seemingly unrelated, reachability problem in \emph{alternating graphs}, as defined by Immerman in~\cite[Definition 3.24]{Immerman99}, we discover that both those problems are decidable in polynomial time for the simulation and the complete simulation preorders, as well as for the ready simulation preorder when the set of actions has constant size (Sections~\ref{Sect:sat-in-P} and~\ref{subsection:primality-simulation}--\ref{sect:primality-LRS}).  
On the other hand,  when the set of actions is unbounded (that is, it is an input of the algorithmic problem at hand), the problems of checking satisfiability and primality for formulae in the logic characterizing the ready simulation preorder are \NP-complete (Theorem~\ref{prop:sat-rs-ts-2s-np-complete}) and \conp-complete (Proposition~\ref{prop:decide-prime-rs-infinite-actions-hard}), respectively.  We also show that deciding whether a formula is characteristic in that setting is \us-hard~\cite{BlassG82} (that is, it is at least as hard as the problem of deciding whether a given Boolean formula has exactly one satisfying truth assignment) and belongs to \dpc, which is the class of languages that are the intersection of one language in \NP and of one in \conp~\cite{PapadimitriouY84}.\footnote{The class \dpc contains both \NP and {\conp}, and is contained in the class of problems that can be solved in polynomial time with an \NP oracle.} (See Corollary~\ref{cor:char-rs-unbounded-complexity}.) 
These negative results are in stark contrast with the positive results for the simulation and the complete simulation preorder, and indicate that augmenting the logic characterizing the simulation preorder with formulae that state that a process cannot perform a given action suffices to make satisfiability and primality checking computationally hard. 

%In passing, w
We also prove that, in the presence of at least two actions, 
%(1) 
for the logics characterizing the trace simulation and 2-nested simulation preorders, satisfiability and primality checking are \NP-complete and \conp-hard respectively, and deciding whether a formula is characteristic is \us-hard; for the logic characterizing the 2-nested simulation preorder, primality is actually \conp-complete, and therefore deciding whether a formula is characterisitic is in \dpc. (See Proposition~\ref{cor:char-ts-complexity} and Corollary~\ref{cor:2S-char-dp}.) 
%\textcolor{orange}{I think we do not include this, so I plan to remove it: 
For the logic that characterizes the trace simulation preorder, deciding whether a formula is characteristic is fixed-parameter tractable~\cite{DowneyF95}, with the modal depth of the input formula as the parameter, when the size of the action set is a constant. (See Theorem~\ref{thm:ts-bounded-depth}.)
%, and }
Finally, deciding whether a formula is characteristic in the modal logic for the $n$-nested simulation preorder~\cite{GV92} is \pspace-complete when $n\geq 3$. The proof of the lower bound for the last result relies on `simulating' Ladner’s reduction proving the \pspace-hardness of satisfiability for modal logic~\cite{ladner1977computational} using the limited alternations of modal operators allowed by the logic for the 3-nested simulation preorder. For the upper bound, we use algorithms based on computationally bounded games, which we present in Section~\ref{Sect:primality-nested-semantics}.

For the logic characterizing the trace simulation preorder, we conjecture that deciding whether a formula is characteristic is in {\dpc}. However, we currently have no tight upper bound for the complexity of this problem. %\textcolor{orange}{am I right?}

To the best of our knowledge, this paper presents the first complexity results for satisfiability and primality checking for the fragments of classic Hennessy-Milner logic that characterize the simulation-based preorders in van Glabbeek's spectrum---with the exception of full Hennessy-Milner logic.
%are the first of their kind.
%\textcolor{orange}{does this read ok?}

We also study the complexity of deciding whether a formula is characteristic modulo the equivalence relations induced by the preorders considered in this article (Section~\ref{section:char-equivalences}). It turns out that the logics characterizing the simulation, complete simulation, and ready simulation preorders have very weak expressive power when it comes to defining characteristic formulae modulo the kernels of those preorders (Proposition~\ref{prop:S-CS-RS-equiv-char}). However, we provide a polynomial-time reduction from the validity problem for all the logics we study, apart from the one that characterizes the simulation preorder, to deciding characteristic formulae with respect to the equivalence relations they induce over processes (Theorem~\ref{thm:reduction-validity-char}). We use that result to establish that deciding whether a formula is characteristic modulo ready simulation equivalence, in the presence of an unbounded action set, and modulo the kernels of the trace simulation and the $2$-nested simulation preorders, in the presence of at least two actions,  is \conp-hard (Corollary~\ref{cor:char-equiv-hard}). We also show that, in the presence of at least two actions, deciding whether a formula  is characteristic for a process modulo the equivalences induced by the $n$-nested simulation preorders, $n\geq 3$, is \pspace-complete (Corollary~\ref{cor:n-char-equiv}).

This paper presents results that were announced in the conference articles \cite{AcetoACI25} and \cite{Aceto_2025}. In particular, it focuses on the results on the complexity of deciding whether formulae are characteristic that were presented, without detailed proofs, in those studies. In~\cite{AcetoACI25}, we also offered results on the complexity of constructing characteristic formulae for processes. To keep the length of this article manageable, we will provide an extended account of those contributions in a separate article. 
%added the Aceto_2025 ref that poins to the GandALF paper

\section{Preliminaries}\label{sec:preliminaries}

To make the paper self-contained and for ease of reference, this section collects the background notions and results used in the remainder of this study. 

%\subsection{Concurrency theory and logic}\label{sect:conc-logics}
\subsection{Labelled transition systems and behavioural relations}\label{sect:LTS}
In this paper, we model processes as finite, loop-free \emph{labelled transition systems} (LTS). An LTS is a triple $\mathS=(P,\act,\longrightarrow)$, where $P$ is a non-empty set of states (or processes), \act is a finite, non-empty set of actions, and ${\longrightarrow}\subseteq P\times \act\times P$ is a transition relation. An LTS is \emph{finite} if so is its set of states $P$. Note that the transition relation of a finite LTS is also finite. We define the \emph{size}  of a finite LTS $\mathS=(P,\act,\longrightarrow)$, denoted  $|\mathS|$, to be $|P|+|{\longrightarrow}|$.  

As usual, we use $p\myarrowa q$ instead of $(p,a,q)\in {\longrightarrow}$. For each $t\in \act^*$, we write $p\myarrowtau q$ to mean that there is a sequence of transitions labelled with $t$ starting from $p$ and ending at $q$. An LTS is \emph{loop-free} iff $p\myarrowtau p$ holds only when $t$ is the empty trace $\varepsilon$. A process $q$ is \emph{reachable} from $p$ if $p\myarrowtau q$, for some $t\in \act^*$.  
\iffalse
We define the \emph{size}  of a finite LTS $\mathS=(P,\act,\longrightarrow)$, denoted  $|\mathS|$, to be $|P|+|{\longrightarrow}|$. 

The \emph{size of a process} $p\in \proc$, denoted  $|p|$, is the cardinality of $\reach(p)=\{q~|~ q \text{ is reachable from } p\}$ plus the cardinality of the set $\longrightarrow$ restricted to $\reach(p)$.
\fi

The set of \emph{initials} of $p$, denoted $I(p)$, is $\{a \in \act \mid  p\myarrowa p' \text{ for some } p'\in \proc\}$. We write $p\myarrowa$ if $a\in I(p)$, $p\notmyarrowa$ if $a\not\in I(p)$, and $p\not\rightarrow$ if $I(p) =\emptyset$. A sequence of actions $t\in \act^*$ is a \emph{trace} of $p$ if there is some $q$ such that $p\myarrowtau q$. We denote 
the set of traces of $p$ by $\mathrm{traces}(p)$.  
The \emph{depth} of a finite, loop-free process $p$, denoted by $\depth(p)$, is the length of a longest trace $t$ of $p$.

Throughout this study, we assume the existence of an ambient LTS $(\proc,\act,\longrightarrow)$ that contains exactly the finite, loop-free LTSs as sub-LTSs (see also Remark~\ref{rem:proc}).  The \emph{size of a process} $p\in \proc$, denoted  $|p|$, is the cardinality of $\reach(p)=\{q~|~ q \text{ is reachable from } p\}$ plus the cardinality of the set $\longrightarrow$ restricted to $\reach(p)$.

In what follows, we shall often describe finite, loop-free processes using the fragment of Milner's CCS~\cite{Milner89} given by the following grammar:
\[
p ::= \mathtt{0} ~\mid~ a.p  ~\mid~  p+p ,
\]
where $a \in \act$. For each action $a$ and terms $p,p'$, we write $p \myarrowa p'$ iff
\begin{enumerate}[(i)]
    \item $p =a.p'$ or
    \item $p =p_1+p_2$, for some $p_1,p_2$, and $p_1\myarrowa p'$ or $p_2\myarrowa p'$ holds.
\end{enumerate}

% \textcolor{orange}{suggestion: [previous: Throughout the paper, unless we specify otherwise, we assume that our processes come from some fixed LTS, $\mathS=(\proc,\act,\longrightarrow)$.]
% [current: 1. use $P$ instead of $proc$ above, 2. remove ``finite'' from the definition of an LTS and define that an LTS is finite when P is finite, 3. say that we assume an LTS with a set of processes proc, such that every finite loop-free $p$ is bisimilar to a process in proc. But then, we need to define bisimilarity first, which is not optimal (or is it?). The set of states of an LTS is just a finite set, so claiming that proc contains all that seems problematic from a set-theoretic point of view. We could say that proc includes the CCS processes defined above and that is enough to describe all finite loop-free processes, and leave it at that.] Sorry again, this has been bugging me.}

In this paper, we consider the following branching-time relations in van Glabbeek's spectrum: the simulation, complete simulation, ready simulation, trace simulation and $n$-nested simulation ($n\geq 2$) preorders,  and bisimilarity. Their definitions are given below.

\begin{defi}[\cite{Milner89,Glabbeek01,AcetoMFI19}]\label{Def:beh-preorders}
We define each of the following preorders as the largest binary relation over $\proc$ that satisfies the corresponding condition.
\begin{enumerate}[(a)]
    \item \emph{Simulation preorder (S):} $p\curle_{S} q$ iff for all $p\myarrowa p'$ there exists some $q\myarrowa q'$ such that $p'\curle_S q'$.
    \item \emph{Complete simulation  preorder (CS):} $p\curle_{CS} q$ iff
    \begin{enumerate}[(i)]
        \item for all $p\myarrowa p'$ there exists some $q\myarrowa q'$ such that $p'\curle_{CS} q'$, and
        \item $I(p)=\emptyset$ iff $I(q)=\emptyset$.        
    \end{enumerate}
     \item \emph{Ready simulation  preorder (RS):} $p\curle_{RS} q$ iff
    \begin{enumerate} [(i)]
        \item for all $p\myarrowa p'$ there exists some $q\myarrowa q'$ such that $p'\curle_{RS} q'$, and
        \item $I(p)=I(q)$.        
    \end{enumerate}
    \item \emph{Trace simulation  preorder (TS):} $p\curle_{TS} q$ iff
   \begin{enumerate}[(i)]
    \item for all $p\myarrowa p'$ there exists some  $q\myarrowa q'$ such that $p'\curle_{TS} q'$, and
    \item $\mathrm{traces}(p)=\mathrm{traces}(q)$.
    \end{enumerate}
\end{enumerate}
%\item \emph{$n$-Nested simulation  
%preorder ($n$S)},
The \emph{$n$-nested simulation  preorder ($n$S)}, where $n\geq 1$, is defined inductively as follows: The $1$-nested simulation preorder $\curle_{1S}$ is $\curle_S$, and the $n$-nested simulation preorder $\curle_{nS}$ for $n > 1$ is the largest relation such that $p\curle_{nS} q$ iff
    \begin{enumerate}[(i)]
        \item for all $p\myarrowa p'$ there exists some $q\myarrowa q'$ such that $p'\curle_{nS} q'$, and
        \item $q\curle_{(n-1)S} p$. 
        %and $p\curle_{i-1} q$.        
    \end{enumerate}
 %\item  \emph{Bisimilarity (BS):} 
 \emph{Bisimilarity (BS)} $\curle_{BS}$ is the largest symmetric relation satisfying the condition defining the simulation preorder.
%  is the largest relation $\sim$ over $P$ satisfying the following conditions for every $p,q\in P$:
% $p\sim q$ $\Leftrightarrow$
%     \begin{enumerate} [(i)]
%         \item for all $p\myarrowa p'$ there exists some $q\myarrowa q'$ such that $p'\sim q'$, and
%         \item for all $q\myarrowa q'$ there exists some $p\myarrowa p'$ such that $p'\sim q'$.    
%     \end{enumerate}
%\end{enumerate}
\end{defi}

It is well-known that bisimilarity is an equivalence relation and all the other relations are preorders~\cite{Glabbeek01,Milner89}. We sometimes write $p\sim q$ instead of $p\curle_{BS} q$. Moreover, for each $n>2$, we have that ${\sim}\subsetneq{\curle_{nS}}\subsetneq
{\curle_{(n-1)S}}\subsetneq {\curle_{TS}}\subsetneq {\curle_{RS}}\subsetneq {\curle_{CS}}\subsetneq {\curle_{S}}$---see~\cite{Glabbeek01}.

\begin{rem}\label{rem:preorders-preserved-under-plus}
 The  preorders defined in Definition~\ref{Def:beh-preorders} are preserved by action prefixing and by the operator $+$~\cite{Glabbeek01}, i.e., for each $X\in\{CS,RS,TS,nS,BS\}$, $n\geq 1$, 
 \begin{itemize}
     \item if $p\curle_X q$, then $a.p\curle_X a.q$, and 
     \item if $p_1\curle_X q_1$ and $p_2\curle_X q_2$, then $p_1+p_2\curle_X q_1+q_2$.
 \end{itemize}
\end{rem}

\begin{rem}\label{rem:proc}
    It is not hard to see that every finite, loop-free process is bisimilar to the LTS associated with a term in the  fragment of CCS that we introduced above. Therefore, we can define the ambient set of processes $\proc$ to consist exactly of these CCS processes. Thus, $\proc$ is well-defined, and it includes all finite, loop-free processes, up to bisimilarity.
\end{rem}

% \begin{rem}\label{rem:preorders-preserved-under-plus}
%  \textbf{Aggela. I suggest that this goes to the appendix.}   As is well-known~\cite{Glabbeek01}, the aforementioned preorders are preserved by action prefixing, i.e.\ if $p\curle_X q$, then $a.p\curle_X a.q$, and by the operator $+$, i.e.\ if $p_1\curle_X q_1$ and $p_2\curle_X q_2$, then $p_1+p_2\curle_X q_1+q_2$.
% \end{rem}

\begin{defi}[Kernels of the preorders]
    For each $X\in\{CS,RS,TS,nS\}$, $n\geq 1$,  the kernel $\equiv_X$ of $\curle_X$ is the equivalence relation defined thus: for every $p,q\in \proc$, $p\equiv_X q$ iff $p\curle_X q$ and $q\curle_X p$. 
\end{defi}

\subsection{Modal logics}\label{Sect:modal-logic}

Each relation $\curle_X$, where $X\in\{CS,RS,TS,nS,BS\}$, where $n\geq 1$, is characterized by a fragment $\mathL_X$ of Hennessy-Milner logic, \hml, defined as follows~\cite{Glabbeek01,AcetoMFI19}.

\begin{defi}\label{def:mathlx}
For $X\in\{CS,RS,TS,nS,BS\}$, $n\geq 1$, $\mathL_X$ is defined by the corresponding grammar given below ($a\in \act$):
\begin{enumerate}[(a)] 
\item $\mathL_S$: 
$\varphi_S::= ~ \true ~ \mid ~ \ff ~ \mid ~ \varphi_S\wedge \varphi_S~ \mid ~\varphi_S\vee \varphi_S~ \mid ~ \langle a \rangle\varphi_S.$
\item $\mathL_{CS}$:
$\varphi_{CS}::= ~ \true ~ \mid ~ \ff ~ \mid ~ \varphi_{CS}\wedge \varphi_{CS} ~ \mid ~\varphi_{CS}\vee \varphi_{CS}
~ \mid ~ \langle a \rangle\varphi_{CS} ~ \mid ~ \zero$,
where $\zero=\bigwedge_{a\in \act} [a]\ff$.
\item $\mathL_{RS}$:
$\varphi_{RS}::= ~ \true ~ \mid ~ \ff ~ \mid ~ \varphi_{RS}\wedge \varphi_{RS} ~ \mid ~\varphi_{RS}\vee \varphi_{RS}
~ \mid ~ \langle a \rangle\varphi_{RS} ~ \mid ~ [a]\ff.$
\item $\mathL_{TS}$: $\varphi_{TS}::= ~ \true ~ \mid ~ \ff ~ \mid ~ \varphi_{TS}\wedge \varphi_{TS}~ \mid ~\varphi_{TS}\vee \varphi_{TS}~ \mid ~ \langle a \rangle\varphi_{TS} ~ \mid ~ \psi_{TS}$, where $\psi_{TS}::=\ff ~ \mid ~ [a]\psi_{TS}$.
\item $\mathL_{nS}$, $n\geq 2$: 
$\varphi_{nS}::= ~ \true ~ \mid ~ \ff ~ \mid ~ \varphi_{nS}\wedge \varphi_{nS} ~ \mid ~\varphi_{nS}\vee \varphi_{nS}
~ \mid ~ \langle a \rangle\varphi_{nS} ~ \mid ~ \neg\varphi_{(n-1)S}.$
\item $\hml$ ($\mathL_{BS}$):
    $\varphi_{BS}::= ~ \true ~ \mid ~ \ff ~ \mid ~ \varphi_{BS}\wedge \varphi_{BS} ~ \mid ~\varphi_{BS}\vee \varphi_{BS}~ \mid ~ \langle a \rangle\varphi_{BS} ~ \mid ~ [a]\varphi_{BS} ~ \mid ~ \neg\varphi_{BS}.$
\end{enumerate}
\end{defi}

Note that the explicit use of negation in the grammar for $\mathL_{BS}$ is unnecessary. However, we included the negation operator explicitly so that $\mathL_{BS}$ extends syntactically each of the other modal logics presented in Definition~\ref{def:mathlx}. 
{Furthermore, we usually identify the negation of a formula with its equivalent formula in negation normal form: $\neg\true=\ff$, $\neg\ff=\true$, $\neg (\varphi\wedge\psi)=\neg\varphi\vee\neg\psi$, $\neg (\varphi\vee\psi)=\neg\varphi\wedge\neg\psi$, $\neg [a]\varphi=\langle a\rangle \neg \varphi$, $\neg \langle a\rangle\varphi=[a] \neg \varphi$, and $\neg\neg\varphi = \varphi$.} For $\mathL \subseteq \mathL_{BS}$, we define the dual fragment of $\mathL$ to be $\compL=\{\varphi\mid \neg\varphi\in\mathL\}$.

Henceforth, we write $\sub(\varphi)$ for the set of subformulae of formula $\varphi\in \mathL_{BS}$.
Let $S$ be a set of formulae. We write 
$\bigwedge S$ for $\bigwedge_{\varphi\in S}\varphi$, when $S$ is finite, and
 $\sub(S)$ for $\{\varphi ~ | ~\varphi\in\sub(\psi) \text{ for some } \psi\in S\}$. Note that $\sub(S)$ is finite, when so is $S$.
 
%Given a formula $\varphi\in\mathL_{BS}$, the \emph{modal depth} of $\varphi$, denoted by $\md(\varphi)$, is the maximum nesting of modal operators in $\varphi$.

\begin{defi}\label{def:modal_depth}
Given a formula $\varphi\in\mathL_{BS}$, the \emph{modal depth} of $\varphi$, denoted by $\md(\varphi)$, is the maximum nesting of modal operators in $\varphi$ and is defined inductively as follows:
\begin{itemize}
    \item $\md(\true)=\md(\ff)=0$,
    \item $\md(\langle a\rangle\varphi)=\md([a]\varphi)=\md(\varphi)+1$,
    \item $\md(\varphi_1\wedge\varphi_2)=\md(\varphi_1\vee\varphi_2)=\max\{\md(\varphi_1),\md(\varphi_2)\}$,
    \item $\md(\neg\varphi)=\md(\varphi)$.
\end{itemize}
\end{defi}

Truth of formulae in states of an  LTS $\mathS= (P,\act,\longrightarrow)$ is defined via the satisfaction relation $\models$ as follows:
$$\begin{aligned}
    &p\models\true \text{ and } p\not\models\ff;\\
    &p\models \neg\varphi \text{ iff } p\not\models \varphi;\\
    &p\models \varphi\wedge \psi \text{ iff both } p\models \varphi \text{ and } p\models \psi;\\
    &p\models \varphi\vee \psi \text{ iff } p\models \varphi \text{ or } p\models \psi;\\
    &p\models \langle a\rangle\varphi \text{ iff there is 
 some } p\myarrowa q \text{ such that } q\models\varphi;\\
    &p\models [a]\varphi \text{ iff for all } p\myarrowa q \text{ it holds that } q\models\varphi  .
\end{aligned}$$
If $p\models \varphi$, we say that $\varphi$ is true, or satisfied, in $p$. Given a process $p$ and $\mathL \subseteq \mathL_{BS}$, we define  $\mathL(p)=\{\varphi\in\mathL \mid p\models\varphi\}$.

If a formula $\varphi$ is satisfied in every process in every LTS, we say that $\varphi$ is valid.  
Formula $\varphi_1$ entails $\varphi_2$, denoted by $\varphi_1\models \varphi_2$, if every process
that satisfies  $\varphi_1$ also satisfies   $\varphi_2$. Moreover, $\varphi_1$ and $\varphi_2$ are logically equivalent, denoted by $\varphi_1\equiv\varphi_2$, if $\varphi_1\models \varphi_2$ and $\varphi_2\models \varphi_1$.
A formula $\varphi$ is \emph{satisfiable} if there is a process that satisfies $\varphi$. 
%Finally, $\sub(\varphi)$ denotes the set of subformulae of formula $\varphi$.
%
We call the problem of deciding whether a given formula in $\mathL \subseteq \mathL_{BS}$ is satisfiable (respectively, valid) the \emph{satisfiability (respectively, validity) problem} for $\mathL$.

% , where $\neg\true=\ff$, $\neg\ff=\true$, $\neg (\varphi\wedge\psi)=\neg\varphi\vee\neg\psi$, $\neg (\varphi\vee\psi)=\neg\varphi\wedge\neg\psi$, $\neg [a]\varphi=\langle a\rangle \neg \varphi$, $\neg \langle a\rangle\varphi=[a] \neg \varphi$, and $\neg\neg\varphi = \varphi$. It is not hard to see that  $p\models \neg\varphi$ iff $p\not\models\varphi$, for every process $p$.

A simplification of the Hennessy-Milner theorem gives a modal characterization of bisimilarity over processes in our ambient LTS $\mathS= (\proc,\act,\longrightarrow)$. An analogous result is true for every preorder examined in this paper.

\begin{thm}[Hennessy-Milner theorem~\cite{HennessyM85}]\label{Thm:HMT}
    For all processes $p,q\in\proc$, $p\sim q$ iff $\mathL_{BS}(p)=\mathL_{BS}(q)$.
\end{thm}

\begin{prop}[\cite{Glabbeek01,AcetoMFI19,DBLP:journals/iandc/GrooteV92}]\label{logical_characterizations}
    Let $X\in\{CS,RS,TS,nS\}$, $n\geq 1$. Then  $p\curle_X q$ iff $\mathL_X(p)\subseteq \mathL_X(q)$, for all $p,q\in \proc$.
\end{prop}

\begin{rem}
Neither $\ff$ nor disjunction are needed in several of the modal characterizations presented in the above result. The reason for adding those constructs to all the logics is that doing so makes our subsequent results more general and uniform. For example, having $\ff$ and disjunction in all logics allows us to provide algorithms that determine whether a formula in a logic $\mathL$ is prime with respect to a sublogic.
\end{rem}

\begin{defi}[\cite{BoudolL92,AFEIP11}]\label{def:prime-formula}
Let $\mathL\subseteq \mathL_{BS}$. 
A formula $\varphi\in \mathL_{BS}$ is \emph{prime in $\mathL$} if for all $\varphi_1,\varphi_2\in \mathL$, $\varphi\models\varphi_1 \vee \varphi_2$ implies $\varphi\models \varphi_1$ or  $\varphi\models \varphi_2$. 
%%%%%%
\iffalse
A formula $\varphi$ is \emph{prime in $\mathL$} if for any finite non-empty set of formulae $\Psi\subseteq\mathL$, it holds that $\varphi\models\bigvee \Psi$ implies $\varphi\models \psi$ for some $\psi\in\Psi$.
\fi
%%%%%%%%%%%%%%%%%
\end{defi}

When the logic $\mathL$ is clear from the context, we simply say that $\varphi$ is prime. Note that every unsatisfiable formula is trivially prime in $\mathL$, for every $\mathL \subseteq \mathL_{BS}$.

\begin{exa}\label{ex:prime}
    The formula $\langle a\rangle \true$ is prime in $\mathL_S$. Indeed, let $\varphi_1,\varphi_2\in\mathL_S$ and assume that $\langle a\rangle \true\models\varphi_1\vee\varphi_2$. Since $a.\mathtt{0} \models\langle a\rangle \true$, without loss of generality, we have that $a.\mathtt{0} \models\varphi_1$. We claim that $\langle a\rangle\true\models\varphi_1$. To see this, let $p$ be some process such that $p\models\langle a\rangle\true$---that is, a process such that $p \myarrowa p'$ for some $p'$. It is easy to see that $a.\mathtt{0}\curle_S p$. Since $a.\mathtt{0} \models \varphi_1$, Proposition~\ref{logical_characterizations} yields that $p\models\varphi_1$, proving our claim and the primality of $\langle a\rangle\true$. On the other hand, the formula $\langle a\rangle\true\vee\langle b\rangle\true$ is not prime in $\mathL_S$. Indeed, $\langle a\rangle\true\vee\langle b\rangle\true\models \langle a\rangle\true\vee\langle b\rangle\true$, but neither $\langle a\rangle\true\vee\langle b\rangle\true \models\langle a\rangle\true$  nor $\langle a\rangle\true\vee\langle b\rangle\true \models\langle b\rangle\true$ hold. 
\end{exa}

We call the problem of deciding whether a formula in $\mathL\subseteq \mathL_{BS}$ is prime in $\mathL$ the \emph{formula primality problem} for $\mathL$.

The notion of a prime formula in \mathL is closely related to that of a characteristic formula, which we now proceed to define. 
%The definition of a characteristic formula within logic \mathL is given next.

\begin{defi}[\cite{AILS07,GrafS86a,SteffenI94}]\label{def:characteristic}
Let $\mathL\subseteq \mathL_{BS}$. 
A formula  $\varphi\in\mathL$ is \emph{characteristic  for $p\in \proc$ within $\mathL$} iff, for all $q \in \proc$, it holds that $q \models \varphi\Leftrightarrow\mathL(p) \subseteq \mathL(q)$. We denote by $\chi(p)$ the unique characteristic formula for $p$ %with respect to 
modulo logical equivalence.
\end{defi}

% \textcolor{orange}{I am starting to think that the use of proc that I am suggesting is problematic. In the above definition, proc cannot be a finite LTS, but it has to be an inclusive one. Should we then perhaps define LTSs, finite LTSs using P, and then use Proc to include all processes (carefully worded)?}

\begin{rem}\label{Remark:charforms}
    Let $X\in\{CS,RS,TS,nS,BS\}$, $n\geq 1$. In light of Theorem~\ref{Thm:HMT} and Proposition~\ref{logical_characterizations}, a formula $\varphi\in \mathL_X$  is characteristic for $p$ within $\mathL_X$ iff, for all $q \in \proc$, it holds that $q \models \varphi\Leftrightarrow p\curle_X q$. This property is often used as an alternative definition of characteristic formula for process $p$ modulo $\curle_X$. In what follows, we shall employ the two definitions interchangeably. 
\end{rem}

In~\cite[Table~1 and Theorem~5]{AcetoMFI19}, Aceto, Della Monica, Fabregas, and Ing\'olfsd\'ottir presented characteristic formulae for each of the semantics we consider in this paper, and showed that characteristic formulae are exactly the satisfiable and prime ones. 

\begin{prop}[\cite{AcetoMFI19}]\label{prop:charact-via-primality}
For every $X\in\{CS,RS,TS,nS\}$ and $n\geq 1$, $\varphi\in \mathL_X$ is characteristic for some process within $\mathL_X$ iff $\varphi$ is satisfiable and prime in $\mathL_X$.
\end{prop}

\begin{rem}
Proposition~\ref{prop:charact-via-primality} is the only result we use from~\cite{AcetoMFI19}  and we employ it as a `black box'. The (non-trivial) methods used in the proof of that result given in that reference do not play any role in our technical developments.

We note, in passing, that the article~\cite{AcetoMFI19} does not deal explicitly with $nS$, $n\geq 3$. However, its results apply to all the $n$-nested simulation preorders.
\end{rem}

We can also consider characteristic formulae modulo equivalence relations as follows.

\begin{defi}\label{def:characteristic-equivalence}
Let $X\in\{CS,RS,TS,nS,BS\}$, where $n\geq 1$. 
    A formula $\varphi\in \mathL_X$ is characteristic for $p\in \proc$ modulo $\equiv_X$  iff for all $q\in \proc$, it holds that $q\models \varphi\Leftrightarrow \mathL_X(p)=\mathL_X(q)$.\footnote{The above definition can also be phrased as follows: A formula $\varphi\in \mathL_X$ is characteristic for $p$ modulo $\equiv_X$ iff, for all $q \in \proc$, it holds that $q \models \varphi\Leftrightarrow p\equiv_X q$. This version of the definition is used, in the setting of bisimilarity, in references such as~\cite{AcetoAFI20,IngolfsdottirGZ87}.}  
    %%%%
    \iffalse\footnote{The definition of characteristic formulae modulo $\sim$ in~\cite{AcetoAFI20} uses $p\sim q$ instead of $\mathL_{BS}(p)=\mathL_{BS}(q)$. }.
    \fi
\end{defi} 

The following proposition holds for characteristic formulae modulo equivalence relations.

\begin{prop}\label{prop:char-mod-equiv}
  Let $X\in\{CS,RS,TS,nS,BS\}$, where $n\geq 1$. 
   A formula $\varphi\in \mathL_X$ is characteristic for a process modulo $\equiv_X$ iff $\varphi$ is satisfiable and for every $p,q\in\proc$ such that $p\models\varphi$ and $q\models\varphi$, $p\equiv_X q$ holds.
\end{prop}

  In the technical developments to follow, we extensively use the disjunctive normal form (DNF) of formulae. To transform a formula $\varphi\in\mathL_{TS}$ into DNF, we distribute conjunctions and $\langle a\rangle$ over disjunctions for every $a\in\act$. 

\begin{lem}\label{lem:DNF-equiv}
    Let $\varphi\in\mathL_X$, where $X\in\{S,CS,RS,TS\}$. Then, the DNF of $\varphi$ is logically equivalent to $\varphi$.
\end{lem}
\begin{proof} 
The proof of the lemma is immediate from the following facts: 
   $\langle a\rangle (\varphi_1\vee\varphi_2)\equiv\langle a\rangle\varphi_1\vee \langle a\rangle\varphi_2$,
$\varphi_1\wedge(\varphi_2\vee\varphi_3)\equiv(\varphi_1\wedge \varphi_2) \vee (\varphi_1\wedge \varphi_3)$, and 
for every $\varphi,\psi,\chi\in\mathL_X$, $X\in\{S,CS,RS,TS\}$, if $\psi\equiv \chi$, then $\varphi[\psi/\chi]\equiv\varphi$.
\end{proof}

The following three basic lemmas  are true for formulae in $\mathL_{BS}$.

\begin{lem}\label{lem:DF-equiv}
    Let $\varphi\in\mathL_{BS}$ and $\varphi_\vee$ be the result of distributing conjunctions over disjunctions in $\varphi$. Then, $\varphi\equiv\varphi_\vee$. 
\end{lem}

\begin{lem}\label{lem:disjunction_lemma}
For every $\varphi_1,\varphi_2,\psi\in \mathL_{BS}$, $\varphi_1\vee \varphi_2\models \psi$ iff $\varphi_1 \models\psi$ and $\varphi_2 \models\psi$.
\end{lem}
\begin{proof}
We prove the two implications separately. 

\noindent 
($\Rightarrow$) Let $p_1\models\varphi_1$. Then, $p_1\models\varphi_1\vee \varphi_2$, and so $p_1\models\psi$. The same argument is true for any $p_2$ that satisfies $\varphi_2$. Thus, $\varphi_1 \models\psi$ and $\varphi_2 \models\psi$.\\
($\Leftarrow$) Let $p\models\varphi_1\vee \varphi_2$. Then, $p\models\varphi_1$ or $p\models\varphi_2$. In either case, $p\models\psi$. So, $\varphi_1\vee \varphi_2\models\psi$.
\end{proof}

\begin{lem}\label{lem:diamond_lemma}
For every $\varphi,\psi\in \mathL_{BS}$, $\langle a\rangle\varphi\models \langle a\rangle \psi$ iff $\varphi \models\psi$.
\end{lem}
\begin{proof}
We prove the two implications separately. 

\noindent 
($\Rightarrow$) Suppose that $\varphi\not\models\psi$ and let $p$ be a process such that $p\models\varphi$ and $p\not\models\psi$. Consider process $q$ such that $q\myarrowa p$ and there is no other $p'$ such that $q\myarrowb p'$, $b\in \act$. Then $q\models\langle a\rangle\varphi$ and $q\not\models\langle a\rangle \psi$, contradiction.\\
($\Leftarrow$) Let $p\models\langle a\rangle\varphi$. Then, there is $p\myarrowa p'$ such that $p'\models \varphi$ and so $p'\models\psi$. Hence, $p\models\langle a\rangle \psi$.
\end{proof}

The following corollaries are immediate from the definitions of prime and characteristic formulae respectively.

\begin{cor}\label{cor:char-disj-form}
    Let $\varphi\in\mathL_{BS}$ be prime and $\varphi\equiv\bigvee_{i=1}^m\varphi_i$, $m\in\mathbb{N}$. Then, there is $1\leq j\leq m$ such that $\varphi\equiv\varphi_j$ and $\varphi_j$ is prime.
\end{cor}

\begin{cor}\label{cor:characteristic}
    Let $\varphi$ be characteristic for $p$ within $\mathL$. For every $\psi\in\mathL$, if $p\models\psi$, then $\varphi\models\psi$.
\end{cor}

%\textcolor{orange}{It might be worth splitting the section here (earlier?) and have a section/subsection called something along the lines of ``the tools that we use'' and then sub(sub) sections instead of paragraphs. Perhaps also move the contents of ``complexity and games'' up to \Cref{prop:pspace-ap} before this new (sub)section.}

\subsection{Complexity and games}\label{Sect:complexity-games}

We introduce 
two complexity classes that play an important role in pinpointing the complexity of deciding whether a formula $\varphi$ is characteristic, which we shall often shorten to `deciding characteristic formulae' in what follows. The first class is $\dpc=\{L_1\cap L_2 \mid L_1\in\NP \text{ and } L_2\in \conp\}$~\cite{PapadimitriouY84} and the second one is  \us~\cite{BlassG82}, which 
is defined thus:
A language $L\in\us$ iff there is a non-deterministic Turing machine $T$ such that, for every instance $x$ of $L$, $x\in L$ iff $T$ has \emph{exactly one} accepting path on input $x$. The problem $\textsc{UniqueSat}$, viz.~the problem of deciding whether a given Boolean formula has exactly one satisfying truth assignment, is \us-complete. Note that $\us\subseteq\dpc$~\cite{BlassG82}.

For formulae in $\mathL_{BS}$, model checking is tractable and satisfiability is \pspace-complete.

\begin{prop}[\cite{HalpernM92}]\label{model-checking-complexity}
    Given a formula $\varphi\in\mathL_{BS}$ and a finite process $p$, there is an algorithm to check if $p$ satisfies $\varphi$ that runs in time $\mathcal{O}(|p|\cdot |\varphi|)$.
\end{prop}

\begin{prop}[\cite{ladner1977computational,HalpernM92}]\label{hml-sat}
    Satisfiability for the modal  logics \logikk and \hml is \pspace-complete.
\end{prop}

Moreover equivalence checking for $\equiv_{2S}$ is also tractable.

\begin{prop}[\cite{DBLP:journals/iandc/GrooteV92, ShuklaRHS96}]\label{prop:2s-equiv-polytime}
    Given two processes $p,q\in\proc$, checking whether $p\equiv_{2S} q$ holds can be done in polynomial time.
\end{prop}

An \emph{alternating Turing machine} is a non-deterministic Turing machine whose set of states is partitioned into existential and universal states. An existential state is considered accepting if at least one of its transitions leads to an accepting state. In contrast, a universal state is accepting only if all of its transitions lead to accepting states. The machine as a whole accepts an input if its initial state is accepting. The complexity class \ap is the class of languages accepted by polynomial-time alternating Turing machines. 
An \emph{oracle Turing machine} is a Turing machine that has access to an oracle---a `black box' capable of solving a specific computational problem in a single operation. An oracle Turing machine can perform all of the usual operations of a Turing machine, and can also query the oracle to obtain a solution to any instance of the computational problem for that oracle. We use $\textsf{C}^{\textsf{C'}[\text{poly}]}$ to denote the complexity class of languages decidable by an algorithm in class $\textsf{C}$ that makes polynomially many oracle calls to a language in $\textsf{C'}$. Note, for example, that $\pspace^{\pspace[\text{poly}]}=\pspace$, since a polynomial-space oracle Turing machine can simulate any \pspace oracle query by solving the problem itself within polynomial space.

\begin{prop}
\label{prop:pspace-ap}
\hfill
\begin{enumerate}[(a)]
    \item  \textnormal{(\cite{DBLP:journals/jacm/ChandraKS81})\textbf{.}} $\ap=\pspace$.
    \item $\ap^{\pspace[\text{poly}]}=\pspace$.
\end{enumerate}
\end{prop}

\noindent 
Consider now two-player games that have the following characteristics: they are \emph{zero sum} (that is, player one's gain is equivalent to player two's loss), \emph{perfect information} (meaning that, at every point in the game, each player is fully aware of all events that have previously occurred), \emph{polynomial depth} (that is, the games proceed for a number of rounds that is polynomial in the input size), and \emph{computationally bounded} (that is, at each round, the computation performed by a player can be simulated by a Turing machine operating within polynomial time in the input size). For two-player games that have all four characteristics described above, there is a polynomial-time alternating Turing machine that decides whether one of the players has a winning strategy~\cite{Feigenbaum1998}. The two-player games we will introduce in Subsection~\ref{subsection:primality-nS} are zero sum, perfect information, and polynomial depth, but they are \emph{not} computationally bounded: in each round, at most a polynomial number of problems in \pspace have to be solved. We call these games zero sum, perfect information, polynomial depth \emph{with a \pspace oracle}. Then, the polynomial-time alternating Turing machine that determines whether one of the players has a winning strategy for such a game has to use a polynomial number of oracle calls to \pspace problems in order to correctly simulate the game. For these games, we can still decide whether a player has a winning strategy in polynomial space because of Proposition~\ref{prop:pspace-ap}(b).

\begin{cor}\label{cor:pspace-games}
   For a two-player, zero-sum, perfect-information, polynomial-depth game with a \pspace oracle, we can decide whether a player has a winning strategy in polynomial space.
\end{cor}

\subsection{Two pervasive tools}\label{Sect:tools}

In the technical developments to follow, we will make repeated use of the reachability problem on alternating graphs to study the complexity of the formula primality problem for various logics and of tableau-based techniques for {\hml}, which we now introduce. 

\paragraph{The reachability problem on alternating graphs}
The definitions of an alternating graph and of the reachability problem on alternating graphs are provided below. They stem from~\cite[Definition~3.24]{Immerman99} and~\cite[pp.~53--54]{Immerman99}. 

\begin{defi}\label{def:alternating-graph}
An \emph{alternating graph} $G=(V,E,A,s,t)$ is a finite directed graph whose vertices are either existential or  universal---that is, they are labelled with $\exists$ or $\forall$, respectively. $V$ and $E$ are the sets of vertices and edges, respectively, $A$ is the set of universal vertices, and $s$ and $t$ are two vertices that are called \emph{source} and \emph{target}, respectively.  
\end{defi}

\begin{defi}
\label{def:alternating_reachability}
Let $G=(V,E,A,s,t)$ be an alternating graph. Let $P^G$ be the smallest binary relation on $V$ 
%$V\times V$ 
that satisfies the following clauses:
\begin{enumerate}
    \item $P^G(x,x)$, for each $x\in V$.
    \item If $x$ is existential and for some $(x,z)\in E$ it holds that $P^G(z,y)$, then $P^G(x,y)$.
    \item If $x$ is universal, there is at least one edge leaving $x$, and $P^G(z,y)$ holds for all edges $(x,z)$, then $P^G(x,y)$.
\end{enumerate}
If $P^G(x,y)$, we say that there is an alternating path from $x$ to $y$.

We define 
$\reacha=\{G ~ | ~ G \text{ is an alternating graph and }P^G(s,t)\}.
$
\end{defi}
The problem $\reacha$ can be solved in linear time~\cite[Algorithm~3.25]{Immerman99} and is $\textsf{P}$-complete via first-order reductions~\cite[Theorem~3.26]{Immerman99}. 

\iffalse
\textcolor{orange}{I suppose we need the upper bound, right? The (joint) definition of and alternating graph and of $\reacha$ is given in \cite{Immerman99}, Def 3.24. Then, Algorithm 3.25 claims linear time to solve it. The subsequent Theorem 3.26 claims P-completeness via first-order reductions. In the historical notes (next page 55, conveniently), Immermann gives two references for the proof of completeness (his papers), but none for the upper bound, which seems to be claimed to be a direct result of P=ASPACE[$\log n$] (which seems fair). In turn, that is (a restriction of) Theorem 2.25 in the book, and Immermann cites Kozen, and Chandra and Stockmeyer \cite{DBLP:journals/jacm/ChandraKS81}. I suggest citing that last reference.}
\fi

\paragraph{The \hml tableau}\label{subsection:hml-tableau}

Tableau constructions will play an important role in our study of the satisfiability and primality problems for several of the logics mentioned above. Hence, we end this section by introducing them together with some classic results on their connections with {\hml}.

\begin{defi}\label{def:proptableau}
Let $T$ be a set of \hml formulae. 
\begin{enumerate}[(a)]
    \item $T$ is \emph{propositionally inconsistent} if  
    $\ff\in T$, or $\psi\in T$ and $\neg\psi\in T$ for some formula $\psi$. Otherwise, $T$ is \emph{propositionally consistent}.
    \item  $T$ is a \emph{propositional tableau} if the following conditions are met:
    \begin{enumerate} [(i)]
        \item if $\psi\wedge\psi'\in T$, then $\psi,\psi'\in T$,
        \item if $\psi\vee\psi'\in T$, then either $\psi\in T$ or $\psi'\in T$, and
        \item $T$ is propositionally consistent. 
    \end{enumerate}
    \item $T$ is  \emph{fully expanded} if for every $\psi\in\sub(T)$, either $\psi\in T$ or $\neg\psi\in T$.
\end{enumerate}
\end{defi}

\begin{defi}\label{def:hmltableau}
Let $\act=\{a_1,\dots,a_k\}$. An \hml tableau is a tuple 
\[
T=(S,L,R_{a_1},\dots,R_{a_k}), 
\]
where $S$ is a finite set of states, $L$ is a labelling function that maps every $s\in S$ to a set $L(s)$ of formulae, and  $R_{a_i}\subseteq S\times S$, for every $1\leq i\leq k$, such that
\begin{enumerate}[(i)]
    \item $L(s)$ is a propositional tableau for every $s\in S$,
    \item if $[a_i] \psi\in L(s)$ and $(s,t)\in R_{a_i}$, then $\psi\in L(t)$, and
    \item if $\langle a_i\rangle\psi\in L(s)$, then there is some $t$ such that $(s,t)\in R_{a_i}$ and $\psi\in L(t)$.
\end{enumerate}
\end{defi}

\noindent An \hml tableau for $\varphi$ is an \hml tableau such that $\varphi\in L(s)$ for some $s\in S$.

The \emph{size} of an \hml tableau $(S,L,R_{a_1},\dots,R_{a_k})$ is $|S|+|R_{a_1}|+\cdots +|R_{a_k}|$ and its length is the length $r$ of the longest path $s_1 s_2 \cdots s_r$, such that for every $i < r$, $s_i R_{a_j} s_{i+1}$ for some $a_j \in \act$.

\begin{prop}[\cite{HalpernM92}]\label{tableausat}
An \hml formula $\varphi$ is satisfiable iff there is an \hml tableau for $\varphi$.
\end{prop}

\begin{rem}\label{rem:tableausat}
    The proof of the `right-to-left' direction of Proposition~\ref{tableausat} constructs an LTS corresponding to a process satisfying $\varphi$ from an \hml tableau for $\varphi$ in a straightforward way: given an \hml tableau $T=(S,L,R_{a_1},\dots,R_{a_k})$ for $\varphi$, define the LTS $\mathS=(P,\myarrowasubone,\dots,\myarrowasubk)$, where $P=S$ and every $\myarrowasubi$ coincides with $R_{a_i}$. Note that $T$ and $\mathS$ have the same size and depth. 
\end{rem}

\paragraph{The tableau construction for $\varphi\in\hml$}

If a set $T$ of formulae is not a propositional tableau, then $\psi$ is a witness to this if $\psi\in T$ and one of clauses (b)(i)--(iii) in Definition~\ref{def:proptableau} does not apply to $\psi$. Similarly, if $T$ is not fully expanded, $\psi$ is a witness to this if $\psi\in\sub(T)$ and neither $\psi\in T$ nor $\neg\psi\in T$. We assume that formulae are ordered in some way, so when there is a witness to one of the above facts, we can choose the least witness.

Let $\varphi\in \hml$. The following procedure constructs a labelled tree from which an \hml tableau for $\varphi$ can be extracted when $\varphi$ is satisfiable.
The \hml tableau construction for $\varphi$ consists of the following steps:
\begin{enumerate}
    \item Construct the node $s_0$ with $L(s_0)=\{\varphi\}$ which is called `the root'.
    \item Repeat (a)--(d) until none of them applies:
    \begin{enumerate}[a.]
        \item \emph{Forming a propositional tableau:} if $s$ is a leaf, $L(s)$ is propositionally consistent, and $L(s)$ is not a propositional tableau and $\psi$ is the least witness to this fact, then:
        \begin{enumerate}[i.]
           % \item if $\psi=\neg\neg\psi'$ for some $\psi'$, then add a successor $s'$ of $s$ (i.e.\ add a node $s'$ to the tree and an edge from $s$ to $s'$) and set $L(s')=L(s)\cup\{\psi'\}$,
            \item if $\psi=\psi_1\wedge\psi_2$ for some $\psi_1,\psi_2$, then add a successor $s'$ of $s$ and set $L(s')=L(s)\cup\{\psi_1,\psi_2\}$,
            \item if $\psi=\psi_1\vee\psi_2$ for some $\psi_1,\psi_2$, then add two successors $s_1$ and $s_2$ of $s$ and set $L(s_i)=L(s)\cup\{\psi_i\}$, $i=1,2$.
        \end{enumerate}
        \item \emph{Forming a fully expanded propositional tableau:} if $s$ is a leaf, $L(s)$ is propositionally consistent, and $L(s)$ is not a fully expanded propositional tableau and $\psi$ is the least witness to this fact, then add two successors $s_1$ and $s_2$ of $s$ and set $L(s_1)=L(s)\cup\{\psi\}$, $L(s_2)=L(s)\cup\{\neg\psi\}$.
        \item \emph{Adding $i$-successor nodes:} if $s$ is a leaf, $L(s)$ is propositionally consistent, and $L(s)$ is a fully expanded propositional tableau, then for every formula of the form $\langle a_i\rangle\psi\in L(s)$ add an $i$-successor node $s'$ (i.e.\ add a node $s'$ to the tree and an edge from $s$ to $s'$ labelled $i$) and let $L(s')=\{\psi\}\cup\{\phi ~ | ~ [a_i]\phi\in L(s)\}$. 
        \item \emph{Marking nodes `satisfiable':} if $s$ is not marked  `satisfiable', then mark $s$ `satisfiable' if one of the following conditions is met:
        \begin{enumerate}[i.]
            \item $L(s)$ is not a fully expanded propositional tableau and $s'$ is marked `satisfiable' for some successor $s'$ of $s$, 
            \item $L(s)$ is a fully expanded propositional tableau, $L(s)$ is propositionally consistent, and it does not contain any formula of the form $\langle a_i\rangle \psi$,
            \item $L(s)$ is a fully expanded propositional tableau, $s$ has successors, and all of them are marked `satisfiable'.
        \end{enumerate}
    \end{enumerate}
\end{enumerate}

\begin{prop}[\cite{HalpernM92}]\label{prop:tableau-construction}
    There is an \hml tableau for $\varphi$ iff the tableau construction for $\varphi$ marks the root $s_0$ `satisfiable'.
\end{prop}

We just describe here how an \hml tableau $T=(S,L,R_{a_1},\dots,R_{a_k})$ for $\varphi$ can be extracted from the tree $T'=(S',L',\longrightarrow,\myarrowone,\dots,\myarrowk)$ constructed by the tableau construction in the case that the root of $T'$ is marked `satisfiable'. The tableau $T$ is defined as follows:
\begin{itemize}
    \item $s\in S$ if $s\in S'$, $s$ is marked `satisfiable',  and $L'(s)$ is a fully expanded propositional tableau;
    \item $(s,t)\in R_{a_i}$ if $s\myarrowi t$ (that is,  $t$ is an $i$-successor of $s$ in $T'$);
    \item $L(s)=L'(s)$,  for every $s\in S$.
\end{itemize}

\section{A bird's eye view of our results}\label{Sect:summary-of-our-results}

In the rest of this paper, we study the complexity of deciding characteristic formulae in the logics $\mathL_{S}$, $\mathL_{CS}$, $\mathL_{RS}$, $\mathL_{TS}$, and $\mathL_{nS}$, $n\geq 2$.
As characteristic formulae are exactly the satisfiable and prime ones (Proposition~\ref{prop:charact-via-primality}), we study both the complexity of satisfiability for these logics and the complexity of deciding whether a formula is prime. Our results are summarized in Tables~\ref{table:sat}--\ref{table:characteristic}. Note that the aforementioned logics characterize the respective preorders $\curle_S$, $\curle_{CS}$, $\curle_{RS}$, $\curle_{TS}$, and $\curle_{nS}$, $n\geq 2$---see Proposition~\ref{logical_characterizations}. We also investigate the complexity of deciding characteristic formulae modulo the kernels of these preorders, namely $\equiv_S$, $\equiv_{CS}$, $\equiv_{RS}$, $\equiv_{TS}$, and $\equiv_{nS}$, $n\geq 2$. Table~\ref{table:characteristic-equiv} summarizes the results in that setting.

\begin{table}[t]
\begin{center}
Satisfiability\\
\vspace{2mm}
\begin{tabular}{ | m{0.7cm} | m{0.7cm} | m{1.8cm} | m{2.2cm} | m{1.5cm} | m{1.5cm} | m{2cm} | m{1.5cm}| } 
\hline
$\mathL_{S}$ & $\mathL_{CS}$ & \multicolumn{2}{ c|}{$\mathL_{RS}$} &  \hspace{2mm} $\mathL_{TS}$ & \hspace{1mm} $\mathL_{2S}$ &  \hspace{1mm} $\mathL_{nS}, n\geq 3$ & \cellcolor[gray]{0.9}\hspace{1mm}  $\hml$\\  
\hline
  &  & \footnotesize $\act \text{ bounded }$ & \footnotesize $\act \text{ unbounded } $  & \footnotesize \hspace{1mm} $|\act|\geq 2$  & \footnotesize \hspace{1mm}$|\act|\geq 2$ & \footnotesize \hspace{3mm}$|\act|\geq 2$ & \cellcolor[gray]{0.9} \\
\hline
\textcolor{teal}{\textsf{P}}  & \textcolor{teal}{\textsf{P}}  & \textcolor{teal}{\textsf{P}}  & \footnotesize \textcolor{purple}{\textsf{NP}-complete} & \textcolor{purple}{\footnotesize \textsf{NP}-complete} &  \textcolor{purple}{\footnotesize  \textsf{NP}-complete} & \footnotesize \textcolor{blue}{\textsf{PSPACE}-complete} & \cellcolor[gray]{0.9}\footnotesize \textcolor{blue}{\textsf{PSPACE}-complete} \\
\hline
\end{tabular}
\end{center}
\caption{The complexity of satisfiability for the fragments of \hml that characterize the preorders considered in this paper. Results shown in white cells are proven in this work, whereas that in light gray is  from~\cite{ladner1977computational,HalpernM92}.}
\label{table:sat}
\end{table}

\begin{table}[t]
\begin{center}
Formula primality\\
\vspace{2mm}
\begin{tabular}{ | m{0.7cm} | m{0.7cm} | m{1.8cm} | m{2.2cm} | m{1.6cm} | m{1.4cm} | m{2cm} | m{1.5cm}| } 
\hline
$\mathL_{S}$ & $\mathL_{CS}$ & \multicolumn{2}{ c|}{$\mathL_{RS}$} &  \hspace{2mm} $\mathL_{TS}$ & \hspace{1mm} $\mathL_{2S}$ &  \hspace{1mm} $\mathL_{nS}, n\geq 3$ & \cellcolor[gray]{0.9}\hspace{1mm}  $\hml$\\  
\hline
   &  & \footnotesize $\act \text{ bounded }$ & \footnotesize $\act \text{ unbounded } $  & \footnotesize \hspace{1mm} $|\act|\geq 2$  & \footnotesize \hspace{1mm}$|\act|\geq 2$ & \footnotesize \hspace{3mm}$|\act|\geq 2$ & \cellcolor[gray]{0.9}  \\
\hline
\textcolor{teal}{\textsf{P}} & \textcolor{teal}{\textsf{P}}  & \textcolor{teal}{\textsf{P}}  & \textcolor{purple}{\footnotesize \textsf{coNP}-complete} & \textcolor{purple}{\footnotesize \textsf{FPT}~\&~}\hspace{3mm}\textcolor{purple}{\footnotesize \textsf{coNP}-hard} & \textcolor{purple}{\footnotesize \textsf{coNP}-complete} & \textcolor{blue}{\footnotesize \textsf{PSPACE}-complete} & \cellcolor[gray]{0.9}\textcolor{blue}{\footnotesize \textsf{PSPACE}-complete} \\
\hline
\end{tabular}
\end{center}
\caption{The complexity of deciding whether a formula is prime in $\mathL_X$, where $X\in\{CS,RS,TS,nS,BS\}$ for $n\geq 1$. As in the previous table, results shown in white cells are established in this paper, whereas that in light gray is from~\cite{AcetoAFI20}.}
\label{table:prime}
\end{table}

\begin{table}[t]
\begin{center}
Deciding characteristic formulae within $\mathL_X$\\
\vspace{2mm}
\begin{tabular}{ | m{0.7cm} | m{0.7cm} | m{1.8cm} | m{2.2cm} | m{1.5cm} | m{1.5cm} | m{2cm} | m{1.5cm}| } 
\hline
$\mathL_{S}$ & $\mathL_{CS}$ & \multicolumn{2}{ c|}{$\mathL_{RS}$} &  \hspace{2mm} $\mathL_{TS}$ & \hspace{1mm} $\mathL_{2S}$ &  \hspace{1mm} $\mathL_{nS}, n\geq 3$ & \cellcolor[gray]{0.9}\hspace{1mm}  $\hml$\\  
\hline
  &  & \footnotesize $\act \text{ bounded }$ & \footnotesize $\act \text{ unbounded } $  & \footnotesize \hspace{1mm} $|\act| \geq 2$   & \footnotesize \hspace{1mm}$|\act|\geq 2$ & \footnotesize \hspace{3mm}$|\act|\geq 2$ & \cellcolor[gray]{0.9} \\
\hline
\textcolor{teal}{\textsf{P}} & \hspace{2mm}\textcolor{teal}{\textsf{P}}  & \hspace{5mm}\textcolor{teal}{\textsf{P}}  & \textcolor{purple}{\footnotesize \textsf{DP}~\&~}\textcolor{purple}{\footnotesize\textsf{US}-hard} & \textcolor{purple}{\footnotesize \textsf{FPT}~\&~}\hspace{3mm}\footnotesize\textcolor{purple}{\textsf{US}-hard} & \textcolor{purple}{\footnotesize \textsf{DP}~\&~}\hspace{3mm}\textcolor{purple}{\footnotesize\textsf{US}-hard} & \textcolor{blue}{\footnotesize \textsf{PSPACE}-complete} & \cellcolor[gray]{0.9} \textcolor{blue}{\footnotesize \textsf{PSPACE}-complete} \\
\hline
\end{tabular}
\end{center}
\caption{The complexity of deciding whether a formula is characteristic within $\mathL_X$, where $X\in\{CS,RS,TS,nS,BS\}$,  $n\geq 1$. The color convention is the same as in the previous tables: white cells denote results proven here, and the light gray cell denotes a result from~\cite{AcetoAFI20}.}
\label{table:characteristic}
\end{table}

\begin{table}[t]
\begin{center}
Deciding characteristic formulae modulo $\equiv_X$\\
\vspace{2mm}
\begin{tabular}{ | m{1.2cm} | m{2.2cm} | m{1.8cm} | m{2.5cm} | m{2.6cm} | m{2.6cm}| } 
\hline
%\vspace{2mm}
\hspace{2mm}$\equiv_{S}$  & \hspace{6mm}$\equiv_{RS}$ &  \hspace{3mm} $\equiv_{TS}$ & \hspace{7mm} $\equiv_{2S}$ &  \hspace{1mm} $\mathL_{nS}, n\geq 3$ & \cellcolor[gray]{0.9}\hspace{4mm}  $\hml$\\  
\hline
  &   \footnotesize $\act \text{ unbounded }$ & \footnotesize \hspace{1mm} $|\act| \geq 2$   & \footnotesize \hspace{5mm}$|\act|\geq 2$ & \footnotesize \hspace{4mm}$|\act|\geq 2$ & \cellcolor[gray]{0.9} \\
\hline
\textcolor{teal}{\textsf{trivial}} & \textcolor{purple}{\footnotesize \textsf{\conp-hard}} & \textcolor{purple}{\footnotesize \textsf{FPT}~\&~}\hspace{3mm}\footnotesize\textcolor{purple}{\conp-hard} & \textcolor{purple}{\footnotesize \textsf{DP}~\&~}\textcolor{purple}{\footnotesize\textsf{\conp}-hard} & \textcolor{blue}{\footnotesize \textsf{PSPACE}-complete} & \cellcolor[gray]{0.9} \textcolor{blue}{\footnotesize \textsf{PSPACE}-complete} \\
\hline
\end{tabular}
\end{center}
\caption{The complexity of deciding whether a formula is characteristic modulo $\mathL_X$, where $X\in\{RS,TS,nS,BS\}$,  $n\geq 1$. The result in light gray is from~\cite{AcetoAFI20}.}
\label{table:characteristic-equiv}
\end{table}

\paragraph{The structure of the rest of the paper.} In Section~\ref{section:deciding-satisfiability}, we prove the results presented in Table~\ref{table:sat} on the complexity of the satisfiability problem for $\mathL_X$, where $X$ is one of $\{S,CS,RS,TS,nS\}$, $n\geq 2$. Sections~\ref{section:deciding-primality} and~\ref{Sect:primality-nested-semantics} are devoted to studying the complexity of the formula primality problem in the logics $\mathL_S$, $\mathL_{CS}$, $\mathL_{RS}$, $\mathL_{nS}$, $n\geq 2$. The results we present in  Sections~\ref{section:deciding-primality} and~\ref{Sect:primality-nested-semantics} are summarized in Table~\ref{table:prime} and, to our mind, are the main technical contributions of the paper. Their proofs are based on the tools we introduced in Section~\ref{Sect:tools}.  By combining the results of Sections~\ref{section:deciding-satisfiability},~\ref{section:deciding-primality} and~\ref{Sect:primality-nested-semantics}, and introducing some additional reductions and observations, we obtain the complexity results for deciding  characteristic formulae in Sections~\ref{section:char-preorders} and~\ref{section:char-equivalences}. Section~\ref{section:char-preorders} studies characteristic formulae within the logics characterizing preorders in van Glabbeek's branching spectrum, while Section~\ref{section:char-equivalences} considers characteristic formulae modulo the corresponding equivalence relations. We end with a discussion of our results and avenues for future research in Section~\ref{section:conclusions}.

\iffalse
\textcolor{red}{I suggest that we describe the structure, compare the paper with the conference papers and explain that we give only some ideas for the primality for CS and RS with a bounded action set and for the conp membership for RS with an unbounded action set because otherwise the paper becomes very long. We mention that we have appendices for the formula primality problem for LCS, LRS with a bounded action set and the conp membership for LRS with an unbounded action set. I have added references to the appendices at the points where the respective results are given. If you wish you can erase whatever you want.}
\fi

\section{The complexity of satisfiability}\label{section:deciding-satisfiability}

In this section, we establish the results shown in Table~\ref{table:sat}. To the best of our knowledge, these are the first results concerning the satisfiability of these fragments of \hml.

\subsection{Satisfiability for \texorpdfstring{$\mathL_S$}{LS}, \texorpdfstring{$\mathL_{CS}$}{LCS} and \texorpdfstring{$\mathL_{RS}$}{LRS} with a bounded action set.}\label{Sect:sat-in-P}

To address the complexity of the satisfiability problem  in $\mathL_S$, $\mathL_{CS}$, or $\mathL_{RS}$, we associate a set $I(\varphi)\subseteq 2^\act$ to every formula $\varphi\in\mathL_{RS}$. Intuitively, $I(\varphi)$ describes all possible sets of initial actions that a process $p$ can have, when $p\models \varphi$.

\begin{defi}\label{def:cs-rs-I(phi)}
    Let $\varphi\in\mathL_{RS}$. We define $I(\varphi)$ inductively as follows:
    \begin{enumerate}[(a)]
        \item $I(\true)=2^\act$,
        \item $I(\ff)=\emptyset$,
        \item $I([a]\ff)= \{X ~\mid ~ X\subseteq \act \text{ and } a\not\in X\}$,
        \item $I(\langle a\rangle \varphi)=\begin{cases}
           \emptyset, &\text{if } I(\varphi)=\emptyset,\\
           \{X ~\mid ~ X\subseteq \act \text{ and } a\in X\}, &\text{otherwise}
        \end{cases}$
        \item $I(\varphi_1\vee\varphi_2)=I(\varphi_1)\cup I(\varphi_2)$,
        \item $I(\varphi_1\wedge\varphi_2)=I(\varphi_1)\cap I(\varphi_2)$.
    \end{enumerate}
    Note that $I(\zero)=\{\emptyset\}$.
\end{defi}

\begin{exa}
     For example, let $\act=\{a,b\}$ and $\varphi=(\langle a\rangle\zero\wedge [b]\ff)\vee(\langle b\rangle\zero\wedge [a]\ff)\vee\zero$. Then, $I(\varphi)=(\{\{a\},\{a,b\}\}\cap\{\emptyset,\{a\}\})\cup(\{\{b\},\{a,b\}\}\cap\{\emptyset,\{b\}\})\cup\{\emptyset\}=\{\{a\},\{b\},\emptyset\}$. Indeed a process that satisfies $\varphi$ can  either have only $a$ as an initial state or have only $b$ or be a deadlocked state.
\end{exa}

\begin{lem}
\label{lem:I(phi)-property}
For every $\varphi\in\mathL_{RS}$, the following statements hold:
\begin{enumerate}[(a)]
    \item  for every $S\subseteq \act$, $S\in I(\varphi)$ iff there is a process $p$ such that $I(p)=S$ and $p\models\varphi$.
    \item  $\varphi$ is unsatisfiable iff $I(\varphi)=\emptyset$.
\end{enumerate}
\end{lem}

\noindent When the number of actions is constant, $I(\varphi)$ can be computed in linear time for every $\varphi\in\mathL_{RS}$. For $\mathL_{CS}$, we need even less information; indeed, it is sufficient to define $I(\varphi)$  so that it encodes whether $\varphi$ is unsatisfiable, or is satisfied only in deadlocked states (that is, states with an empty set of initial actions), or is satisfied only in processes that are not deadlocked, or is satisfied both in some deadlocked and non-deadlocked states. This information can be computed in linear time for every $\varphi\in\mathL_{CS}$, regardless of the size of the action set. The details for the complexity of satisfiability for $\mathL_{CS}$ can be found in Appendix~\ref{subsection:cs-satisfiability}.

\begin{cor}\label{cor:sat-s-cs-rs-poly}\hfill
    \begin{enumerate}[(a)]
     \item Satisfiability of formulae in $\mathL_{CS}$ and $\mathL_S$ is  decidable in linear time.
     \item Let $|\act|=k$, where $k\geq 1$ is a constant. Satisfiability of formulae in $\mathL_{RS}$ is  decidable in linear time.
    \end{enumerate}
\end{cor}

\subsection{Satisfiability for \texorpdfstring{$\mathL_{RS}$}{LRS} with an unbounded action set, \texorpdfstring{$\mathL_{TS}$}{LTS}, and \texorpdfstring{$\mathL_{2S}$}{L2S}.} 

On the other hand, if we can use an unbounded number of actions, the duality of $\langle a\rangle$ and $[a]$ can be employed to define a polynomial-time reduction from  {\SAT}, the satisfiability problem for propositional logic, to satisfiability in $\mathL_{RS}$. Moreover, if we are allowed to nest $[a]$ modalities ($a\in \act$) and have at least two actions, we can encode $n$ propositional literals using formulae of $\log n$ size and reduce \SAT to satisfiability in  $\mathL_{TS}$ in polynomial time. Finally, satisfiability in $\mathL_{2S}$ is in \NP, which can be shown by an appropriate tableau construction. These statements are proven below.

\begin{prop}\label{prop:sat-RS-TS}\hfill
\begin{enumerate}[(a)]
\item Let $\act$ be unbounded. Satisfiability in $\mathL_{RS}$ is $\NP$-hard. 
\item Let $|\act|\geq 2$ and $X\in\{TS,2S\}$. Satisfiability of formulae in $\mathL_X$ is $\NP$-hard. 
\end{enumerate}
\end{prop}
\begin{proof} \hfill
\begin{enumerate}[(a)]
\item Consider a propositional formula $\psi$ in conjunctive normal form (CNF) with variables $x_1,\dots, x_n$. Construct formula $\psi'$ by replacing each literal $x_i$ with $\langle a_i\rangle\true$ and each literal $\neg x_i$ with $[a_i]\ff$. Then, it is not hard to see that $\psi$ is satisfiable iff there is a process that satisfies $\psi'$. 
\item We show the statement for $TS$. Then it also holds for $2S$, since $\mathL_{TS}\subseteq \mathL_{2S}$.  Assume that $\act=\{0,1\}$. Consider a propositional formula $\psi$ in CNF with variables $x_1,\dots, x_n$. We use $\mathL_{TS}$ formulae of logarithmic size to encode literals of $\psi$. 
We associate a positive literal $x_i$, $i=1,\dots,n$, with the binary representation of $i$, that is,  ${b_i}_1\dots{b_i}_k$, where every ${b_i}_j\in\{0,1\}$ and $k=\lceil\log n\rceil$. The binary string ${b_i}_1\dots {b_i}_k$ can now be mapped to formula 
$\enc(x_i)= \langle {b_i}_1\rangle \langle {b_i}_2\rangle\dots \langle {b_i}_k\rangle\true$.
We map a negative literal $\neg x_i$ to $\enc(\neg x_i)=[{b_i}_1] [{b_i}_2]\dots [{b_i}_k]\ff$. We construct formula $\varphi\in\mathL_{TS}$ by starting with $\psi$ and replacing every literal $\ell$ with $\enc(\ell)$ in $\psi$. It is not hard to see that the resulting formula $\varphi$ is satisfiable iff $\psi$ is satisfiable.\qedhere
\end{enumerate}
\end{proof}

\begin{thm}\label{prop:sat-rs-ts-2s-np-complete}
Let either $X=RS$ and $\act$ be unbounded or $X\in\{TS,2S\}$ and $|\act|\geq 2$. Satisfiability of formulae in $\mathL_X$ is \NP-complete.
\end{thm}
\begin{proof}
    \NP-hardness is immediate from Proposition~\ref{prop:sat-RS-TS}. To show membership in \NP for the three logics, it suffices to prove it for $2S$ since $\mathL_{RS}\subseteq \mathL_{TS}\subseteq \mathL_{2S}$. We can decide whether $\varphi\in\mathL_{2S}$ is satisfiable in a standard way by constructing a tableau for $\varphi$ and checking whether its root is satisfiable as described in Subsection~\ref{subsection:hml-tableau}.  Consider the non-deterministic polynomial-time Turning machine $M$ such that each path of $M$ constructs a branch of a tableau for $\varphi$ as follows. It starts from the node $r$ of the tableau that corresponds to $\varphi$ labelled with $L(r)=\{\varphi\}$. Let $s$ be a node of the tableau that has been already created by $M$ labelled with $L(s)$. If one successor $s'$ of a node $s$ must be created because of a formula $\psi\wedge\psi'\in L(s)$, then $M$ creates $s'$. If for every $\langle a_i\rangle\varphi'\in L(s)$, an $i$-successor of $s$ must be created, then $M$ creates all these successors. If two different successors $s'$ and $s''$ of $s$ that are not $i$ successors of $s$ must be created, then $M$ non-deterministically chooses to create one of them. In this last case, $s'$ and $s''$ correspond either to some $\psi\vee\psi'\in L(s)$ or some $\psi\in\sub(\varphi')$, where $\varphi'\in L(s)$, such that $\psi\not\in L(s)$ and $\neg\psi\in L(s)$. To decide if the root $r$ is marked satisfiable, we need to know whether at least one of $s$, $s''$ is marked satisfiable and this will be done by $M$ using non-determinism. After constructing this branch of the tableau, $M$ propagates information from the leaves to the root and decides whether the root must be marked `satisfiable'. It accepts iff $r$ is marked `satisfiable'. It holds that $\varphi$ is satisfiable iff there is some branch that makes $r$ satisfiable iff $M$ has an accepting path. Since $\varphi\in\mathL_{2S}$, the formula has no diamond operators $\langle a_i\rangle$ in the scope of box operators $[a_i]$. So, the longest path of a branch is polynomial in the number of nested diamond operators occurring in $\varphi$ and a branch has size polynomial with respect to the number of diamond operators in $\varphi$; moreover, a branch can be computed in polynomial time.
\end{proof}

\subsection{Satisfiability for \texorpdfstring{$\mathL_{nS}$}{LnS}, \texorpdfstring{$n\geq 3$}{n>=3}.}

Moving from $2S$ to $3S$ makes the satisfiability problem even harder. Deciding satisfiability in $\mathL_{3S}$ when $|\act|\geq 2$ turns out to be \pspace-complete, which matches the complexity of the problem in $\mathL_{BS}$.
Let $|\act|\geq 2$. We show that $\compL_{2S}$-satisfiability is \pspace-complete. 
In what follows $\mathL_{\Box\Diamond}$ denotes $\compL_{2S}$, i.e.\ the dual fragment of $\mathL_{2S}$, which consists of all \hml formulae that have no box subformulae in the scope of a diamond operator.

\begin{thm}\label{thm:2s-validity}
Let $|\act|\geq 2$. Satisfiability of formulae in $\compL_{2S}$ is \pspace-complete.
\end{thm}
\begin{proof}
    That $\compL_{2S}$-satisfiability is in \pspace\ is a direct result of Proposition~\ref{hml-sat} and $\compL_{2S}\subseteq\hml$.
To prove the \pspace-hardness of $\compL_{2S}$-satisfiability, we consider 
    $\mathL_{\Box\Diamond x}$ to be the extension of $\mathL_{\Box\Diamond}$ with literals, i.e.\ with propositional variables and their negation.
    We can interpret variables and their negation in the usual way by introducing a labelling of each process in an LTS with a set of propositional variables (see \cite{HalpernM92} for instance).   
    We observe that Ladner's reduction in the proof for the \pspace-hardness of \textbf{K}-satisfiability from \cite{ladner1977computational} constructs a one-action formula in $\mathL_{\Box\Diamond x}$, and therefore $\mathL_{\Box\Diamond x}$-satisfiability is \pspace-hard, even with only one action.
 
    We now give a reduction from $\mathL_{\Box\Diamond x}$-satisfiability to $\mathL_{\Box\Diamond}$-satisfiability by encoding literals with formulae that have no box modalities.
    Let $\mathL_{[a]\langle a \rangle^d x}^k$ be the fragment of $\mathL_{\Box\Diamond x}$ that includes the formulae of modal depth up to $d$ that use only action $a$ and $k$ propositional variables, $x_1,x_2,\ldots ,x_k$. 
    We write the negation of a variable $x$ as $\Bar{x}$. Let $b \neq a$ be an action.
    We now describe how to encode each $x_i$ and $\Bar{x_i}$.
    For each $0 \leq i \leq k$ and $0\leq j <  \lceil\log k \rceil$, let $\alpha(i,j) = a$, if position $j$ in  the binary representation of $i$ (using $\lceil\log k \rceil$ bits) has bit $1$, and $\alpha(i,j) = b$ otherwise.
    Let $e (x_i)  = \langle b \rangle\langle \alpha(i,0) \rangle\langle \alpha(i,1) \rangle\cdots \langle \alpha(i,\lceil\log k \rceil) \rangle\langle a \rangle \true$, and 
    $e (\Bar{x_i})  = \langle b \rangle\langle \alpha(i,0) \rangle\langle \alpha(i,1) \rangle\cdots \langle \alpha(i,\lceil\log k \rceil) \rangle\langle b \rangle \true$.
    The negations of those formulae are defined as
    expected thus: 
    \begin{eqnarray*}
     \neg e( x_i ) & = & [ b ][ \alpha(i,0) ][ \alpha(i,1) ]\cdots [ \alpha(i,\lceil\log k \rceil) ] [a]  \ff   , \text{ and } \\
     \neg e( \Bar{x_i} ) & = & [ b ][ \alpha(i,0) ][ \alpha(i,1) ]\cdots [ \alpha(i,\lceil\log k \rceil) ] [b]  \ff . 
     \end{eqnarray*}
    The formula $e(x_i)$ asserts that a process has a trace of the form $bt_ia$, where $t_i$ encodes $i$ in binary, using $a$ to stand for $1$ and $b$ for $0$. Similarly, $e(\Bar{x_i})$ asserts that a process has a trace of the form $bt_ib$. 
    Notice that $t_i \neq t_j$, for $i \neq j$.  Therefore, every conjunction that can be formed from formulae of the above-mentioned types is satisfiable, unless it contains a formula of the form $e(x_i)$ or $e(\Bar{x_i})$ and its negation. 
    
    For each $d,k\geq 0$, 
    let $\varphi_d^k = \bigwedge_{j=0}^d [a]^j \bigwedge_{i=1}^k \left(
    (e(x_i) \land \neg e(\Bar{x_i}))
    \lor 
    (\neg e(x_i) \land  e(\Bar{x_i}))
    \right)$. The formula $\varphi_d^k$ asserts that for every process that can be reached via sequences of $a$-transitions of length up to $d$, for each $1 \leq i \leq k$, exactly one of $e(x_i)$ and $e(\Bar{x_i})$ must be true.

    For each formula $\varphi \in \mathL_{[a]\langle a \rangle x}$ of modal depth $d$ and on variables $\{x_1,x_2,\ldots, x_k\}$, let $\varphi_{-x} = \varphi' \land \varphi_d^k$, where $\varphi'$ is the result of replacing each positive occurrence of $x_i$ in $\varphi$ with $e(x_i)$ and each occurrence of $\Bar{x_i}$ in $\varphi$ with $e(\Bar{x_i})$.
    For each $1 \leq i \leq k$, let $p_i$ be the process that only has $t_i a$ and its prefixes as traces; and let $\neg p_i$ be the process that only has $t_i b$ and its prefixes as traces.
    For each labelled LTS $\mathS$ with only $a$-transitions, we  can define the LTS $\mathS_e$
    that additionally includes the processes $p_i$ and $\neg p_i$ and for every $p$ in $\mathS$ and $1 \leq i \leq k$, it includes an $a$-transition to $p_i$, if $x_i$ is in the labelling of $p$, and an $a$-transition to $\neg p_i$, otherwise.
    It is easy to see that for every $p$ in $\mathS$, $p$ satisfies $\varphi_d^k$; also by straightforward induction on $\varphi$, $p$ satisfies $\varphi$ in $\mathS$ if and only if 
    $p$ satisfies $\varphi'$ in $\mathS_e$.
    Therefore, if $\varphi$ is $\mathL_{\Box\Diamond x}$-satisfiable, then $\varphi_{-x}$ is $\mathL_{\Box\Diamond }$-satisfiable.

    Let $\mathS$ be an LTS; we define $\mathS_x$ to be the labelled LTS that results from labelling each $p$ in $\mathS$ with $\{ x_i \mid b t_i a \in \mathrm{traces}(p) \}$.
    It is then not hard to use induction on $\varphi$ to prove that for every process $p$ in $\mathL$, if $p$ satisfies $\varphi_{-x}$ in $\mathL$, then $p$ satisfies $\varphi$ in $\mathL_x$.
    Therefore, if $\varphi_{-x}$ is $\mathL_{\Box\Diamond }$-satisfiable, then $\varphi$ is $\mathL_{\Box\Diamond x}$-satisfiable.
\end{proof}

\begin{cor}\label{prop:3-s-sat}
    Let $|\act|\geq 2$. Satisfiability of formulae in $\mathL_{3S}$ is \pspace-complete.
\end{cor}
\begin{proof}
    \pspace-hardness is a corrolary of Theorem~\ref{thm:2s-validity} and the inclusion $\compL_{2S}\subseteq\mathL_{3S}$. The problem belongs to \pspace because of Proposition~\ref{hml-sat} and $\mathL_{3S}\subseteq\hml$.
\end{proof}

\begin{cor}\label{prop:n-s-sat}
     Let $|\act|\geq 2$. Satisfiability of formulae in $\mathL_{nS}$ for $n\geq 3$ is \pspace-complete.
\end{cor}

As a corollary of Theorem~\ref{thm:2s-validity}, we also obtain the following result, which will be used later to prove $\pspace$-hardness of the formula primality problem for $\mathL_{nS}$, $n \geq 3$.

\begin{cor}\label{cor:2s-validity}
    Let $|\act|\geq 2$. Validity for $\mathL_{2S}$ is \pspace-complete.
\end{cor}

\section{The complexity of formula primality}\label{section:deciding-primality}

 In this section, we determine the complexity of the formula primality problem for the logics $\mathL_S$, $\mathL_{CS}$, $\mathL_{RS}$, and $\mathL_{TS}$. Combined with the results for the logics characterizing the $n$-nested simulation preorders ($n\geq 2$), we offer in Section~\ref{Sect:primality-nested-semantics}, these contributions yield the complexity bounds shown in Table~\ref{table:prime}. 
 %characterizing the branching-time semantics in van Glabbeek's spectrum and prove the results summarized in Table~\ref{table:prime}. 
 
\subsection{The formula primality problem for \texorpdfstring{$\mathL_S$}{LS}.} \label{subsection:primality-simulation}

Our order of business in this section is to prove that the formula primality problem for $\mathL_S$ is decidable in polynomial time. We shall do so by providing a polynomial-time reduction from that problem to the reachability problem in alternating graphs (see Definitions~\ref{def:alternating-graph} and~\ref{def:alternating_reachability}). As key intermediate steps in our technical developments, we will offer necessary and sufficient conditions for primality for arbitrary formulae in $\mathL_S$ (Proposition~\ref{prop:primality-LS}) and for formulae that do not contain $\ff$ (see Proposition~\ref{prop:primality-LS-2} and Corollary~\ref{cor:primality-LS-3}). 

We first provide some simple lemmas. 

\begin{lem}\label{lem:conjunction_lemma_simulation}
 For every $\varphi_1,\varphi_2,\psi\in \mathL_{S}$, $\varphi_1\wedge\varphi_2\models \langle a\rangle \psi$ iff $\varphi_1\models \langle a\rangle \psi   \text{ or }\varphi_2\models \langle a\rangle \psi$.
\end{lem}
\begin{proof}
We prove the two implications separately. 

\noindent 
($\Leftarrow$) If $\varphi_1\models \langle a\rangle \psi   \text{ or }\varphi_2\models \langle a\rangle \psi$, then $\varphi_1\wedge\varphi_2\models \langle a\rangle \psi$ follows immediately. In fact, this implication holds for all {\hml} formulae $\varphi_1,\varphi_2,\psi$.

\noindent 
($\Rightarrow$) Assume that $\varphi_1\wedge\varphi_2\models \langle a\rangle \psi$ and, without loss of generality, that $\varphi_1\not\models \langle a\rangle \psi$. We  show that $\varphi_2\models \langle a\rangle \psi$. Since  $\varphi_1\not\models \langle a\rangle \psi$, there is some $p_1$ such that $p_1\models\varphi_1$ and $p_1\not\models \langle a\rangle \psi$. If $\varphi_2$ is not satisfiable, then $\varphi_2\models\langle a\rangle \psi$ trivially holds. Assume now that $\varphi_2$ is satisfiable and let $p_2\models\varphi_2$. Observe that 
$p_1 \curle_S p_1+p_2$ and $p_2 \curle_S p_1+p_2$.  Then, $p_1+p_2\models \varphi_1\wedge\varphi_2$ by Proposition~\ref{logical_characterizations}, and therefore $p_1+p_2\models\langle a\rangle \psi$. This means that either $p_1\myarrowa p'$ or $p_2\myarrowa p'$ for some $p'$ such that $p'\models \psi$. Since $p_1\not\models \langle a\rangle \psi$, we have that $p_2\myarrowa p'$, and so $p_2\models \langle a\rangle \psi$, which was to be shown.
\end{proof}

Lemma~\ref{lem:prime-no-ff-disj} below states that all formulae in $\mathL_S$ that do not contain $\ff$ and disjunctions are prime. To show that result, we first associate a process $p_\varphi$ to a formula $\varphi$ of that form and prove that $\varphi$ is characteristic within $\mathL_S$ for $p_\varphi$.

\begin{defi}\label{def:simulation-associated-process}
   Let $\varphi\in\mathL_{S}$ be given by the grammar $\varphi::=\true ~|~ \langle a \rangle \varphi ~|~ \varphi\wedge\varphi$. We define process $p_\varphi$ inductively as follows.
\begin{itemize}
    \item If $\varphi=\true$, then $p_\varphi=\mathtt{0}$.
     \item If $\varphi=\langle a\rangle \varphi'$, then $p_\varphi=a. p_{\varphi'}$.
     \item If $\varphi=\varphi_1\wedge \varphi_2$, then $p_\varphi=p_{\varphi_1}+p_{\varphi_2}$.
\end{itemize}
\end{defi}

\begin{lem}\label{lem:prime-no-ff-disj}
Let $\varphi$ be a formula given by the  grammar $\varphi::=\true~|~ \langle a\rangle \varphi ~|~ \varphi\wedge\varphi$. Then, $\varphi$ is prime. In particular, $\varphi$ is characteristic within $\mathL_S$ for $p_\varphi$. 
\end{lem}
\begin{proof}  We prove that $\varphi$ is characteristic within $\mathL_S$ for $p_\varphi$. Then, from Proposition~\ref{prop:charact-via-primality}, $\varphi$ is also prime.
Let $p\models\varphi$.  From Remark~\ref{Remark:charforms}, it suffices to show that $p_\varphi\curle_S p$. We show this claim by induction on the structure of $\varphi$.
\begin{description}
    \item[Case $\varphi=\true$] The claim follows 
    %It holds $p_\varphi=\mathtt{0}
    %\curle_S p$, 
    since $\mathtt{0}\curle_S p$ holds for every process $p$.
    \item[Case $\varphi=\langle a\rangle \varphi'$] As $p\models \langle a\rangle\varphi'$, there is some transition $p\myarrowa p'$ such that $p'\models\varphi'$. By the inductive hypothesis, $p_{\varphi'}\curle_S p'$. Thus, $p_\varphi = a.p_{\varphi'}\curle_S p$. 
    \item[Case $\varphi=\varphi_1\wedge \varphi_2$] Since $p\models\varphi_1\wedge\varphi_2$, we have that $p\models\varphi_1$ and $p\models\varphi_2$. By the inductive hypothesis, $p_{\varphi_1}\curle_S p$ and $p_{\varphi_2}\curle_S p$. As noted in Remark~\ref{rem:preorders-preserved-under-plus}, $\curle_S$ is preserved by the  $+$ operation. Therefore,  $p_{\varphi_1}+p_{\varphi_2}\curle_S p+p\sim p$. Since the  $+$ operation is idempotent, $p_\varphi\curle_S p$ follows. 
\end{description}
The converse implication, namely that $p_\varphi\curle_{S} p$ implies $p\models\varphi$, can be easily shown following similar lines by induction on the structure of $\varphi$.
\end{proof}

The next  statement generalization of of the above result follows immediately from Lemma~\ref{lem:prime-no-ff-disj}, the fact that unsatisfiable formulae are prime, and the observation that each $\mathL_S$-formula containing $\ff$ but no occurrence of $\lor$ is not satisfiable.

\begin{cor}\label{cor:prime-no-disj}
   Let $\varphi\in\mathL_S$ be a formula given by the  grammar $\varphi::=\true~|~ \ff~|~ \langle a\rangle \varphi ~|~ \varphi\wedge\varphi$. Then, $\varphi$ is prime. 
\end{cor}

Proposition~\ref{prop:primality-LS} provides a necessary and sufficient condition for primality in $\mathL_S$. 

\begin{prop}\label{prop:primality-LS}
Let $\varphi\in \mathL_{S}$ and $\bigvee_{i=1}^k \varphi_i$ be $\varphi$ in DNF. Then, $\varphi$ is prime iff $\varphi\models\varphi_j$ for some $1\leq j\leq k$.
\end{prop}
 \begin{proof} We prove the two implications separately. 

\noindent 
($\Rightarrow$) Assume that $\varphi$ is prime. From Lemma~\ref{lem:DNF-equiv}, $\varphi\equiv\bigvee_{i=1}^k \varphi_i$. By the definition of primality, $\varphi\models\varphi_j$, for some $1\leq j\leq k$. \\
($\Leftarrow$) From Lemma~\ref{lem:DNF-equiv}, it suffices to  show the claim for $\bigvee_{i=1}^k\varphi_i$. Assume that $\bigvee_{i=1}^k\varphi_i\models \varphi_j$, for some $1\leq j\leq k$.  Let $\bigvee_{i=1}^k\varphi_i\models \bigvee_{l=1}^m \phi_l$. From Lemma~\ref{lem:disjunction_lemma}, $\varphi_i\models\bigvee_{l=1}^m \phi_l$, for every $1\leq i\leq k$, and, in particular, $\varphi_j\models\bigvee_{l=1}^m \phi_l$. Since $\varphi_j$ does not contain disjunctions,  Corollary~\ref{cor:prime-no-disj} guarantees that it is prime. Thus, $\varphi_j\models \phi_s$, for some $1\leq s\leq m$, and since $\bigvee_{i=1}^k\varphi_i\models \varphi_j$, it holds that $\bigvee_{i=1}^k\varphi_i\models \phi_s$. 
\end{proof}

To give necessary and sufficient conditions for the primality of formulae in $\mathL_S$ without $\ff$, we present some 
%examine and state
results regarding the DNF of such formulae.

\begin{lem}\label{lem:disjuncts-of-DNF-no-ff}
    Let $\varphi$ be given by the grammar $\varphi::= \true ~|~ \langle a\rangle \varphi ~|~ \varphi\wedge\varphi  ~|~ \varphi\vee\varphi$; let also $\bigvee_{i=1}^k \varphi_i$ be $\varphi$ in DNF. Then, for every $1\leq i\leq k$, $\varphi_i$ is characteristic for $p_{\varphi_i}$.
\end{lem}
\begin{proof}
    Every $\varphi_i$ does not contain disjunctions and so it is characteristic for $p_{\varphi_i}$ from Lemma~\ref{lem:prime-no-ff-disj}.
\end{proof}

\begin{lem}\label{lem:common-divisor-pairs}
    Let $\varphi$ be given by the grammar $\varphi::= \true ~|~ \langle a\rangle \varphi ~|~ \varphi\wedge\varphi  ~|~ \varphi\vee\varphi$;
    let also $\bigvee_{i=1}^k \varphi_i$ be $\varphi$ in DNF. If for every pair $p_{\varphi_i},p_{\varphi_j}$, $1\leq i,j\leq k$, there is some process $q_{ij}$ such that $q_{ij}\curle_S p_{\varphi_i}$, $q_{ij}\curle_S p_{\varphi_j}$, and $q_{ij}\models\varphi$, then there is some process $q$ such that $q\curle_S p_{\varphi_i}$ for every $1\leq i\leq k$, and $q\models\varphi$. 
\end{lem}
\begin{proof}  We prove that, under the assumptions of the lemma, for all processes $p_{\varphi_{i_1}}, \dots,p_{\varphi_{i_m}}$, with $2\leq m\leq k$, there is some process $q$ such that $q\curle_S p_{\varphi_{i_1}},\dots, q\curle_S p_{\varphi_{i_m}}$ and $q\models\varphi$. The proof is by strong induction on $m$.
\begin{description}
    \item[Base case.] Let $m=2$. This is true from the hypothesis of the lemma.
    \item[Inductive step.] Assume that the claim is true for every $m\leq n-1$. We show that it is true for $m=n$. Let $p_{\varphi_{i_1}}, \dots,p_{\varphi_{i_n}}$ be processes associated to the disjunction-free formulae $\varphi_{i_1}, \dots,\varphi_{i_n}$, respectively. Consider the pairs 
    \[
(p_{\varphi_{i_1}},p_{\varphi_{i_2}}), (p_{\varphi_{i_3}},p_{\varphi_{i_4}}), \ldots, (p_{\varphi_{i_{n-1}}},p_{\varphi_{i_n}}) , 
\]
when $n$ is even, and the pairs 
\[
(p_{\varphi_{i_1}},p_{\varphi_{i_2}}), (p_{\varphi_{i_3}},p_{\varphi_{i_4}}), \ldots, (p_{\varphi_{i_{n-2}}},p_{\varphi_{i_{n-1}}}) , 
(p_{\varphi_{i_{n}}},p_{\varphi_{i_{n}}}) , 
\]
when $n$ is odd. In what follows, we consider only the case that $n$ is even since the case that $n$ is odd is similar.  By the assumptions of the lemma, there are processes $q_1,\dots, q_{n/2}$ such that $q_1\curle_S p_{\varphi_{i_1}}, p_{\varphi_{i_2}}$, $q_2\curle_S p_{\varphi_{i_3}}, p_{\varphi_{i_4}}$, $\dots$, $q_{n/2}\curle_S p_{\varphi_{i_{n-1}}},p_{\varphi_{i_n}}$, and $q_i\models\varphi$ for every $1\leq i\leq n/2$. Thus, for every $1\leq i\leq n/2$, there is some  $1\leq j_i\leq k$ such that $q_i\models \varphi_{j_i}$. From Lemma~\ref{lem:disjuncts-of-DNF-no-ff}, every $\varphi_{j_i}$ is characteristic for $p_{\varphi_{j_i}}$ and from Remark~\ref{Remark:charforms}, $p_{\varphi_{j_i}}\curle_S q_i$ for every $1\leq i\leq n/2$ and $1\leq j_i\leq k$. By the assumptions of the lemma and the inductive hypothesis, there is some process $q$ such that $q\curle_S p_{\varphi_{j_1}},\dots,p_{\varphi_{j_{n/2}}}$ and $q\models\varphi$. By transitivity of $\curle_S$, $q\curle_S p_{\varphi_{i_1}}, \dots,p_{\varphi_{i_n}}$, and we are done.  \qedhere
\end{description}
\end{proof}

\begin{rem}
    Note that if we consider processes $p_1,\dots, p_k$ and, for every pair $p_i$ and $p_j$, there is some $q\neq\mathtt{0}$ such that $q\curle_S p_i$ and $q\curle_S p_j$, then it may be the case that there is no $q\neq\mathtt{0}$ such that $q\curle_S p_i$ for every $1\leq i\leq k$. For example, this is true for processes $p_1$, $p_2$, and $p_3$, where $p_1\myarrowa \mathtt{0}$, $p_1\myarrowb \mathtt{0}$, $p_1\myarrowc \mathtt{0}$, $p_2\myarrowc \mathtt{0}$, $p_2\myarrowd \mathtt{0}$, $p_2\myarrowe \mathtt{0}$, $p_3\myarrowe \mathtt{0}$, $p_3\myarrowf \mathtt{0}$, and $p_3\myarrowa \mathtt{0}$.
\end{rem}

\begin{cor}\label{cor:common-divisor-pairs}
    Let $\varphi$ be given by the grammar $\varphi::= \true ~|~ \langle a\rangle \varphi ~|~ \varphi\wedge\varphi  ~|~ \varphi\vee\varphi$;
    let also $\bigvee_{i=1}^k \varphi_i$ be $\varphi$ in DNF. If for every pair $p_{\varphi_i},p_{\varphi_j}$, $1\leq i,j\leq k$, there is some process $q$ such that $q\curle_S p_{\varphi_i}$, $q\curle_S p_{\varphi_j}$, and $q\models\varphi$, then there is some $1\leq m\leq k$, such that $p_{\varphi_m}\curle_S p_{\varphi_i}$ for every $1\leq i\leq k$. 
\end{cor}
\begin{proof}
    From Lemma~\ref{lem:common-divisor-pairs}, there is $q$ such that $q\curle_S p_{\varphi_i}$ for every $1\leq i\leq k$, and $q\models\varphi$. Consequently, $q\models\varphi_m$ for some $1\leq m\leq k$, so $p_{\varphi_m}\curle_S q$. By transitivity of $\curle_S$, $p_{\varphi_m}\curle_S p_{\varphi_i}$ for every $1\leq i\leq k$. Since, $p_{\varphi_m}\models\varphi_m$, we have that $p_{\varphi_m}\models\varphi$ as well.
\end{proof}

Proposition~\ref{prop:primality-LS-2} provides a necessary and sufficient condition for primality in $\mathL_S$ for formulae that do not contain $\ff$.

\begin{prop}\label{prop:primality-LS-2}
    Let $\varphi$ be given by the grammar $\varphi::= \true ~|~ \langle a\rangle \varphi ~|~ \varphi\wedge\varphi  ~|~ \varphi\vee\varphi$;
    let also $\bigvee_{i=1}^k \varphi_i$ be $\varphi$ in DNF. Then, $\varphi$ is prime iff for every pair $p_{\varphi_i},p_{\varphi_j}$, $1\leq i,j\leq k$, there is some process $q$ such that $q\curle_S p_{\varphi_i}$, $q\curle_S p_{\varphi_j}$, and $q\models \varphi$.
\end{prop}
\begin{proof}
We prove the two implications separately. 

\noindent 
    $(\Leftarrow)$ From Corollary~\ref{cor:common-divisor-pairs}, there is some $1\leq m\leq k$, such that $p_{\varphi_m}\curle_S p_{\varphi_i}$ for every $1\leq i\leq k$. From the fact that $\varphi_i$ is characteristic for $p_{\varphi_i}$ and Remark~\ref{Remark:charforms}, $\varphi_i\models \varphi_m$ for every $1\leq i\leq m$. From Proposition~\ref{prop:primality-LS}, $\varphi$ is prime. 
    % From Lemma~\ref{lem:common-divisor-pairs}, there is a process $q$ such that $q\curle_S p_{\varphi_i}$ for every $1\leq i\leq k$, and $q\models \varphi$. So $q\models \varphi_j$ for some $1\leq j\leq k$. From Lemma~\ref{lem:prime-no-ff-disj} and Corollary~\ref{cor:characteristic}, $p_{\varphi_j}\curle_S q$. By transitivity of $\curle_S$, $p_{\varphi_j}\curle_S p_{\varphi_i}$ for every $1\leq i\leq k$. From Proposition~\ref{logical_characterizations}, $\mathL_S(p_{\varphi_j})\subseteq \mathL_S(p_{\varphi_i})$, and so $p_{\varphi_i}\models\varphi_j$ for every $1\leq i\leq k$. Let $p\models \varphi_i$, then from Definition~\ref{def:characteristic}, $\mathL_S(p_{\varphi_i})\subseteq \mathL_S(p)$, and so $p\models\varphi_j$. This means that for every $1\leq i\leq k$, $\varphi_i\models\varphi_j$. From Proposition~\ref{prop:primality-LS}, $\varphi$ is prime.

    \noindent
    $(\Rightarrow)$ Let $\varphi$ be prime. From Lemma~\ref{lem:prime-no-ff-disj}, there is some $1\leq m
    \leq k$ such that $\varphi\models\varphi_m$. From Lemmas~\ref{lem:DNF-equiv} and~\ref{lem:disjunction_lemma}, $\varphi_i\models\varphi_m$ for all $1\leq i\leq k$. From Remark~\ref{Remark:charforms}, $p_{\varphi_m}\curle_S p_{\varphi_i}$ for every $1\leq i\leq k$. As a result, for every pair $p_{\varphi_i},p_{\varphi_j}$, it holds that $p_{\varphi_m}\curle_S p_{\varphi_i}$, $p_{\varphi_m}\curle_S p_{\varphi_j}$, and $p_{\varphi_m}\models\varphi_m$, which implies that $p_{\varphi_m}\models\varphi$.
    % From Lemma~\ref{lem:disjuncts-of-DNF-no-ff}, $\varphi_i$ is characteristic for $p_{\varphi_i}$. So, for every $1\leq i\leq k$, $p_{\varphi_i}\models\varphi_i$. Since $\varphi_i\models\varphi_m$, we have that $p_{\varphi_i}\models\varphi_m$, and from Corollary~\ref{cor:characteristic}, $p_{\varphi_m}\curle_S p_{\varphi_i}$. As a result, for every pair $p_{\varphi_i},p_{\varphi_j}$, it holds that $p_{\varphi_m}\curle_S p_{\varphi_i}$, $p_{\varphi_m}\curle_S p_{\varphi_j}$, and $p_{\varphi_m}\models\varphi_m$, which implies that $p_{\varphi_m}\models\varphi$.
\end{proof}

\begin{cor}\label{cor:primality-LS-3}
    Let $\varphi$ be given by the grammar $\varphi::= \true ~|~ \langle a\rangle \varphi ~|~ \varphi\wedge\varphi  ~|~ \varphi\vee\varphi$;
    let also $\bigvee_{i=1}^k \varphi_i$ be $\varphi$ in DNF. Then, $\varphi$ is prime iff for every pair $\varphi_i$, $\varphi_j$ there is some $1\leq m \leq k$ such that $\varphi_i\models \varphi_m$ and $\varphi_j\models\varphi_m$.
\end{cor}
\begin{proof}
We prove the two implications separately. 

\noindent
  $(\Rightarrow)$ From Proposition~\ref{prop:primality-LS}, if $\varphi$ is prime, then there is some $1\leq m \leq k$ such that $\varphi_i\models \varphi_m$ for every $1\leq i \leq k$. 
  
  \noindent 
  $(\Leftarrow)$ If for every $\varphi_i$, $\varphi_j$ there is some $1\leq m \leq k$ such that $\varphi_i\models \varphi_m$ and $\varphi_j\models\varphi_m$, then from the fact that $\varphi_i,\varphi_j$ are characteristic for $p_{\varphi_i}$ and $p_{\varphi_j}$, and Remark~\ref{Remark:charforms}, $p_{\varphi_m}\curle_S p_{\varphi_i}$ and $p_{\varphi_m}\curle_S p_{\varphi_j}$. It also holds that  $p_{\varphi_m}\models\varphi$, so from Proposition~\ref{prop:primality-LS-2}, $\varphi$ is prime.
\end{proof}

%  In other words, we are interested in the existence of an alternating path from the source to the target.

% \end{defi}\label{def:alternating-path}
%     Let $G=(V,E,A,s,t)$ be an alternating graph. An alternating path $P_a=(V_a,E_a)$ from $s$ to $t$ is a DAG such that:
%     \begin{enumerate}
%         \item $s\in V_a$.
%         \item If $x\in V_a$, $x\neq t$, and $x$ is existential, then there is at least one $(x,z)\in E$ such that $z\in V_a$, $(x,z)\in E_a$ and $P^G(z,t)$.
%         \item If $x\in V_a$, $x\neq t$, and $x$ is universal, then for every $(x,z)\in E$, it holds that $z\in V_a$, $(x,z)\in E_a$, and $P^G(z,t)$.
%         \item If $x\in V_a$ and $x=t$, then there is no $(x,z)\in E_a$.
%     \end{enumerate}
% \end{defi}
%
% \begin{cor}\label{cor:alternating-reachability}
% Let $G=(V,E,A,s,t)$ be an alternating graph. $G\in\reacha$ iff $P^G(s,t)$ iff there is an alternating path from $s$ to $t$.
% \end{cor}

\iffalse
To prove correctness of algorithm $\algos$, we need the following lemmas.
\fi

We are now ready to prove that the formula primality problem for $\mathL_S$ is decidable in polynomial time. 

\begin{prop}\label{prop:sim-primality}
    Let $\varphi\in\mathL_S$ such that $\ff\not\in\sub(\varphi)$. Deciding whether $\varphi$ is prime is in \cP.
\end{prop}
\begin{proof}
We describe algorithm $\algos$ that, on input  $\varphi$, decides primality of $\varphi$. 
%, prove its correctness and analyze its complexity.
$\algos$ constructs a rooted directed acyclic graph, denoted by $G_\varphi$, from the formula $\varphi$ as follows. Every vertex of the graph is either of the form $\varphi_1,\varphi_2\Rightarrow\psi$---where $\varphi_1$, $\varphi_2$ and $\psi$ are sub-formulae of $\varphi$---, or \textsc{True}. The algorithm starts from vertex $x=(\varphi,\varphi\Rightarrow\varphi)$ and applies some rule in Table~\ref{tab:S-rules} to $x$ in top-down fashion to generate one or two new vertices that are given at the bottom of the rule. These vertices are the children of $x$ and the vertex $x$ is labelled with either $\exists$ or $\forall$, depending on which one is displayed at the bottom of the applied rule. If $x$ has only one child, \algos labels it with $\exists$. The algorithm recursively continues this procedure on the children of $x$. If no rule can be applied on a vertex, then this vertex has no outgoing edges. For the sake of clarity and consistency, we assume that right rules, i.e.\ (R$\vee$) and (R$\wedge$),  are applied before the left ones, i.e.\ (L$\vee_i$) and (L$\wedge_i$), $i=1,2$, by the algorithm.
The graph generated in this way 
is an \emph{alternating graph}, as defined by Immerman in~\cite[Definition 3.24]{Immerman99} (see Definition~\ref{def:alternating-graph}). In $G_\varphi$,  the source vertex $s$ is  $\varphi,\varphi\Rightarrow\varphi$, and the target vertex $t$ is \textsc{True}. Algorithm \algos solves the {\reacha} problem on input $G_\varphi$, where \reacha is the \reachability problem on alternating graphs (see  Definition~\ref{def:alternating_reachability}).
It accepts $\varphi$ iff \reacha accepts $G_\varphi$. 
 Since the size of $G_\varphi$ is $\mathcal{O}(|\varphi|^3)$ (see Lemma~\ref{lem:graph-size} to follow) and  \reacha can be solved in linear time~\cite[Algorithm~3.25]{Immerman99}, $\algos$ runs  in $\mathcal{O}(|\varphi|^3)$. 
The correctness of algorithm $\algos$ is shown in Lemmas~\ref{lem:sim-algo-correct-1} and~\ref{lem:sim-algo-correct-2} to follow.
\end{proof}

\begin{cor}\label{cor:primality-in-S}
    The formula primality problem for $\mathL_S$ is in \cP.
\end{cor}
\begin{proof}
    Given a formula $\varphi\in\mathL_S$ it can be checked whether it is satisfiable in polynomial time from Corollary~\ref{cor:sat-s-cs-rs-poly}(a). If $\varphi$ is unsatisfiable, then it is prime. In case $\varphi$ is satisfiable there is a polynomial-time algorithm that returns $\varphi'$ such that (a) $\varphi\equiv\varphi'$, and (b)  $\ff\not\in\sub(\varphi')$: the algorithm just repeatedly applies the rules $\langle a\rangle\ff\ruleff \ff$, $\ff\vee\psi\ruleff\psi$, $\psi\vee\ff\ruleff\psi$, $\ff\wedge\psi\ruleff \ff$, and $\psi\wedge\ff\ruleff \ff$ on $\varphi$ until no rule can be applied, and returns the resulting formula.  We can decide whether $\varphi'$ is prime because of Proposition~\ref{prop:sim-primality}.
\end{proof}

We now provide the proofs of the lemmas that were used in the proof of Proposition~\ref{prop:sim-primality} to establish the correctness of algorithm $\algos$ and its polynomial-time complexity.

\begin{lem}
    Let $\varphi\in\mathL_S$ such that $\ff\not\in\sub(\varphi)$. For every formula $\psi$ that appears in vertices of $G_\varphi$, it holds that $\ff\not\in\sub(\psi)$.
\end{lem}

\begin{table}
\fbox{\begin{minipage}{0.45 \textwidth}
\begin{prooftree}
    \AxiomC{$\varphi_1\vee\varphi_2,\varphi\Rightarrow \psi$ }
    \RightLabel{\scriptsize(L$\vee_1$)}
    \UnaryInfC{$\varphi_1,\varphi\Rightarrow \psi~|_{\forall}~ \varphi_2,\varphi\Rightarrow \psi$}
\end{prooftree}
\vspace{2mm}
\begin{prooftree}
    \AxiomC{$\varphi_1\wedge\varphi_2,\varphi\Rightarrow \langle a\rangle\psi$}
    \RightLabel{\scriptsize(L$\wedge_1$)}
    \UnaryInfC{$\varphi_1,\varphi\Rightarrow \langle a\rangle \psi~|_{\exists}~ \varphi_2,\varphi\Rightarrow \langle a\rangle \psi$}
\end{prooftree}
\vspace{2mm}
\begin{prooftree}
    \AxiomC{$\varphi_1,\varphi_2\Rightarrow\psi_1\wedge \psi_2$ }
    \RightLabel{\scriptsize(R$\wedge$)}
    \UnaryInfC{$\varphi_1,\varphi_2\Rightarrow \psi_1~|_{\forall}~\varphi_1,\varphi_2\Rightarrow \psi_2$}
\end{prooftree}
\vspace{2mm}
\begin{prooftree}
    \AxiomC{$\langle a\rangle \varphi_1, \langle a\rangle \varphi_2 \Rightarrow \langle a\rangle \psi$ }
    \RightLabel{\scriptsize($\diamond$)}
    \UnaryInfC{$\varphi_1,\varphi_2\Rightarrow \psi$}
\end{prooftree}
\end{minipage}
\hfill
\begin{minipage}{0.45\textwidth}
\begin{prooftree}
    \AxiomC{$\varphi,\varphi_1\vee\varphi_2\Rightarrow \psi$ }
    \RightLabel{\scriptsize(L$\vee_2$)}
    \UnaryInfC{$\varphi_1,\varphi\Rightarrow \psi~|_{\forall}~ \varphi_2,\varphi\Rightarrow \psi$}
\end{prooftree}
\vspace{2mm}
\begin{prooftree}
    \AxiomC{$\varphi,\varphi_1\wedge\varphi_2\Rightarrow \langle a\rangle\psi$}
    \RightLabel{\scriptsize(L$\wedge_2$)}
    \UnaryInfC{$\varphi_1,\varphi\Rightarrow \langle a\rangle \psi~|_{\exists}~ \varphi_2,\varphi\Rightarrow \langle a\rangle \psi$}
\end{prooftree}
\vspace{2mm}
\begin{prooftree}
    \AxiomC{$\varphi_1,\varphi_2\Rightarrow\psi_1\vee \psi_2$ }
    \RightLabel{\scriptsize(R$\vee$)}
    \UnaryInfC{$\varphi_1,\varphi_2\Rightarrow \psi_1~|_{\exists}~\varphi_1,\varphi_2\Rightarrow \psi_2$}
\end{prooftree}
\vspace{2mm}
\begin{prooftree}
    \AxiomC{$\varphi_1,\varphi_2\Rightarrow\true$ }
    \RightLabel{\scriptsize(tt)}
    \UnaryInfC{\textsc{True}}
\end{prooftree}
\end{minipage}}   
\vspace{2mm}
    \caption{Rules for the simulation preorder. If $|_{\forall}$ is displayed  in the conclusion of a rule, then the rule is called universal. Otherwise, it is called existential.}
    \label{tab:S-rules}
\end{table}

\begin{lem}\label{lem:sim-algo-correct-1}
Let $\varphi\in\mathL_S$ such that $\ff\not\in\sub(\varphi)$. If there is an alternating path in $G_\varphi$ from $(\varphi,\varphi\Rightarrow\varphi)$ to \textsc{True}, then $\varphi$ is prime.
\end{lem}
\begin{proof}
Let $\varphi_1,\varphi_2\Rightarrow \psi$ be a vertex in an alternating path from $(\varphi,\varphi\Rightarrow\varphi)$ to \textsc{True} and $\bigvee_{i=1}^{k_1}\varphi_{1}^i$, $\bigvee_{i=1}^{k_2}\varphi_2^i$, and $\bigvee_{i=1}^{k_3}\psi^{i}$ be $\varphi_1$, $\varphi_2$, and $\psi$ in DNF, respectively. We show that $\varphi_1,\varphi_2 \text{ and } \psi$ satisfy the following property $P_1$: 
\begin{quote}
    `For every $\varphi_1^i,\varphi_2^j$ there is $\psi^k$ such that $\varphi_1^i\models\psi^k$ and $\varphi_2^j\models\psi^k$.'
\end{quote}
Edges of $G_\varphi$ correspond to the application of some rule of Table~\ref{tab:S-rules}. In an alternating path from $(\varphi,\varphi\Rightarrow\varphi)$ to \textsc{True}, the edges connect the top vertex of a rule to one or two vertices that correspond to the bottom of the same rule: if a rule is universal, the top vertex connects to two children. Otherwise, it connects to one child. We show that $P_1$ is true for every vertex in the alternating path from $s$ to $t$ by induction on the type of rules, read from their conclusions to their premise. 
\begin{description}
    \item[Case (tt)] $P_1$ is trivial for $\varphi_1$, $\varphi_2$, and $\true$, since for every $\varphi_1^i$, $\varphi_2^j$, it holds that $\varphi_1^i\models\true$ and $\varphi_2^j\models\true$.
    \item[Case (L$\vee_1$)] Since rule (L$\vee_1$) is universal, assume that $P_1$ is true for both $\varphi_1,\varphi,\psi$ and $\varphi_2,\varphi,\psi$. Let $\bigvee_{i=1}^{k_{12}}\varphi_{12}^i$  be $\varphi_1\vee\varphi_2$  in DNF. Then, $\bigvee_{i=1}^{k_{12}}\varphi_{12}^i=\bigvee_{i=1}^{k_1}\varphi_1^i\vee\bigvee_{i=1}^{k_2}\varphi_2^i$. So, $P_1$ holds for $\varphi_1\vee\varphi_2,\varphi,\psi$ as well. Case (L$\vee_2$) is completely analogous.
    \item[Case (L$\wedge_1$)] Since rule (L$\wedge_1$) is existential, assume that $P_1$ is true for either $\varphi_1,\varphi,\langle a\rangle\psi$ or $\varphi_2,\varphi,\langle a\rangle\psi$. Let $\bigvee_{i=1}^{k_{12}}\varphi_{12}^i$ and $\bigvee_{i=1}^k\varphi^i$ be $\varphi_1\wedge\varphi_2$ and $\varphi$ in DNF, respectively. Then, every $\varphi_{12}^i$ is $\varphi_1^j\wedge\varphi_2^k$ for some $1\leq j\leq k_1$ and $1\leq k\leq k_2$. Property $P_1$ for $\varphi_1\wedge\varphi_2,\varphi,\langle a\rangle\psi$ is as follows: for every $\varphi_1^j\wedge\varphi_2^k, \varphi^i$ there is $\langle a\rangle \psi^m$ such that $\varphi_1^j\wedge\varphi_2^k\models\langle a\rangle\psi^m$ and $\varphi^i\models\langle a \rangle\psi^m$. This is true since $\varphi_1^j\wedge\varphi_2^k\models\langle a \rangle\psi^m$ is equivalent to $\varphi_1^j\models\langle a\rangle \psi^m$ or $\varphi_2^k\models\langle a \rangle\psi^m$ from Lemma~\ref{lem:conjunction_lemma_simulation}. Case (L$\wedge_2$) is completely analogous.
    \item[Case (R$\wedge$)] Let $\bigvee_{i=1}^{m_1}\psi_1^i$ and $\bigvee_{i=1}^{m_2}\psi_2^i$, and $\bigvee_{i=1}^{m}\psi_{12}^i$ be $\psi_1$, $\psi_2$, and $\psi_1\wedge\psi_2$ in DNF, respectively. Assume $P_1$ is true for $\varphi_1,\varphi_2,\psi_1$ and $\varphi_1,\varphi_2,\psi_2$, which means that for every  $\varphi_1^i,\varphi_2^j$ there are $\psi_1^{k_1},\psi_2^{k_2}$ such that $\varphi_1^i\models\psi_1^{k_1},\psi_2^{k_2}$ and $\varphi_2^j\models\psi_1^{k_1},\psi_2^{k_2}$. So,  $\varphi_1^i\models\psi_1^{k_1}\wedge\psi_2^{k_2}$ and $\varphi_2^j\models\psi_1^{k_1}\wedge\psi_2^{k_2}$. Since every $\psi_{12}^i$ in the DNF of $\psi_1\wedge\psi_2$ is $\psi_1^j\wedge\psi_2^k$ for some $1\leq j\leq m_1$ and $1\leq k\leq m_2$, $P_1$ is also true for $\varphi_1,\varphi_2,\psi_1\wedge\psi_2$.
    \item[Case (R$\vee$)] Let $\bigvee_{i=1}^{m_1}\psi_1^i$ and $\bigvee_{i=1}^{m_2}\psi_2^i$, and $\bigvee_{i=1}^{m}\psi_{12}^i$ be $\psi_1$, $\psi_2$, and $\psi_1\vee\psi_2$ in DNF, respectively. Assume $P_1$ is true for $\varphi_1,\varphi_2,\psi_1$ or $\varphi_1,\varphi_2,\psi_2$, which means that for every  $\varphi_1^i,\varphi_2^j$ there is $\psi_n^{k}$  such that $\varphi_1^i\models\psi_n^{k}$ and $\varphi_2^j\models\psi_n^{k}$, where $n=1$ or $n=2$. Since every $\psi_{12}^i$ in the DNF of $\psi_1\vee\psi_2$ is either some $\psi_1^j$, $1\leq j\leq m_1$, or $\psi_2^k$, $1\leq k\leq m_2$, $P_1$ is also true for $\varphi_1,\varphi_2,\psi_1\vee\psi_2$.
    \item[Case ($\diamond$)] Assume that $P_1$ is true for $\varphi_1,\varphi_2,\psi$, so for every $\varphi_1^i,\varphi_2^j$, there is some $\psi^k$ such that $\varphi_1^i\models\psi^k$ and $\varphi_2^j\models\psi^k$. The DNFs of $\langle a\rangle\varphi_1$, $\langle a\rangle\varphi_2$, and $\langle a\rangle\psi$ are $\bigvee_{i=1}^{k_1}\langle a\rangle \varphi_1^i$, $\bigvee_{i=1}^{k_2}\langle a\rangle \varphi_2^i$, and $\bigvee_{i=1}^{k_3}\langle a\rangle \psi^i$, respectively.  From Lemma~\ref{lem:diamond_lemma}, $\langle a\rangle\varphi_1^i\models\langle a\rangle\psi^k$ and $\langle a\rangle\varphi_2^j\models\langle a\rangle\psi^k$ hold and 
    $P_1$ is also true for $\langle a\rangle\varphi_1,\langle a\rangle \varphi_2,\langle a\rangle\psi$.
\end{description}
Consequently, if  $(\varphi_1,\varphi_2\Rightarrow\psi)$ is a vertex in $G_\varphi$, $P_1$ is true for $\varphi_1,\varphi_2,\psi$. In particular, $(\varphi,\varphi\Rightarrow\varphi)$ is a vertex in $G_\varphi$. Thus, for every $\varphi^i,\varphi^j$ there is $\varphi^k$ such that $\varphi^i\models\varphi^k$ and $\varphi^j\models\varphi^k$. From Corollary~\ref{cor:primality-LS-3}, $\varphi$ is prime.
\end{proof}

\begin{lem}\label{lem:sim-algo-correct-2}
 Let $\varphi\in\mathL_S$ such that $\ff\not\in\sub(\varphi)$. If $\varphi$ is prime, then there is an alternating path in $G_\varphi$ from $(\varphi,\varphi\Rightarrow\varphi)$ to \textsc{True}.
\end{lem}
\begin{proof}
   Assume that $\varphi$ is prime. Let $\varphi_1,\varphi_2,\psi\in\mathL_S$, such that they do not contain $\ff$, and $\bigvee_{i=1}^{k_1}\varphi_{1}^i$, $\bigvee_{i=1}^{k_2}\varphi_2^i$, and $\bigvee_{i=1}^{k_3}\psi^{i}$ be their DNFs, respectively. We say that $\varphi_1,\varphi_2,\psi$ satisfy property $P_2$ if there is $\psi^k$ such that $\varphi_1\models\psi^k$ and $\varphi_2\models\psi^k$. We prove Claims~\ref{claim-one} and~\ref{claim-two}.
  % \begin{description}
        \begin{clm}\label{claim-one}
        For every vertex $x=(\varphi_1,\varphi_2\Rightarrow \psi)$  in $G_\varphi$ such that $\psi\neq\true$ and $\varphi_1,\varphi_2,\psi$ satisfy $P_2$, the following are true: 
        \begin{enumerate}[(a)]
            \item One of the rules from Table~\ref{tab:S-rules} can be applied on $x$.
            \item If an existential rule is applied on $x$, then there is some $z=(\varphi_1',\varphi_2'\Rightarrow \psi')$ such that $(x,z)\in E$ and $\varphi_1',\varphi_2',\psi'$ satisfy $P_2$. 
            \item If a universal rule is applied on $x$, then for all $z=(\varphi_1',\varphi_2'\Rightarrow \psi')$ such that $(x,z)\in E$, $\varphi_1',\varphi_2',\psi'$ satisfy $P_2$.
        \end{enumerate}
        \end{clm}
        \begin{proof}
         We first prove statement (a). To this end, suppose that $(\varphi_1,\varphi_2\Rightarrow \psi)$ is a vertex such that no rule from Table~\ref{tab:S-rules} can be applied and  $\varphi_1,\varphi_2,\psi$ satisfy $P_2$. Then, it must be the case that $\varphi_1=\langle a\rangle \varphi_1'$, $\varphi_2=\langle b\rangle \varphi_2'$, and $\psi=\langle c\rangle \psi'$, where $a=b=c$ is not true. Assume that $a=c\neq b$. Hence, there is $\psi^k$ such that $\langle a\rangle \varphi_1'\models\psi^k$ and $\langle b\rangle \varphi_2'\models\psi^k$. However, $\psi^k=\langle a\rangle \psi'$ for some $\psi'$, and $\langle b\rangle \varphi_2'\not\models\langle a\rangle \psi'$, contradiction. The other cases that make $a=b=c$ false can be proven analogously.

We prove parts (b) and (c) of the claim by induction on the type of the rules. 
\begin{description}
    \item[Case (L$\vee_1$)] Assume there is $\psi^k$ such that $\varphi_1\vee\varphi_2\models\psi^k$ and $\varphi\models\psi^k$. From Lemma~\ref{lem:disjunction_lemma}, it holds that $\varphi_1\models\psi^k$ and $\varphi_2\models\psi^k$, and so $P_2$ is true for both $\varphi_1,\varphi,\psi$ and $\varphi_2,\varphi,\psi$. Case (L$\vee_2$) is similar.
    \item[Case (L$\wedge_1$)] Assume there is $\langle a\rangle \psi^k$ such that $\varphi_1\wedge\varphi_2\models\langle a\rangle\psi^k$ and $\varphi\models\langle a\rangle\psi^k$. From Lemma~\ref{lem:conjunction_lemma_simulation}, it holds that $\varphi_1\models\langle a\rangle\psi^k$ or $\varphi_2\models\langle a\rangle\psi^k$, and so $P_2$ is true for either $\varphi_1,\varphi,\langle a\rangle\psi$ or $\varphi_2,\varphi,\langle a\rangle\psi$. Case (L$\wedge_2$) is similar.
    \item[Case (R$\wedge$)] Let $\bigvee_{i=1}^m \psi_{12}^i$ be the DNF of $\psi_1\wedge\psi_2$. Assume there is $\psi_{12}^k$ such that $\varphi_1\models\psi_{12}^k$ and $\varphi_2\models\psi_{12}^k$. Since every $\psi_{12}^k$ is $\psi_1^i\wedge\psi_2^j$ for some $1\leq i\leq k_1$ and $1\leq j\leq k_2$, it holds that $P_2$ is true for both $\varphi_1,\varphi_2,\psi_1$ and $\varphi_1,\varphi_2,\psi_2$.
    \item[Case (R$\vee$)] Let  $\bigvee_{i=1}^m \psi_{12}^i$ denote the DNF of $\psi_1\vee\psi_2$. Then, $\bigvee_{i=1}^m \psi_{12}^i=\bigvee_{i=1}^{k_1}\varphi_1^i\vee\bigvee_{i=1}^{k_2}\varphi_2^i$. This immediately implies that if $P_2$ is true for $\varphi_1,\varphi_2, \psi_1\vee\psi_2$, then $P_2$ is true for $\varphi_1,\varphi_2, \psi_1$ or $\varphi_1,\varphi_2, \psi_2$.
    \item[Case ($\diamond$)] If there is $\langle a \rangle \psi^k$ such that $\langle a\rangle \varphi_1\models \langle a \rangle \psi^k$ and $\langle a\rangle \varphi_2\models \langle a \rangle \psi^k$, then from Lemma~\ref{lem:diamond_lemma}, $\varphi_1\models\psi^k$ and $\varphi_2\models\psi^k$. From the fact that the DNFs of $\langle a\rangle\varphi_1$, $\langle a\rangle\varphi_2$, and $\langle a\rangle\psi$ are $\bigvee_{i=1}^{k_1}\langle a\rangle \varphi_1^i$, $\bigvee_{i=1}^{k_2}\langle a\rangle \varphi_2^i$, and $\bigvee_{i=1}^{k_3}\langle a\rangle \psi^i$, respectively, we have that $P_2$ is true for $\varphi_1, \varphi_2, \psi$.\qedhere
    \end{description}
        \end{proof}
        \begin{clm}\label{claim-two} If $x$ is a vertex $(\varphi_1,\varphi_2\Rightarrow\psi)$ in $G_\varphi$ such that $\varphi_1,\varphi_2,\psi$ satisfy $P_2$, then there is an alternating path from $x$ to \textsc{True}.
        \end{clm}
\begin{proof} Let $x=(\varphi_1,\varphi_2\Rightarrow\psi)$ be a vertex in $G_\varphi$ such that $\varphi_1,\varphi_2,\psi$ satisfy $P_2$. We prove the claim by induction on the form of $x$.
\begin{description}
   \item[Case $x=(\varphi_1,\varphi_2\Rightarrow\true)$] In this case, $\varphi_1,\varphi_2,\psi$ satisfy $P_2$ and $P^G(x,\textsc{True})$ trivially holds. 
   \item[Case $x=(\varphi_1\vee\varphi_2\Rightarrow\psi)$] In this case, $x$ is universal and from Claim~\ref{claim-one}(c), all $z$ such that $(x,z)\in E$, i.e.\ $z_1=(\varphi_1,\varphi\Rightarrow\psi)$ and $z_2=(\varphi_2,\varphi\Rightarrow\psi)$, are vertices such that  $\varphi_1,\varphi,\psi$ and $\varphi_2,\varphi,\psi$ satisfy $P_2$, respectively. By  inductive hypothesis, $P^G(z_1,\textsc{True})$ and $P^G(z_2,\textsc{True})$. As a result, $P^G(x,\textsc{True})$.
\end{description}
Similarly to the case $x=(\varphi_1\vee\varphi_2\Rightarrow\psi)$ and using Claim~\ref{claim-one}(b)--(c), Claim~\ref{claim-two} can be proven for all the other cases that correspond to the top of the rules in Table~\ref{tab:S-rules}. From Claim~\ref{claim-one}(a), we know that these are the only forms that vertex $x$ can have. Finally, primality of $\varphi$ and Proposition~\ref{prop:primality-LS} imply that $P_2$ is true for $\varphi,\varphi,\varphi$, and so there is an alternating path from $(\varphi,\varphi\Rightarrow\varphi)$ to \textsc{True}. 
\end{proof}
The lemma follows from the aforementioned claims.
\end{proof}

The polynomial-time complexity of algorithm $\algos$ on $\varphi$ derives from the polynomial size of $G_\varphi$ and linear-time complexity of \reacha.

\begin{lem}\label{lem:graph-size}
    Given a formula $\varphi\in\mathL_S$ such that $\ff\not\in\sub(\varphi)$, the size of $G_\varphi$ is  $\mathcal{O}(|\varphi|^3)$.
\end{lem}
\begin{proof}
 Let $|\varphi|=n$. To construct $G_\varphi$, we start from $(\varphi,\varphi\Rightarrow\varphi)$ and apply repeatedly rules from Table~\ref{tab:S-rules} until no rule can be applied. Let $x=(\varphi_1,\varphi_2\Rightarrow\psi)$ be a vertex in $G_\varphi$. Apart from (tt), every rule generates new vertices by replacing at least one of $\varphi_1,\varphi_2,\psi$ with one of its subformulae. 
 %%and generates at most two new vertices. 
 Thus, every vertex of the form $(\varphi_1',\varphi_2'\Rightarrow\psi')$ is such that all $\varphi_1',\varphi_2',\psi'\in\sub(\varphi)$. Since $|\sub(\varphi)|=\mathcal{O}(n)$, the number of different vertices is at most $\mathcal{O}(n^3)$.
\end{proof}

Given a characteristic formula within $\mathL_S$ it is also tractable to generate a process for which that formula is characteristic.

\begin{cor}\label{cor:find-p-algo}
Let $\varphi\in\mathL_S$ such that $\ff\not\in\sub(\varphi)$. If $\varphi$ is prime, there is a polynomial-time algorithm that constructs a process for which  $\varphi$ is characteristic within $\mathL_S$.
\end{cor}
\begin{proof}
    Let $\varphi$ be satisfiable and prime and $p_\varphi$ be a process for which $\varphi$ is characteristic within $\mathL_S$. From Proposition~\ref{prop:primality-LS}, there is $1\leq j\leq k$, such that $\varphi\models\varphi_j$. If $p_j$ denotes a process for which $\varphi_j$ is characteristic within $\mathL_S$, then from Remark~\ref{Remark:charforms}, $p_j\curle_S p_{\varphi}$ and so $p_j\equiv_S p_\varphi$. Consider now algorithm \algos described in the proof of Proposition~\ref{prop:sim-primality} (in the main body of the paper). When \algos checks whether there is an alternating path in $G_\varphi$ from $s$ to $t$, it can also find an alternating path, denoted here by $\mathP_a$. As we move from the starting vertex $s=(\varphi,\varphi\Rightarrow\varphi)$ to the descendants of $s$ along $\mathP_a$, formula $\varphi$ on the right-hand side of $\Rightarrow$ gets deconstructed to give some $\varphi_j$ such that $\varphi\models\varphi_j$. We construct a process $p_j$ for which $\varphi_j$ is characteristic within $\mathL_S$, by following $\mathP_a$ bottom-up, i.e.\ from $t$ to $s$, and associating a process $p$ to every vertex $x$ in $\mathP_a$. Process $p$ depends only on the right-hand side of $\Rightarrow$ that appears in vertex $x$.  At the end, the process corresponding to $s$ is $p_j$.
    \begin{itemize}
        \item If $x=(\varphi_1,\varphi_2\Rightarrow \true)$ belongs to $\mathP_a$, then  $p=\mathtt{0}$ corresponds to $x$.
        \item If $p$ corresponds to $x=(\varphi_1,\varphi_2\Rightarrow\psi)$ and $y=(\langle a\rangle\varphi_1,\langle a\rangle\varphi_2\Rightarrow \langle a\rangle\psi)$ is the parent of $x$ in $\mathP_a$, then $q=a.p$ corresponds to $y$.
        \item If $p_1$ corresponds to $x_1=(\varphi_1,\varphi_2\Rightarrow \psi_1)$, $p_2$ corresponds to $x_2=(\varphi_1,\varphi_2\Rightarrow \psi_2)$, and $y=(\varphi_1,\varphi_2\Rightarrow \psi_1\wedge\psi_2)$ is the parent of $x_1$ and $x_2$ in $\mathP_a$, then $p_1+p_2$ corresponds to $y$.
        \item If $p\in \proc$ corresponds to $x=(\varphi_1,\varphi_2\Rightarrow \psi_1)$ and $y=(\varphi_1,\varphi_2\Rightarrow \psi_1\vee\psi_2)$ is the parent of $x$ in $\mathP_a$, then $p$ corresponds to $y$.
        \item If $p\in \proc$ corresponds to $x=(\varphi_1,\varphi\Rightarrow \langle a\rangle\psi)$ and $y=(\varphi_1\wedge\varphi_2,\varphi\Rightarrow \langle a\rangle\psi)$ (or $y=(\varphi,\varphi_1\wedge\varphi_2\Rightarrow \langle a\rangle\psi)$) is the parent of $x$ in $\mathP_a$, then $p$ corresponds to $y$.
         \item If $p_1\in \proc$ corresponds to $x_1=(\varphi_1,\varphi\Rightarrow \psi)$, $p_2\in \proc$ corresponds to $x_2=(\varphi_2,\varphi\Rightarrow \psi)$ and $y=(\varphi_1\vee\varphi_2,\varphi\Rightarrow \psi)$ (or $y=(\varphi,\varphi_1\vee\varphi_2\Rightarrow \psi)$) is the parent of $x_1$ and $x_2$ in $\mathP_a$, then w.l.o.g.\ $p_1$ corresponds to $y$.\qedhere
    \end{itemize}
\end{proof}

\subsection{The formula primality problem for \texorpdfstring{$\mathL_{CS}$}{LCS}.}\label{sect:primality-LCS}

Note that, in the case of $\mathL_{CS}$, the rules in Table~\ref{tab:S-rules} do not work any more because, unlike $\mathL_{S}$, the logic $\mathL_{CS}$ can express some `negative information' about the behaviour of processes. For example, let $\act=\{a\}$ and $\varphi=\langle a\rangle \true$. Then, \algos accepts $\varphi$, even though  $\varphi$ is not prime in $\mathL_{CS}$. Indeed, $\varphi\models\langle a\rangle\langle a\rangle \true\vee\langle a\rangle\zero$, but $\varphi\not\models\langle a\rangle\langle a\rangle \true$ and $\varphi\not\models\langle a\rangle\zero$. However, we can overcome this problem as described in the proof sketch of Proposition~\ref{prop:cs-primality} below.

\begin{restatable}{prop}{csprime}\label{prop:cs-primality}
    Let $\varphi\in\mathL_{CS}$ be a formula such that every $\psi\in\sub(\varphi)$ is satisfiable. Deciding whether $\varphi$ is prime is in \cP.
\end{restatable}
\begin{proof}
   Consider the algorithm that first computes the formula $\varphi^\diamond$ by applying rule $\langle a\rangle\true\rulediam\true$, and rules $\true\vee\psi\rulett\true$ and $\true\wedge\psi\rulett\psi$ modulo commutativity on $\varphi$. It holds that $\varphi$ is prime iff $\varphi^\diamond$ is prime and $\varphi^\diamond\models\varphi$. Next, the algorithm decides primality of $\varphi^\diamond$ by solving reachability on a graph constructed as in the case of simulation using the rules in Table~\ref{tab:S-rules}, where rule (tt) is replaced by rule (0), whose premise is $\zero,\zero\Rightarrow\zero$ and whose  conclusion is \textsc{True}. To verify $\varphi^\diamond\models\varphi$, the algorithm computes a process $p$ for which $\varphi^\diamond$ is characteristic within $\mathL_{CS}$ and checks whether $p\models\varphi$. In fact, the algorithm has also a preprocessing phase during which it applies a set of rules on $\varphi$ and obtains an equivalent formula with several desirable properties. See Appendix~\ref{subsection:cs-primality-appendix} for full details, where Remark~\ref{rem:type-ordering-rules} comments on the type and ordering of the rules applied.
\end{proof}

\begin{cor}\label{cor:primality-in-CS}
    The formula primality problem for $\mathL_{CS}$ is in \cP.
\end{cor}

\subsection{The formula primality problem for \texorpdfstring{$\mathL_{RS}$}{LRS}.}\label{sect:primality-LRS}
The presence of formulae of the form $[a]\ff$ in $\mathL_{RS}$ means that a prime formula $\varphi\in\mathL_{RS}$ has at least to describe which actions are necessary or forbidden for any process that satisfies $\varphi$. For example, let $\act=\{a,b\}$. Then, $\langle a\rangle \zero$ is not prime, since $\langle a\rangle \zero\models (\langle a\rangle \zero\wedge [b]\ff)\vee (\langle a\rangle \zero\wedge \langle b\rangle\true)$, and $\langle a\rangle \zero$ entails neither $\langle a\rangle \zero\wedge [b]\ff$ nor $\langle a\rangle \zero\wedge \langle b\rangle\true$. Intuitively, we call a formula $\varphi$ \emph{saturated} if $\varphi$ describes exactly which actions label the outgoing edges of any process $p$ such that $p\models\varphi$. Formally, $\varphi$ is saturated iff $I(\varphi)$, which we introduced in Definition~\ref{def:cs-rs-I(phi)}, is a singleton.

If the action set is bounded by a constant, given $\varphi$, we can efficiently construct a formula $\varphi^s$ such that 
\begin{enumerate}
    \item $\varphi^s$ is saturated and for every $\langle a\rangle\varphi'\in\sub(\varphi^s)$, $\varphi'$ is saturated,
    \item $\varphi^s\models\varphi$, 
    \item $\varphi$ is prime iff $\varphi^s$ is prime, and
    \item the primality of $\varphi^s$ can be efficiently reduced to $\reacha(G_{\varphi^s})$.
\end{enumerate}
\iffalse
(1) $\varphi^s$ is saturated and for every $\langle a\rangle\varphi'\in\sub(\varphi^s)$, $\varphi'$ is saturated, (2) $\varphi$ is prime iff $\varphi^s$ is prime and $\varphi^s\models\varphi$, and (3) primality of $\varphi^s$ can be efficiently reduced to $\reacha(G_{\varphi^s})$. 
\fi
The proofs of Proposition~\ref{prop:rs-primality} and Corollary~\ref{cor:primality-in-RS} are included in Appendix~\ref{subsection:rs-primality-bounded-appendix}.

\begin{prop}\label{prop:rs-primality}
    Let $|\act|=k$, where $k\geq 1$ is a constant, and $\varphi\in\mathL_{RS}$ be such that if $\psi\in\sub(\varphi)$ is unsatisfiable, then $\psi=\ff$ and $\psi$ occurs in the scope of some $[a]$. Deciding whether $\varphi$ is prime is in \cP.
\end{prop}

\begin{cor}\label{cor:primality-in-RS}
The formula primality problem for $\mathL_{RS}$ with a bounded action set is in  \cP.   
\end{cor}

As the following result indicates, primality checking for formulae in $\mathL_{RS}$ becomes computationally hard when $|\act|$ is not a constant.

\begin{restatable}{prop}
{rsprimehard}\label{prop:decide-prime-rs-infinite-actions-hard}
The formula primality problem for $\mathL_{RS}$ with an unbounded action set is  \conp-complete.
\end{restatable}
\begin{proof}
We give a polynomial-time reduction from \SAT to deciding whether a formula in $\mathL_{RS}$ is not prime. Let $\varphi$ be a propositional formula over $x_0,\dots,x_{n-1}$. We set $\varphi'=(\varphi\wedge \neg x_n)\vee (x_n\wedge \bigwedge_{i=1}^{n-1} \neg x_i)$ and let $\varphi''$ be the modal formula in $\mathL_{RS}$ that is obtained from $\varphi'$ by replacing $x_i$ with $\langle a_i \rangle \zero$ and $\neg x_i$ with $[a_i] \ff$,
where $\act=\{a_0,\dots, a_n\}$. As $\varphi''$ is satisfied in $a_n.\mathtt{0}$, it is a satisfiable formula, and so $\varphi''$ is prime in $\mathL_{RS}$ iff $\varphi''$ is characteristic within $\mathL_{RS}$.

    We show that $\varphi$ is satisfiable iff $\varphi''$ is \emph{not} characteristic  within $\mathL_{RS}$. Let $\varphi$ be satisfiable and let $s$ denote a satisfying assignment for $\varphi$. Consider $p_1,p_2\in \proc$ such that: 
    \begin{itemize}
        \item $p_1\myarrowasubi \mathtt{0}$ iff $s(x_i)=\text{true}$, for $0\leq i\leq n-1$, and $p_1\notmyarrowan$, and
        \item $p_2\myarrowan \mathtt{0}$ and $p_2\notmyarrowa$ for every $a\in \act\setminus\{a_n\}$.
    \end{itemize}  
    We have that $p_i\models \varphi''$, $i=1,2$, 
    %It also holds that 
    $p_1\notcurle_{RS} p_2$, and $p_2\notcurle_{RS} p_1$. Suppose that there is a process $q$, 
    %different from $p_1$ and $p_2$, 
    such that $\varphi''$ is characteristic for $q$ within $\mathL_{RS}$. If $q\myarrowan$, then $q\notcurle_{RS} p_1$. On the other hand, if $q\notmyarrowan$, then $q\notcurle_{RS} p_2$. So, both cases lead to a contradiction, which means that $\varphi''$ is not characteristic within $\mathL_{RS}$. For the converse implication, assume that $\varphi$ is unsatisfiable. This implies that there is no process satisfying the first disjunct of $\varphi''$. Thus, $\varphi''$ is characteristic for $p_2$, described above, within $\mathL_{RS}$.
    
    Proving the matching upper bound is non-trivial. There is a \conp algorithm that uses some properties of prime formulae and the rules in Table~\ref{tab:S-rules}, carefully adjusted to the case of ready simulation. Details can be found in Appendix~\ref{subsection:rs-primality-unbounded-appendix}.
\end{proof}

\subsection{The formula primality problem for \texorpdfstring{$\mathL_{TS}$}{LTS}}

We now show that, in the case of $\mathL_{TS}$, the formula primality problem is \conp-hard if $\act$ contains at least two actions. Furthermore, the problem is fixed-parameter tractable (FPT) \cite{DowneyF95} when parameterized by the  modal depth of the formula and $|\act|$.

\begin{prop}\label{prop:decide-prime-ts-infinite-actions-hard}
 Let $|\act|\geq 2$. The formula primality problem for $\mathL_{TS}$ is \conp-hard.
\end{prop}
\begin{proof}
    We show that the complement of the problem is \NP-hard. To this end, we describe a polynomial-time reduction from \SAT to deciding non-prime formulae within $\mathL_{TS}$, which is based on the proofs of Propositions~\ref{prop:sat-RS-TS}(b) and~\ref{prop:decide-prime-rs-infinite-actions-hard}. However, we have to encode the literals more carefully here since the set of actions is fixed.
    
    Let $\act=\{0,1\}$ and $\varphi$ be an instance of \SAT over the variables $x_0,\dots, x_{n-1}$. Similarly to the proof of Proposition~\ref{prop:sat-RS-TS}(b), the initial idea is to associate every variable $x_i$ with the binary representation of $i$. In more detail, for every $0\leq i\leq n-1$, we associate  $x_i$ with $0\,{b_i}_1\dots{b_i}_k$, where every ${b_i}_j\in\{0,1\}$ and $k=\lceil\log n\rceil$.
    This means that $x_i$ is associated with the binary string ${b_i}_0{b_i}_1\dots{b_i}_k$, where ${b_i}_0=0$ and ${b_i}_1\dots{b_i}_k$ is the binary representation of $i$. The binary string ${b_i}_0{b_i}_1\dots {b_i}_k$ can now be mapped to formula $\enc(x_i)= [\overline{{b_i}_0}]\ff
    \wedge \langle {b_i}_0\rangle( [\overline{{b_i}_1}]\ff\wedge \langle {b_i}_1\rangle (\dots ([\overline{{b_i}_k}]\ff\wedge \langle {b_i}_k\rangle\zero)\dots))$, where 
    $\overline{b}$ stands for the complement of $b$. 
    \iffalse
    $\overline{b}=\begin{cases}
        1, &\text{if } b=0,\\
        0, &\text{if } b=1
    \end{cases}$. 
    \fi 
    We map a negative literal $\neg x_i$ to $\enc(\neg x_i)=[{b_i}_1] [{b_i}_2]\dots [{b_i}_k]\ff$. 
    
    Define formula $\varphi'$ to be $\varphi$ where every literal $l$ has been replaced by $\enc(l)$ and formula $\varphi''=(\varphi'\wedge [1]\ff)\vee (\langle 1\rangle ([0]\ff\wedge [1]\ff)\wedge [1][0]\ff\wedge[1][1]\ff\wedge[0]\ff)$. As in the proof of Proposition~\ref{prop:decide-prime-rs-infinite-actions-hard}, we can prove that if $\varphi$ is satisfiable, then $\varphi''$ is not prime and not characteristic within $\mathL_{TS}$. If $\varphi$ is not satisfiable, then $\varphi''$ is prime and characteristic for $p=a.\mathtt{0}$ within $\mathL_{TS}$, where $a=1$.
\end{proof}

To prove that the formula primality problem for $\mathL_{TS}$ is FPT, we define $P^d \subseteq \proc$ to be the set of processes of depth at most $d$, up to bisimilarity.
We also define the tower function $\operatorname{tow}(k,d)$ recursively on $d \geq 0$ thus: $\operatorname{tow}(k,0)=1$ and $\operatorname{tow}(k,d+1) = k \cdot 2^{\operatorname{tow}(k,d)}$.

\begin{lem}
    For each $d \geq 0$, $|P^d| \leq \operatorname{tow}(|\act|,d)$. Furthermore, for each $p \in P^d$ and $d>0$, $|p| \leq \operatorname{tow}(|\act|,d-1) + 1$.
\end{lem}
\begin{proof}
    The proof for $|P^d| \leq \operatorname{tow}(|\act|,d)$ is by straightforward induction on $d$ and the expected combinatorial arguments. See also~\cite{halpern1995effect} and~\cite[Theorem 1]{achilleos2012parameterized}.
    To bound the size of $p$, notice that for every process $q$ reachable from $p$, either $q = p$ or $q \in P^{d-1}$.
\end{proof}

The modal depth of a formula limits the space of processes that we need to check for satisfiability and primality.

\begin{lem}\label{lem:TS_d_bounds_search}
    Let $\varphi \in \mathL_{TS}$ and let $d$ be the modal depth of $\varphi$. Then, $\varphi$ is satisfiable and prime iff there is some $p\in P^d$ such that $p\models\varphi$ and for every $q\in P^{d}$, $q\models\varphi\implies p\curle_{TS} q$.
\end{lem}
\begin{proof}
 Assume that $\varphi$ is satisfiable and prime. Since $\varphi$ is satisfiable, then it must be satisfied by a process with depth less than or equal to $d$---one can see this by the tableau construction in Subsection~\ref{subsection:hml-tableau}, which returns tableaux with depth bounded by the formula's modal depth, Remark~\ref{rem:tableausat} and Proposition~\ref{prop:tableau-construction}. (See also Proposition~\ref{prop:sat-rs-ts-2s-np-complete}.) 
 So, there is some $q$ that satisfies $\varphi$ and $\depth(q)=m\leq d$. 
 Since $\varphi$ is prime, and so characteristic within $\mathL_{TS}$, there is some $p$ such that for every $p'$, $p'\models\varphi\Leftrightarrow p\curle_{TS} p'$. 
 In particular, 
 %for every $p'\models\varphi$, 
 we have that 
 %$p\curle_{TS} p'$ and 
 $p\curle_{TS} q$, and since $\mathrm{traces}(p) = 
 %\mathrm{traces}(p')=
 \mathrm{traces}(q)$, it holds that $\depth(p) = 
 %\depth(p')=
 \depth(q)=m$,
 and therefore 
 %So, 
 %$\depth(p')
 $p
 %,p'
 \in P^d$. 
 
 Conversely, assume that there is some $p\in P^d$ that satisfies $\varphi$ and for every $q\in P^{d}$, $q\models\varphi\implies p\curle_{TS} q$. Then, $\depth(p)=m$, for some $m\leq d$, by definition of $P^d$. This implies that if $q\notin P^{d}$ with $\depth(q) \geq d+1$, then $q\not \models \varphi$, because if such a $q$ satisfies $\varphi$, $p\curle_{TS} q$ and $\mathrm{traces}(p)\neq\mathrm{traces}(q)$, which is impossible. 
 Consequently, for every $q\in \proc$, if $\depth(q)\geq d+1$, then $q\not\models\varphi$, since $\md(\varphi)=d$. This, in turn, implies that for every $q\in \proc$, $q\models \varphi\implies p\curle_{TS} q$. For the other direction, if $p\curle_{TS} q$, then $q\models \varphi$ from Proposition~\ref{logical_characterizations}. As a result $\varphi$ is characteristic within $\mathL_{TS}$ for $p$.
\end{proof}

\begin{thm}\label{thm:ts-bounded-depth}
  Let $|\act|= k$ and $\varphi\in\mathL_{TS}$ with $\md(\varphi)=d$.
  %, where $k,d\geq 1$ are constants. 
  Then, there is an algorithm that decides whether $\varphi$ is prime in time 
  %linear in $|\varphi|$.
   $\mathcal{O}(\operatorname{tow}(k,d)^3\cdot |\varphi|)$.
\end{thm}

\begin{proof}
%\textcolor{orange}{not ready yet}
    We describe an algorithm that decides prime formulae within $\mathL_{TS}$ and analyze its complexity.
    The algorithm first checks whether $p \curle_{TS} q$ for all processes $p,q \in P^d$ and stores the results. This can be done by iterating through all $p,q \in P^d$ in $\mathcal{O}(\operatorname{tow}(k,d)^2)$ time, and then through all $p',q' \in P^{d-1}$, such that $p \myarrowa p'$ and $q \myarrowa q'$ for some $\alpha \in \act$. Overall, this step takes up to $\mathcal{O}(\operatorname{tow}(k,d)^2 \cdot \operatorname{tow}(k,d-1)^2)$ time.
    %\textcolor{orange}{does this make sense to everyone?}
    The algorithm then computes $P^d_{sat} = \{ p \in P^d \mid p \models \varphi \}$---checking if $p \models \varphi$ can be done in $\mathcal{O}(|p| \cdot |\varphi|) = \mathcal{O}(\operatorname{tow}(k,d-1) \cdot |\varphi|)$ time. 
    If $P^d_{sat} = \emptyset$, then the algorithm accepts the input.
    The algorithm then iterates through all processes in $P^d_{sat}$ twice: the first time it searches for a $\curle_{TS}$-smallest process $r$ and the second time it verifies that $r$ is indeed a $\curle_{TS}$-smallest process in $P^d_{sat}$. If one such process $r$ exists, then it accepts. Otherwise, it rejects.

    From the above, the algorithm runs in \[\mathcal{O}(\operatorname{tow}(k,d)^2\cdot \operatorname{tow}(k,d-1)^2 + \operatorname{tow}(k,d) \cdot \operatorname{tow}(k,d-1) \cdot |\varphi| + \operatorname{tow}(k,d) ) = \mathcal{O}(\operatorname{tow}(k,d)^3  \cdot |\varphi|  ).\]
    Its correctness follows from Lemma \ref{lem:TS_d_bounds_search}.
\end{proof}

\begin{rem}\label{rem:rambling-about-parameter-dependency-FPT}
    We note that in the proof of Theorem~\ref{thm:ts-bounded-depth}, we have not calculated the cost of looking up the calculated value of $p \curle_{TS} q$, but for a reasonable implementation it should take at most $\mathcal{O}(\operatorname{tow}(k,d-1))$ time, which does not affect the analysis above.
    We also expect that the rather steep parameter dependency of $\operatorname{tow}(k,d)^3$ can be improved, but it is unclear by how much. 
    On the one hand, a similar parameter dependency for general modal satisfiability is known to be tight~\cite{achilleos2012parameterized}, but on the other hand, the satisfiability problem for $\mathL_{TS}$ is in \NP~(Theorem~\ref{prop:sat-rs-ts-2s-np-complete}) and thus not expected to be \pspace-hard.
\end{rem}

\begin{rem}\label{rem:what-is-FPT}
    Theorem~\ref{thm:ts-bounded-depth} means that primality checking for $\mathL_{TS}$ is fixed-parameter tractable (\fpt) with the modal depth of the formula as the parameter.
     Let $L\subseteq \Sigma^*\times\Sigma^*$ be a parameterized problem---in our case, the input formula $\varphi$ can be written as $(\varphi,\md(\varphi))$ to indicate that $\md(\varphi)$ is the parameter. 
     % We denote by $L_y$ the associated fixed-parameter problem $L_y=\{x ~|~ (x,y)\in L\}$, where $y$ is the parameter. 
     Then, $L\in\fpt$ (or $L$ is fixed-parameter tractable) if there are a constant $\alpha$ and an algorithm to determine if $(x,y)$ is in $L$ in time $f(|y|)\cdot |x|^\alpha$, where $f$ is a computable function. See, for instance,~\cite{DowneyF13,FlumG06} for textbook accounts of parameterised complexity theory.
     We also note that Corollaries~\ref{cor:ts-bounded-depth} and \ref{cor:char-ts-equiv-bounded-depth} also show that the corresponding problems (checking if a $\mathL_{TS}$-formula is characteristic for $\curle_{TS}$ and $\equiv_{TS}$, respectively) are \fpt.
\end{rem}

\section{The formula primality problem for the nested-simulation-preorder logics}\label{Sect:primality-nested-semantics}

We will now study the complexity of the formula primality problem for the logics $\mathL_{nS}$, $n\geq 2$, that characterize the nested-simulation preorders. 

So far, we have presented our results according to the inclusion order $\mathL_S\subseteq\mathL_{CS}\subseteq\mathL_{RS}\subseteq \mathL_{TS}$. However, we organize the presentation of our results on the complexity of the formula primality problem for $\mathL_{2S}$ and $\mathL_{nS}$ with $n\geq 3$, differently.  We begin with $\mathL_{nS}$, $n\geq 3$, in this Section~\ref{subsection:primality-nS} and then consider $\mathL_{2S}$ in Section~\ref{subsec:prime-2S}. We chose this %The reverse 
order in presenting our results on the $n$-nested simulation semantics because the results for $\mathL_{nS}$, $n\geq 3$, help our readers to follow the arguments for $\mathL_{2S}$ that appear later.

\subsection{The formula primality problem for \texorpdfstring{$\mathL_{nS}$}{LnS}, \texorpdfstring{$n\geq 3$}{n>=3}}\label{subsection:primality-nS}

The goal of this section is to prove that the formula primality problem for {$\mathL_{nS}$} is \pspace-complete, when $n\geq 3$. To this end, we first establish that the problem is \pspace-hard via a reduction from the validity problem for $\mathL_{2S}$ to the formula primality problem for $\mathL_{nS}$, $n\geq 3$. We then show that the formula primality problem for $\mathL_{nS}$ can be solved in polynomial space for each $n\geq 3$. Perhaps surprisingly, the proof of that complexity upper bound is much more involved than the one for the lower bound stated in the following theorem. 

%The problem is \pspace-hard for $\mathL_{nS}$, $n\geq 3$.

\begin{thm}\label{thm:2scompl-prime}
    The formula primality problem for $\mathL_{nS}$, where $n\geq 3$, is \pspace-hard.
\end{thm}
\begin{proof}
    Let $n\geq 3$. We reduce the validity problem for $\mathL_{2S}$ to the formula primality problem for $\mathL_{nS}$. Then the theorem holds because of Corollary~\ref{cor:2s-validity}.
    Let $\varphi \in \mathL_{2S}$.
    The reduction will return a formula $\varphi'$, such that $\varphi$ is $\mathL_{2S}$-valid if and only if $\varphi'$ is prime in $\mathL_{nS}$.
    If $\mathtt{0} \not\models \varphi$, then let $\varphi' = \true$; in this case, $\varphi$ is not valid and $\true$ is not prime in $\mathL_{nS}$.
    Otherwise, let $\varphi' = \zero \vee \neg \varphi$.
    If $\varphi$ is valid, then $\varphi' \equiv \zero$ and therefore $\varphi'$ is prime in $\mathL_{nS}$.
    On the other hand, if $\varphi$ is not valid, then there is some process $p \models \neg \varphi$. 
    From $\mathtt{0} \models \varphi$, for each such $p$, it holds that $p\myarrowa$ for some $a\in\act$.
    Then, $\varphi'\models\zero\vee\bigvee_{a\in \act} \langle a\rangle\true$, but $\varphi'\not\models\zero$ and $\varphi'\not\models\bigvee_{a\in \act} \langle a\rangle\true$. Therefore $\varphi'$ is not prime in $\mathL_{nS}$.
\end{proof}

Next, we present a polynomial-space algorithm that solves the formula primality problem for $\mathL_{nS}$, $n\geq 3$, matching the lower bound from Theorem~\ref{thm:2scompl-prime}. Our order business in the rest of this subsection is to prove the following theorem.

\begin{restatable}{thm}{nSpspacecomp}\label{prop:3S-char-in-pspace}
 The formula primality problem for $\mathL_{nS}$, $n\geq 3$, is  \pspace-complete.
\end{restatable}

To prove the above result, we introduce two families of games: the \emph{char-for-n-nested-simulation-equivalence} game, referred to as the \nsimeq game, for every $n\geq 1$, and the \emph{prime-for-n-nested-simulation-preorder} game, referred to as the \nsimpre game, for every $n\geq 3$.

Assume that $\varphi$ is a satisfiable formula in $\mathL_{nS}$. All of the games are played between players $A$ and $B$. If a game is initiated on $\varphi$, it starts with two or three states each of which has a label equal to $\{\varphi\}$.  As the game proceeds, the players extend the already existing structures and explore (two or three) tableaux for $\varphi$ that satisfy some additional, game-specific conditions. 

Player $A$ has a winning strategy for the \nsimeq game iff every two processes that satisfy $\varphi$ are equivalent modulo $\equiv_{nS}$---that is, $\varphi$ is a characteristic formula modulo $\equiv_{nS}$---as stated below.

\begin{restatable}{prop}{charmodns}\label{prop:twosimul-game}
  Let $\varphi\in\mathL_{\ell S}$, where $\ell\geq n$, be a satisfiable formula. Player $A$ has a winning strategy for the \nsimeq game on $\varphi$ iff for every two processes $r_1,r_2$ that satisfy $\varphi$, $r_1\equiv_{nS} r_2$.
\end{restatable}

The existence of a winning strategy for player $A$ in the \nsimpre game on $\varphi$ is equivalent to the primality of $\varphi$ in $\mathL_{nS}$.

\begin{restatable}{prop}{gameprime}\label{prop:winning-strategy-primality}
  Let $\varphi\in\mathL_{nS}$, where $n\geq 3$, be satisfiable. Assume that $A$ has a winning strategy for the \nsimpre game on $\varphi$. Then, $\varphi$ is prime in $\mathL_{nS}$.
\end{restatable}

A difference between the games \nsimeq and \nsimpre is that the former can be initiated on a satisfiable formula that belongs to $\mathL_{\ell S}$, where $\ell\geq n$, whereas the latter is only started on a satisfiable formula that is in $\mathL_{nS}$. When $A$ and $B$ play one of the games \nsimeq or \nsimpre, at some point, they have to play the \nmosimeq 
game initiated on states labelled with possibly different finite subsets of $\mathL_{nS}$ formulae. This is why the \nsimeq game is generalized to start with such labelled states. 

For the presentation of the games, let $\act=\{a_1,\dots,a_k\}$. The basic moves that $A$ and $B$ can play are presented in Table~\ref{tab:moves}.

\begin{table}%[t]
\begin{center}
\begin{tabular}{ | m{2.3cm} | m{11cm} | } 
\hline
\textbf{Move name}  & \hspace{4cm}\textbf{Move description}\\  
 \hline
Pl($\wedge$) & For every $\psi_1\wedge\psi_2\in L_i(p)$,  $Pl$ replaces $\psi_1\wedge\psi_2$ with both $\psi_1$ and $\psi_2$ in $L_i(p)$. \\  
 \hline
Pl($\vee$) & For every $\psi_1\vee\psi_2\in L_i(p)$,  $Pl$  chooses $\psi\in\{\psi_1,\psi_2\}$ and replaces $\psi_1\vee\psi_2$ with $\psi$ in $L_i(p)$.\\  
 \hline
Pl($\Diamond$) & For every $\langle a_j\rangle \psi\in L_i(p)$,  $Pl$ adds a new state $p'$ to $S_i$, $(p,p')$ to $R_{a_j}^i$, and sets $L_i(p')=\{\psi\}\cup\{\psi'\mid [a_j]\psi'\in L_i(p)\}$.\\  
 \hline
B($\square$) & $B$ chooses between doing nothing and picking some $1\leq j\leq k$. In the latter case, $B$ adds a new state $p'$ to $S_i$, $(p,p')$ to $R^i_{a_j}$, and  sets $L_i(p')=\{\psi ~ | ~ [a_j]\psi\in L_i(p)\}$.  \\  
 \hline
 A(sub) & For every $\psi\in\sub(\varphi)$, $A$ chooses between adding or not adding $\psi$ to $L_i(p)$.  \\  
 \hline
 A(rem) & For every $j$-successor $p'$ of $p$, $A$ removes $p'$ from $T_i$ if there is a $j$-successor $p''$ of $p$, such that $p'\neq p''$ and $L_i(p')\subseteq L_i(p'')$. \\  
 \hline
\end{tabular}
\end{center}
\caption{Basic moves that players $A$ and $B$ can play in any game initiated on formula $\varphi$. The description is for player $Pl\in\{\text{A,B}\}$ who plays on state $p\in S_i$, where $i\in\{1,2,3\}$, and action $a_j\in\act$.
}
\label{tab:moves}
\end{table}

\paragraph{The \nsimeq game, \texorpdfstring{$n\geq 1$}{n>=1}} We present the first family of games.  We begin by describing the \simequiv game, followed by the \nsimeq game for $n\geq 2$. Let $\varphi$ be a satisfiable formula in $\mathL_{\ell S}$, where $\ell\geq n$. The games are defined so  
%We show by induction on $n$ 
that player $A$ has a winning strategy for the \nsimeq game played on $\varphi$ iff every two processes satisfying $\varphi$ are $n$-nested-simulation equivalent: we prove this statement for the \simequiv game, and assuming that this is true for the \nmosimeq game, we show the statement for the \nsimeq game.

\paragraph{The \simequiv game}
We first introduce the \simequiv game started on $\varphi$. During the game, $B$ constructs two labelled trees $T_1$ and $T_2$ that correspond to two arbitrary processes $p_1$ and $p_2$ satisfying $\varphi$ and challenges $A$ to construct a simulation relation between the states of $T_1$ and $T_2$ showing that $p_1\curle_S p_2$. The labelled trees constructed by $B$ are denoted $T_1=(S_1,L_1,R^1_{a_1},\dots,R^1_{a_k})$ and $T_2=(S_2,L_2,R^2_{a_1},\dots,R^2_{a_k})$, and the game starts with $S_1=\{p_0^1\}$ and $S_2=\{p_0^2\}$, $L_1(p_0^1)=L_2(p_0^2)=\{\varphi\}$, and $R^i_{a_j}=\emptyset$, for every $i=1,2$ and $1\leq j\leq k$. We describe the $l$-th round of the game, where $l\geq 1$, in Table~\ref{tab:simulation-game}. States $p_1,p_2$ are $p_0^1,p_0^2$ respectively, if $l\in\{1,2\}$, or the two states that $B$ and $A$ respectively chose at the end of round $l-1$, if $l>2$. For two states $p,p'$ such that $(p,p')\in R_{a_j}$, we say that $p'$ is a $j$-successor of $p$. We use $p$, $p_1$, $p_2$, etc.~to denote both processes and states of the labelled trees; the intended meaning will be clear from the context.
\begin{table}%[t]
\begin{center}
\begin{tabular}{ | p{15cm} | } 
\hline
\textbf{$\mathbf{1}^{\text{st}}$ round.} 
 $B$ plays moves $B(\wedge)$ and $B(\vee)$ on $p_i$, for both $i=1,2$, until no formula can be replaced in $L_i(p_i)$. 
 If $\bigwedge L_i(p_i)$ becomes unsatisfiable, then $B$ loses.

\textbf{$\mathbf{l}^{\text{th}}$ round, $\mathbf{l\geq 2}$.} 
\begin{enumerate}
    \item For every $a_j\in \act$, $B$ plays as follows. He plays move $B(\Diamond)$ on $p_i$ for both $i=1,2$, and move $B(\Box)$ only on $p_1$. Then, for both $i=1,2$, $B$ plays moves $B(\wedge)$ and $B(\vee)$  on every $p_i'$ such that $(p_i,p_i')\in R_{a_j}^i$ until no formula can be replaced in $L_i(p_i')$. If $\bigwedge L_i(s)$ becomes unsatisfiable for some $i=1,2$ and $s\in S_i$, then $B$ loses. 
    \item $B$ chooses a $1\leq j\leq k$ and a $j$-successor $p_1'$ of $p_1$. If $p_1$ has no $j$-successors, then $B$ loses.
    \item $A$ chooses a $j$-successor $p_2'$  of $p_2$. If $p_2$ has no $j$-successors, then $A$ loses.
    \item The $l+1$-th round starts on $p_1'$, $p_2'$.
\end{enumerate}  \\
 \hline
\end{tabular}
\end{center}
\caption{The \simequiv game initiated on a satisfiable $\varphi\in\mathL_{\ell S}$, where $\ell\geq 1$.}
\label{tab:simulation-game}
\end{table}

\begin{exa}\label{ex:simulation-game}
   (a) Consider the formula $\varphi=\langle a_1\rangle \mathbf{0}$. Note that both the processes $r_1=a_1.\mathtt{0}$ and $r_2=a_1.\mathtt{0}+a.2.\mathtt{0}$ satisfy $\varphi$, and $r_1\not\equiv_S  r_2$. Therefore, player $B$ should have a winning strategy for the \simequiv game on $\varphi$, which is true as $B$ can play as follows. At the first round, he can make no replacement in $L_i(p_i)$ for both $i=1,2$. At step 1 of the second round, he generates states $p_1'$ and $p_2'$ that are $1$-successors of $p_1$ and $p_2$ respectively, when he plays move B($\Diamond$). Then, when $B$ plays move B($\square$) on $p_1$, he chooses to generate state $p_1''$ that is a $2$-successor of $p_1$, adds $(p_1,p_1'')$ to $R^1_{a_2}$, and  sets $L_1(p_1'')=\emptyset$. He applies move B($\wedge$) on $p_i'$  to obtain $L_i(p_i')=\{[a_1]\ff,\dots,[a_k]\ff\}$ for $i=1,2$. At step 2, $B$ chooses $p_1''$ and since $p_2$ has no $2$-successors, $A$ loses at step 3.

    (b) On the other hand, player $A$ has a winning strategy for the \simequiv game initiated on $\psi=\langle a_1\rangle \mathbf{0}\wedge \bigwedge_{i=2}^k [a_i]\ff$. (Note that the process $r_1=a_1.\mathtt{0}$ is the unique process modulo $\equiv_S$ that satisfies $\psi$.) After completing the first round, $B$ generates two states $p_1'$ and $p_2'$ which are $1$-successors of $p_1$ and $p_2$ respectively, and sets $L_1(p_1')=L_2(p_2')=\{\mathbf{0}\}$ when applying move B($\Diamond$). If he chooses to generate a $j$-successor $p_1''$ of $p_1$, where $j\neq 1$, when he plays move B($\square$), then he loses, since $L_1(p_1'')=\{\ff\}$ is unsatisfiable. So, he chooses to do nothing at move B($\square$) and picks $p_1'$ at step 2. Then, $A$ picks $p_2'$ at step 3. In round 3, $B$ either generates a $j$-successor of $p_1'$ for some $1\leq j\leq k$ when applying move B($\square$) and loses because the label set of the new state is unsatisfiable or generates no successors and loses at step 2. 
    \end{exa}

The labelled trees $T_1, T_2$, constructed during the \simequiv game on $\varphi$, form partial tableaux for $\varphi$. This is because some states are abandoned during the game, which may result in $T_1, T_2$ failing to satisfy condition (iii) of Definition~\ref{def:hmltableau}. The \simequiv game can be generalized so that it starts with  $S_1=\{s_1\}$, $S_2=\{s_2\}$, $R_{a_j}^i$ being empty for every $i=1,2$ and $1\leq j\leq k$, and $L_1(s_1)=U_1$, $L_2(s_2)=U_2$, where $U_1,U_2$ are finite subsets of $\mathL_{\ell S}$, $\ell\geq 1$. We denote by $\simab(U_1,U_2)$ the \simequiv game that starts from the configuration just described. In particular, $\simab(\{\varphi\},\{\varphi\})$ is called the \simequiv game on $\varphi$.

Let $p\in S_i$, where $i=1,2$. We denote by $\inlabeli(p)$ the initial label of $p$ before moves B($\wedge$) and B($\vee$) are applied on $p$ and $\finlabeli(p)$ the final label of $p$ after moves B($\wedge$) and B($\vee$) have been applied on $p$ (until no formula can be replaced in $L_i(p)$).
As shown in Example~\ref{ex:simulation-game}, in the \simequiv game on $\varphi$, player $B$ consistently plays $T_i$, $i=1,2$, on a process $r$  satisfying $\varphi$. Intuitively, $T_i$ represents part or all of $r$, viewing states in $S_i$ as processes reachable from $r$ and $R^i_{a_j}$ as transitions. The formal definition follows.
\begin{defi}\label{def:B-plays-consistently}
    Assume that the \simequiv game is played on $\varphi\in\mathL_{\ell S}$, $\ell\geq 1$. We say that $B$ plays $T_i$, where $i\in\{1,2\}$, consistently on a process $r$ if there is a mapping $map:S_i\rightarrow \proc$ such that the following conditions are satisfied:
  \begin{enumerate}
      \item for every $p\in S_i$, $map(p)\models \bigwedge \finlabeli(p)$,
      \item for every $(p,p')\in R^i_{a_j}$, $map(p)\myarrowasubj map(p')$, and 
      \item $map(p_0^i)=r$, where $p_0^i$ is the initial state of $T_i$.
  \end{enumerate}
\end{defi}

We first prove that if every two processes that satisfy $\varphi$ are equivalent with respect to $\equiv_S$, then $A$ has a winning strategy for the \simequiv game on $\varphi$. 

\begin{lem}\label{lem:steps-Bab-first}
    Let $p$ be a state of $T_i$, $i=1,2$. The formulae in $\finlabeli(p)$ are either $\true$, $\ff$, or formulae that start with either $\langle a\rangle$ or $[a]$, $a\in \act$.
    %If $B$ applies steps B(a)--(b) on $p$ until no formula can be replaced in $L_i(p)$, then the formulae in $L_i(p)$ are either $\true$, $\ff$, or formulae that start with either $\langle a\rangle$ or $[a]$, $a\in \act$.
\end{lem}
\begin{proof}
  This is immediate from the definition of moves B($\wedge$) and B($\vee$).
\end{proof}

\begin{defi}\label{def:disj_form}
    Let $\varphi\in\mathL_{\ell S}$, $\ell\geq 1$. The disjunctive form of $\varphi$, denoted  $DF(\varphi)$, is the formula obtained by applying distributivity of conjunction over disjunction to all outermost conjunctions in $\varphi$. A conjunction is said to be outermost if it is not within the scope of any $\langle a\rangle$ or $[a]$, $a\in\act$.
\end{defi}

 For example, $DF(\langle a\rangle (\langle b\rangle \true \wedge (\langle c\rangle \true\vee \langle a\rangle \true)) \wedge ([b]\ff \vee \langle b\rangle \true))=(\langle a\rangle (\langle b\rangle \true \wedge (\langle c\rangle \true\vee \langle a\rangle \true)) \wedge [b]\ff) \vee (\langle a\rangle (\langle b\rangle \true \wedge (\langle c\rangle \true\vee \langle a\rangle \true)) \wedge \langle b\rangle \true)$.

\begin{lem}\label{lem:steps-Bab-second}
Let $p$ be a state of $T_i$, where $i=1,2$, and $\bigvee_{1\leq i\leq m}\varphi_i$ be the disjunctive form of $\bigwedge \inlabeli(p)$. Then $\bigwedge \finlabeli(p)=\varphi_j$ for some $1\leq j\leq m$. Conversely, for every $\varphi_j$, $1\leq j\leq m$, there is a sequence of choices in the application of B($\wedge$) and B($\vee$) such that $\bigwedge \finlabeli=\varphi_j$.
% Let $p$ be a state of $T_i$, where $i=1,2$, and $\bigvee_{1\leq i\leq m}\varphi_i$ be the disjunctive form of $\bigwedge L_i(p)$. If steps B(a)--(b) are applied on $p$ until no formula can be replaced in $L_i(p)$, then $\bigwedge L_i(p)=\varphi_j$ for some $1\leq j\leq m$. Inversely, for every $\varphi_j$, $1\leq j\leq m$, there is a sequence of choices in the application of B(a)--(b) such that after no formula can be replaced in $L_i(p)$, $\bigwedge L_i(p)=\varphi_j$.
\end{lem}
\begin{proof}
    This is immediate from the definition of moves B($\wedge$) and B($\vee$).
\end{proof}

\begin{lem}\label{lem:final-initial-label}
Let $p$ be a state of $T_i$, where $i=1,2$. For every $r\in\proc$, if $r\models\bigwedge\finlabeli(p)$, then $r\models \bigwedge \inlabeli(p)$.
\end{lem}
\begin{proof}
    Let $r\in\proc$ and $\bigvee_{1\leq i\leq m}\varphi_i=DF(\bigwedge \inlabeli(p))$. Lemma~\ref{lem:DF-equiv} implies that $\bigwedge \inlabeli(p)\equiv \bigvee_{1\leq i\leq m}\varphi_i$. Moreover, from Lemma~\ref{lem:steps-Bab-second}, $\bigwedge \finlabeli(p)=\varphi_j$ for some $1\leq j\leq m$. If $r\models \varphi_j$, then $r\models\bigvee_{1\leq i\leq m}\varphi_i$, and equivalently $r\models\bigwedge \inlabeli(p)$.
\end{proof}

% \begin{defi}\label{def:B-plays-consistently}
%     Assume that the \simequiv game is played on $\varphi\in\mathL_{\ell S}$, $\ell\geq 1$. We say that $B$ plays $T_i$, where $i\in\{1,2\}$, consistently on a process $r$ if there is a mapping $map:S_i\rightarrow \proc$ such that the following conditions are satisfied:
%   \begin{enumerate}
%       \item for every $p\in S_i$, $map(p)\models \bigwedge \finlabeli(p)$,
%       %where $L_i(p)$ is the label set of $p$ after steps B(a)--(b) have been applied on $p$,
%       \item for every $(p,p')\in R^i_{a_j}$, $map(p)\myarrowasubj map(p')$, and 
%       \item $map(p_0^i)=r$, where $p_0^i$ is the initial state of $T_i$.
%   \end{enumerate}
% \end{defi}

\begin{defi}\label{def:mapB}
  Assume that the \simequiv game is played on $\varphi\in\mathL_{\ell S}$, $\ell\geq 1$, and $B$ constructs $T_i$, $i=1,2$, such that $\bigwedge \finlabeli(p)$ is satisfiable for every state $p$ added to $S_i$, $i=1,2$. We inductively define the mapping $map_B:S_1\cup S_2\rightarrow \proc$ such that:
  \begin{itemize}
      \item $map_B(p)=\sum_{(p,p')\in R^1_{a_j}} a_j.map_B(p')$, if $p\in S_1$ and $B$ chooses $p$ at step 2 of some round;
      \item $map_B(p)=\sum_{(p,p')\in R^2_{a_j}} a_j.map_B(p')$, if $p\in S_2$ and $A$ chooses $p$ at step 3 of some round;
      \item $map_B(p)=r_p$, where $r_p$ is the least process (w.r.t.\ to depth and size) such that $r_p\models\bigwedge \finlabeli(p)$, otherwise.
  \end{itemize}
\end{defi}

Note that in Definition~\ref{def:mapB}, $map_B$ is well-defined because $\bigwedge \finlabeli(p)$ is satisfiable for every $p\in S_i$ and $i=1,2$.

\begin{lem}\label{lem:mapping-r-simulation}
 Assume that the \simequiv game is played on  $\varphi\in\mathL_{\ell S}$, $\ell\geq 1$ and $B$ constructs $T_i$, $i=1,2$, such that $\bigwedge \finlabeli(p)$ is satisfiable for every state $p$ added to $S_i$, $i=1,2$. Then, $B$ plays $T_i$ consistently on $map_B(p_0^i)$ and $map_B(p_0^i)\models\varphi$, where $i=1,2$.
\end{lem}
\begin{proof}
We prove the lemma for $i=1$; the case  $i=2$ follows by analogous arguments. To this end, we prove that $B$ plays $T_1$ consistently on $map_B(p_0^1)$ by showing that conditions 1 and 2 of Definition~\ref{def:B-plays-consistently} hold for $map_B$ and every $p\in S_1$. Note that condition 3 is trivially true. Let $p\in S_1$. 
\begin{description}
    \item[Condition 2] For every $(p,p')\in R^1_{a_j}$, condition 2 is satisfied by the definition of $map_B(p)$.
    \item[Condition 1] We show by induction on the structure of $map_B(p)$ that for every formula $\phi\in \finlabelone(p)$,  $map_B(p)\models\phi$.  If $B$ does not choose $p$ at step 2 of some round, then this is immediate from the definition of $map_B$. Let $p$ be a state chosen by $B$ at step 2 of some round. Since $B$ does not create any state $s$ such that $\bigwedge \finlabelone(s)$ is unsatisfiable, $\ff\not\in \finlabelone(p)$.
    %after steps B(a)--(b) have been applied on $p$. 
    From Lemma~\ref{lem:steps-Bab-first}, every $\phi\in \finlabelone(p)$ is of the form $\true$, $\langle a\rangle \psi$, or $[a]\psi$, $a\in \act$.
    %after steps B(a)--(b) have been applied on $p$.
    Let $\langle a_j\rangle \psi\in \finlabelone(p)$. Then, there is some $p'$ such that $(p,p')\in R^1_{a_j}$ and $\inlabelone(p')=\{\psi\}\cup\{\psi'\mid [a_j]\psi'\in \finlabelone(p)\}$.
    %before steps B(a)--(b) are applied on $p'$. 
    By the inductive hypothesis, $map_B(p')\models\finlabelone(p')$, and from Lemma~\ref{lem:final-initial-label}, $map_B(p')\models\inlabelone(p')$, which implies that $map_B(p')\models \psi$. Therefore, $map_B(p)\models \langle a_j\rangle\psi$.
    % Let $\Psi$ denote  $\bigwedge \inlabelone(p')$ and
    % %Let $S_{p'}$ denote $\inlabelone(p')$ and $\Psi=\bigwedge S_{p'}$;
    % $DF(\Psi)=\bigvee_{i=1}^m\Psi_i$. From Lemma~\ref{lem:steps-Bab-second}, 
    % %after B(a)--(b) is applied on $p'$ (until no formula can be replaced in $L_1(p')$), 
    % $\bigwedge \finlabelone(p')=\Psi_j$ for some $1\leq j\leq m$. This means that 
    % %after steps B(a)--(b) have been applied on $p'$, 
    % $\bigwedge \finlabelone(p')\models\Psi$ and by the inductive hypothesis, $map_B(p')\models\bigwedge \finlabelone(p')$. Therefore, $map_B(p')\models\Psi$ and so $map_B(p')\models \psi$. Thus, $map_B(p)\models \langle a_j\rangle\psi$.
    For $[a_j]\psi\in L_1(p)$, we can prove in a similar way that $map_B(p)\models [a_j]\psi$. It trivially holds that $map_B(p)\models\true$.
\end{description}
Finally, we show that $map_B(p_0^1)\models\varphi$. We have that $map_B(p_0^1)\models \bigwedge \finlabelone(p_0^1)$ because condition 1 holds for $p_0^1$. From Lemma~\ref{lem:final-initial-label}, $map_B(p_0^1)\models \bigwedge \inlabelone(p_0^1)$, where $\inlabelone(p_0^1)=\{\varphi\}$.
% Let $DF(\varphi)=\bigvee_{i=1}^m \varphi_i$. From Lemma~\ref{lem:steps-Bab-second}, $\bigwedge L_1(p_0^1)=\varphi_i$ for some $1\leq i\leq m$, which implies that $\bigwedge L_1(p_0^1)\models\varphi$ and therefore, $map_B(p_0^1)\models\varphi$.
\end{proof}

\begin{lem}\label{lem:bounded-depth}
    Let $\varphi\in\mathL_{\ell S}$, $\ell\geq 1$, and $p_1,p_2$ be such that $p_1\models\varphi$ and $p_2\models\varphi$. If for every $p,q$, $p\models\varphi$ and $q\models\varphi\Rightarrow p\equiv_S q$, then $\depth(p_1)=\depth(p_2)\leq\md(\varphi)$.
\end{lem}
\begin{proof}
    % Let $\varphi$ be as in the hypothesis of the lemma. Then, from Propositions~\ref{tableausat} and~\ref{prop:tableau-construction} and the tableau construction described in Subsection~\ref{subsec:tableau-construction}, there is a finite process that satisfies $\varphi$. Let $p_1,p_2$ be two finite processes that satisfy $\varphi$. Using the definition of $\curle_S$, it is not hard to see that  $\depth(p_1)>\depth(p_2)\Rightarrow p_1\not\curle_S p_2$, contradiction.
    We start with the following claim, which is straightforward from the definition of $\curle_S$.
    \begin{clm}\label{claim:not-sim}
      If $\depth(r_1)>\depth(r_2)$, then $r_1\notcurle_S r_2$.
    \end{clm}
    \begin{proof}
      Immediate from the definition of $\curle_S$.
    \end{proof}
    Thus, $\depth(p_1)=\depth(p_2)$ follows immediately from the hypothesis of the lemma and Claim~\ref{claim:not-sim}. Assume, towards a contradiction,  that $\depth(p_i)>\md(\varphi)$ for both $i=1,2$. In the sequel, we need to be able to trim a process $r$ up to depth $d$, that is, we discard the processes reachable from $r$ that have depth greater than $d$, where $d$ is to be determined. To this end, we define process $\trim(r,d)$ for every process $r$ and $d\in\mathbb{N}$, as follows.
   $$\trim(r,d)=\begin{cases}
       r, &\text{if } d\geq \depth(r),\\
       \mathtt{0}, &\text{if } d=0,\\
       \sum_{\substack{a\in\act\\r\myarrowa r'}} a.\trim(r',d-1), &\text{otherwise.}
   \end{cases}$$
  We prove the following claims and use them to show the lemma.
  \begin{clm}\label{claim:trim-depth}
   If $\depth(r) > d$, then $\depth(\trim(r,d))=d$.
  \end{clm}
  \begin{proof}
      Immediate from the definition of $\trim(r,d)$.
  \end{proof}
  \begin{clm}\label{claim:trim-models-phi}
      If $r\models\varphi$, then $\trim(r,d)\models\varphi$, for every $d\geq \md(\varphi)$.
  \end{clm}
  \begin{proof}
     If $\depth(r)\leq d$, then by definition, $\trim(r,d)=r$ and the claim is true. Let $\depth(r)> d$. We prove the lemma by induction on $\varphi$.
      \begin{itemize}
          \item If $\varphi=\true$, then $\trim(r,d)\models\varphi$.
          \item If $\varphi=[a]\varphi'$ for some $a\in\act$, then 
          \begin{itemize}
              \item If $\trim(r,d)\notmyarrowa$, then $\trim(r,d)\models [a]\varphi'$.
              \item Let $\trim(r,d)\myarrowa p$. Then, $p=\trim(r',d-1)$, for some $r\myarrowa r'$, and $d-1\geq \md(\varphi')$. From the inductive hypothesis, $\trim(r',d-1)\models \varphi'$. Therefore, $\trim(r,d)\models [a]\varphi'$.
          \end{itemize}
          \item If $\varphi=\langle a\rangle\varphi'$ for some $a\in\act$, then the proof is similar to the case $\varphi=[a]\varphi'$.
          \item If $\varphi=\varphi_1\wedge\varphi_2$ or $\varphi=\varphi_1\vee\varphi_2$, the claim follows immediately from the inductive hypothesis. \qedhere
      \end{itemize}
      \end{proof}
  From Claims~\ref{claim:trim-depth} and~\ref{claim:trim-models-phi}, $\trim(p_1,\md(\varphi))\models\varphi$ and $\depth(\trim(p_1,\md(\varphi)))=\md(\varphi)$.  This implies that $\depth(p_1)\neq \depth(\trim(p_1,\md(\varphi)))$, which contradicts the fact that $p_1\equiv_S \trim(p_1,\md(\varphi))$ and so $\depth(p_1)$ $= \depth(\trim(p_1,\md(\varphi)))$ from Claim~\ref{claim:not-sim}.
\end{proof}

\begin{lem}\label{lem:B-stops}
     Let $\varphi\in\mathL_{\ell S}$, $\ell\geq 1$, be satisfiable and assume that $p_1\equiv_S p_2$ holds for all processes $p_1,p_2$ that satisfy $\varphi$. Then,  $B$ stops generating states after  at most $\md(\varphi)+1$ rounds
     %a finite number of rounds 
     in the \simequiv game on $\varphi$.
\end{lem}
\begin{proof}
   If $B$ generates a state $p\in S_i$, $i=1,2$, such that $\bigwedge \finlabelone(p)$ is unsatisfiable, the game stops and the lemma is trivially true. If $B$ does not generate any state $p\in S_i$ with an unsatisfiable $\bigwedge \finlabelone(p)$, then the lemma is immediate from Lemmas~\ref{lem:mapping-r-simulation} and~\ref{lem:bounded-depth}.
\end{proof}

\begin{prop}\label{prop:simulation-right-to-left}
    Let $\varphi\in\mathL_{\ell S}$, $\ell\geq 1$. Assume that for every $p_1,p_2$ such that $p_1\models\varphi$ and $p_2\models\varphi$, it holds that $p_1\equiv_S p_2$. Then, $A$ has a winning strategy for the \simequiv game on $\varphi$.
\end{prop}
\begin{proof}
  Assume that the \simequiv game is played on $\varphi$ and consider the processes $map_B(p_0^1)$ and $map_B(p_0^2)$. From Lemma~\ref{lem:mapping-r-simulation}, $B$ plays $T_i$ consistently on $map_B(p_0^i)$ and $map_B(p_0^i)\models\varphi$, for both $i=1,2$. We prove that $A$ can play in such a way that the following condition is true for every round $l\geq 1$.
    \begin{quote}
        \textit{Condition C:} `The round starts with $p_1$, $p_2$ such that $map_B(p_1)\curle_S map_B(p_2)$ and unless $B$ loses, it ends with $p_1'$, $p_2'$ such that $map_B(p_1')\curle_S map_B(p_2')$.'
    \end{quote}
    Note that $map_B(p_0^i)\models\varphi$ for both $i=1,2$, and so, from the hypothesis of the proposition, $map_B(p_0^1)\equiv_S map_B(p_0^2)$. The first round starts and ends with $p_0^1$ and $p_0^2$ so condition C is true for $l=1$. Let $l\geq 2$ and assume that condition C is true for round $l-1$. Then, round $l$ starts with the states that round $l-1$ ended with, so the first part of the condition is true for round $l$. If $B$ does not lose at round $l$, then he successfully chooses a state $p_1'$ such that $(p_1,p_1')\in R^1_{a_j}$ at step 2. From Lemma~\ref{lem:mapping-r-simulation}, $map_B(p_1)\myarrowasubj map_B(p_1')$. Since $map_B(p_1)\curle_S map_B(p_2)$, there is $r_{p_2}$ such that $map_B(p_2)\myarrowasubj r_{p_2}$ and $map_B(p_1')\curle_S r_{p_2}$. From Definition~\ref{def:mapB}, there is $p_2'\in S_2$ such that $(p_2, p_2')\in R^2_{a_j}$ and $map_B(p_2')=r_{p_2}$. Player $A$ chooses $p_2'$ and the round ends as the condition describes.

    Since $A$ can play so that every round satisfies condition C, player $A$ can play in such a way that the game stops only if $B$ loses. From Lemma~\ref{lem:B-stops}, after at most $\md(\varphi)+1$ rounds,
    %a finite number of rounds 
    $B$ will not generate any successors at moves B($\Diamond$) and B($\square$) and so he will lose.
\end{proof}

Next, we prove that if there are two processes $r_1, r_2$ such that $r_1\models\varphi$, $r_2\models\varphi$, and $r_1\not\curle_S r_2$, then $B$ has a winning strategy for the \simequiv game on $\varphi$.

\begin{lem}\label{lem:B-can-play-consistently}
    Assume that the \simequiv game is played on $\varphi\in\mathL_{\ell S}$, $\ell\geq 1$. Let $r$ be a process that satisfies $\varphi$ and $i\in\{1,2\}$. Then, there is a strategy of $B$ that allows him to play $T_i$ consistently on $r$.
\end{lem}
\begin{proof} 
When we refer to conditions 1--3, we mean conditions 1--3 in  Definition~\ref{def:B-plays-consistently}. Let $map:S_i\rightarrow \proc$ be such that $map(p_0^i)=r$, and so  condition 3 is satisfied. We prove that $B$ can play so that (a) condition 1 is true for  $p_0^i$ and (b) if condition 1 is true for $p\in S_i$, then for every $(p,p')\in R^i_{a_j}$ conditions 1--2 are true for $p'$.
    \begin{enumerate}[(a)]
        \item Condition 1 holds for $p_0^i$: When the game starts, $\inlabeli(p_0^i)=\{\varphi\}$. Let $DF(\varphi)=\bigvee_{i=1}^m\varphi_i$, $m\in\mathbb{N}$. From Lemma~\ref{lem:DF-equiv}, $\varphi\equiv\bigvee_{i=1}^m\varphi_i$, and so $r\models\varphi_j$ for some $1\leq j\leq m$. From Lemma~\ref{lem:steps-Bab-second}, $B$ can apply moves B($\wedge$) and B($\vee$) on $p_0^i$ such that
        %after no formula can be replaced in $L_i(p_0^i)$, 
        $\bigwedge \finlabeli(p_0^i)=\varphi_j$. Then, $map(p_0^i)=r\models\bigwedge \finlabeli(p_0^i)$.
        %where $L_i(p_0^i)$ is the label set of $p_0^i$ after steps B(a)--(b) have been applied on $p_0^i$. 
        \item Assume that $p\in S_i$ and that condition 1 holds for $p$. Then, $B$  generates successors of $p$ only when he applies move B($\Diamond$) on $p$ and not during move B($\square$) on $p$.
        %at step 1(c) and chooses not to generate any successor at step 1(d). 
        We show that for every $(p,p')\in R^i_{a_j}$, $map(p)\myarrowasubj map(p')$ and condition 1 is true for $p'$. Let $(p,p')\in R^i_{a_j}$. Then, there is some formula $\langle a_j\rangle \psi\in \finlabeli(p)$, 
        %where $L_i(p)$ is the label set of $p$ after steps B(a)--(b) have been applied on $p$ (or just before the successors of $p$ are generated by $B$), 
        such that $\inlabeli(p')=\{\psi\}\cup\{\psi'\mid [a_j]\psi'\in \finlabeli(p)\}$.
        %when $p'$ is generated.
        Let
        %$\psi\wedge\bigwedge_{[a_j]\psi'\in L_2(p)}\psi'$
        $DF(\bigwedge\inlabeli(p'))=\bigvee_{i=1}^n\Psi_i$. From Lemma~\ref{lem:DF-equiv}, we know that $\bigwedge\inlabeli(p')\equiv \bigvee_{i=1}^n\Psi_i$. Since condition 1 holds for $p$, $map(p)\models \bigwedge \finlabeli(p)$, which implies that $map(p)\models\langle a_j\rangle\bigwedge\inlabeli(p')$, and so there is $map(p)\myarrowasubj r_p$ such that $r_p\models \bigwedge\inlabeli(p')$. Therefore,  $r_p\models\Psi_j$ for some $1\leq j\leq n$. Then, $B$ can apply moves B($\wedge$) and B($\vee$) on $p'$ in such a way that $\bigwedge \finlabeli(p')=\Psi_j$ 
        %after no formula can be replaced in $L_i(p')$. 
        Thus, we can set $map(p')=r_p$ and condition 1 holds for $p'$.
    \end{enumerate}
From (a) and (b) above, $B$ can play in such a way that there is a mapping $map:S_i\rightarrow \proc$ which satisfies conditions 1--3.
\end{proof}

\begin{lem}\label{lem:B-on-first-process}
 Assume that the \simequiv game is played on $\varphi\in\mathL_{\ell S}$, $\ell\geq 1$. Let $r$ be a process that satisfies $\varphi$ and $r\myarrowasubone r^1\myarrowasubtwo r^2\dots\myarrowasubm r^m$, $m\in\mathbb{N}$, be a trace. Then, there is a strategy of $B$ that allows him to choose state $p_l^1\in S_1$ at step 2 of round $l+1$, for every $1\leq l\leq m$, such that 
there is a mapping $g:\{p_0^1,\dots,p_m^1\}\rightarrow \proc$ satisfying the following conditions:
\begin{enumerate}[i.]
    \item  $g(p_0^1)=r$ and  $g(p_l^1)=r^l$ for every $1\leq l\leq m$, 
    \item $(p_{l-1}^1,p_l^1)\in R_{a_l}$ for every $1\leq l\leq m$, and
    \item  $g(p_l^1)\models \bigwedge \finlabeli(p_l^1)$ for every $0\leq l\leq m$.
    %where $L_1(p_l^1)$ is the label set of $p_l^1$ after steps B(a)--(b) are applied on $p_l^1$.
\end{enumerate}
We say that the choices of $B$ (at the first $m+1$ rounds) are consistent with trace $r\myarrowasubone r^1\myarrowasubtwo r^2\dots\myarrowasubm r^m$.
\end{lem}
\begin{proof}
  Let $DF(\varphi)=\bigvee_{i=1}^n\varphi_i$. Then, $r\models\varphi_j$ for some $1\leq j\leq n$. From Lemma~\ref{lem:steps-Bab-second}, $B$ can apply moves B($\wedge$) and B($\vee$) such that 
  %after no formula can be replaced in $L_1(p_0^1)$, it holds that 
  $\bigwedge \finlabelone(p_0^1)=\varphi_j$, and so conditions 1 and 2 are true for $p_0^1$. Assume that the choices of $B$ at the first $k+1$ rounds are consistent with the trace $r\myarrowasubone r^1\dots\myarrowasubk r^k$, $k<m$. We show that $B$ can play such that his choices (at the first $k+2$ rounds) are consistent with the trace $r\myarrowasubone r^1\dots\myarrowasubkplusone r^{k+1}$. When $B$ applies move B($\square$) on $p_1^k$ at round $k+2$, he chooses to generate an $i_{(k+1)}$-successor of $p_1^k$ with $\inlabelone(p^1_{k+1})=\{\psi\mid [a_{i_{(k+1)}}]\psi\in \finlabelone(p^1_k)\}$. Let $DF(\bigwedge \inlabelone(p_{k+1}^1))=\bigvee_{i=1}^{n'}\Psi_i$. From Lemma~\ref{lem:DF-equiv}, we know that $\bigwedge \inlabelone(p_{k+1}^1)\equiv \bigvee_{i=1}^{n'}\Psi_i$.  From the inductive hypothesis $g(p_k^1)=r^k\models\bigwedge \finlabelone(p_k^1)$ and since $r^k\myarrowasubkplusone r^{k+1}$, we have that $r^{k+1}\models\bigwedge \inlabelone(p_{k+1}^1)$, and so $r^{k+1}\models\Psi_j$ for some $1\leq j\leq n'$.
  %and so  $\bigwedge \inlabelone(p_{k+1}^1)$ is satisfiable.
  %from Corollary~\ref{cor:prop-consistent-sat}. 
  %Let $S$ denote $L_1(p_{k+1}^1)$ and $\Psi$ denote $\bigwedge S$; 
  % Then, $r^k\models [a_{i_{(k+1)}}]\psi$ for every $[a_{i_{(k+1)}}]\psi\in \finlabelone(p_k^1)$, which implies that $r^{k+1}\models\psi$ for every  $[a_{i_{(k+1)}}]\psi\in L_1(p_k^1)$ or equivalently, for every $\psi\in S$. Consequently, $r^{k+1}\models \Psi$ and so $r^{k+1}\models\Psi_j$ for some $1\leq j\leq n'$.
  $B$ can apply moves B($\wedge$) and B($\vee$) on $p_{k+1}^1$ such that 
  %after no formula can be replaced in $L_1(p_{k+1}^1)$, it holds that 
  $\bigwedge \finlabelone(p_{k+1}^1)=\Psi_j$. Thus, $B$ can choose $p_{k+1}^1$ at step 2 of round $k+2$. The mapping $g$ can be extended such that $g(p_{k+1}^1)=r^{k+1}$ so that conditions i--iii  are satisfied for $p_{k+1}^1$.
 \end{proof}

 \begin{lem}\label{lem:finite_processes}
     Let $\varphi\in\mathL_{\ell S}$, $\ell\geq 1$, such that there are two processes $p_1,p_2$ that both satisfy $\varphi$ and $p_1\not\curle_S p_2$. Then, there are $q_1,q_2\in\proc$ such that $q_1\not\curle_S q_2$, $q_i\models\varphi$, and $\depth(q_i)\leq\md(\varphi)+1$ for both $i=1,2$.
 \end{lem}
 \begin{proof}
  Let $p_1,p_2$ be as described in the lemma. If $\depth(p_i)\leq \md(\varphi)+1$ for both $i=1,2$, then the lemma holds. Assume that some $p_i$ has depth greater than $\md(\varphi)+1$ for some $i=1,2$.
  To prove the lemma, we make use of the definition of $\trim(r,d)$ for every process $r$ and $d\in\mathbb{N}$, introduced in the proof of Lemma~\ref{lem:bounded-depth}, and the claims following it.
  
  We distinguish the following cases.
   \begin{description}
      \item[Case 1: $\depth(p_i)\geq \md(\varphi)+1$ for both $i=1,2$]  Let $q_1=\trim(p_1,\md(\varphi)+1)$ and $q_2=\trim(p_2,\md(\varphi))$. We have that $\depth(q_1)=\md(\varphi)+1$ and $\depth(q_2)=\md(\varphi)$ from the definition of $\trim(r,d)$ and Claim~\ref{claim:trim-depth}. Moreover, $q_1\not\curle_S q_2$ because of Claim~\ref{claim:not-sim}. Finally, $q_i\models\varphi$, for both $i=1,2$, from Claim~\ref{claim:trim-models-phi}.
      \item[Case 2: $\depth(p_1)\leq \md(\varphi)+1$ and $\depth(p_2)>\md(\varphi)+1$] Let $q_1=\trim(p_1,\md(\varphi))$ and $q_2=\trim(p_2,$ $\md(\varphi)+1)$. Similarly to the previous case, we can show that the lemma holds for $q_1$ and $q_2$.
      \item[Case 3: $\depth(p_1)>\md(\varphi)+1$ and $\depth(p_2)\leq \md(\varphi)+1$] Let $q_1=\trim(p_1,\md(\varphi)+1)$ and $q_2=\trim(p_2,\md(\varphi))$. Again, we can show similarly to the first case, that $q_1$ and $q_2$ satisfy the lemma.\qedhere
  \end{description}
  %  We distinguish the following cases.
  % \begin{description}
  %     \item[Case 1: $p_i$ is infinite for both $i=1,2$.]  Let $q_1=\trim(p_1,\md(\varphi)+1)$ and $q_2=\trim(p_2,\md(\varphi)$. We have that $\depth(q_1)=\md(\varphi)+1$ and $\depth(q_2)=\md(\varphi)$ from  Claim~\ref{claim:trim-depth}. Therefore, $q_1\not\curle_S q_2$ because of Claim~\ref{claim:not-sim}. Finally, $q_i\models\varphi$, for both $i=1,2$, from Claim~\ref{claim:trim-models-phi}. 
  %     % Finally, from Claim~\ref{claim:trim-path}, $q_1=\trim(p_1^0,m+1)\myarrowasubone \trim(p_1^1,m)\dots\myarrowasubk \trim(p_1^k,m+1-k)\myarrowa$ and $q_2=\trim(p_2^0,m)\myarrowasubone \trim(p_2^1,m-1)\dots\myarrowasubk \trim(p_2^k,m-k)\notmyarrowa$, since $m\geq k+1$.
  %     \item[Case 2: $p_1$ is infinite and $p_2$ is finite.] Let $q_1=\trim(p_1,\md(\varphi)+1)$ and $q_2=\trim(p_2,\md(\varphi))$. Similarly to the previous case, we can show that the lemma holds for $q_1$ and $q_2$.
  %     \item[Case 3: $p_1$ is finite and $p_2$ is infinite.] Let $q_1=\trim(p_1,\md(\varphi))$ and $q_2=\trim(p_2,\md(\varphi)+1)$. Again, we can show, similarly to the first case, that $q_1$ and $q_2$ satisfy the lemma.\qedhere
  % \end{description}   
 \end{proof}

\begin{prop}\label{prop:simulation-left-to-right}
    Let $\varphi\in\mathL_{\ell S}$, $\ell\geq 1$. Assume that there are $r_1,r_2$ such that $r_1\models \varphi$, $r_2\models\varphi$, and $r_1\not\curle_S r_2$. Then, $B$ has a winning strategy for the \simequiv game on $\varphi$.
\end{prop}
\begin{proof}
    Let $r_1, r_2$ be two processes that satisfy $\varphi$ and $r_1\not\curle_S r_2$. From Lemma~\ref{lem:finite_processes}, we can assume that $\depth(r_i)\leq\md(\varphi)+1$, $i=1,2$. Let $B$ play as follows. $B$  plays $T_2$ consistently on $r_2$, which is possible from Lemma~\ref{lem:B-can-play-consistently}. Regarding $T_1$, $B$ makes the choices that would allow him to play $T_1$ consistently on $r_1$, described in the proof of Lemma~\ref{lem:B-can-play-consistently}, and, in addition, he applies move B($\square$) and picks states at step 2 so that after every round $k$, his choices are consistent with a trace $r_1\myarrowasubone r_1^1\cdots\myarrowasubkminusone r_1^{k-1}$. Moreover, if $p_1^1,\dots,p_{k-1}^1$ are the choices of $B$ and $p_1^2,\dots,p_{k-1}^2$ are the choices of $A$ at steps 2 and 3, respectively, of the first $k$ rounds, then $g(p_i^1)\not\curle_S map(p_i^2)$ for every $0\leq i\leq k-1$. 
    
    Before round 2 starts, $B$ can play such that there is a mapping $map$ that satisfies conditions 1--3 of Definition~\ref{def:B-plays-consistently} for $p_0^2$ because of Lemma~\ref{lem:B-can-play-consistently}. From the proof of Lemma~\ref{lem:B-on-first-process}, he can play such that there is a mapping $g$ which satisfies conditions i--iii of Lemma~\ref{lem:B-on-first-process} for $p_0^1$. Then, $g(p_0^1)\not\curle_S map(p_0^2)$ since $g(p_0^1)=r_1$, $map(p_0^2)=r_2$, and $r_1\not\curle_S r_2$. Assume that the following condition is true for some $k\in\mathbb{N}$.
     \begin{quote}
      \textit{Condition C:} At the first $k$ rounds, player $B$ can follow 
       a strategy that allows him to play $T_2$ consistently on $r_2$, make  choices that are consistent with the trace $r_1\myarrowasubone r_1^1\cdots\myarrowasubkminusone r_1^{k-1}$ and $g(p_i^1)\not\curle_S map(p_i^2)$ for every $0\leq i\leq k-1$, where  $p_1^1,\dots,p_{k-1}^1$ are the choices of $B$ at step 2 and $p_1^2,\dots,p_{k-1}^2$ are the choices of $A$ at step 3 of rounds $2,\dots,k$, respectively.   
     \end{quote}
We show that condition C is also true for $k+1$ unless $A$ loses at round $k+1$. $B$ plays the first $k$ rounds such that condition C is satisfied for $k$ and he plays as follows at round $k+1$. He continues playing $T_2$ consistently on $r_2$ as described in the proof of Lemma~\ref{lem:B-can-play-consistently}. Since $g(p_{k-1}^1)\not\curle_S map(p_{k-1}^2)$, there is $a_{i_k}\in\act$ and $r_k^1$ such that
\begin{itemize}
    \item either $g(p_{k-1}^1)\myarrowasubk r_k^1$ and $map(p_{k-1}^2)\notmyarrowasubk$, 
    \item or $g(p_{k-1}^1)\myarrowasubk r_k^1$ and for every $map(p_{k-1}^2)\myarrowasubk r_k^2$, $r_k^1\not\curle_S r_k^2$.
\end{itemize}  
Then, $B$ can make choices that are consistent with the trace $r_1\myarrowasubone r_1^1\cdots\myarrowasubkminusone r_1^{k-1}\myarrowasubk r_k^1$ from Lemma~\ref{lem:B-on-first-process}. As described in the proof of  Lemma~\ref{lem:B-on-first-process}, $B$ generates an $i_k$-successor $p_k^1$ of $p_{k-1}^1$ when he plays move B($\square$) on $p_{k-1}^1$ such that $g$ can be extended by mapping $p_k^1$ to $r_1^k$ and makes a sequence of choices when applying moves B($\wedge$) and B($\vee$) such that conditions i--iii of Lemma~\ref{lem:B-on-first-process} are satisfied. Then, $B$ chooses $p_k^1$ at step 2. If $A$ successfully chooses an $i_k$-successor $p_k^2$ of $p_{k-1}^2$, then from Definition~\ref{def:B-plays-consistently}, $map(p_{k-1}^2)\myarrowasubk map(p_k^2)$ because $(p_{k-1}^2,p_k^2)\in R^2_{a_{i_k}}$. This means that $r_k^1\not\curle_S map(p_k^2)$ because $r_k^1$ was specifically chosen so that it is not simulated by any of the $i_k$-successors of  $map(p_{k-1}^2)$. Since $g(p_k^1)=r_k^1$, $g(p_k^1)\not\curle_S map(p_k^2)$.

If $B$ follows the strategy described above, the game can only stop if $A$ loses. Since $\depth(r_i)\leq\md(\varphi)+1$ for both $i=1,2$, and $r_1\not\curle_S r_2$, there is some $m\leq\md(\varphi)$ such that the choices of $B$ are consistent with $r_1\myarrowasubone r_1^1\myarrowasubtwo\cdots\myarrowasubm r_1^{m}$ and there is some $a\in \act$ such that  $r_1^m\myarrowa$, $map(p_m^2)\notmyarrowa$. Then, $B$ can easily play such that $A$ loses at step 3 of round $m+2$.
\end{proof}

\begin{prop}\label{prop:simul-game}
     Let $\varphi\in\mathL_{\ell S}$, where $\ell\geq 1$, be a satisfiable formula. Player $A$ has a winning strategy for the \simequiv game on $\varphi$ iff  $r_1\equiv_S r_2$, for every two processes $r_1,r_2$ that satisfy $\varphi$.
    \end{prop}
\begin{proof}
   Note that the proofs of Propositions~\ref{prop:simulation-right-to-left} and~\ref{prop:simulation-left-to-right} imply that the game always terminates 
   %no later than the 
   %($\md(\varphi)+2$)-th round. 
   in at most $\md(\varphi)+2$ rounds. By induction, it can be proven that either $A$ has a winning strategy or $B$ has a winning strategy, which is a standard proof for zero-sum, two-player games. Then, the proposition is immediate from Propositions~\ref{prop:simulation-right-to-left} and~\ref{prop:simulation-left-to-right}.  
\end{proof}

 \begin{prop}\label{prop:sim-win-algo}
    Let $\varphi\in\mathL_{\ell S}$, $\ell\geq 1$, be a satisfiable formula. Deciding whether every two processes $p_1,p_2$ that satisfy $\varphi$ are equivalent modulo $\equiv_S$ can be done in polynomial space.
    \end{prop}
     \begin{proof}
     From Proposition~\ref{prop:simul-game}, it suffices to show that determining whether player $A$ has a winning strategy for the \simequiv game on $\varphi$ can be done in polynomial space. The \simequiv game is a zero-sum and perfect-information game. It is also of polynomial depth, since it stops after at most $\md(\varphi)+2$ rounds. Finally, in every round, the satisfiability of $\bigwedge L_i(s)$ has to be checked a polynomial number of times. If $\varphi\in\mathL_S$, then Corollary~\ref{cor:sat-s-cs-rs-poly}(a) yields that the game is computationally bounded. If $\varphi\in\mathL_{\ell S}$,  for some $\ell\geq 2$, then the \simequiv game is a game with a \pspace oracle by  Corollaries~\ref{cor:sat-s-cs-rs-poly}(a) and~\ref{prop:n-s-sat}. The desired conclusion then follows from Corollary~\ref{cor:pspace-games}.
     % The proposition follows from Proposition~\ref{prop:pspace-ap} and Corollary~\ref{cor:ap-pspace-oracle}. We provide a brief sketch of how the proposition can be established.
     % There is an alternating algorithm $\onesimwin(\{\varphi\},\{\varphi\})$ that determines whether $A$ has a winning strategy for the \simequiv game on $\varphi$ by simulating $B$'s choices universally and $A$'s choices existentially. The algorithm needs polynomial time to simulate each choice of the players, and uses an oracle to decide whether $\bigwedge L_i(s)$ is satisfiable when this is needed. From the proof of Proposition~\ref{prop:simul-game}, the game stops after at most $\mod(\varphi)+2$ rounds, i.e.\ this is a game of polynomial depth.
     % Consequently, algorithm $\onesimwin(\{\varphi\},\{\varphi\})$ can be implemented by an alternating polynomial-time Turing machine that uses polynomially many calls to a \pspace oracle, and therefore, by a polynomial-space Turing machine.    
     \end{proof}

     By determining whether $A$ has a winning strategy for the game $\simab(U_1,U_2)$, where $U_1,U_2$ are two finite subsets of formulae, we can decide whether every process that satisfies $\bigwedge U_1$ is simulated by any process that satisfies $\bigwedge U_2$. Therefore, the latter problem also lies in \pspace. 

     \begin{cor}\label{cor:simul-game}
   Let $U_1,U_2$ be finite subsets of $\mathL_{\ell S}$, $\ell\geq 1$. Player $A$ has a winning strategy for $\simab(U_1,U_2)$ iff $p_1\curle_S p_2$, for every $p_1,p_2$ such that $p_1\models\bigwedge U_1$ and $p_2\models\bigwedge U_2$.
    \end{cor}

\begin{cor}\label{cor:simul-game-complexity}
    Let $U_1,U_2$ be finite subsets of $\mathL_{\ell S}$, $\ell\geq 1$. Deciding whether $p_1\curle_S p_2$ is true for every two processes $p_1,p_2$ such that $p_1\models\bigwedge U_1$ and $p_2\models\bigwedge U_2$ can be done in polynomial space.
    \end{cor}
\begin{proof}
From Corollaries~\ref{cor:sat-s-cs-rs-poly} and~\ref{prop:n-s-sat}, we can decide whether $\bigwedge U_1$ and $\bigwedge U_2$ are satisfiable in polynomial space. If one of them is unsatisfiable, then the statement of the corollary is trivially true. If both are satisfiable, let the \simequiv game start with $S_1=\{s_1\}$, $S_2=\{s_2\}$, $L_1(s_1)=U_1$,  $L_2(s_2)=U_2$, and $R^i_{a_j}$ empty for every $i=1,2$ and $1\leq j\leq k$. By Corollary~\ref{cor:simul-game}, player $A$ has a winning strategy  for $\simab(U_1,U_2)$ iff for every two processes $p_1,p_2$ such that $p_1\models\bigwedge U_1$ and $p_2\models\bigwedge U_2$, $p_1\curle_S p_2$ holds. We can determine whether $A$ has a winning strategy for $\simab(U_1,U_2)$ in polynomial space by Corollary~\ref{cor:pspace-games}, since the game has all the characteristics required in the corollary.
\end{proof}

\paragraph{The \nsimeq game, \texorpdfstring{$n\geq 2$}{n>=2}}
Let $n\geq 2$. We denote by $\isimab(U_1,U_2)$, where $i\geq 2$, the \isimeq  that starts with  $S_1=\{s_1\}$, $S_2=\{s_2\}$, $R_{a_j}^t$ being empty for every $t=1,2$ and $1\leq j\leq k$, and $L_1(s_1)=U_1$, $L_2(s_2)=U_2$ with $U_1,U_2$ being finite subsets of $\mathL_{\ell S}$, $\ell\geq i$. Specifically, $\isimab(\{\varphi\},\{\varphi\})$ is called the \isimeq game on $\varphi$. We say that the \isimeq game is correct if Propositions~\ref{prop:simul-game} and~\ref{prop:sim-win-algo} and Corollaries~\ref{cor:simul-game} and~\ref{cor:simul-game-complexity} hold, when $\curle_S$ and $\equiv_S$ are replaced by $\curle_{iS}$ and $\equiv_{iS}$, respectively, $\mathL_{\ell S}$, with $\ell\geq 1$, by $\mathL_{\ell S}$, with $\ell\geq i$, and $\simab$ by $\isimab$. Assume that the \nmosimeq game has been defined so that it is played by players $A$ and $B$ and it is correct.

We can now describe the \nsimeq game on $\varphi$. 
%and prove that it is also correct.
Each round of the \nsimeq game on $\varphi$ follows the steps of the respective round of the \simequiv game on $\varphi$ and includes some additional steps. Analogously to the \simequiv game, if $A$ wins the \nsimeq game on $\varphi$, then the labelled trees $T_1, T_2$, constructed during the game, will correspond to two processes  $p_1,p_2$ such that $p_i\models \varphi$ for both $i=1,2$, and $p_1\curle_{nS} p_2$. By the definition of $\curle_{nS}$, a necessary condition for $p_1\curle_{nS}p_2$ is $p_1\equiv_{(n-1)S} p_2$. This fact is the intuition behind the step preceding the first round and steps 5--6 of the game described below.
%after $B$ and $A$ have chosen two states $p_1',p_2'$ among the ones that $B$ has just generated, they play the \nmosimeq game twice, once starting with state $s_1$ labelled $L_1(p_1')$, and $s_2$ labelled $L_2(p_2')$ and a second time starting with $s_1$ labelled $L_2(p_2')$ and $s_2$ labelled $L_1(p_1')$. If $A$ wins both these sub-games, then the \nsimeq game continues on $p_1',p_2'$, otherwise $A$ loses the \nsimeq game. 
%Moreover, when the \nsimeq game on $\varphi$ starts and before the first round begins, $A$ and $B$ play the \nmosimeq game on $\varphi$. If $A$ loses, then she loses the \nsimeq game. Otherwise, the game resumes. 

The \nsimeq game on $\varphi$ starts with $A$ and $B$ playing the \nmosimeq game on $\varphi$. If $A$ wins, the \nsimeq game resumes. Otherwise, $A$ loses the \nsimeq game on $\varphi$. During the game, $B$ constructs two labelled trees, denoted  $T_1=(S_1,L_1,R^1_{a_1},\dots,R^1_{a_k})$ and $T_2=(S_2,L_2,R^2_{a_1},\dots,R^2_{a_k})$. The first round starts with $S_1=\{p_0^1\}$, $S_2=\{p_0^2\}$, $L_1(p_0^1)=L_2(p_0^2)=\{\varphi\}$, and all $R^i_{a_j}$ being empty, $i=1,2$, and is the same as the first round of the \simequiv game. For $l\geq 2$, the $l$-th round of the game includes steps 1--4 of the $l$-th round of the \simequiv game together with the following steps.
\begin{enumerate}\setcounter{enumi}{4}
    \item $A$ and $B$ play two versions of the \nmosimeq game:  $\nmosimab(L_1(p_1'),L_2(p_2'))$ and $\nmosimab(L_2(p_2'),$ $L_1(p_1'))$.
    \item If $A$ wins both versions of the \nmosimeq game at step 5,  round $l+1$ of the \nsimeq game starts on $p_1',p_2'$. Otherwise, $A$ loses.
\end{enumerate}

The following propositions and corollaries are adaptations of Propositions~\ref{prop:simul-game} and~\ref{prop:sim-win-algo}, as well as Corollaries~\ref{cor:simul-game} and~\ref{cor:simul-game-complexity}, to the case of \nsimeq game. The \nsimeq game is correct since these statements hold.

\charmodns*
\noindent\textcolor{darkgray}{\textit{\textbf{Proof sketch.}}}
    Because of the correctness of the \nmosimeq game, if every $r_1,r_2$ that satisfy $\varphi$ are equivalent modulo $\equiv_{nS}$, and so equivalent modulo $\equiv_{(n-1)S}$, then $A$ has a winning strategy for the \nmosimeq game on $\varphi$ that is played at the beginning of the game. Similarly to the \simequiv game on $\varphi$, in every round, $A$ can always choose a state $p_2'$ such that the process corresponding to $p_1'$ is $n$-nested-simulated by the process corresponding to $p_2'$ where these processes are defined through the mapping $map_B$ exactly as in the proof of Lemma~\ref{lem:mapping-r-simulation}. From the assumption that the \nmosimeq game is correct, player $A$ also has a winning strategy for the two versions of the \nmosimeq game played at step 5 of each round. So, the game continues until $B$ loses. For the converse, analogously to the proof of Proposition~\ref{prop:simulation-left-to-right}, if there are $r_1,r_2$ that satisfy $\varphi$ and $r_1\not\curle_{nS} r_2$, we can show that $B$ has a strategy that allows him, in every round, to choose a state $p_1'$  such that the process corresponding to $p_1'$ is not $n$-nested-simulated by the process corresponding to $p_2'$, until $A$ loses.
\hfill \qed %$\blacktriangleleft$

\begin{prop}\label{prop:twosim-win-algo}
Let $\varphi\in\mathL_{\ell S}$, where $\ell\geq n$. Deciding whether every two processes $p_1,p_2$ that satisfy $\varphi$ are equivalent modulo $\equiv_{nS}$ can be done in polynomial space.   
\end{prop}
\begin{proof} 
Similarly to the \simequiv game on $\varphi$, the \nsimeq game on $\varphi$ is a two-player, perfect-information, and polynomial-depth game. In each round, the satisfiability problem for $\mathL_{\ell S}$, $\ell\geq n$, is checked a polynomial number of times, and additionally, it is checked twice whether player $A$ has a winning strategy for $\nmosimab$ on some input. It is also checked, once before the first round begins, whether $A$ has a winning strategy for $\nmosimab(\{\varphi\},\{\varphi\})$; this can be viewed as a separate initial round of the game. Determining whether $A$ has a winning strategy for $\nmosimab$ on any input is in \pspace since the \nmosimeq game is correct, and satisfiability of formulae in $\mathL_{\ell S}$, where $\ell\geq 2$, is in \pspace from Theorem~\ref{prop:sat-rs-ts-2s-np-complete} and Corollary~\ref{prop:n-s-sat}. Therefore, Corollary~\ref{cor:pspace-games} implies that we can decide whether $A$ has a winning strategy for the \nsimeq game on $\varphi$ in polynomial space, and thus, from Proposition~\ref{prop:twosimul-game}, we can decide whether every two processes that satisfy $\varphi$ are equivalent modulo $\equiv_{nS}$ in polynomial space.
\end{proof}

\begin{cor}\label{cor:twosimul-game}
 Let $U_1,U_2$ be finite subsets of $\mathL_{\ell S}$, $\ell\geq n$. Player $A$ has a winning strategy for $\nsimab(U_1,U_2)$ iff for every $p_1,p_2$ such that $p_1\models\bigwedge U_1$ and $p_2\models \bigwedge U_2$, we have that $p_1\curle_{2S} p_2$.
\end{cor}

\begin{cor}\label{cor:two-sim-game-complexity}
    Let $U_1,U_2$ be finite subsets of $\mathL_{\ell S}$, $\ell\geq n$. Deciding whether $p_1\curle_{nS} p_2$ is true for every two processes $p_1,p_2$ such that $p_1\models\bigwedge U_1$ and $p_2\models\bigwedge U_2$, can be done in polynomial space.
\end{cor}

\paragraph{The \nsimpre game, $n\geq 3$}\label{subsec:the-n-prime-game}

Let $n\geq 3$. We use the \nsimpre game on a satisfiable $\varphi\in\mathL_{nS}$ to check whether $\varphi$ is prime in $\mathL_{nS}$, and thus characteristic for a process within $\mathL_{nS}$.
To this end, the \nsimpre game is developed so that $A$ has a winning strategy iff for every two processes $p_1,p_2$ satisfying $\varphi$ there is a process $q$ satisfying $\varphi$ and $q\curle_{nS} p_i$ for both $i=1,2$. We then show that the latter statement is equivalent to $\varphi$ being characteristic within $\mathL_{nS}$; that is, there is a process $q$ satisfying $\varphi$ such that for all processes $p$ satisfying $\varphi$, %we have
$q\curle_{nS} p$.

The game is presented in Table~\ref{tab:n-prime-game}. Player $B$ constructs two labelled trees, denoted  $T_1=(S_1,L_1,R^1_{a_1},\dots,R^1_{a_k})$ and $T_2=(S_2,L_2,R^2_{a_1},\dots,R^2_{a_k})$, and  player $A$ builds a third labelled tree, denoted  $T_3=(S_3,L_3,R^3_{a_1},\dots,R^3_{a_k})$. The game starts with $S_1=\{p^1_0\}$, $S_2=\{p^2_0\}$, $S_3=\{q_0\}$, $L_1(p^1_0)=L_2(p^2_0)=L_3(q_0)=\{\varphi\}$, and all $R^i_{a_j}$ being empty, where $i=1,2,3$. We describe the $l$-th round of the game for $l\geq 1$. States $p_1,p_2$, and $q$ are  equal to $p^1_0,p^2_0$, and $q_0$, respectively, if $l\in\{1,2\}$ or $p_1,p_2$ are the states that $A$ chose at the end of round $l-1$ and $q$ is the state that $B$ chose at the end of round $l-1$, if $l>2$.
 \begin{table}%[h]
\begin{center}
\begin{tabular}{ | p{15cm} | } 
\hline
\textbf{$\mathbf{1}^{\text{st}}$ round.} 
\begin{enumerate}
    \item $A$ and $B$ play $\nmosimab(\{\varphi\},\{\varphi\})$. If $A$ wins, they continue playing the \nsimpre game. Otherwise, $B$ wins the \nsimpre game. 
    \item  $B$ plays moves B($\wedge$) and B($\vee$) on $p_i$, for both $i=1,2$, until no formula can be replaced in $L_i(p_i)$.
    If $\bigwedge L_i(p_i)$ becomes unsatisfiable, then $B$ loses.
    \item $A$ plays move A(sub) once on $q$ and then she plays moves A($\wedge$) and A($\vee$) on $q$ until no formula can be replaced in $L_3(q)$. 
    If $\bigwedge L_3(q)$ becomes unsatisfiable, then $A$ loses.
\end{enumerate} 

\textbf{$\mathbf{l}^{\text{th}}$ round, $\mathbf{l\geq 2}$.}
\begin{enumerate}
    \item For every $a_j\in \act$ and both $i=1,2$, $B$ plays as follows. He plays move B($\Diamond$) on $p_i$. Then, $B$ plays moves B($\wedge$) and B($\vee$)  on every $p_i'$ such that $(p_i,p_i')\in R_{a_j}^i$ until no formula can be replaced in $L_i(p_i')$.
    If, for some $s\in S_i$, $\bigwedge L_i(s)$ is unsatisfiable, then $B$ loses.
    \item  For every $a_j\in \act$, $A$ plays as follows. She plays move A($\Diamond$) on $q$. Then, for every $j$-successor $q'$ of $q$, $A$ plays move A(sub) once and moves A($\wedge$) and A($\vee$) on $q'$ until no formula can be replaced in $L_3(q')$. Finally, $A$ plays move A(rem) on $q$. If, for some $s\in S_3$, $L_3(s)$ is unsatisfiable, then $A$ loses.
    \item $B$ chooses a $j\in\{1,\dots,k\}$ and a $j$-successor $q'$ of $q$. If $q$ has no $j$-successors, then $B$ loses.
    \item $A$ chooses two states $p_1'$ and $p_2'$ that are  $j$-successors of $p_1$ and  $p_2$ respectively. If some of $p_1$ and $p_2$ has no $j$-successors, then $A$ loses.
     \item $A$ and $B$ play the following four games: (i) $\nmosimab(L_3(q'),L_1(p_1'))$, (ii) $\nmosimab(L_1(p_1'),L_3(q'))$, (iii) $\nmosimab(L_3(q'),L_2(p_2'))$, and (iv) $\nmosimab(L_2(p_2'),L_3(q'))$. If $A$ loses any of (i)--(iv), then $A$ loses.
     %Otherwise, the game continues.
    \item If $l=\md(\varphi)+2$, the game ends and $B$ wins. If $l\leq \md(\varphi)+1$, the $l+1$-th round  starts on $p_1'$, $p_2'$, and $q'$. 
\end{enumerate} \\
 \hline
\end{tabular}
\end{center}
\caption{The \nsimpre game, where $n\geq 3$, initiated on a satisfiable $\varphi\in\mathL_{nS}$.}
\label{tab:n-prime-game}
\end{table}

We use the same notation introduced for the \simequiv game. For every $p\in S_i$, where $i=1,2$, we denote by $\inlabeli(p)$ and $\finlabeli(p)$ the initial and final labels of $p$ before and after moves B($\wedge$) and B($\vee$) are applied on $p$ respectively. For every $p\in S_3$, we denote by $\inlabelthree(p)$ the initial label of $p$ before move A(sub) is applied on $p$, $\starlabelthree(p)$ the label of $p$ after move A(sub) has been applied on $p$, and $\finlabelthree(p)$ the final label of $p$ after moves A($\wedge$) and A($\vee$) have been applied on $p$.

All moves in the \nsimpre game align with those in the \nsimeq game, except for the A(sub) and A(rem) moves. Below, we provide the intuition behind these two specific moves. Let $\act=\{a,b\}$ and consider a formula $\varphi$ of the form $\langle a\rangle\psi_1\wedge\langle a\rangle \psi_2\wedge [a]\psi\wedge [b]\ff$, which is characteristic within  $\mathL_{nS}$. Assume that $\psi_1\wedge\psi$ is not characteristic within $\mathL_{nS}$. Then, $\psi_2\wedge\psi$ must be characteristic and must entail $\psi_1\wedge \psi$, i.e.\ $\psi_2\wedge\psi\models\psi_1\wedge\psi$. This implies that removing $\langle a\rangle\psi_1$ from $\varphi$ yields a logically equivalent formula. Now, let $s\in S_3$ be 
a state in the tree constructed by $A$, such that $\finlabelthree(s)=\{\varphi\}$. When $A$ generates two states $s_1$ and $s_2$ with $L_3(s_i)=\{\psi_i,\psi\}$ for $i=1,2$, she can choose to apply move A(sub) to add all required formulae to $L_3(s_2)$, so that, in the end, $\finlabelthree(s_1)\subseteq \finlabelthree(s_2)$ holds. Thus, when $A$ plays A(rem), she can remove $s_1$. Furthermore, she can ensure that $\bigwedge\finlabelthree(s_2)$ remains characteristic. By applying A(sub) and A(rem) according to this strategy, $B$ is forced, at step 3, to choose a state whose label set corresponds to a characteristic formula. This, in turn, allows $A$ to complete each round without losing.

First, we  show that if $\varphi$ is satisfiable and prime, then player $A$ has a winning strategy for the \nsimpre game on $\varphi$.

\begin{lem}\label{lem:extend-prime-set}
 Let $\varphi\in\mathL_{nS}$ be a prime formula.  If $\varphi\models\psi$, where $\psi\in\mathL_{nS}$, then $\varphi\wedge\psi$ is prime.
\end{lem}
\begin{proof} If $\varphi\models\psi$, then $\varphi$ is logically equivalent to $\varphi\wedge \psi$, and so the latter is also prime. 
% If $\varphi$ is not satisfiable, then the lemma trivially holds. If $\varphi$ is satisfiable, let $p_{min}$ be a process for which $\varphi$ is characteristic within $\mathL_{3S}$. To show that $\varphi\wedge\psi$ is prime, let $\varphi\wedge \psi\models\psi_1\vee\psi_2$. Then $p_{min}\models\varphi\wedge\psi$ because $\varphi\models\psi$. Thus, $p_{min}\models\psi_1\vee\psi_2$, which in turn implies that $p_{min}\models \psi_i$ for some $i=1,2$. From Corollary~\ref{cor:characteristic}, $\varphi\models \psi_i$ and $\varphi\wedge \psi\models\psi_i$. 
\end{proof}

\begin{lem}\label{lem:step-Aastar-prime}
Assume that the \nsimpre game is played on $\varphi\in\mathL_{nS}$ and $q$ is a state of $T_3$ such that $\bigwedge \inlabelthree(q)$ is prime. Let $A$ apply move A(sub) on $q$ such that for every $\psi\in\sub(\varphi)$, she adds $\psi$ to $L_3(q)$ if $\bigwedge L_3(q)\models \psi$. Then, $\bigwedge \starlabelthree(q)$ is prime.
%after A($a^*$) is applied.
\end{lem}
\begin{proof}
 Let $q$ be as described in the lemma. From Lemma~\ref{lem:extend-prime-set}, if $\bigwedge L_3(q)\models\psi$, then $\bigwedge L_3(q)\wedge\psi$ is prime. Therefore, if $\bigwedge L_3(q)$ is prime, after the addition of some $\psi$ to $L_3(q)$ as described in the lemma, $\bigwedge L_3(q)$ remains prime.
\end{proof}

\begin{lem}\label{lem:steps-Aab-first}
    Let $q$ be a state of $T_3$. Then the formulae in $\finlabelthree(q)$  are either $\true$, $\ff$, or formulae that start with either $\langle a\rangle$ or $[a]$, $a\in \act$.
    %If $A$ applies steps A(a)--(b) on $q$ until no formula can be replaced in $L_3(q)$, then the formulae in $L_3(q)$  are either $\true$, $\ff$, or formulae that start with either $\langle a\rangle$ or $[a]$, $a\in \act$.
\end{lem}
\begin{proof}
  This is immediate from the definition of moves A($\wedge$) and A($\vee$).
\end{proof}

\begin{lem}\label{lem:steps-Aab-second}
    Let $q$ be a state of $T_3$ and $\bigvee_{1\leq i\leq m}\varphi_i$, $m\in\mathbb{N}$, be the disjunctive form of $\bigwedge \starlabelthree(q)$. 
    %If steps A(a)--(b) are applied on $q$ until no formula can be replaced in $L_3(q)$, 
    Then $\bigwedge \finlabelthree(q)=\varphi_j$ for some $1\leq j\leq m$. Conversely, for every $\varphi_j$, $1\leq j\leq m$, there is a sequence of choices in the application of A($\wedge$) and A($\vee$) on $q$ such that
    %after no formula can be replaced in $L_3(q)$, 
    $\bigwedge \finlabelthree(q)=\varphi_j$.
\end{lem}
\begin{proof}
    This is immediate from the definition of moves A($\wedge$) and A($\vee$).
\end{proof}

\begin{rem}\label{rem:steps-Bab}
The results of Lemmas~\ref{lem:steps-Aab-first} and~\ref{lem:steps-Aab-second} hold for a state $p_i$ in $T_i$, where $i=1,2$, after moves B($\wedge$) and B($\vee$) are applied on $p_i$. For Lemma~\ref{lem:steps-Aab-second}, $\starlabelthree(q)$ has to be replaced by $\inlabeli(p_i)$ in this case.
\end{rem}

\begin{lem}\label{lem:final-initial-label-two}
Let $p$ be a state of $T_i$, where $i=1,2,3$. For every $r\in\proc$, if $r\models\bigwedge\finlabeli(p)$, then $r\models \bigwedge \inlabeli(p)$.
\end{lem}

\begin{lem}\label{lem:steps-Aab-prime}
 Let $q$ be a state of $T_3$ such that $\bigwedge \starlabelthree(q)$ is prime. There is a sequence of choices that $A$ can make when applying moves A($\wedge$) and A($\vee$) on $q$ such that $\bigwedge \finlabelthree(q)$ is prime.
 %after no other formula can be replaced in $L_3(q)$.
\end{lem}
\begin{proof}
    Let $\bigvee_{1\leq i\leq m}\varphi_i$, $m\in\mathbb{N}$, be the disjunctive form of $\bigwedge \starlabelthree(q)$. From Lemma~\ref{lem:DF-equiv}, we have that $\bigwedge \starlabelthree(q)\equiv \bigvee_{1\leq i\leq m}\varphi_i$. From Corollary~\ref{cor:char-disj-form}, there is some $1\leq j\leq m$ such that $\varphi_j$ is prime and $\bigwedge \starlabelthree(q)\equiv\varphi_j$.   From Lemma~\ref{lem:steps-Aab-second}, when $A$ applies A($\wedge$) and A($\vee$) on $q$, there is a sequence of choices that she can make such that $\bigwedge \finlabelthree(q)=\varphi_j$.
    %after no formula can be replaced in $L_3(q)$.
\end{proof}

\begin{cor}\label{cor:steps-Aastarab-prime}
   Assume that the \nsimpre game is played on $\varphi\in\mathL_{nS}$ and $q\in S_3$ such that $\bigwedge \inlabelthree(q)$ is characteristic within $\mathL_{3S}$. Let player $A$ apply moves A(sub), A($\wedge$), and A($\vee$) on $q$ as follows. 
    \begin{itemize}
        \item $A$ applies move A(sub) on $q$ such that for every $\psi\in\sub(\varphi)$, she adds $\psi$ to $L_3(q)$ if $\bigwedge L_3(q)\models \psi$.
        \item Let $DF(\starlabelthree(q))=\bigvee_{1\leq i\leq m}\phi^*_i$, $m\in\mathbb{N}$. Player $A$ applies moves A($\wedge$) and A($\vee$) on $q$ in such a way that $\bigwedge \finlabelthree(q)=\phi^*_j$, for some $1\leq j\leq m$ such that $\phi^*_j$ is prime and $\starlabelthree(q)\equiv\phi^*_j$. 
    \end{itemize}
    Then,  $\bigwedge\inlabelthree(q)\models\bigwedge \finlabelthree(q)$ and $\bigwedge \finlabelthree(q)$ is characteristic within $\mathL_{nS}$.
\end{cor}

\begin{lem}\label{lem:step-Ad-prime}
Let $q$ be a state of $T_3$ such that $\bigwedge \finlabelthree(q)$ is characteristic within $\mathL_{nS}$, and $\finlabelthree(q)$ contains only $\true$ or formulae that start with either $\langle a\rangle$ or $[a]$, $a\in \act$. If player $A$ can construct some $j$-successor of $q$ at move A($\Diamond$), she can also construct at least one $j$-successor $q'$ of $q$ such that $\bigwedge \inlabelthree(q')$ is characteristic within $\mathL_{nS}$.
\end{lem}
\begin{proof}  Let $\langle a_j \rangle\phi_1,\dots, \langle a_j \rangle\phi_{n_j}$  and $[a_j] \psi_1,\dots, [a_j]\psi_{m_j}$ be all the formulae in $\finlabelthree(q)$ that start with $\langle a_j\rangle$ and $[a_j]$ respectively. Let also $\Phi_i=\phi_i\wedge\bigwedge_{k=1}^{m_j}\psi_k$, $1\leq i\leq n_j$. Then, $A$ generates one $j$-successor $q_i$ of $q$ for each $\Phi_i$ and $\bigwedge \inlabelthree(q_i)=\Phi_i$. It is not hard to see that $\langle a_j\rangle\phi_i\wedge\bigwedge_{k=1}^{m_j} [a_j]\psi_k\models \langle a_j\rangle (\phi_i\wedge\bigwedge_{k=1}^{m_j}\psi_k)$ or equivalently, $\langle a_j\rangle\phi_i\wedge\bigwedge_{k=1}^{m_j} [a_j]\psi_k\models \langle a_j\rangle \Phi_i$. This means that if some $\Phi_i$ is not satisfiable, then $\bigwedge \finlabelthree(q)$ is not satisfiable, which contradicts the hypothesis of the lemma. Thus, every $\Phi_i$ is satisfiable, and we show that at least one of the $\Phi_i$'s is prime. Assume, towards a contradiction, that every $\Phi_i$, $1\leq i\leq n_j$, is not prime. Let $p_{min}$ be a process for which $\bigwedge \finlabelthree(q)$ is characteristic within $\mathL_{nS}$. From the fact that $\langle a_j\rangle\phi_i\wedge\bigwedge_{k=1}^{m_j} [a_j]\psi_k\models \langle a_j\rangle \Phi_i$ for every $1\leq i\leq n_j$, there is $p_{min}\myarrowasubj p_i$ such that $p_i\models\Phi_i$ for every $1\leq i\leq n_j$. Let $\Psi_i$ be the characteristic formula for $p_i$ within $\mathL_{nS}$. We prove the following three claims.
     \begin{clm} \label{claim:psi-phi-one}
     For every $1\leq i\leq n_j$, $\Psi_i\models\Phi_i$.
     \end{clm}
     \begin{proof}
         It holds that $p_{min}\models \langle a_j\rangle\Psi_i$ and so, from Corollary~\ref{cor:characteristic}, $\bigwedge \finlabelthree(q)\models \langle a_j\rangle\Psi_i$. Moreover, for every process $p$ such that $p\models\Psi_i$, $\mathL_{nS}(p_i)\subseteq\mathL_{nS}(p)$, which implies $p\models \Phi_i$. So, $\Psi_i\models\Phi_i$. 
     \end{proof}
     \begin{clm} \label{claim:psi-phi-two}
     For every $1\leq m\leq n_j$ there is some $1\leq i\leq n_j$ such that $\Phi_i\models\Psi_m$.
     \end{clm}
     \begin{proof}
        Suppose that there is some $\Psi_{m_0}$ such that  $\Phi_i\not\models \Psi_{m_0}$, for every $1\leq i\leq n_j$. Then, there are processes $q_1,\dots,q_{n_j}$  such that  $q_i\models\Phi_i$ and $q_i\not\models\Psi_{m_0}$, for every $1\leq i\leq n_j$. Define the process $q =\sum_{i=1}^{n_j} a_j.q_i$. It is not hard to see that $q\models\bigwedge \finlabelthree(q)$ and $q\not\models\langle a_j\rangle \Psi_{m_0}$. However, since  $p_{min}\models \Psi_i$ for every $1\leq i\leq n_j$, we have from Corollary~\ref{cor:characteristic} that $\bigwedge \finlabelthree(q)\models\langle a_j\rangle \Psi_i$ for every $1\leq i\leq n_j$, and so $q\models\langle a_j\rangle\Psi_i$ for every $1\leq i\leq n_j$, which contradicts the fact that $q\not\models\langle a_j\rangle \Psi_{m_0}$. 
       \end{proof}
      \begin{clm} \label{claim:psi-phi-three}
          There is at least one $1\leq i\leq n_j$ such that $\Phi_i\models\Psi_i$.
      \end{clm}
      \begin{proof}
      Let $i_1\in\{1,\dots,n_j\}$. From Claim~\ref{claim:psi-phi-one}, $\Psi_{i_1}\models\Phi_{i_1}$ and from Claim~\ref{claim:psi-phi-two}, there is some $i_2\in\{1,\dots,n_j\}\setminus\{i_1\}$ such that $\Phi_{i_2}\models\Psi_{i_1}$. By repeating this argument we can take the following logical implications:\\
    $$\begin{aligned}
        \Phi_{i_2}\models&\Psi_{i_1}\models \Phi_{i_1}, ~ i_2\neq i_1\\
        \Phi_{i_3}\models&\Psi_{i_2}\models \Phi_{i_2}, ~ i_3\neq i_2\\
                         &\vdots\\
        \Phi_{i_{(n_j+1)}}\models&\Psi_{i_{n_j}}\models\Phi_{i_{n_j}}, ~ i_{(n_j+1)}\neq i_{n_j}
    \end{aligned}$$
    If $\Phi_{i_2},\dots,\Phi_{i_{(n_j+1)}}$ are pairwise distinct, then  $\Phi_{i_{(n_j+1)}}=\Phi_{i_1}$. In this case, $\Phi_{i_{(n_j+1)}}\models\Psi_{i_{n_j}}\models\Phi_{i_{n_j}}\models\Psi_{i_{(n_j-1)}}\models\Phi_{i_{(n_j-1)}}\models\dots\models\Psi_{i_1}\models\Phi_{i_1}$, which implies that $\Phi_{i_1}\models\Psi_{i_1}$. If there are $i_m,i_k\in\{i_2,\dots,i_{(n_{j+1})}\}$ with $k>m$ such that $\Phi_{i_m}=\Phi_{i_k}$, then $\Phi_{i_k}\models\Psi_{i_{(k-1)}}\models\Phi_{i_{(k-1)}}\models\dots\models\Phi_{i_{(m+1)}}\models\Psi_{i_{m}}\models\Phi_{i_m}$, which implies that $\Phi_{i_m}\models\Psi_{i_m}$.
      \end{proof} 
    From Claims~\ref{claim:psi-phi-one} and~\ref{claim:psi-phi-three}, there is some $1\leq i\leq n_j$ such that $\Phi_i\equiv \Psi_i$, which is impossible as $\Psi_i$ is prime and $\Phi_i$ is not prime. Consequently, there is at least one $\Phi_i$ that is prime.
\end{proof}

\begin{lem}\label{lem:step-Ad-prime-second}
   Assume that $q\in S_3$ such that $\bigwedge\finlabelthree(q)$ is characteristic within $\mathL_{nS}$ and $A$ has just applied move A($\Diamond$). Then, for every $j$-successor $q'$ of $q$ such that $\bigwedge \inlabelthree(q')$ is not prime, there is a $j$-successor $q''$ of $q$ such that $\bigwedge \inlabelthree(q'')\models\bigwedge \inlabelthree(q')$ and $\bigwedge \inlabelthree(q'')$ is characteristic within $\mathL_{nS}$, where $1\leq j\leq k$.
\end{lem}
\begin{proof}
     The proof is similar to the proof of Lemma~\ref{lem:step-Ad-prime}.  Let $\langle a_j \rangle\phi_1,\dots, \langle a_j \rangle\phi_{n_j}$  and $[a_j] \psi_1,\dots,$ $ [a_j]\psi_{m_j}$ be all the formulae in $\finlabelthree(q)$ that start with $\langle a_j\rangle$ and $[a_j]$ respectively. Let also $\Phi_i=\phi_i\wedge\bigwedge_{k=1}^{m_j}\psi_k$, $1\leq i\leq n_j$. Then, $A$ generates one $j$-successor $q_i$ of $q$ for each $\Phi_i$ and $\bigwedge \inlabelthree(q_i)=\Phi_i$. We also see that every $\Phi_i$ is satisfiable as we observed in the proof of Lemma~\ref{lem:step-Ad-prime}. Define $X =\{\Phi_i \mid \not\exists~ \Phi_j \text{ such that } \Phi_j \text{ is prime and } \Phi_j\models\Phi_i\}=\{\Phi_1,\dots,\Phi_t\}$, where $t\leq n_j$. Note that $X$ contains non-prime formulae, and it suffices to show that $X$ is empty. Let $p_{min}$ be a process for which $\bigwedge \finlabelthree(q)$ is characteristic within $\mathL_{nS}$. From the fact that $\langle a_j\rangle\phi_i\wedge\bigwedge_{k=1}^{m_j} [a_j]\psi_k\models \langle a_j\rangle \Phi_i$ for every $1\leq i\leq n_j$, there is $p_{min}\myarrowasubj p_i$ such that $p_i\models\Phi_i$ for every $1\leq i\leq n_j$. Let $\Psi_i$ be the characteristic formula for $p_i$ within $\mathL_{nS}$.  Claims~\ref{claim:psi-phi-one} and~\ref{claim:psi-phi-two} allow us to take the following logical implications.
     $$\begin{aligned}
        \Phi_{j_1}\models&\Psi_1\models \Phi_1, \\
        \Phi_{j_2}\models&\Psi_2\models \Phi_2,\\
                         &\vdots\\
        \Phi_{j_t}\models&\Psi_t\models\Phi_t.
    \end{aligned}$$
     Note that if $\Phi_{j_i}\not\in X$ for some $1\leq i\leq t$, then there is some $\Phi_k$ such that $\Phi_k$ is prime and $\Phi_k\models \Phi_{j_i}\models\Psi_i\models\Phi_i$, which contradicts the fact that $\Phi_i\in X$. As a result, $\Phi_{j_i}\in X$ for every $1\leq i\leq t$. By applying the same argument as in the proof of Lemma~\ref{claim:psi-phi-three}, we can show that there is some $1\leq i\leq t$ such that $\Phi_i\equiv \Psi_i$, which contradicts the fact that $\Phi_i$ is not prime and $\Psi_i$ is  prime. Consequently, $X$ must be empty.
\end{proof}

\begin{lem}\label{lem:step-Adastarabe-prime}
    Assume that $q\in S_3$ such that $\bigwedge L_3(q)$ is characteristic within $\mathL_{nS}$ and $A$ has just applied move A($\Diamond$). Let $1\leq j\leq k$. There is a strategy of $A$ such that for every $j$-successor $q'$ of $q$,
    \begin{itemize}
        \item if $\bigwedge \inlabelthree(q')$ is not prime, $q'$ is removed from $T_3$ when move A(rem) is applied on $q$, and
        \item if $\bigwedge \inlabelthree(q')$ is prime, then $\bigwedge \finlabelthree(q')$ is also prime.
    \end{itemize} 
\end{lem}
\begin{proof}
    To prove the first statement, let  $q'$ be a $j$-successor of $q$ such that $\bigwedge \inlabelthree(q')$ is not prime. From Lemma~\ref{lem:step-Ad-prime-second}, there is a $j$-successor $q''$ of $q$ such that $\bigwedge \inlabelthree(q'')\models\bigwedge \inlabelthree(q')$ and $\bigwedge \inlabelthree(q'')$ is characteristic within $\mathL_{nS}$.  It suffices to describe a sequence of choices that $A$ can make when applying A(sub), A($\wedge$), and A($\vee$) on $q',q''$ such that $\finlabelthree(q')\subseteq \finlabelthree(q'')$. Then, $q'$ will be removed from $T_3$ when A(rem) is applied on $q$.  Assume that the \nsimpre game is played on $\varphi\in\mathL_{nS}$. 
    \begin{description}
        \item[Application of A(sub), A($\wedge$), and A($\vee$) on $q''$]  The choices of $A$ on $q''$ are the ones described in Corollary~\ref{cor:steps-Aastarab-prime}. At move A(sub), for every $\psi\in\sub(\varphi)$, $A$ adds $\psi$ to $L_3(q'')$ if $\bigwedge L_3(q'')\models\psi$. Let $DF(\bigwedge \starlabelthree(q''))=\bigvee_{i=1}^l\varphi^{q''}_i$, $l\in\mathbb{N}$. $A$ applies A($\wedge$) and A($\vee$) on $q''$ such that $\bigwedge \finlabelthree(q'')=\varphi^{q''}_j$ for some $1\leq j\leq l$, such that $\varphi^{q''}_j$ is prime and $\bigwedge \starlabelthree(q'')\equiv\varphi^{q''}_j$. This is possible from Lemma~\ref{lem:steps-Aab-prime}. 
        \item[Application of A(sub), A($\wedge$), and A($\vee$) on $q'$] When $A$ applies A(sub) on $q'$, she does not add any $\psi\in\sub(\varphi)$ to $\inlabelthree(q')$. Note that after A(sub) is applied on $q''$, $\inlabelthree(q')\subseteq\starlabelthree(q'')$, since $\bigwedge\inlabelthree(q'')\models \psi$ for every $\psi\in\inlabelthree(q')$. When $A$ applies moves A($\wedge$) and A($\vee$) on $q'$, for every $\psi\in L_3(q')$, she makes the same choices she made for formula $\psi$, when she applied A($\wedge$) and A($\vee$) on $q''$. In this way, $\finlabelthree(q')\subseteq\finlabelthree(q'')$.
         \end{description}
    If for every $j$-successor $q''$ of $q$ such that $\bigwedge\inlabelthree(q'')$ is prime,  moves A(sub), A($\wedge$), and A($\vee$) are applied as described above for $q''$, then $\bigwedge\finlabelthree(q'')$ is prime and the second statement also holds.
\end{proof}

\begin{lem}\label{lem:2s-equiv-processes}
    Let $\varphi,\psi\in\mathL_{nS}$ such that $\psi$ is prime and $\varphi\models\psi$. Then for every two processes $p,q$, if $p\models\varphi$ and $q\models\psi$, then $p\equiv_{(n-1)S} q$.
\end{lem}
\begin{proof}
   Let $p,q$ be two processes such that $p\models\varphi$ and $q\models\psi$. Then, $\psi$ is satisfiable and hence, characteristic for a process $p_{min}$ within $\mathL_{nS}$. From $p\models \varphi$ and $\varphi\models\psi$, it holds that $p\models \psi$. From Definition~\ref{def:characteristic} and Proposition~\ref{logical_characterizations}, $p_{min}\curle_{nS} p$ and $p_{min}\curle_{nS} q$, and so by the definition of $\curle_{nS}$, $p_{min}\equiv_{(n-1)S} p\equiv_{(n-1)S} q$.
   %note: if $\varphi,\psi$ are satisfiable, then $\varphi$ is prime.
\end{proof}

\begin{lem}\label{lem:modal-depth-prime}
    Let $\varphi\in\mathL_{nS}$ be characteristic within $\mathL_{nS}$. Then, all processes that satisfy $\varphi$ have the same depth $d'$, where $d'\leq \md(\varphi)$.
\end{lem}
\begin{proof}
    Let $q_1,q_2$ be two processes that satisfy $\varphi$. From Definition~\ref{def:characteristic} and Proposition~\ref{logical_characterizations}, there is $p$ such that $p\curle_{nS} q_i$ for both $i=1,2$. As a result, $q_1\equiv_S q_2$. From Lemma~\ref{lem:bounded-depth}, $\depth(q_1)=\depth(q_2)\leq\md(\varphi)$. 
\end{proof}

\begin{prop}\label{prop:right-to-left}
     Let $\varphi\in\mathL_{nS}$ be characteristic within $\mathL_{nS}$. Then, player $A$ has a winning strategy for the \nsimpre game on $\varphi$.
\end{prop}
\begin{proof}
Let $d$ denote the modal depth of $\varphi$. We will show that $A$ has a strategy that allows her to continue the game until $B$ loses at round $m$, for some $1\leq m\leq d+2$. Consider the following condition.
\begin{quote}
    \textit{Condition C:} `In the beginning of the round,  $\bigwedge \finlabelthree(q)$  is characteristic within  $\mathL_{3S}$ and  $\bigwedge \finlabeli(p_i)\models \bigwedge \finlabelthree(q)$, for both $i=1,2$.'
\end{quote}
We show that player $A$ can play so that the second round satisfies condition C, and if round $l\geq 2$ satisfies condition C, then she can complete the $l$-th round without losing. In case player $B$ does not lose in the $l$-th round, the $l+1$-th round satisfies condition C. Moreover, we prove that the game does not last more than $d+1$ rounds.
Note that the first round starts with $L_1(p_1)=L_2(p_2)=L_3(q)=\{\varphi\}$, where $\varphi$ is characteristic within $\mathL_{nS}$.

\begin{clm}\label{claim:cond-c}
    Player $A$ can make a sequence of choices so that the second round satisfies condition C. Moreover, if round $l$, where $2\leq l\leq d$, satisfies condition C, then player $A$ can make a sequence of choices so that round $l+1$ satisfies condition C.
\end{clm}
\begin{proof}
\begin{description}
    \item[The second round can start satisfying condition C] Player $A$ can apply moves A(sub), A($\wedge$), and A($\vee$) as described in Corollary~\ref{cor:steps-Aastarab-prime} so that $\bigwedge \finlabelthree(q)$ is prime and $\varphi\models\bigwedge \finlabelthree(q)$ at the end of the round. Thus, $\bigwedge \finlabelthree(q)$ will be satisfiable. Regardless of  $B$'s choices, $\bigwedge \finlabeli(p_i)\models\varphi$ at the end of the round from Lemma~\ref{lem:final-initial-label-two}.
    %from Lemmas~\ref{lem:steps-Aab-second} and~\ref{lem:disjunction_lemma}.
    Therefore,  $\bigwedge \finlabeli(p_i)\models\bigwedge \finlabelthree(q)$ for both $i=1,2$, and the second round starts with the label sets of $p_i$'s and $q$ satisfying condition C.
    \item[The $(l+1)$-th round, $l\geq 2$, can start satisfying condition C] Let the $l$-th round satisfy condition C. In round $l$, $A$ generates $q_1,\dots,q_{n_j}$. From Lemmas~\ref{lem:step-Ad-prime-second} and~\ref{lem:step-Adastarabe-prime}, there is a sequence of choices that $A$ can make when applying moves A(sub), A($\wedge$), and A($\vee$) on $q_i$'s such that if $\bigwedge \inlabelthree(q_i)$ is not prime, then $q_i$ is removed from $T_3$, and for every $q_i$ that remains in $T_3$, $\bigwedge\finlabelthree(q_i)$ is prime. Therefore, if $B$ chooses some $q'$ at step 3 of the round, then $\bigwedge \finlabelthree(q')$ is characteristic within $\mathL_{nS}$. When $q'$ was created by $A$, $\inlabelthree(q')=\{\phi\}\cup\{\phi'\mid [a_j]\phi'\in \finlabelthree(q)\}$ for some $\langle a_j\rangle\phi\in \finlabelthree(q)$ and  $\bigwedge \finlabelthree(q)\models\langle a_j\rangle\bigwedge \inlabelthree(q')$. After moves A(sub), A($\wedge$), and A($\vee$) are applied on $q'$ as described in Lemma~\ref{lem:step-Adastarabe-prime} and Corollary~\ref{cor:steps-Aastarab-prime}, $\bigwedge\inlabelthree(q')\models \bigwedge \finlabelthree(q')$,
    %$\phi\wedge\bigwedge_{[a_j]\phi'\in L_3(q)}\phi'\models \bigwedge L_3(q')$, 
    so $\bigwedge \finlabelthree(q)\models \langle a_j\rangle \bigwedge \finlabelthree(q')$. From this last fact
    %fact that $\bigwedge L_3(q)\models \langle a_j\rangle \bigwedge L_3(q')$ 
    and since round $l$ satisfies condition C, in the beginning of round $l$, $\bigwedge  \finlabeli(p_i)\models\langle a_j\rangle \bigwedge \finlabelthree(q')$, for both $i=1,2$. From Lemma~\ref{lem:steps-Aab-first} and Remark~\ref{rem:steps-Bab}, it holds that $\finlabeli(p_i)$ contains formulae that start either with $\langle a\rangle$ or $[a]$ (and maybe formula $\true$) and so there is some $\langle a_j\rangle\psi_i\in \finlabeli(p_i)$ such that $\psi_i\wedge\bigwedge_{[a_j]\psi\in \finlabeli(p_i)} \psi\models \finlabelthree(q')$. Consequently, there is a $j$-successor $p_i'$ of $p_i$ such that $\bigwedge \inlabeli(p_i')\models\bigwedge \finlabelthree(q')$. 
    %From Lemmas~\ref{lem:steps-Aab-first} and~\ref{lem:steps-Aab-second} and Remark~\ref{rem:steps-Bab}, 
    From Lemma~\ref{lem:final-initial-label-two}, $\bigwedge\finlabeli(p_i')\models\bigwedge\inlabeli(p_i')$, and so 
    %after steps B(a)--(b) are applied on $p_i'$, 
    $\bigwedge \finlabeli(p_i')\models\bigwedge \finlabelthree(q')$. Thus, given a state $q'$ chosen by $B$ at step 3, $A$ chooses these two states $p_1',p_2'$, whose existence was just proven, and $A$ does not lose at step 4. From Lemma~\ref{lem:2s-equiv-processes}, for every two processes $p,q$ such that $p\models\bigwedge \finlabeli(p_i')$ and $q\models\bigwedge \finlabelthree(q')$, $p\equiv_{2S} q$. From Corollary~\ref{cor:twosimul-game}, $A$ wins all versions of the \nmosimeq game at step 5. The round $l+1$ starts with $p_1',p_2'$, and $q'$, which implies that it satisfies condition C.\qedhere
\end{description} 
\end{proof}
Finally,  we prove that if $A$ plays according to the strategy described above, after at most $d+1$ rounds, she will not create $j$-successors at step 2, and $B$ will lose at step 3. Let $T_3$ be constructed by $A$ according to the strategy described in the proof of Claim~\ref{claim:cond-c}. From Proposition~\ref{tableausat} and Remark~\ref{rem:tableausat}, we can construct a process  $p$ that satisfies $\varphi$ and has the same depth as $T_3$. Then, from Lemma~\ref{lem:modal-depth-prime}, the depth of  $T_3$ can be at most $d$ and from the definition of the game, $d+1$ rounds suffice to construct any path of $T_3$. After that, at round $d+2$, $A$ will not create any new $j$-successors of $q$ and $B$ will lose. 
%In the case that round $d+2$ is completed without $B$ loosing, it means that $\varphi$ is not prime and the game ends with $B$ winning.
\end{proof}

%left-to-right-direction

Let $\varphi\in\mathL_{nS}$ be satisfiable. Next, we prove that if $A$ has a winning strategy for the \nsimpre game on $\varphi$, then $\varphi$ is prime.
%First, recall 
(Recall Definition~\ref{def:B-plays-consistently}, where we defined when $B$ plays $T_i$  consistently on a process $r$, where $i=1,2$.)

\begin{lem}\label{lem:Bs-play}
  Assume that the \nsimpre game is played on $\varphi\in\mathL_{nS}$. Let $r_1,r_2$ be two processes that satisfy $\varphi$. Then, there is a strategy for player $B$ that allows him to play $T_i$ on $r_i$, where $i=1,2$.
\end{lem}
\begin{proof}
  The proof is exactly the same as the proof of Lemma~\ref{lem:B-can-play-consistently}.
  
  We show the lemma for $r_1$. When we refer to conditions 1--3, we mean conditions 1--3 of  Definition~\ref{def:B-plays-consistently}. Let $map:S_1\rightarrow \proc$ be such that $map(p_0^1)=r_1$. We prove that $B$ can play such that (a) condition 1 is true for $p_0^1$ and (b) if condition 1 is true for $p\in S_1$, then for every $(p,p')\in R^1_{a_j}$, conditions 1--2 are true for $p'$. 
    \begin{enumerate}[a.]
        \item Condition 1 holds for $p_0^1$: When the game starts, $\inlabelone(p_0^1)=\{\varphi\}$. Let $DF(\varphi)=\bigvee_{i=1}^m\varphi_i$, $m\in\mathbb{N}$. From Lemma~\ref{lem:DF-equiv}, $\varphi\equiv \bigvee_{i=1}^m\varphi_i$, and so $r_1\models\varphi_j$ for some $1\leq j\leq m$. From Lemma~\ref{lem:steps-Aab-second} and Remark~\ref{rem:steps-Bab}, $B$ can apply moves B($\wedge$) and B($\vee$) on $p_0^1$ such that 
        %after no formula can be replaced in $L_1(p_0^1)$,
        $\bigwedge \finlabelone(p_0^1)=\varphi_j$. Then, $map(p_0^1)=r_1\models\bigwedge \finlabelone(p_0^1)$.
        %where $L_1(p_0^1)$ is the label set of $p_0^1$ after steps B(a)--(b) have been applied on $p_0^1$. 
        \item Assume that $p\in S_1$ for which condition 1 holds. We show that for every $(p,p')\in R^1_{a_j}$, $map(p)\myarrowasubj map(p')$ and condition 1 is true for $p'$. Let $(p,p')\in R^1_{a_j}$. Then, there is some formula $\langle a_j\rangle \psi\in \finlabelone(p)$,
        %where $L_1(p)$ is the label set of $p$ after steps B(a)--(b) have been applied on $p$ (or just before the successors of $p$ are generated by $B$), 
        and $\inlabelone(p')=\{\psi\}\cup\{\psi'\mid [a_j]\psi'\in L_1(p)\}$. Let  $DF(\bigwedge \inlabelone(p'))=\bigvee_{i=1}^n\Psi_i$. Since $map(p)\models \bigwedge \finlabelone(p)$, $map(p)\models\langle a_j\rangle\bigwedge\inlabelone(p')$, and so there is $map(p)\myarrowasubj r_p$ such that $r_p\models \bigwedge\inlabelone(p')$, which implies that $r_p\models\Psi_j$ for some $1\leq j\leq n$, since $\bigwedge \inlabelone(p')\equiv\bigvee_{i=1}^n\Psi_i$ from Lemma~\ref{lem:DF-equiv}.  Then, $B$ can apply moves B($\wedge$) and B($\vee$) on $p'$ in such a way that $\finlabelone(p')=\Psi_j$. Therefore, we can set $map(p')=r_p$ and condition 1 holds for $p'$.
    \end{enumerate}
From (a) and (b) above, $B$ can play in such a way that there is a mapping $map:S_1\rightarrow \proc$ that satisfies conditions 1--3.
The proof for $r_2$ is completely analogous.
\end{proof}

\begin{lem}\label{lem:As-play}
Assume that $A$ has a winning strategy for the \nsimpre game on $\varphi$. Then, for every two processes $r_1,r_2$ that satisfy $\varphi$, there is a process $t$ such that 
\begin{itemize}
    \item $t\models\varphi$,
    \item $t\curle_{nS} r_1$ and $t\curle_{nS} r_2$,
    \item $|t|\leq (2m+1)^{m+1}$, where $m=|\varphi|$.
\end{itemize} 
\end{lem}
\begin{proof}
    Let $A$ play according to her winning strategy and $r_1, r_2$ be two processes that satisfy $\varphi$. From Lemma~\ref{lem:Bs-play}, we can assume that $B$ plays $T_1, T_2$ consistently on $r_1,r_2$, respectively; let also $map:S_1\cup S_2\rightarrow\proc$ be the mapping described in Definition~\ref{def:B-plays-consistently}. Assume that when the $l$-th round starts, where $l\geq 2$, $A$ plays on $q$ and $B$ plays on $p_1,p_2$.
    %Since the first and second round start with the same states, it is sufficient to consider $l>1$.
    Let $B$ generate states $p_1^i,\dots,p_{k_i}^i$ that are $j^i_1,\dots,j^i_{k_i}$-successors of $p_i$, $i=1,2$, when he applies move B($\Diamond$) on $p_i$, and $A$ generate $q_1,\dots,q_{k_3}$, which are $j_1,\dots, j_{k_3}$-successors of $q$, when she applies move A($\Diamond$) on $q$. We inductively define  process $t(q,p_1,p_2)$ as follows:
    \begin{itemize}
        \item $t(q,p_1,p_2)=\mathtt{0}$ if $A$ generates no $j$-successor of $q$ when she plays move A($\Diamond$) on $q$.
        \item $t(q,p_1,p_2)=\sum_{i=1}^{k_3}a_{j_i}.t(q_i,p_1(q_i),p_2(q_i))$, where for every $1\leq i\leq k_3$, $p_1(q_i)$, $p_2(q_i)$ are the $j_i$-successors of $p_1,p_2$, respectively, that $A$ chooses at step 4 if $B$ has chosen $q_i$ at step 3.
    \end{itemize}
    Consider the states $q_0$ and $p_0^1,p_0^2$ that $A$ and $B$, respectively, start the game. Let $dt$ denote the depth of $t(q_0,p_0^1,p_0^2)$ and $m$ be the size of $\varphi$. By definition, $dt$ is the length of the maximum trace $q_0, q_1,\dots, q_s$ (for every $1\leq i\leq s$ there is a $1\leq j\leq k$ such that $q_i$ is a $j$-successor of $q_{i-1}$) that $A$ constructs during a play of the game where $B$ plays $T_1, T_2$ consistently on $r_1,r_2$, respectively, and $A$ follows her winning strategy. Since $A$ always wins, we know that $B$ does not win at step 6, and so round $\md(\varphi)+2$ never reaches step 6. Moreover, $A$ starts round $2$ with state $q_0$ and at every round $l$, $2\leq l\leq \md(\varphi)+2$, the length of the path constructed by $A$ is increased by 1. At the end, $A$ constructs a path of length at most $\md(\varphi)+1\leq m+1$. Therefore, the depth of $t(q_0,p_0^1,p_0^2)$ is at most $m+1$.
    
    We show that 
    \begin{itemize}
        \item $t(q,p_1,p_2)\models \bigwedge \finlabelthree(q)$ (Claim~\ref{claim:process-t-one}),
        \item $t(q,p_1,p_2)\curle_{nS} map(p_1)$ and $t(q,p_1,p_2)\curle_{nS} map(p_2)$ (Claim~\ref{claim:process-t-two}), and 
        \item $|t(q,p_1,p_2)|\leq (2m+1)^{dt}$ (Claim~\ref{claim:process-t-three}).
    \end{itemize}  
    \begin{clm}\label{claim:process-t-one}
        $t(q,p_1,p_2)\models \bigwedge \finlabelthree(q)$.
    \end{clm}
    \begin{proof}
    From Lemma~\ref{lem:steps-Aab-first} and the fact that $A$ plays according to her winning strategy, $\finlabelthree(q)$ contains only formulae that start with either $\langle a\rangle$ or $[a]$, $a\in \act$ (and maybe formula $\true$). We show the claim by induction on the depth of $t(q,p_1,p_2)$.
    \begin{itemize}
        \item If $t(q,p_1,p_2)=\mathtt{0}$, then $q$ has no $j$-successors for any $1\leq j \leq k$, which means that all formulae in $\finlabelthree(q)$ start with $[a]$, where $a\in \act$, or they are the formula $\true$. It holds that $\mathtt{0}\models [a]\psi$ for every $a\in \act$ and $\psi\in\mathL_{nS}$, and $\mathtt{0}\models\true$. Therefore, $t(q,p_1,p_2)\models\bigwedge \finlabelthree(q)$.
        \item Let $t(q,p_1,p_2)=\sum_{i=1}^{k_3}a_{j_i}.t(q_i,p_1(q_i),p_2(q_i))$ and $\langle a_j\rangle \psi\in L_3(q)$ for some $1\leq j\leq k$ and $\psi\in\mathL_{nS}$. Then, there is some $q\myarrowasubj q_i$ such that $\psi\in \inlabelthree(q_i)$. By the inductive hypothesis, $t(q_i,p_1(q_i),$ $p_2(q_i))$ $\models\bigwedge \finlabelthree(q_i)$ and so from Lemma~\ref{lem:final-initial-label-two}, $t(q_i,p_1(q_i),p_2(q_i))\models\bigwedge \inlabelthree(q_i)$. This implies that $t(q_i,p_1(q_i),p_2(q_i))\models\psi$. Therefore, $t(q,p_1,p_2)\models \langle a_j\rangle\psi$. Let  $[a_j]\psi\in \finlabelthree(q)$ for some $1\leq j\leq k$ and $\psi\in\mathL_{nS}$. Then, for every $q\myarrowasubj q_i$ it holds that $\psi\in \inlabelthree(q_i)$ and again, by the inductive hypothesis and Lemma~\ref{lem:final-initial-label-two}, $t(q_i,p_1(q_i),p_2(q_i))\models\psi$. Thus, $t(q,p_1,p_2)\models [a_j]\psi$. Consequently, $t(q,p_1,p_2)\models \bigwedge \finlabelthree(q)$.\qedhere
    \end{itemize}
    \end{proof}
    \begin{clm}\label{claim:process-t-two}
        $t(q,p_1,p_2)\curle_{nS} map(p_1)$ and $t(q,p_1,p_2)\curle_{nS} map(p_2)$.
    \end{clm}
    \begin{proof}
       We show by induction on the depth of $t(q,p_1,p_2)$ that if $A$ has a winning strategy on $q,p_1,p_2$, then $t(q,p_1,p_2)\curle_{nS} map(p_1)$ and $t(q,p_1,p_2)\curle_{nS} map(p_2)$. 
    \begin{itemize}
        \item If $t(q,p_1,p_2)=\mathtt{0}$, then $q$ has no $j$-successors for any $1\leq j\leq k$. We have that  $map(p_i)\models\bigwedge \finlabeli(p_i)$, $i=1,2$, from Lemma~\ref{def:B-plays-consistently}(1) and $t(q,p_1,p_2)\models \bigwedge \finlabelthree(q)$ from Claim~\ref{claim:process-t-one}. Since $A$ has a winning strategy on $q,p_1,p_2$, she can play such that she wins all versions of the \nmosimeq game played on $q$, $p_1$, $p_2$ at step 5 of round $l-1$ (or step 1 of the first round if $l=2$) and so, from Corollary~\ref{cor:twosimul-game}, $t(q,p_1,p_2)\equiv_{(n-1)S} map(p_1)\equiv_{(n-1)S} map(p_2)$. This implies that $map(p_i)=\mathtt{0}$ for $i=1,2$, which in turn implies that $t(q,p_1,p_2)\curle_{nS} map(p_1)$ and $t(q,p_1,p_2)\curle_{nS} map(p_2)$.
        \item Let $t(q,p_1,p_2)\myarrowasubj t'$ for some $1\leq j\leq k$. As in the case of $t(q,p_1,p_2)=0$, $t(q,p_1,p_2)\equiv_{(n-1)S} map(p_1)\equiv_{(n-1)S} map(p_2)$. By the definition of $t(q,p_1,p_2)$, there is $q\myarrowasubj q_i$ such that $t'=t(q_i,p_1(q_i),p_2(q_i))$.  By the definition of $p_1(q_i),p_2(q_i)$, it also holds that $p_1\myarrowasubj p_1(q_i)$ and $p_2\myarrowasubj p_2(q_i)$, which together with Definition~\ref{def:B-plays-consistently}(2) implies that $map(p_1)\myarrowasubj map(p_1(q_i))$ and $map(p_2)\myarrowasubj map(p_2(q_i))$. Moreover, by the definition of $p_1(q_i),p_2(q_i)$, $A$ has a winning strategy on $q_i,p_1(q_i),$ $p_2(q_i)$. By the inductive hypothesis, $t(q_i,p_1(q_i),p_2(q_i))\curle_{nS} map(p_1(q_i))$ and $t(q_i,p_1(q_i),p_2(q_i))$ $\curle_{nS} map(p_2(q_i))$, which completes the proof of this case. \qedhere
    \end{itemize}
    \end{proof}
    \begin{clm}  \label{claim:process-t-three}
 $|t(q,p_1,p_2)|\leq (2m+1)^{dt}$, where $dt$ denotes the depth of $t(q,p_1,p_2)$. 
 \end{clm}
\begin{proof}  By induction on $dt$. If $t(q,p_1,p_2)=\mathtt{0}$, the claim is trivial. If $t(q,p_1,p_2)=\sum_{i=1}^{k_3}a_{j_i}.t(q_i,$ $p_1(q_i),p_2(q_i))$, then $|t(q,p_1,p_2)|\leq m\cdot \max_{1\leq i\leq k_3}\{|t(q_i,p_1(q_i),p_2(q_i))|\}+1+m$, where we bounded the number of successors of $q$ by $m$, the plus $1$ is due to the root, and the plus $m$ comes from adding the number of edges to the size.  
Then, $|t(q,p_1,p_2)|\leq (2m+1)\cdot \max_{1\leq i\leq k_3}\{|t(q_i,p_1(q_i),p_2(q_i))|\}$ is an immediate implication of the above inequality. Therefore, we have that $|t(q,p_1,p_2)|\leq (2m+1)^{dt}$.
\end{proof}

Claim~\ref{claim:process-t-two} implies that $t(q_0,p_0^1,p_0^2)\curle_{nS} map(p_0^1)$ and $t(q_0,p_0^1,p_0^2)\curle_{nS} map(p_0^2)$. Moreover, $map(p_0^i)=r_i$, $i=1,2$, from Definition~\ref{def:B-plays-consistently}(3), and so $t(q_0,p_0^1,p_0^2)\curle_{nS} r_1$ and $t(q_0,p_0^1,p_0^2)\curle_{nS} r_2$. Claim~\ref{claim:process-t-one} says that $t(q_0,p_0^1,p_0^2)\models \bigwedge \finlabelthree(q_0)$. From Lemma~\ref{lem:final-initial-label-two}, $t(q_0,p_0^1,p_0^2)\models \bigwedge \inlabelthree(q_0)$, and since $\inlabelthree(q_0)=\{\varphi\}$, $t(q_0,p_0^1,p_0^2)\models \varphi$.
Finally, from Claim~\ref{claim:process-t-three} and the observation that $dt\leq m+1$, we have that $|t(q_0,p_0^1,p_0^2)|\leq (2m+1)^{m+1}$. We conclude that process $t(q_0,p_0^1,p_0^2)$ satisfies all three conditions of the lemma.
\end{proof}

\begin{defi}\label{def:expphi}
 We say that a process $p$ is in \expphi if $p\models\varphi$ and $|p|\leq (2m+1)^{m+1}$, where $m=|\varphi|$.
\end{defi}

\begin{cor}\label{cor:for-every-two-there-is}
 Assume that $A$ has a winning strategy for the \nsimpre game on $\varphi\in\mathL_{nS}$. Then, for every two processes $r_1,r_2\in\expphi$ there is a process $t\in\expphi$ such that $t\curle_{nS} r_1$ and $t\curle_{nS} r_2$.
\end{cor}
\begin{proof}
     Immediate from Lemma~\ref{lem:As-play}.
\end{proof}

\begin{cor}\label{cor:for-every-there-is-small}
 Assume that $A$ has a winning strategy for the \nsimpre game on $\varphi\in\mathL_{nS}$. Then, for every process $r$ that satisfies $\varphi$, there is some $t\in\expphi$ such that $t\curle_{nS} r$.
\end{cor}
\begin{proof}
 Immediate from Lemma~\ref{lem:As-play}.
\end{proof}

\begin{lem}\label{lem:there-is-for-every-small}
    Assume that $A$ has a winning strategy for the \nsimpre game on $\varphi\in\mathL_{nS}$. Then, there is a process $t\in \expphi$ such that for every process $r\in\expphi$, $t\curle_{nS} r$.
\end{lem}
\begin{proof}
      Let $m\in\{2,\dots,|\expphi|\}$. We prove that for every $m$ processes $r_1, \dots,r_m\in\expphi$ there is some process $t\in\expphi$ such that $t\curle_{nS} r_1,\dots,r_m$. The proof is by strong induction on $m$.
\begin{description}
    \item[Base case] For $m=2$, the claim follows from Corollary~\ref{cor:for-every-two-there-is}.
    \item[Inductive step] Assume that the claim is true for every $2\leq m\leq \ell-1$, where $\ell\geq 3$. We show that it is also true for $m=\ell$. Assume, without loss of generality, that $\ell$ is even. Let $r_1,\dots,r_\ell$ be processes in \expphi. Consider the pairs $(r_1,r_2)$, $(r_3,r_4),$ $\dots,$
    $(r_{\ell-1},r_\ell)$. From the inductive hypothesis, there are $t_1,\dots, t_{\ell/2}\in\expphi$ such that $t_1\curle_{nS} r_1,r_2$, $t_2\curle_{nS} r_3,r_4$, $\dots$, $t_{\ell/2}\curle_{nS} r_{\ell-1},r_\ell$. From the inductive hypothesis, there is some $t\in\expphi$ such that $t\curle_{nS} t_1,\dots,t_{\ell/2}$. By transitivity of $\curle_{nS}$, it follows that $t\curle_{nS} r_1,\dots,r_\ell$, and we are done.  \qedhere
\end{description}
\end{proof}

\begin{cor}\label{cor:there-is-for-every}
    Assume that $A$ has a winning strategy for the \nsimpre game on $\varphi\in\mathL_{nS}$. Then, there is some $t\in\expphi$ such that $t\curle_{nS} r$, for every process $r$ that satisfies $\varphi$.
\end{cor}
\begin{proof}
    Let $t\in\expphi$ be such that for every $r\in\expphi$, $t\curle_{nS} r$, the existence of which is guaranteed by Lemma~\ref{lem:there-is-for-every-small}. Let $r'$ be a process that satisfies $\varphi$. From Corollary~\ref{cor:for-every-there-is-small}, there is $r''\in\expphi$ such that $r''\curle_{nS} r'$. As a result, $t\curle_{nS} r''\curle_{nS} r'$ which was to be shown.
\end{proof}

\begin{prop}\label{prop:left-to-right}
  Let $\varphi\in\mathL_{nS}$ be satisfiable. Assume that $A$ has a winning strategy for the \nsimpre game on $\varphi$. Then, $\varphi$ is characteristic for some process within $\mathL_{nS}$.
\end{prop}
\begin{proof}
   From Definition~\ref{def:characteristic}, it suffices to show that there is a process $t$ such that for every process $r$, $r\models \varphi$ iff $\mathL_{nS}(t)\subseteq\mathL_{nS}(r)$. Let $t$ be as described in Corollary~\ref{cor:there-is-for-every} and let $r$ be any process. If $r\models\varphi$, then from Corollary~\ref{cor:there-is-for-every}, $t\curle_{nS} r$, which together with Proposition~\ref{logical_characterizations} implies that $\mathL_{nS}(t)\subseteq\mathL_{nS}(r)$. If  $\mathL_{nS}(t)\subseteq\mathL_{nS}(r)$, then $r\models\varphi$ because $t\models\varphi$.
    \end{proof}

\begin{prop}\label{prop:A-wins-prime}
Let $\varphi\in\mathL_{nS}$, where $n\geq 3$, be satisfiable. Then, $A$ has a winning strategy for the \nsimpre game on $\varphi\in\mathL_{nS}$ iff $\varphi$ is characteristic for some process within $\mathL_{nS}$.
\end{prop}
\begin{proof}
     Immediate from Propositions~\ref{prop:right-to-left} and~\ref{prop:left-to-right}. 
\end{proof}

\gameprime*
\begin{proof}
The proposition follows from Proposition~\ref{prop:charact-via-primality} and~\ref{prop:A-wins-prime}.    
\end{proof}

\nSpspacecomp*
\begin{proof}
  The problem is \pspace-hard from Theorem~\ref{thm:2scompl-prime}. Let $n\geq 3$ and $\varphi\in\mathL_{nS}$. We describe a Turing machine $M$ that runs in polynomial space and decides whether $\varphi$ is prime. First, $M$ checks whether $\varphi$ is satisfiable, which can be done in polynomial space from Corollary~\ref{prop:3-s-sat}. If $\varphi$ is not satisfiable, then $M$ accepts. If $\varphi$ is satisfiable, than $M$ simulates a polynomial-space algorithm to decide whether $A$ has a winning strategy for the \nsimpre game on $\varphi$. Such an algorithm exists from Corollary~\ref{cor:pspace-games}, since the \nsimpre is a two-player, perfect-information, polynomial-depth game with a \pspace oracle.
\end{proof}

\subsection{The formula primality problem for  \texorpdfstring{$\mathL_{2S}$}{L2S}}\label{subsec:prime-2S}

We now turn our attention to the formula primality problem for {$\mathL_{2S}$}. We will show that the problem is \conp-complete for $\mathL_{2S}$. As was the case for $\mathL_{nS}$, $n\geq 3$, the most challenging part of the proof is to establish the \conp upper bound. 

\begin{prop}\label{prop:2s-prime-conp-hard}
    The formula primality problem for $\mathL_{2S}$ is \conp-hard.
\end{prop}
\begin{proof}
    The proof is analogous to the proof of Proposition~\ref{prop:decide-prime-ts-infinite-actions-hard}.
\end{proof}

We will now prove that the formula primality problem for $\mathL_{2S}$ belongs to \conp,  in contrast to the \pspace upper bound for the same problem in $\mathL_{nS}$, $n\geq 3$. This difference in complexity is mainly because, for a satisfiable formula $\varphi\in\mathL_{2S}$, there is always a tableau for $\varphi$---and so a corresponding process satisfying $\varphi$---of polynomial size. Regarding the satisfiability problem for the logic, an execution of the standard non-deterministic tableau construction~\cite{HalpernM92} must result in a tableau for $\varphi$
(and a corresponding process that satisfies $\varphi$)
and, therefore, we obtain an \NP algorithm, as was shown in Theorem~\ref{prop:sat-rs-ts-2s-np-complete}. In contrast, for the formula primality problem, we accept the formula $\varphi$ under the following conditions: 
\begin{enumerate}
    \item[(i)] all executions of the non-deterministic tableau construction fail---implying that $\varphi$ is unsatisfiable and hence prime;  or
    \item[(ii)] we run the tableau construction twice in parallel, and for each pair of executions that return two tableaux for $\varphi$, corresponding to two processes $p_1,p_2$ satisfying $\varphi$, we check whether there is a process $q$ that also satisfies $\varphi$ and is 2-nested-simulated by both $p_1$ and $p_2$.
\end{enumerate} 
Note that a winning strategy for $A$ in the \nsimpre game on $\varphi$ is equivalent to the second condition for some $\varphi\in\mathL_{nS}$, $n\geq 3$. When we implement the procedure outlined above---see Algorithm~\ref{alg:prime-twosim}---each execution runs in polynomial time and, since we universally quantify over all such executions, the problem lies in \conp. In the rest of this subsection we prove the following theorem.

\begin{restatable}{thm}{twosprimeconp}\label{prop:2S-primality-conpc}
The formula primality problem for $\mathL_{2S}$ is \conp-complete.
\end{restatable}

Given two processes $p,q$, their maximal lower bound is defined below.

\begin{defi}\label{def:gcd}
    Let $p,q\in\proc$. We say that $g$ is  
    %greatest simulated process
    a maximal lower bound of $p$ and $q$ with respect to a preorder $\leq$
    %denoted  $\grcd_\leq(p,q)$,
    if $g\leq p$ and $g\leq q$, and for every $g'$ such that  $g'\leq p$ and $g'\leq q$, it holds that $g'\leq g$. 
\end{defi}

\begin{rem}\label{rem:gcd-uniqueness}
 Let $p\equiv q$ iff $p\leq q$ and $q\leq p$. Then, for two processes $p,q$, either $p, q$ do not have a maximal lower bound with respect to $\leq$ or a maximal lower bound of $p$ and $q$ with respect to $\leq$ exists and is unique up to $\equiv$. In the latter case, we write $\grcd_\leq(p,q)$ for the unique  maximal lower bound of $p$ and $q$ modulo $\equiv$.
\end{rem}

\begin{lem}\label{lem:sim-equiv-processes}
    Let $p,q\in\proc$  such that $p\equiv_S q$. Then,
    \begin{enumerate}[(a)]
        \item $I(p)=I(q)$ and for every $a\in I(p)$ there are $p^*,q^*$ such that  $p\myarrowa p^*$, $q\myarrowa q^*$, and $p^*\equiv_S q^*$. 
        \item For every $p\myarrowa p'$, either there is some $q\myarrowa q'$ and $p'\equiv_S q'$, or there are $p^*,q^*$ such that  $p\myarrowa p^*$, $q\myarrowa q^*$, $p^*\equiv_S q^*$, and $p'\curle_S p^*$.
    \end{enumerate}
\end{lem}
\begin{proof}
    (b) Let $p\myarrowa p^{(1)}$ and assume that there is no $q\myarrowa q'$ such that $p^{(1)}\equiv_S q'$; let also $n=|\{p'\mid p\myarrowa p'\}|$, i.e.\ $n$ is the number of processes reachable from $p$ through an $a$-transition. Since $p\equiv_S q$, we have that there are $q^{(1)},p^{(2)},q^{(2)},\dots,p^{(n)},q^{(n)}, p^{(n+1)}$ such that, for every $2\leq i\leq n+1$ and $1\leq j\leq n$,
    $p\myarrowa p^{(i)}$, $q\myarrowa q^{(j)}$, and 
    \[
    p^{(1)}\curle_S q^{(1)}\curle_S p^{(2)}\curle_S q^{(2)}\curle_S\cdots\curle_S p^{(n)}\curle_S q^{(n)}\curle_S p^{(n+1)}. 
    \]
    From the pigeonhole principle, there are some $m_1$ and $m_2$ such that $1 \leq m_1 < m_2 \leq n+1$ and $p^{(m_1)}=p^{(m_2)}$.
    Then, $p^{(m_1)}\curle_S q^{(m_1)}\curle_S p^{(m_2)}$, which implies that $p^{(m_1)}\equiv_S q^{(m_1)}$. Moreover, $p^{(1)}\curle_S p^{(m_1)}$. 
    So, $p^*, q^*$ of the lemma are $p^{(m_1)}, q^{(m_1)}$, respectively.

    (a) This is immediate from (b).
\end{proof}

\begin{lem}\label{lem:gcd-2s-poly-time}
Let $p,q\in\proc$. We can decide whether $\grcd_{\curle_{2S}}(p,q)$ exists in polynomial time. Moreover, if $\grcd_{\curle_{2S}}(p,q)$ exists, its size is at most $2|p||q|$\footnote{Here, we mean that there is a process $g$ that satisfies the properties of Definition~\ref{def:gcd} and is of size at most $2|p||q|$. In the sequel, we use the notation $\grcd_{\curle_{2S}}(p,q)$ in the same way.} and it can be computed in polynomial time.
\end{lem}
\begin{proof} 
In case $p\not\equiv_{S} q$, there is no $r\in\proc$ such that $r\curle_{2S} p$ and $r\curle_{2S} q$, and so $\grcd_{\curle_{2S}}(p,q)$ does not exist. For $p,q\in\proc$ such that $p\equiv_S q$, we define $(p,q)$ as follows.
\[
(p,q)=\sum \{ a.(p',q') \mid a\in \act,~p\myarrowa p',~q\myarrowa q'\text{ and } p'\equiv_S q' \}.
\]
%$$(p,q)=\sum_{\substack{a\in \act\\ p\myarrowa %p'\\q\myarrowa q'\\ p'\equiv_S q'}} a.(p',q').$$
Intuitively, for every $p'\in\reach(p)$ and $q'\in\reach(q)$, we form a pair $(p',q')$ only if $p'\equiv_S q'$. Then, we connect two pairs $(p',q')$ and $(p'',q'')$ through transition $a$, if $p'\myarrowa p''$ and $q'\myarrowa q''$. We prove that for every $p,q\in\proc$, process $(p,q)$ satisfies the conditions of Definition~\ref{def:gcd}.
    % Consider the LTSs $\mathS_{p_1}=(P_1,\act,\longrightarrow_1)$ and $\mathS_{p_2}=(P_2,\act,\longrightarrow_2)$ that correspond to $p_1$ and $p_2$, respectively. We construct LTS $\mathS_{p_1\times p_2}=(P,\act,\longrightarrow_{1\times 2})$ by starting with empty $P$ and $\longrightarrow_{1\times 2}$ and applying the following steps.
    % \begin{description}
    %     \item[Step 1: Add states.] For every $p\in P_1$, $q\in P_2$, such that $p\equiv_S q$, add $(p,q)$ to $P$.
    %     \item[Step 2: Add transitions.] For every $(p,q), (p',q')\in P$ and $a\in \act$, if $p\myarrowa p'$ and $q\myarrowa q'$, then add $((p,q),a, (p',q'))$ to $\longrightarrow_{1\times 2}$.
    %     \item[Step 3. Remove non-reachable states.] For every $(p,q)\in P$, if $(p,q)\not\in\reach((p_1,p_2))$, then remove $(p,q)$ from $P$.
    % \end{description}
    %  We show that if $\mathS_{p_1\times p_2}$ is not empty, then it is the unique $\grcd_{\curle_{2S}}(p_1,p_2)$ up to $\equiv_{2S}$. 
     \begin{clm}\label{claim:gcd-one}
     For every $(p,q)$, it holds that $(p,q)\curle_{2S} p$ and $(p,q)\curle_{2S} q$.
     \end{clm}
     \begin{proof}
     We prove the claim by induction on the size of $(p,q)$.
      \begin{description}
        \item[Let $(p,q)=\mathtt{0}$] From the definition of $(p,q)$, we have that $p\equiv_S q$. Assume, towards a contradiction, that $I(p)\neq\emptyset$ and $a\in I(p)$ for some $a\in\act$. Then, Lemma~\ref{lem:sim-equiv-processes}(a) implies that there are $p\myarrowa p^*$ and $q\myarrowa q^*$ such that $p^*\equiv_S q^*$. By the definition of $(p,q)$, we have that $(p,q)\myarrowa (p^*,q^*)$, which contradicts the proviso for this case. 
        %We conclude the same contradiction if we assume %that $I(q)\neq\emptyset$. 
        Therefore, $I(p)=\emptyset$ and, by symmetry, $p=q=\mathtt{0}$. Finally, $(p,q)\curle_{2S} p$ and $(p,q)\curle_{2S} q$. 
        \item[Let $(p,q)\myarrowa (p',q')$ for some $a\in\act$] Then, there are $p',q'\in\proc$, such that $p'\equiv_S q'$, $p\myarrowa p'$, and $q\myarrowa q'$. From the inductive hypothesis, $(p',q')\curle_{2S} p'$ and $(p',q')\curle_{2S} q'$. We also show that $p\curle_S (p,q)$ and $q\curle_S (p,q)$. Let $p\myarrowa p'$. From Lemma~\ref{lem:sim-equiv-processes}(b), one of the following holds.
        \begin{itemize}
            \item There is $q\myarrowa q'$ and $p'\equiv_S q'$, which means that $(p,q)\myarrowa (p',q')$. From the inductive hypothesis, $p'\curle_S (p',q')$.
            \item There are $p^*,q^*$ such that $p\myarrowa p^*$, $q\myarrowa q^*$, $p^*\equiv_S q^*$, and $p'\curle_S p^*$. From the inductive hypothesis, $p^*\curle_S (p^*,q^*)$ and therefore, $p'\curle_S (p^*,q^*)$.
        \end{itemize}
        We can show that $q\curle_S (p,q)$ in an analogous way.\qedhere
        \end{description}
        \end{proof}
   \begin{clm}\label{claim:gcd-two}
       For every $p,q\in\proc$ and $r\in\proc$ such that $r\curle_{2S} p$ and $r\curle_{2S} q$, it holds that $r\curle_{2S} (p,q)$.
   \end{clm}
    \begin{proof}
        Assume that there is a process $r$ such that $r\curle_{2S} p$ and $r\curle_{2S} q$. We prove the claim by induction on $p$ and $q$.
        \begin{itemize}
            \item If $p=q=\mathtt{0}$, then $r=\mathtt{0}$ and the claim trivially holds.
            \item Let $I(p)\neq \emptyset$ or $I(q)\neq\emptyset$. From the hypothesis of the claim, $r\equiv_S p \equiv_S q$.
            From Claim~\ref{claim:gcd-one}, $(p,q)\equiv_S p\equiv_S q$ and therefore, $(p,q)\equiv_S r$. Let $r\myarrowa r'$. From the hypothesis of the claim, there are $p\myarrowa p'$ and $q\myarrowa q'$ such that $r'\curle_{2S} p'$ and $r'\curle_{2S} q'$. Hence, $p'\equiv_S q'$, and from the definition of $(p,q)$, $(p,q)\myarrowa (p',q')$. From the inductive hypothesis, $r'\curle_{2S} (p',q')$.\qedhere
        \end{itemize} 
        \end{proof}
    \begin{clm}\label{claim:gcd-three}
    For every $p,q\in\proc$, $\grcd_{\curle_{2S}}(p,q)$ exists iff $p\equiv_S q$. 
    \end{clm}
    \begin{proof}
        Let $p,q\in\proc$ and $g$ denote $\grcd_{\curle_{2S}}(p,q)$. If $g$ exists, then $g\curle_{2S} p$ and $g\curle_{2S} q$, which implies that $g\equiv_S p\equiv_S q$. Conversely, if $p\equiv_S q$, then from Claim~\ref{claim:gcd-one}, $(p,q)\curle_{2S} p$ and $(p,q)\curle_{2S} q$. From Claim~\ref{claim:gcd-two}, if there is $r\in\proc$ such that $r\curle_{2S} p$ and $r\curle_{2S} q$, then $r\curle_{2S} (p,q)$. So $(p,q)$ satisfies the conditions of Definition~\ref{def:gcd}.
    \end{proof} 
    
Claim~\ref{claim:gcd-three} implies that the existence of $\grcd_{\curle_{2S}}(p,q)$, $p,q\in\proc$, can be decided in polynomial time, since $p\equiv_S q$ can be checked in polynomial time~\cite{HT94, KS90}.
    \begin{clm}\label{claim:gcd-four}
       If $\grcd_{\curle_{2S}}(p,q)$ exists, then $(p,q)\equiv_{2S} \grcd_{\curle_{2S}}(p,q)$.
    \end{clm}
    \begin{proof}
     This is immediate from the proof of Claim~\ref{claim:gcd-three} and Remark~\ref{rem:gcd-uniqueness}.   
    \end{proof} 
    
    Claim~\ref{claim:gcd-four} says that $(p,q)$ is the unique $\grcd_{\curle_{2S}}(p,q)$ modulo $\equiv_{2S}$. From Claim~\ref{claim:gcd-four}, the definition of $(p,q)$,  and the fact that $p\equiv_S q$ can be checked in polynomial time, computing and returning $(p,q)$ can be done in polynomial time. In particular, the number of processes reachable from $(p,q)$ is bounded by the number of pairs $(p',q')$ such that $p'\in\reach(p)$ and $q'\in\reach(q)$. Therefore, $|(p,q)|\leq 2 |p| |q|$, where the factor of 2 accounts for both the pairs reachable from $(p,q)$ and the transitions between them.
    \end{proof}

     Procedure \gcd{$p,q$} in Algorithm~\ref{alg:procedures} computes the unique $\grcd_{\curle_{2S}}(p,q)$ modulo $\equiv_{2S}$ as was described in the proof of Lemma~\ref{lem:gcd-2s-poly-time}, and so \gcd{$p,q$} runs in polynomial time.

% \begin{cor}\label{cor:gcd-existence}
%     Let $p_1,p_2\in\proc$.  $\grcd_{\curle_{2S}}(p_1,p_2)$ exists iff $p_1\equiv_S p_2$.
% \end{cor}
% \begin{proof}
%     This is immediate from the proof of Lemma~\ref{lem:gcd-2s-poly-time}.
% \end{proof}

\begin{lem}\label{lem:gcd-satisfies-phi}
Let $\varphi\in\mathL_{2S}$ and $p,q\in\proc$ such that $p\models\varphi$ and $q\models\varphi$. Then, the following are equivalent.
\begin{enumerate}
    \item $\grcd_{\curle_{2S}}(p,q)$ exists and $\grcd_{\curle_{2S}}(p,q)\models\varphi$.
    \item There is a process $r$ such that $r\models\varphi$, $r\curle_{2S} p$ and $r\curle_{2S} q$.
\end{enumerate}
\end{lem}
\begin{proof}
    Let $g$ denote  $\grcd_{\curle_{2S}}(p,q)$.\\
    $(1)\Rightarrow (2)$ If $g\models\varphi$, then (2) is true by the definition of $g$.\\
    $(2)\Rightarrow (1)$ If there is a process $r$ as described in (2), then by the definition of $g$, $r\curle_{2S} g$, and from Proposition~\ref{logical_characterizations}, 
    $g\models \varphi$.
\end{proof}

We introduce two algorithms, namely $\mathtt{ConPro}$ in Algorithm~\ref{alg:process-construction}, and $\mathtt{Prime_{2S}}$ in Algorithm~\ref{alg:prime-twosim}. Let $\varphi\in\mathL_{2S}$ be an input to the first algorithm. Lines 1--19 of $\mathtt{ConPro}$ are an implementation of the tableau construction for $\varphi$---see Subsection~\ref{subsection:hml-tableau}. If $\varphi$ is unsatisfiable, then $\mathtt{ConPro}(\varphi)$ stops without returning an output because it stops at lines 12 or 18. In the case that $\mathtt{ConPro}(\varphi)$ returns an output, then its output is an LTS corresponding to a process that satisfies $\varphi$. If there are $r_1,r_2$ satisfying $\varphi$ such that $r_1\not\curle_S r_2$, the tableau construction cannot guarantee the generation of two processes that are not simulation equivalent. This is precisely the role of lines 20--30 in $\mathtt{ConPro}$. Given such processes $r_1, r_2$, when run twice, the algorithm can choose two processes $p_1,p_2$ based on $r_1,r_2$. During construction of $p_1$, lines 20--30 can be used to add  to $p_1$ up to $|\varphi|$ states that witness the failure of $r_1\curle_S r_2$. Note that in the case of the \simequiv game, player $B$ could follow a similar strategy by using move B($\square$) and introducing a trace that witnesses $r_1\not\curle_S r_2$. In the case of $\mathL_{2S}$, since the full tableau is constructed, the algorithm needs only to construct a `small' process that serves as a witness to the same fact. 

Algorithm $\mathtt{Prime_{2S}}$ decides whether its input $\varphi\in\mathL_{2S}$ is prime: $\varphi$ is prime iff every execution of $\mathtt{Prime_{2S}}(\varphi)$ accepts. This algorithm runs $\mathtt{ConPro}(\varphi)$ twice. If $\mathtt{ConPro}(\varphi)$ fails to return an output, $\mathtt{Prime_{2S}}(\varphi)$  rejects at line 5---this line deals with unsatisfiability. For every two processes $p_1,p_2$ that satisfy $\varphi$, at line 7, $\mathtt{Prime_{2S}}(\varphi)$ constructs their maximal lower bound, denoted $\grcd_{\curle_{2S}}(p_1,p_2)$, which is a process $g$ that is 2-nested-simulated by both $p_i$'s and $r\curle_{2S} g$, for every process $r$ such that $r\curle_{2S} p_i$. Processes $p_1,p_2$ have a maximal lower bound iff $p_1\equiv_S p_2$.
%and $\grcd_{\curle_{2S}}(p_1,p_2)$ can be constructed in polynomial time. 
In case $\grcd_{\curle_{2S}}(p_1,p_2)$ does not exist, the algorithm discovers two processes satisfying $\varphi$ such that there is no process that is 2-nested-simulated by both of them and it rejects the input---$\varphi$ is not prime. On the other hand, if $\grcd_{\curle_{2S}}(p_1,p_2)$ exists, then there is a process that is 2-nested simulated by both $p_i$'s and it can be constructed in polynomial time. It remains to check whether $\grcd_{\curle_{2S}}(p_1,p_2)$ satisfies $\varphi$. If so, then the second condition described above is met, and the algorithm accepts. If $\grcd_{\curle_{2S}}(p_1,p_2)\not\models\varphi$ it can be shown that there is no process $r$ satisfying $\varphi$ that is 2-nested-simulated by both $p_i$'s, and the algorithm rejects at line 10. This establishes Theorem~\ref{prop:2S-primality-conpc}.

\begin{algorithm}
\caption{Algorithm $\mathtt{ConPro}$ that takes as input $\varphi\in\mathL_{2S}$, and extends the tableau construction for $\varphi$ with lines 20--30. }\label{alg:process-construction}
\DontPrintSemicolon
\KwIn{$\varphi\in\mathL_{2S}$}
{$S\gets\{s_0\}$}\;
{$L(s_0)=\{\varphi\}$, $d(s_0)\gets 0$}\;
{$BoxCount\gets 0$}\;
\lFor{all $a_j\in \act$}{$R_{a_j}\gets\emptyset$}
{$Q.\mathtt{enqueue}(s_0)$}\;
\While{$Q$ is not empty}{
{$s\gets Q.\mathtt{dequeue()}$}\;
\While{$L(s)$ contains $\psi_1\wedge\psi_2$ or $\phi_1\vee\phi_2$}
{{$L(s)\gets L(s)\setminus\{\psi_1\wedge\psi_2\}\cup\{\psi_1\}\cup\{\psi_2\}$}\;
{non-deterministically choose $\phi$ between $\phi_1$ and $\phi_2$}\;
{$L(s)\gets L(s)\setminus\{\phi_1\vee\phi_2\}\cup\{\phi\}$}\;}
\lIf{$\ff\in L(s)$}{stop}
\For{all $\langle a_j\rangle\psi\in L(s)$}
{{$S\gets S\cup\{s'\}$} \Comment{$s'$ is a fresh state}\;
{$L(s')=\{\psi\}\cup\{\phi\mid [a_j]\phi\in L(s)\}$}\;
{$d(s')\gets d(s)+1$}\;
{$R_{a_j}\gets R_{a_j}\cup\{(s,s')\}$}\;
{\lIf{$\ff\in L(s')$}{stop}}
{\lIf{$d(s')<\md(\varphi)+1$}{$Q.\mathtt{enqueue}(s')$}
}}
{Non-deterministically choose to go to line 6 or line 21}\;
{Non-deterministically choose $N\in\{1,\dots,|\varphi|-BoxCount\}$}\;
\For{$i\gets 1$  to $N$}{
{Non-deterministically choose $j\in\{1,\dots,k\}$}\;
{$S\gets S\cup\{s'\}$} \Comment{$s'$ is a fresh state}\;
{$L(s')=\{\phi\mid [a_j]\phi\in L(s)\}$}\;
{$d(s')\gets d(s)+1$}\;
{$R_{a_j}\gets R_{a_j}\cup\{(s,s')\}$}\;
\lIf{$\ff\in L(s')$}{stop}
{\lIf{$d(s')<\md(\varphi)+1$}{$Q.\mathtt{enqueue}(s')$}}
{$BoxCount\gets BoxCount+1$}
}}
{Return $S, R_{a_1},\dots, R_{a_k}$}
\end{algorithm}

\begin{algorithm}
\caption{Algorithm $\mathtt{Prime_{2S}}$ decides whether $\varphi\in\mathL_{2S}$ is prime. State $s_0^i$, $i=1,2$, denotes the first state that is added to $S_i$ by $\mathtt{ConPro}(\varphi)$.
%The algorithm also uses the procedures that are presented in Algorithm~\ref{alg:procedures}
Procedure 
$\mathtt{Process}(S,s_0^i,R_{a_1},\dots,R_{a_k})$ 
computes a process corresponding to the output of $\mathtt{ConPro}$ and 
$\mathtt{MLB}(p_1,p_2)$
%\gcd{$p_1,p_2$} 
returns the $\grcd_{\curle_{2S}}(p_1,p_2)$.}\label{alg:prime-twosim}
\DontPrintSemicolon
\KwIn{$\varphi\in\mathL_{2S}$}
{$(S_1,R^1_{a_1},\dots, R^1_{a_k})\gets \mathtt{ConPro}(\varphi)$}\;
{$p_1\gets \process{$S_1,s_0^{1}, R^1_{a_1},\dots, R^1_{a_k}$}$}\;
{$(S_2,R^2_{a_1},\dots, R^2_{a_k})\gets \mathtt{ConPro}(\varphi)$}\;
{$p_2\gets \process{$S_2,s_0^{2}, R^2_{a_1},\dots, R^2_{a_k}$}$}\;
\lIf{some of the two calls of $\mathtt{ConPro}(\varphi)$ stops without an output}{accept}
\Else{
{$g\gets\gcd{$p_1,p_2$}$}\;
\lIf{$g$ is empty}{reject}
\lIf{$g\models\varphi$}{accept}
\lElse{reject}
}
\end{algorithm}

\begin{lem}\label{lem:phi-sat}
Let $\varphi\in\mathL_{2S}$ be a satisfiable formula. There is a sequence of non-deterministic choices that $\mathtt{ConPro}(\varphi)$ can make such that it outputs some $(S,R_{a_1},\dots,R_{a_k})$. 
\end{lem}
\begin{proof}
   Let $s\in S$ and $L_{12}(s)$ denote the set of formulae that $L(s)$ contains after line 12 of Algorithm~\ref{alg:process-construction} is completed. The following two claims follow directly from the operations performed in lines 8--12 of the algorithm.
   \begin{clm}\label{claim:line-eleven}
       After lines 8--12 of Algorithm~\ref{alg:process-construction} are executed and if the algorithm does not stop at line 12, $L_{12}(s)$ contains formulae that are either the formula $\true$ or start with $\langle a\rangle$ or $[a]$, for some $a\in\act$.
   \end{clm}
   \begin{clm}\label{claim:lines-seven-to-eleven}
       Assume that $\mathtt{ConPro}(\varphi)$ is at line 6 and starts examining a state $s$ and $DF(\bigwedge L(s))=\bigvee_{i=1}^m\Phi_i$. After executing lines 8--11, $\bigwedge L_{12}(s)=\Phi_n$ for some $1\leq n\leq m$. Conversely, for every $1\leq n\leq m$, there is a sequence of non-deterministic choices that $\mathtt{ConPro}(\varphi)$ can make, when executing lines 8--11, such that $\bigwedge L_{12}(s)=\Phi_n$.
   \end{clm}
   We show that there is a sequence of non-deterministic choices of $\mathtt{ConPro}(\varphi)$ such that the algorithm does not stop at lines 12, 18, and 28, and it outputs some $(S, R_{a_1}, \dots, R_{a_k})$. 
  \begin{clm}\label{claim:line-25}
     There is a sequence of non-deterministic choices such that $\mathtt{ConPro}(\varphi)$ does not stop at line 28.
  \end{clm}
  \begin{proof}
      When $\mathtt{ConPro}(\varphi)$ executes line 20, non-deterministically chooses to go to line 6, and so lines 21--30 are not executed.
  \end{proof}
\begin{clm}\label{claim:lines-11-25}
Assume that $\mathtt{ConPro}(\varphi)$ is at line 7 and starts examining a state $s$ such that $\bigwedge L(s)$ is satisfiable. Then, there is a sequence of non-deterministic choices that $\mathtt{ConPro}(\varphi)$ can make at lines 8--11 such that it does not stop at lines 12 and 18. Moreover, for every $1\leq j\leq k$ and $(s,s')$ that is added to $R_{a_j}$ at line 17, $\bigwedge L(s')$ is satisfiable.
\end{clm}
\begin{proof}
    Let $DF(\bigwedge L(s))=\bigvee_{i=1}^m\Phi_i$. From Lemma~\ref{lem:DF-equiv}, $\bigwedge L(s)\equiv\bigvee_{i=1}^m\Phi_i$. Assume that $\mathtt{ConPro}(\varphi)$ examines $s$ and executes lines 8--12. Since $\bigwedge L(s)$ is satisfiable and $\bigwedge L(s)\equiv\bigvee_{i=1}^m\Phi_i$, there is some satisfiable $\Phi_n$, $1\leq n\leq m$. From Claim~\ref{claim:lines-seven-to-eleven}, $\mathtt{ConPro}(\varphi)$ can make such non-deterministic choices while executing lines 8--11 so that $\bigwedge L_{12}(s)=\Phi_n$. Since $\Phi_n$ is satisfiable,  $L_{12}(s)$ does not contain $\ff$ and $\mathtt{ConPro}(\varphi)$ does not stop at line 12. Let $1\leq j\leq k$ and $\langle a_j\rangle\psi_1,\dots,\langle a_j\rangle\psi_t,[a_j]\psi_1',\dots,[a_j]\psi'_{t'}$ be all formulae in $L_{12}(s)$ that start with $\langle a_j\rangle$ or $[a_j]$. Then, since $\Phi_n$ is satisfiable, for every $1\leq i\leq t$, $\langle a_j\rangle\psi_i\wedge\bigwedge_{l=1}^{t'} [a_j]\psi'_l$ is satisfiable. As a result, $\psi_i\wedge\bigwedge_{l=1}^{t'}\psi'_l$ is satisfiable, which implies that for every $(s,s')$ that is added to $R_{a_j}$ at line 17, $\bigwedge L(s')$ is satisfiable. In particular, $L(s')$ does not contain $\ff$ and the algorithm does not stop at line 18.
\end{proof}

\begin{clm}\label{claim:output}
    Assume that $\mathtt{ConPro}(\varphi)$ makes the non-deterministic choices described in Claims~\ref{claim:line-25} and~\ref{claim:lines-11-25}. Then, $\mathtt{ConPro}(\varphi)$ outputs some $(S,R_{a_1},\dots, R_{a_k})$.
\end{clm}
\begin{proof}
    Immediate from Claims~\ref{claim:line-25} and~\ref{claim:lines-11-25} and the fact that the algorithm starts with $L(s_0)=\{\varphi\}$, where $\varphi$ is satisfiable.
\end{proof}

Claim~\ref{claim:output} guarantees that the algorithm returns an output when the input is a satisfiable formula in $\mathL_{2S}$.
\end{proof}

Next, we prove that in case $\mathtt{ConPro}(\varphi)$ returns an output, then $\varphi\in\mathL_{2S}$ is satisfiable. The next lemma uses procedure \process{$S,s,R_{a_1},\dots, R_{a_k}$}, which takes as input a set of states $S$, a designated state $s\in S$, and the  binary relations $R_{a_1},\dots, R_{a_k}$ over $S$, and returns the process that naturally corresponds to its input. It is described in Algorithm~\ref{alg:procedures}.

\begin{lem}\label{lem:model-of-phi-and-phi-unsat}
Let $\varphi\in\mathL_{2S}$, $(S,R_{a_1},\dots,R_{a_k})$ be an output of $\mathtt{ConPro}(\varphi)$, and $r$ denote \process{$S,s_0,R_{a_1},\dots,R_{a_k}$}, where  $s_0$ is the first state added to $S$ by the algorithm. Then, $|r|\leq\md(\varphi)+1$ and $r\models\varphi$.
\end{lem}
\begin{proof}
Note that a state $s$ is added to the queue $Q$ at lines 19 and 29 only if $d(s)\leq \md(\varphi)$ and is examined by the algorithm during a future iteration. When $s$ is examined, if a new state $s'$ is added to $S$, then $d(s')=d(s)+1$. Given that the initial state $s_0$ has $d(s_0)=0$, and a state $s'$ can be also added to $S$ when examining a state $s$ with $d(s)=\md(\varphi)$, it follows that, upon completion of the algorithm, every $s\in S$ satisfies $d(s)\leq\md(\varphi)+1$. It is immediate from the structure of the procedure \process{$\cdot$} described in Algorithm~\ref{alg:procedures}, that \process{$S,s_0,R_{a_1},\dots,R_{a_k}$} is a process of depth bounded by $\md(\varphi)+1$. We show that \process{$S,s_0,R_{a_1},\dots,R_{a_k}$}$\models\varphi$.

\begin{clm}\label{claim:modal-depth}
    For every $s\in S$, $\max\{\md(\psi)\mid \psi\in L_{12}(s)\}=\begin{cases}
        \md(\varphi)-d(s), &\text{if } d(s)\leq\md(\varphi),\\
        0, &\text{otherwise}
    \end{cases}$.
\end{clm}
\begin{proof}
    By an easy induction on $d(s)$.
\end{proof}

  \begin{clm}\label{claim:process}
      For every $s\in S$, \process{$S,s, R_{a_1}, \dots, R_{a_k}$}$\models \bigwedge L_{12}(s)$.
  \end{clm}
  \begin{proof}
    By induction on $\max\{\md(\psi)\mid \psi\in L_{12}(s)\}$. In the base case, $L_{12}(s)$ either contains formula $\true$ or is empty, and so the claim trivially holds. It is not hard to show the inductive step.
    \end{proof}
 The following claim is straightforward from Claim~\ref{claim:process}.
\begin{clm}\label{claim:pzero}
    \process{$S,s_0, R_{a_1}, \dots, R_{a_k}$}$\models\varphi$.
\end{clm}
\begin{proof}
 Let $DF(\varphi)=\bigvee_{i=1}^n\varphi_i$. From Lemma~\ref{lem:DF-equiv}, $\varphi\equiv \bigvee_{i=1}^n\varphi_i$.  From Claim~\ref{claim:process}, when the algorithm examines $s_0$ and after lines 8--11 have been executed, $\bigwedge L_{12}(s_0)=\varphi_j$ for some $1\leq j\leq n$. Consequently, $\bigwedge L_{12}(s_0)\models\varphi$, and, \process{$S,s_0, R_{a_1}, \dots, R_{a_k}$}$\models\varphi$.
\end{proof}
The lemma follows directly from the observations of the first paragraph of this proof and Claim~\ref{claim:pzero}.
\end{proof}

\begin{lem}\label{lem:model-of-phi-size}
  Let $\varphi\in\mathL_{2S}$, $(S,R_{a_1},\dots,R_{a_k})$ be an output of $\mathtt{ConPro}(\varphi)$, and $r$ denote $\mathtt{Process}(S,s_0,R_{a_1},\dots,$ $R_{a_k})$, where  $s_0$ is the first state added to $S$ by the algorithm. Then, $|r|\leq 4|\varphi|$.
\end{lem}
\begin{proof}
 If lines 21--30 of $\mathtt{ConPro}$ are omitted, then the algorithm constructs a tableau for $\varphi$, which is of polynomial size as was shown in~\cite[Proposition 77]{AACI24}. In particular, if the algorithm did not contain lines 21--30, the number of states added to $S$ would be bounded by $|\varphi|$.  Since a new pair is added to $R_{a_j}$, for some $1\leq j\leq k$, every time a fresh state is added to $S$, the number of pairs added to all $R_{a_j}$, $1\leq j\leq k$, would also be at most $|\varphi|$. 
 In the worst case, lines 23--30, i.e.\ the last for-loop of the algorithm, will be executed at most $|\varphi|$ times: variable $BoxCount$ is increased by one every time these lines are completed and the number of iterations of the for-loop, namely $N$, is chosen to be at most $|\varphi|-BoxCount$ at line 21. Consequently, the execution of lines 21--30 can add at most $|\varphi|$ states to $S$ and at most $|\varphi|$ pairs to all $R_{a_j}$. Note also that if a state $s$ is added to $S$ at line 24, then $L(s)$ consists of formulae that do not contain $\langle a \rangle$ for any $a\in \act$, because $\varphi\in\mathL_{2S}$. Consequently, when $s$ is examined, the for-loop starting at line 13 will be skipped. So, state $s$ can only lead to the addition of more states to $S$ at line 24, and we already argued that no more than $|\varphi|$ can be added to $S$ because of that line.
  As a result, there are at most $2|\varphi|$ states in $S$ and $2|\varphi|$ pairs in all $R_{a_j}$ when the algorithm returns the output $(S, R_{a_1},\dots, R_{a_k})$.   
 Given the structure of \process{$\cdot$} and the fact that $(S, R_{a_1}, \dots, R_{a_k})$ forms a directed tree when interpreted as a graph, it follows that \process{$S, s_0, R_{a_1}, \dots, R_{a_k}$} is of size at most $4|\varphi|$.
\end{proof}

%We introduce the non-deterministic polynomial-time algorithm 
We now examine $\mathtt{Prime_{2S}}$, which decides primality in the logic $\mathL_{2S}$. As shown in Algorithm~\ref{alg:prime-twosim}, $\mathtt{Prime_{2S}}(\varphi)$ uses $\mathtt{ConPro}(\varphi)$ as a subroutine, as well as procedures \process{$\cdot$} and \gcd{$\cdot$} that are presented in Algorithm~\ref{alg:procedures}. Previously, in Lemma~\ref{lem:phi-sat}, our focus was on the existence of an execution of $\mathtt{ConPro}(\varphi)$ that returns an output when $\varphi$ is satisfiable. In contrast, our current analysis concerns whether every execution of $\mathtt{Prime_{2S}}(\varphi)$ results in acceptance. This shift in perspective leads us to examine the behavior of all possible executions of $\mathtt{ConPro}$ within $\mathtt{Prime_{2S}}$.

\begin{algorithm}
\caption{Procedure 
$\mathtt{Process}(S,s_0,R_{a_1},\dots,R_{a_k})$
returns the process that naturally corresponds to its input. Procedure 
$\mathtt{MLB}(p,q)$ 
returns  
$\grcd_{\curle_{2S}}(p,q)$.}
\label{alg:procedures}
\DontPrintSemicolon
  \SetKwFunction{process}{Process}
  \SetKwProg{Fn}{procedure}{:}{}
  \Fn{\process{$S,s_0,R_{a_1},\dots,R_{a_k}$}}{
  {return $\displaystyle\sum_{\substack{a\in\act\\(s, s')\in R_{a}}}a.\process{$S,s',R_{a_1},\dots,R_{a_k}$}$}}
  \;
   \SetKwFunction{gcd}{MLB}
  \SetKwProg{Fn}{procedure}{:}{}
  \Fn{\gcd{$p,q$}}{
     \lIf{$p\not\equiv_S q$}{stop}
     %\lIf{$p=q=\mathtt{0}$}{return $\mathtt{0}$}
     \lElse{return $\displaystyle\sum_{\substack{a\in \act\\ p\myarrowa p'\\q\myarrowa q'\\ p'\equiv_S q'}} a.$\gcd{$p',q'$}}}
\end{algorithm}

We show that $\varphi\in\mathL_{2S}$ is prime iff every execution of $\mathtt{Prime_{2S}}(\varphi)$ accepts, thus establishing that the problem lies in \conp.

\begin{prop}\label{prop:prime-implies-accept}
    If $\varphi\in\mathL_{2S}$ is prime, then every sequence of non-deterministic choices that $\mathtt{Prime}_{2S}(\varphi)$ makes leads to acceptance.
\end{prop}
\begin{proof}
    If $\varphi$ is unsatisfiable and prime, then from Lemma~\ref{lem:model-of-phi-and-phi-unsat}, $\mathtt{ConPro}$ does not return an output and $\mathtt{{Prime}_{2S}}(\varphi)$ accepts at line 5. Assume that $\varphi$ is satisfiable and prime and $\mathtt{ConPro}(\varphi)$ returns an output at both lines 1 and 3 of $\mathtt{{Prime}_{2S}}(\varphi)$. From Lemma~\ref{lem:model-of-phi-and-phi-unsat}, $p_1\models\varphi$ and $p_2\models\varphi$. From Proposition~\ref{prop:charact-via-primality}, $\varphi$ is characteristic for a process within $\mathL_{2S}$. Let $p$ denote the process for which $\varphi$ is characteristic within $\mathL_{2S}$. Then, because of Definition~\ref{def:characteristic}, $p\models\varphi$, and $\mathL_{2S}(p)\subseteq \mathL_{2S}(p_i)$, or from Proposition~\ref{logical_characterizations}, $p\curle_{2S} p_i$ for $i=1,2$. Therefore, from the definition of $\curle_{2S}$, $p\equiv_S p_1\equiv_S p_2$ and from Claim~\ref{claim:gcd-three}, $\grcd_{\curle_{2S}}(p_1,p_2)$ exists and \gcd{$p_1,p_2$} returns $\grcd_{\curle_{2S}}(p_1,p_2)$ at line 7. Since there is a process, namely $p$, such that satisfies $\varphi$ and is 2-nested simulated by both $p_1$ and $p_2$, Lemma~\ref{lem:gcd-satisfies-phi} guarantees that $\grcd_{\curle_{2S}}(p_1,p_2)\models\varphi$, and so $\mathtt{Prime_{2S}}$ accepts at line 9.
\end{proof}

Conversely, we show that if $\mathtt{Prime_{2S}}(\varphi)$ always accepts, then $\varphi$ is prime. To this end, we first prove that if every execution of $\mathtt{Prime_{2S}}(\varphi)$ leads to acceptance, then every two processes that satisfy $\varphi$ are equivalent modulo $\equiv_S$.

\begin{defi}\label{def:a-witness}
Let $r_1, r_2$ be two processes such that $r_1\not\curle r_2$. We say that $r_1'$ is an $a$-witness of $r_1\not\curle r_2$ if $r_1\myarrowa r_1'$ and for all $r_2\myarrowa r_2'$, $r_1'\not\curle_S r_2'$. When we write $r_1'=\wit(r_1,r_2,a)$ we mean that $r_1'$ is an $a$-witness of $r_1\not\curle r_2$.
\end{defi}

\begin{lem}\label{lem:wit-existence}
If $r_1\not\curle_S r_2$ and $a\in I(r_1)\Rightarrow a\in I(r_2)$ for every $a\in \act$, then there is $b\in\act$ and $r_1\myarrowb r_1'$ such that $r_1'=\wit(r_1,r_2,b)$.    
\end{lem}
\begin{proof}
    Immediate from the definition of $\curle_S$.
\end{proof}

\begin{defi}\label{def:witness-process}
    Let $r_1, r_2$ be two processes such that $r_1\not\curle r_2$. We inductively define the witness process of $r_1\not\curle_S r_2$, denoted $\witproc(r_1,r_2)$, as follows.
    \begin{itemize}
        \item $\witproc(r_1,r_2)=a.\mathtt{0}$ if $r_1\myarrowb$ and $r_2\notmyarrowb$ for some $b\in\act$, and $a$ is the first such action in $\{a_1,\dots,a_k\}$;
        \item $\displaystyle\witproc(r_1,r_2)= a.\sum_{r_2\myarrowa r_2'} \witproc(r_1',r_2')$, where $a$ is the first action in $\{a_1,\dots,a_k\}$ such that there is an $a$-witness of $r_1\not\curle_S r_2$ and $r_1'=\wit(r_1,r_2,a)$, otherwise.
    \end{itemize}
\end{defi}

Note that $\witproc(r_1,r_2)$ is well-defined when $r_1\not\curle_S r_2$, because of Lemma~\ref{lem:wit-existence}.

\begin{lem}\label{lem:witproc-properties}
    Let $r_1,r_2$ be two processes such that $r_1\not\curle_S r_2$. Then, $\witproc(r_1,r_2)\curle_S r_1$, $\witproc(r_1,r_2)\not\curle_S r_2$, $\depth(\wit(r_1,r_2))\leq \depth(r_1)$, and $|\witproc(r_1,r_2)|\leq |r_2|$.
\end{lem}
\begin{proof}
  It is straightforward from the definition of $\wit(p_1,p_2)$.
\end{proof}

In what follows $S^\square$ (respectively, $S^\Diamond$) denotes the subset of $S$ consisting of the states added when lines 21--30 (respectively, linees 13--19) are executed by $\mathtt{ConPro}$. Similarly, $R_{a_j}^\square$ (respectively, $R_{a_j}^\Diamond$) denotes the subset of $R_{a_j}$ consisting of the pairs added when lines 21--30 (respectively, lines 13--19) are executed by $\mathtt{ConPro}$.
The following lemma is the analogue of Lemma~\ref{lem:B-can-play-consistently}, which guaranteed that player $B$ can play consistently on a process that satisfies $\varphi$ when playing the \simequiv game on $\varphi$.

\begin{lem}\label{lem:algo-chooses-consistently}
    Let $\varphi\in\mathL_{2S}$ and $r\in\proc$ such that $r\models\varphi$. There is a sequence of non-deterministic choices that $\mathtt{ConPro}(\varphi)$ can make such that there is a mapping $map: S^\Diamond\rightarrow \proc$ satisfying the following conditions.
    \begin{enumerate}
        \item for every $s\in S^\Diamond\cup \{s_0\}$, $map(s)\models\bigwedge L_{12}(s)$,
        \item for every $(s,s')\in R_{a_j}^\Diamond$, $map(s)\myarrowasubj map(s')$, and
        \item $map(s_0)=r$, where $s_0$ is the first state added to $S$ by $\mathtt{ConPro}(\varphi)$.
    \end{enumerate}
    In the case that $\mathtt{ConPro}(\varphi)$ makes these non-deterministic choices, we say that $\mathtt{ConPro}(\varphi)$ chooses its output $(S,R_{a_1},\dots,R_{a_k})$ consistently on $r$.
\end{lem}
\begin{proof}
    The proof is completely analogous to the proof of Lemma~\ref{lem:B-can-play-consistently}, where we replace moves B($\wedge$) and B($\vee$) with the execution of lines 8--11 of Algorithm~\ref{alg:process-construction}, move B($\Diamond$) with the execution of lines 13--19, and move B($\square$) with the execution of lines 21--30. Note that no state is added to $S$ during the execution of lines 21--30 when $\mathtt{ConPro}$ makes the non-deterministic choices described in the proof of Lemma~\ref{lem:B-can-play-consistently}. Therefore, $S=\{s_0\}\cup S^\Diamond$ and $R_{a_j}=R^\Diamond_{a_j}$ for every $1\leq j\leq k$.
\end{proof}

\begin{lem}\label{lem:choose-cons-simulation}
 Let $\varphi\in\mathL_{2S}$ and $r\in\proc$ such that $r\models\varphi$. Assume that $\mathtt{ConPro}(\varphi)$ chooses its output $(S,R_{a_1},\dots,R_{a_k})$ consistently on $r$ and $p=$~\process{$S,R_{a_1},\dots,R_{a_k}$}. Then, 
 \begin{enumerate}[(a)]
\item  $p\curle_S r$, and
\item if, in addition, every two processes that satisfy $\varphi$ are simulation-equivalent, then $p\curle_{2S} r$.
 \end{enumerate}
\end{lem}
\begin{proof}
   Immediate from the definitions of $\curle_S$ and $\curle_{2S}$, the existence of a mapping $map:S\rightarrow\proc$ satisfying conditions 1--3 of Lemma~\ref{lem:algo-chooses-consistently}, and the definition of procedure \process{$\cdot$}. The proof of part (b) relies also on Lemma~\ref{lem:model-of-phi-and-phi-unsat}, which implies that $r\equiv_S p$.
\end{proof}

\begin{lem}\label{lem:box-2s}
    Let $S\subseteq\mathL_{2S}$ such that it contains only formula $\true$ and formulae that start with either $\langle a\rangle$ or $[a]$, $a\in\act$. Then, for every  $\phi\in\{\psi\mid [b]\psi\in S \text{ for some } b\in\act\}$, $\neg\phi\in\mathL_S$.
\end{lem}
\begin{proof}
    Immediate from the definitions of $\mathL_S$ and $\mathL_{2S}$.
\end{proof}

\begin{lem}\label{lem:cons-on-subprocess}
    Let $\varphi\in\mathL_{2S}$, $r\in\proc$ such that $r\models \varphi$. Let $p\in\proc$ be a process satisfying the following conditions: (1) $p\curle_S r$, (2) $\depth(p)\leq\md(\varphi)+1$, (3) $|\reach(p)|\leq |\varphi|$ and $|p|\leq 2|\reach(p)|$, and (4) every $p',p''\in\reach(p)$ are connected through exactly one transition. Then, there is a sequence of non-deterministic choices that $\mathtt{ConPro}(\varphi)$ can make such that there is a surjective mapping $\cop:S^\square\cup\{s_0\}\rightarrow \reach(p)$ satisfying the following conditions.
    \begin{enumerate}[i.]
        \item $\cop(s_0)=p$,
        \item $(s,s')\in R^\square_{a_j}\Leftrightarrow \cop(s)\myarrowasubj \cop(s')$,
        \item $\cop(s)\models\bigwedge L_{12}(s)$ for every $s\in S^\square$.
    \end{enumerate}
    If $\mathtt{ConPro}(\varphi)$ makes these non-deterministic choices, we say that $\mathtt{ConPro}(\varphi)$ adds a copy of $p$ to its output $(S,R_{a_1},\dots,R_{a_k})$.
\end{lem}
\begin{proof}
    $\mathtt{ConPro}(\varphi)$ chooses its output consistently on $r$, i.e.\ it makes the non-deterministic choices described in the proof of Lemma~\ref{lem:cons-on-subprocess}. As a result, there is a mapping $map:S^\Diamond\cup\{s_0\}\rightarrow\proc$ as described in Lemma~\ref{lem:cons-on-subprocess}, which implies that $r\models\bigwedge L_{12}(s_0)$. Note that the algorithm always skips the execution of lines 21--30 when it chooses its output consistently on $r$. In this proof, we describe how the algorithm can complement these choices by sometimes executing lines 21--30 such that a surjective mapping $\cop:S^\square\cup\{s_0\}\rightarrow \reach(p)$ satisfying conditions i--iii exists. 
    %We also describe what choices it makes during the execution of lines 21--30 if they are reached. 
    We start by setting $\cop(s_0)=p$, and we will demonstrate that the algorithm can proceed in such a way that $\cop$ can be extended to ultimately satisfy the conditions of the lemma.
    \begin{itemize}
        \item When $\mathtt{ConPro}(\varphi)$ examines state $s_0$ and reaches line 20, it decides to go to line 21. Then, it chooses $N=|\{p'\mid p\myarrowa p' \text{ where } a\in\act\}|$, and it makes one iteration of the for-loop for each $p\myarrowa p'$. Let $p\myarrowasubi p'$ for some $a_i\in\act$ and $p'\in\proc$. During the iteration that corresponds to this $a_i$-transition, the algorithm chooses $j=i$, adds state $s'$ to $S^\square$ and $(s_0,s')$ to $R^\square_{a_i}$ at lines 23, 24, and 27, respectively. We show that the algorithm does not stop at line 28 and $p'\models \bigwedge L_{12}(s')$, and so we can set $\cop(s')=p'$. Recall that $L(s')=\{\psi\mid [a_i]\psi\in L_{12}(s)\}$. Since $p\curle_S r$, there is $r\myarrowasubi r'$ such that $p'\curle_S r'$. Thus, $r\not\models [a_i]\ff$. As we noted above, $r\models \bigwedge L_{12}(s_0)$, which implies $[a_i]\ff\not\in L_{12}(s_0)$, $\ff\not\in L(s')$, and the algorithm does not stop at line 28. From the fact that $r\models\bigwedge L_{12}(s_0)$, we have also that $r'\models\bigwedge L(s')$. Let $DF(\bigwedge L(s'))=\bigvee_{i=1}^m\Psi_i$, $m\in\mathbb{N}$. From Lemma~\ref{lem:DF-equiv}, $\bigwedge L(s')\equiv \bigvee_{i=1}^m\Psi_i$ and so $r'\models\Psi_t$ for some $1\leq t\leq m$. From Claim~\ref{claim:lines-seven-to-eleven}, $\mathtt{ConPro}$ can execute lines 8--11 so that $\bigwedge L_{12}(s')=\Psi_t$. Since $\varphi\in\mathL_{2S}$, Claim~\ref{claim:line-eleven} and Lemma~\ref{lem:box-2s} imply that for every formula $\phi\in L(s')$, $\neg\phi\in\mathL_S$. As a result,$\neg\Psi_t\in\mathL_S$. If $p'\models\neg\Psi_t$, then from Proposition~\ref{logical_characterizations} and the fact that $p'\curle_S r'$, $r'\models\neg\Psi_t$, contradiction. We conclude that $p'\models\Psi_t$, or $p'\models \bigwedge L_{12}(s')$.
        \item Assume that $\mathtt{ConPro}(\varphi)$ examines some $s'\in S^\square$ for which we have set $\cop(s')=p'$ and $p'\models\bigwedge L_{12}(s')$ holds. Then, the algorithm chooses to go to line 21, picks $N=|\{p''\mid p'\myarrowa p'' \text{ where } a\in\act\}|$, and makes one iteration of the for-loop for each $p'\myarrowa p''$. Let $p'\myarrowasubi p''$ for some $a_i\in\act$ and $p''\in\proc$. As in the previous case, during the iteration that corresponds to this $a_i$-transition, the algorithm chooses $j=i$, adds state $s''$ to $S^\square$ and $(s',s'')$ to $R^\square_{a_i}$ at lines 23, 24, and 27, respectively. Similarly to the above, we can show that the algorithm does not terminate at line 28 and that $p''\models\bigwedge L_{12}(s'')$. So, we can set $\cop(s'')=p''$.
        \item In the case that $\mathtt{ConPro}(\varphi)$ examines a state $s\in S^\Diamond$ and reaches line 20, it chooses to go to line 6, and so lines 21--30 are not executed.
    \end{itemize}
    Note that if $\mathtt{ConPro}(\varphi)$ follows the non-deterministic choices described above, it adds exactly $|\reach(p)|$ states to $S^\square$. 
    %This is feasible because  (1) $\depth(p)\leq\md(\varphi)+1$ and for every $s$ added to $S$, $d(s)$ can be up to $\md(\varphi)+1$, and (2) $|\reach(p)| \leq |\varphi|$, and the constraints on $N$ ensure that the for-loop can be executed at most $|\varphi|$ times overall. 
    Given the four conditions satisfied by $p$ and the choices that $\mathtt{ConPro}(\varphi)$ makes which were described in the three cases above, we have that mapping $\cop:S^\square\cup\{s_0\}\rightarrow \reach(p)$ is surjective and satisfies conditions i--iii. Moreover, $\mathtt{ConPro}(\varphi)$ is able to add all states to $S^\square$ and pairs to $R_{a_j}$'s in order to follow these choices.
\end{proof}

\begin{lem}\label{lem:not-equiv-algo-rejects}
    Let $\varphi\in\mathL_{2S}$ be a satisfiable formula. If $\mathtt{Prime_{2S}}(\varphi)$ always accepts, then every two processes $r_1,r_2$ that satisfy $\varphi$ are equivalent with respect to $\equiv_S$.
\end{lem}
\begin{proof} We prove the contrapositive of the lemma. Assume that there are $r_1,r_2\in\proc$ that satisfy $\varphi$ and $r_1\not\curle_S r_2$. From Lemma~\ref{lem:finite_processes}, we assume w.l.o.g.\ that $\depth(r_i)\leq\md(\varphi)+1$ for both $i=1,2$. We show that there is an execution of $\mathtt{Prime_{2S}}(\varphi)$ that rejects. Let $\mathtt{Prime_{2S}}(\varphi)$ play as follows. $\mathtt{ConPro}(\varphi)$ chooses its output $(S_2,R_{a_1}^2,\dots, R_{a_k}^2)$ consistently on $r_2$, when it is called at line 3. From Lemma~\ref{lem:choose-cons-simulation}, $p_2\curle_S r_2$. Consequently, $r_1\not\curle_S p_2$. When $\mathtt{ConPro}(\varphi)$ is called at line 1, it chooses its output $(S_1,R_{a_1}^1,\dots, R_{a_k}^1)$ consistently on $r_1$ and adds a copy of $\witproc(r_1,p_2)$ to its output, which is possible from Lemmas~\ref{lem:witproc-properties}, \ref{lem:algo-chooses-consistently},  \ref{lem:cons-on-subprocess} and the fact that $\witproc(r_1,p_2)$ satisfies all five conditions of Lemma~\ref{lem:cons-on-subprocess}. It is not hard to see that \process{$S_1,R_{a_1}^1,\dots, R_{a_k}^1$} is process $p_1=\witproc(r_1,p_2)+p_1'$, where $p_1'$ is the result of $\mathtt{ConPro}(\varphi)$ choosing its output consistently on $r_1$. From Lemma~\ref{lem:witproc-properties}, $\witproc(r_1,p_2)\not\curle_S p_2$, and so $p_1\not\curle_S p_2$. From Claim~\ref{claim:gcd-three}, $\grcd_{\curle_{2S}}(p_1,p_2)$ does not exist, procedure \gcd{$p_1,p_2$} does not generate a process because of line 5 in Algorithm~\ref{alg:procedures}, and $\mathtt{Prime_{2S}}(\varphi)$ rejects at line 8.
% The proof is similar to the proof of Proposition~\ref{prop:simulation-left-to-right}. We describe the basic idea here, and we refer the reader to the aforementioned proof for the details. We prove the contrapositive of the lemma. Assume that there are $r_1,r_2\in\proc$ that satisfy $\varphi$ and $r_1\not\curle_S r_2$. We show that there is an execution of $\mathtt{Prime_{2S}}(\varphi)$ that rejects. From Lemma~\ref{lem:finite_processes}, we assume w.l.o.g.\ that $\depth(r_i)\leq\md(\varphi)+1$ for both $i=1,2$. There is a sequence of non-deterministic choices that $\mathtt{ConPro}$ can make such that there is a mapping $map:\reach(p_2)\rightarrow\reach(r_2)$ for which conditions 1--3 of Definition~\ref{def:B-plays-consistently} hold. This can be done as described in the proof of Lemma~\ref{lem:B-can-play-consistently}. Moreover, $\mathtt{ConPro}(\varphi)$ can make non-deterministic choices such that $p_1$ at line 2 contains a copy of a trace of $r_1$ that witnesses the fact that $r_1\not\curle_S r_2$ as described in the proofs of Lemma~\ref{lem:B-on-first-process} and Proposition~\ref{prop:simulation-left-to-right}.
% This execution of $\mathtt{Prime_{2S}}(\varphi)$  constructs  $p_1,p_2$ such that $p_1\not\curle_S p_2$. From Claim~\ref{claim:gcd-three}, $\grcd_{\curle_{2S}}(p_1,p_2)$ does not exist, procedure \gcd{$p_1,p_2$} does not generate a process because of line 5 in Algorithm~\ref{alg:procedures}, and $\mathtt{Prime_{2S}}(\varphi)$ rejects at line 8.
\end{proof}

For the following lemmas, let $p\in \mathrm{Bounded}(|\varphi|)$ denote that $p\models\varphi$ and $|p|\leq 16|\varphi|^2$.

\begin{lem}\label{lem:for-every-two-there-is}
    Let $\varphi\in\mathL_{2S}$ be a satisfiable formula and assume that $\mathtt{Prime_{2S}}(\varphi)$ always accepts. Then, for every $r_1,r_2$ that satisfy $\varphi$, there is a process $g$ such that $g\in \mathrm{Bounded}(|\varphi|)$, $g\curle_{2S} r_1$, and $g\curle_{2S} r_2$.
\end{lem}
\begin{proof}
    Let $r_1,r_2\in\proc$ such that $r_i\models\varphi$ for both $i=1,2$; let also the two calls of $\mathtt{ConPro}(\varphi)$ make non-deterministic choices such that the two outputs at lines 1 and 3 are chosen consistently on $r_1,r_2$ respectively. We denote by 
    $map$ the mapping described in Lemma~\ref{lem:algo-chooses-consistently} and $g$ the $\grcd_{\curle_{2S}}(p_1,p_2)$ computed by $\mathtt{Prime_{2S}}(\varphi)$ at line 7. Since $\mathtt{Prime_{2S}}(\varphi)$ always accepts, $g\models\varphi$ because of lines 8 and 9. From Lemma~\ref{lem:gcd-2s-poly-time}, $|g|\leq 2|p_1||p_2|$ and thus, from Lemma~\ref{lem:model-of-phi-size}, $|g|\leq 16|\varphi|^2$. Since $\mathtt{Prime_{2S}}(\varphi)$ always accepts, Lemma~\ref{lem:not-equiv-algo-rejects} implies that every two processes that satisfy $\varphi$ are simulation-equivalent, and so from Lemma~\ref{lem:choose-cons-simulation}(b), we have that $p_1\curle_{2S} r_1$ and $p_2\curle_{2S} r_2$. Since, from Claim~\ref{claim:gcd-one}, $g\curle p_i$ for both $i=1,2$, we have that $g\curle_{2S} r_i$ for both $i=1,2$.
    % We show that $g\curle_{2S} map(s_0^i)$, where $map(s_0^i)=r_i$, $i=1,2$, by induction on $g$. Note that from Lemma~\ref{lem:model-of-phi-and-phi-unsat}, $p_i\models\varphi$ for both $i=1,2$. This implies that $p_1\equiv_S p_2\equiv_S r_1\equiv_S r_2$ because of Lemma~\ref{lem:not-equiv-algo-rejects}. Since $g\curle_{2S} p_i$, we also have that $g\equiv_S r_1\equiv_S r_2$.
    % \begin{itemize}
    %     \item Let $g=\mathtt{0}$. From the fact that $g\equiv_S r_1\equiv_S r_2$, $r_1=r_2=\mathtt{0}$ and $g\curle_{2S} r_i$ for both $i=1,2$.
    %     \item Let $g\myarrowa g'$. Then, from the construction of $g$ there are $p_1',p_2'$ such that $p_1\myarrowa p_1'$, $p_2\myarrowa p_2'$, and from Claims~\ref{claim:gcd-one} and~\ref{claim:gcd-two}, $g'$ is  $\grcd_{\curle_{2S}}(p_1',p_2')$. From condition 2 of Definition~\ref{def:B-plays-consistently}, there are also $map(p_1)\myarrowa map(p_1')$ and $map(p_2)\myarrowa map(p_2')$. From the inductive hypothesis, $g'\curle_{2S} map(p_i')$, $i=1,2$. Therefore, $g\curle_{2S}  map(p_i)$, $i=1,2$.\qedhere
    % \end{itemize} 
\end{proof}

\begin{cor}\label{cor:2S-for-every-two-there-is}
Let $\varphi\in\mathL_{2S}$ be a satisfiable formula and assume that $\mathtt{Prime_{2S}}(\varphi)$ always accepts.  Then, for every two processes $r_1,r_2$ such that $r_i\in \mathrm{Bounded}(|\varphi|)$ for both $i=1,2$, there is a process $g$ such that $g\in \mathrm{Bounded}(|\varphi|)$, $g\curle_{2S} r_1$, and $g\curle_{2S} r_2$.
\end{cor}
\begin{proof}
     Immediate from Lemma~\ref{lem:for-every-two-there-is}.
\end{proof}

\begin{cor}\label{cor:2S-for-every-there-is-small}
 Let $\varphi\in\mathL_{2S}$ be a satisfiable formula and assume that $\mathtt{Prime_{2S}}(\varphi)$ always accepts. Then, for every process $r$ such that $r\models\varphi$, there is $g$  such that $g\in \mathrm{Bounded}(|\varphi|)$ and $g\curle_{2S} r$.
\end{cor}
\begin{proof}
 Immediate from Lemma~\ref{lem:for-every-two-there-is}.
\end{proof}

\begin{lem}\label{lem:2S-there-is-for-every-small}
    Let $\varphi\in\mathL_{2S}$ be a satisfiable formula and assume that $\mathtt{Prime_{2S}}(\varphi)$ always accepts. Then, there is a process $g$ such that $g\in \mathrm{Bounded}(|\varphi|)$ and for every process $r$ such that $r\in \mathrm{Bounded}(|\varphi|)$, $g\curle_{2S} r$.
\end{lem}
\begin{proof}
The proof is similar to the proof of Lemma~\ref{lem:there-is-for-every-small}.
\end{proof}

\begin{cor}\label{cor:2S-there-is-for-every}
    Let $\varphi\in\mathL_{2S}$ be a satisfiable formula and assume that $\mathtt{Prime_{2S}}(\varphi)$ always accepts. Then, there is $g$ such that $g\in \mathrm{Bounded}(|\varphi|)$ and for every process $r$ such that $r\models\varphi$, $g\curle_{2S} r$.
\end{cor}
\begin{proof}
This is immediate from Lemma~\ref{lem:2S-there-is-for-every-small} and Corollary~\ref{cor:2S-for-every-there-is-small}.
\end{proof}

\begin{prop}\label{prop:accept-implies-prime}
   Let $\varphi\in\mathL_{2S}$. If $\mathtt{Prime}_{2S}(\varphi)$ always accepts, then $\varphi$ is prime.
\end{prop}
\begin{proof}
   If  $\mathtt{Prime_{2S}}(\varphi)$ always accepts at line 5, then $\varphi$ is unsatisfiable from Lemma~\ref{lem:phi-sat}, and so it is prime. In the case that there is an execution of $\mathtt{Prime_{2S}}(\varphi)$ that accepts at line 9, two processes that satisfy $\varphi$ are returned at lines 2 and 4, and $\varphi$ is satisfiable from Lemma~\ref{lem:phi-sat}. Then, we can prove that $\varphi$ is prime similarly to the proof of Proposition~\ref{prop:left-to-right} using Corollary~\ref{cor:2S-there-is-for-every}.
\end{proof}
% \begin{prop}\label{prop:accept-implies-prime}
%    Let $\varphi\in\mathL_{2S}$ be satisfable. If $\mathtt{Prime}_{2S}(\varphi)$ always accepts, then $\varphi$ is characteristic within $\mathL_{2S}$.
% \end{prop}
% \begin{proof}
%     The proof is similar to the proof of Proposition~\ref{prop:left-to-right}.
% \end{proof}

\begin{cor}\label{cor:2s-prime-algo-correct}
     Let $\varphi\in\mathL_{2S}$. Then, $\varphi$ is prime iff $\mathtt{Prime_{2S}}(\varphi)$ always accepts.
\end{cor}
\begin{proof}
 Immediate  from Propositions~\ref{prop:prime-implies-accept} and~\ref{prop:accept-implies-prime}.
\end{proof}

\twosprimeconp*
% \begin{restatable}{proposition}{twosprimeconp}\label{prop:2S-primality-conpc}
% The Formula Primality problem for $\mathL_{2S}$ is \conp-complete.
% \end{restatable}
\begin{proof}
    The problem is \conp-hard from Proposition~\ref{prop:2s-prime-conp-hard}. To prove that it is also in \conp, let $\varphi\in\mathL_{2S}$. From Corollary~\ref{cor:2s-prime-algo-correct}, it suffices to show that every execution of $\mathtt{Prime_{{2S}}}(\varphi)$ runs in polynomial time.
     From Lemma~\ref{lem:model-of-phi-size}, every execution of $\mathtt{ConPro}(\varphi)$ needs polynomial time and from Lemma~\ref{lem:gcd-2s-poly-time}, the construction of $\grcd_{\curle_{2S}}(p_1,p_2)$ can be done in polynomial time. Model checking at line 7 also requires polynomial time from Proposition~\ref{model-checking-complexity}. Therefore, deciding prime formulae in $\mathL_{2S}$ is in \conp.
\end{proof}

We also prove the following result, analogous to Proposition~\ref{prop:twosim-win-algo}, for the case of $\equiv_{2S}$.

\begin{prop}\label{prop:2s-equivalent-processes-conp}
    Let $\varphi\in\mathL_{2S}$. The problem of deciding whether every two processes satisfying $\varphi$ are 2-nested-simulation equivalent is in \conp.
\end{prop}
\begin{proof}
    Consider the algorithm $\mathtt{EquivProc_{2S}}$, which on input $\varphi$ proceeds as follows: it executes lines 1--5 of $\mathtt{Prime_{2S}}(\varphi)$. If $\mathtt{EquivProc_{2S}}(\varphi)$ does not accept at line 5, it continues by checking whether $p_1$ and $p_2$ are 2-nested-simulation equivalent.  This equivalence check can be performed in polynomial time, by Proposition~\ref{prop:2s-equiv-polytime}. Using reasoning similar to the correctness proof of the $\mathtt{Prime_{2S}}$ algorithm, we can show that every execution of $\mathtt{EquivProc_{2S}}$ accepts iff every two processes satisfying $\varphi$ are $2$-nested-simulation equivalent. Moreover, each execution of $\mathtt{EquivProc_{2S}}$  needs polynomial time in $|\varphi|$ since $|p_i|\leq 4|\varphi|$ from Lemma~\ref{lem:model-of-phi-size}. Thus, the problem belongs to \conp.
\end{proof}

\section{The complexity of deciding characteristic formulae}\label{section:char-preorders}
%within the logics considered in this work}

Using the results of the previous sections and introducing some additional reductions, we now proceed to establish the results summarized in Table~\ref{table:characteristic}.

\begin{cor}\label{cor:simulation-decide-characteristic}
   Deciding characteristic formulae within $\mathL_S$ is polynomial-time solvable.
\end{cor}
\begin{proof}
   Immediate from Proposition~\ref{prop:charact-via-primality} and Corollaries~\ref{cor:sat-s-cs-rs-poly}(a) and~\ref{cor:primality-in-S}.
   In particular, deciding characteristic formulae in $\mathL_S$ %requires 
   can be done in time $\mathcal{O}(n^3)$.
\end{proof}

\begin{restatable}{cor}{cschar}\label{cor:cs-decide-characteristic}
   Deciding characteristic formulae within $\mathL_{CS}$ and $\mathL_{RS}$ with a bounded action set is polynomial-time solvable.
\end{restatable}
\begin{proof}
    This is a corollary of Proposition~\ref{prop:charact-via-primality} and Corollaries~\ref{cor:sat-s-cs-rs-poly}(a) and~\ref{cor:primality-in-CS} for $\mathL_{CS}$ and Proposition~\ref{prop:charact-via-primality} and Corollaries~\ref{cor:sat-s-cs-rs-poly}(b) and~\ref{cor:primality-in-RS} for $\mathL_{RS}$.
\end{proof}

The problem belongs to \dpc and is \us-hard for $\mathL_{RS}$ with an unbounded action set. 

\begin{cor}\label{cor:char-rs-unbounded-complexity}
    Let $\act$ be unbounded. Deciding characteristic formulae within $\mathL_{RS}$ (a) is \us-hard, and (b) belongs to \dpc.   
\end{cor}
\begin{proof}
(a)  Let $\varphi$ be an instance of $\textsc{UniqueSat}$. We construct $\varphi'\in\mathL_{RS}$, which is obtained from $\varphi$ by replacing $x_i$ with $\langle a_i\rangle \zero$ and $\neg x_i$  with $[a_i]\ff$. Note that $\varphi$ has exactly one satisfying assignment iff $\varphi'$ is characteristic within $\mathL_{RS}$. Indeed, let $s$ denote the unique satisfying assignment of $\varphi$. A process for which $\varphi'$ is characteristic within $\mathL_{RS}$ is the process $\displaystyle p=\sum_{i \,:\,s(x_i)=\mathrm{true}}\langle a_i\rangle \mathtt{0}$. Conversely, let $\varphi'$ be characteristic for a process $p$ within $\mathL_{RS}$. The truth assignment that maps to true exactly the $x_i$'s for which $a_i\in I(p)$ is a unique satisfying assignment for $\varphi$.\\
(b) This claim is immediate from Proposition~\ref{prop:charact-via-primality} and membership of the satisfiability and the formula primality  problems for $\mathL_{RS}$ in \NP and \conp, respectively.
\end{proof}

\us-hardness holds for $\mathL_{TS}$ and $\mathL_{2S}$ as well.

\begin{prop}\label{cor:char-ts-complexity}
 Let $|\act|\geq 2$. Deciding characteristic formulae within $\mathL_{TS}$ or $\mathL_{2S}$ is \us-hard.
\end{prop}
\begin{proof}
    Let $|\act|=2$. There is a polynomial-time reduction from \textsc{UniqueSat} to deciding characteristic formulae within $\mathL_{TS}$. Given an instance $\varphi$ of $\textsc{UniqueSat}$, we construct $\varphi''\in\mathL_{TS}$ such that $\varphi$ has a unique satisfying assignment iff $\varphi''$ is characteristic within $\mathL_{TS}$. We first define $\varphi'\in\mathL_{TS}$ to be $\varphi$ where every literal $l$ is substituted with $\enc(l)$ as described in the proof of Proposition~\ref{prop:decide-prime-ts-infinite-actions-hard}. We denote $\lceil \log n \rceil$ by $k$. Formula $\varphi'$ is one conjunct of $\varphi''$. To complete $\varphi''$, the idea is that in the case that $\varphi$ has a unique satisfying assignment, any process satisfying $\varphi''$ must have only traces which correspond to variables that are set to true in the satisfying assignment (and perhaps prefixes of such traces). So, we have to forbid all other traces. For every sequence of actions $a_1\dots a_l\in\act$, where $2\leq l\leq k$, we define $\displaystyle\mathrm{fb}^{a_1\dots a_l}:=\bigvee_{t\in \act^{k+1-l}} \langle a_1\rangle \dots \langle a_l\rangle t\vee[a_1]\dots [a_l]\ff$. Intuitively, $\mathrm{fb}^{a_1\dots a_l}$ says that if $a_1\dots a_l$ has not been imposed by a variable that was set to true, then it is forbidden. For a specific sequence $a_1\dots a_l$, formula $\mathrm{fb}^{a_1\dots a_l}$ contains $2^{k+1-l}+1$ disjuncts and there are $2^{l-1}$ different sequences of length $l$ that we have to take care of since the first symbol is always $0$ (see the definition of $\enc(l)$ for a literal $l$). Thus, we need to include $n+2^{l-1}$ disjuncts for every $1\leq l\leq k$. Considering all literals $l$, we get $\mathcal{O}(n \log n)$ disjuncts of length at most $\log n$ each. Finally, we define $\displaystyle\varphi'':=\varphi'\wedge \bigwedge_{2\leq l\leq k}\bigwedge_{a_1\dots a_l}\mathrm{fb}^{a_1\dots a_l} \wedge [1]\ff$. From the analysis above, $\varphi''$  is of polynomial size. Moreover, $\varphi$ has a unique satisfying assignment iff $\varphi''$ is characteristic for $p$, where $p$ is the process whose traces are the ones corresponding to the variables set to true in the unique satisfying assignment.

    The reduction can be modified to work for deciding characteristic formulae within $\mathL_{2S}$ as well. In this case, formulae of the form $[a_1]\dots[a_l] ([0]\ff\vee [1]\ff)$ must be included as conjuncts in $\varphi''$, for every sequence of actions $a_1\dots a_l$ and  every $2\leq l\leq k$.
\end{proof}

For the logic $\mathL_{2S}$ we show membership in \dpc below.

\begin{cor}\label{cor:2S-char-dp}
   Let $|\act|\geq 2$.  Deciding whether a formula in $\mathL_{2S}$ is characteristic for a process within $\mathL_{2S}$ is in \dpc.
 \end{cor}
 \begin{proof}
     This follows from the definition of the class \dpc, Proposition~\ref{prop:charact-via-primality}, and the fact that satisfiability for $\mathL_{2S}$ is in \NP by Theorem~\ref{prop:sat-rs-ts-2s-np-complete}, and the formula primality problem for $\mathL_{2S}$ is in \conp by Theorem~\ref{prop:2S-primality-conpc}.
 \end{proof}

\begin{cor}\label{cor:ts-bounded-depth}
  Let $|\act|= k$ and $\varphi\in\mathL_{TS}$ with $\md(\varphi)=d$.
  %, where $k,d\geq 1$ are constants. 
  Then, there is an algorithm that decides whether $\varphi$ is characterisitic in time 
  %linear in $|\varphi|$.
   $\mathcal{O}(\operatorname{tow}(k,d)^3\cdot |\varphi|)$.
\end{cor}

\begin{proof}
    This is effectively the same algorithm as in the proof of Theorem~\ref{thm:ts-bounded-depth}, except that if $P^d_{sat} = \emptyset$, then the algorithm rejects the input.
\end{proof}

 Finally the problem becomes \pspace-complete for $\mathL_{nS}$, $n\geq 3$.

\begin{cor}\label{cor:3s-char-pspacec}
  Let $|\act|\geq 2$.  Deciding whether a formula in $\mathL_{nS}$, $n\geq 3$, is characteristic for a process within $\mathL_{nS}$ is \pspace-complete.
\end{cor}
\begin{proof}
   The proof of Theorem~\ref{thm:2scompl-prime} implies that deciding whether $\varphi\in\mathL_{3S}$ is characteristic within $\mathL_{3S}$ is \pspace-hard.
   Since the characteristic formulae within $\mathL_{nS}$ are the satisfiable and prime ones, the problem is in \pspace from Corollary~\ref{prop:n-s-sat} and Theorem~\ref{prop:3S-char-in-pspace}.
\end{proof}

\section{Deciding characteristic formulae modulo equivalence relations}\label{section:char-equivalences}

So far, we have studied the complexity of deciding characteristic formulae in the modal logics that characterize the simulation-based preorders in van Glabbeek's spectrum. As shown in~\cite{AcetoMFI19}, those logics are powerful
enough to describe characteristic formulae for each finite, loop-free process up to the preorder they characterize. It is therefore natural to wonder whether they can also express characteristic formulae modulo the kernels of those preorders. The following result indicates that the logics $\mathL_X$, where $X\in\{S,CS,RS\}$, have very weak expressive power when it comes to defining characteristic formulae modulo $\equiv_X$. 

\begin{prop}\label{prop:S-CS-RS-equiv-char}
\hfill
   \begin{enumerate}[(a)]
       \item No formula in $\mathL_S$ is characteristic for some process $p$ with respect to $\equiv_S$.
       \item A formula $\varphi$ is characteristic for some process $p$ with respect to $\equiv_{CS}$ or $\equiv_{RS}$ iff it is logically equivalent to $\bigwedge_{a\in \act} [a] \ff$. 
   \end{enumerate}   
\end{prop}
\begin{proof}
   (a) Assume, towards contradiction, that there is a formula $\varphi_c^S$ in $\mathL_S$ that is characteristic for some process $p$ with respect to $\equiv_S$. Let $\ell$ be the depth of $p$ and $a\in \act$. Define process $q = p + a^{\ell + 1}\mathtt{0}$---that is, $q$ is a copy of $p$ with an additional path that has exactly $\ell+1$ $a$-transitions. It is easy to see that $p \curle_S q$, but $q \not\curle_S p$.  Since $p\models\varphi_c^S$, it holds that $q\models \varphi_c^S$. However, $q\not\equiv_S p$, which contradicts our assumption that $\varphi_c^S$ is characteristic for $p$ with respect to $\equiv_S$. \\
    (b) In the case of $\equiv_{CS}$ and $\equiv_{RS}$, note that a formula $\varphi$  is logically equivalent to $\bigwedge_{a\in \act} [a] \ff$ iff it is satisfied only by processes without outgoing transitions, and so it is characteristic for 
    any such process modulo $\equiv_X$. To prove that no formula is  characteristic for some process $p$ with positive depth  modulo  $\equiv_{CS}$ or $\equiv_{RS}$, a similar argument to the one for $\equiv_S$ can be used. For $\equiv_{RS}$, the action $a$ should be chosen such that $p\myarrowa p'$ for some $p'$.
    \end{proof}

    For $TS$ and $2S$, there  are non-trivial characteristic formulae modulo $\equiv_{TS}$ and $\equiv_{2S}$, respectively. For example, if $\act=\{a,b\}$, the formula $\varphi_a = \langle a\rangle([a]\ff \wedge [b]\ff)\wedge [b]\ff\wedge [a][a]\ff\wedge [a][b]\ff$ is satisfied only by processes that are equivalent, modulo those equivalences, to process $p_a = a.\mathtt{0}$ that has a single transition labelled with $a$. Thus, $\varphi_a$ is characteristic for $p_a$ modulo both $\equiv_{TS}$ and $\equiv_{2S}$. 

We can use the following theorem as a tool to prove hardness of deciding characteristic formulae modulo some equivalence relation.
Theorem~\ref{thm:reduction-validity-char} below is an extension of~\cite[Theorem 26]{AcetoAFI20}, so that it holds for every $X$ such that a characteristic formula modulo $\equiv_X$  exists, namely $X\in\{CS,RS,TS,nS,BS\}$, where $n\geq 2$.

\begin{thm}\label{thm:reduction-validity-char}
Let $X\in\{CS,RS,TS,2S,3S, BS\}$. Validity in $\compL_X$ reduces in polynomial time to deciding characteristic formulae with respect to $\equiv_X$.
\end{thm}
\begin{proof}
Given $\varphi\in\compL_X$, where $X\in\{CS,RS,TS,nS,BS\}$, $n\geq 2$, we construct a formula $\varphi'\in\mathL_X$ such that $\varphi$ is valid if and only if  $\varphi'$ is characteristic modulo $\equiv_X$ for some $p$. Let $\varphi_{ch}$ be a characteristic formula modulo $\equiv_X$ for process $p_{ch}$ and $\varphi_{nch}\in\mathL_X$ be a formula that is not characteristic modulo $\equiv_X$. For each $X\in\{CS,RS,TS,2S\}$, the formulae $\varphi_{ch}$ and $\varphi_{nch}$ exist by Proposition~\ref{prop:S-CS-RS-equiv-char} and the discussion in the paragraph following its proof. In the same way, we can show that characteristic formulae modulo $\equiv_{3S}$ or $\equiv_{BS}$ exist. Given $\varphi\in\compL_X$, it can be determined in linear time whether $p_{ch}\models\varphi$. We distinguish the following two cases:
\begin{itemize}
    \item Assume that $p_{ch}\not\models\varphi$. Then $\varphi$ is not valid and we set $\varphi'=\varphi_{nch}$.
    \item Assume that $p_{ch}\models\varphi$. In this case, we set $\varphi'=\neg \varphi \vee \varphi_{ch}$. We show that $\varphi$ is valid if and only if $\varphi'$ is characteristic modulo $\equiv_X$ for some process $p$. 
    \begin{itemize}
        \item For the implication from left to right, assume that $\varphi$ is valid. Then $\neg \varphi \vee \varphi_{ch}$ is equivalent to $\varphi_{ch}$, which is characteristic for $p_{ch}$ modulo $\equiv_X$.
        \item Conversely, assume that $\neg \varphi \vee \varphi_{ch}$ is characteristic for some process $p$ modulo $\equiv_X$. Let $q$ be any process. We show that $q\models \varphi$ and therefore that $\varphi$ is valid. If $p_{ch}\equiv_X q$, then $\mathL_X(p_{ch})=\mathL_X(q)$ and, since $p_{ch}\models \varphi$ by assumption, it holds that $q\models\varphi$. Suppose now that $p_{ch}\not\equiv_X q$.  Since we have that $p_{ch}\models \neg\varphi\vee \varphi_{ch}$ and $\neg\varphi\vee \varphi_{ch}$ is characteristic for $p$ modulo $\equiv_X$, it follows that $\varphi_{ch} \equiv\neg\varphi\vee \varphi_{ch}$.  Therefore, $q\not\models\neg\varphi\vee \varphi_{ch}$, which implies that $q\models\varphi$, and we are done.\qedhere
    \end{itemize}
\end{itemize}
\end{proof}

\noindent Note that, from the results of Section~\ref{section:deciding-satisfiability}, validity in $\compL_{RS}$ with an unbounded action set, $\compL_{TS}$  with $|\act|\geq 2$, and $\compL_{2S}$ with $|\act|\geq 2$ is \conp-complete, whereas  
validity in $\compL_{nS}$, $n\geq 3$, with $|\act|\geq 2$ is \pspace-complete. Consequently, by applying Theorem~\ref{thm:reduction-validity-char}, we obtain the following result.

\begin{cor}\label{cor:char-equiv-hard}\hfill
\begin{enumerate}[(a)]
    \item Deciding whether a formula is characteristic modulo $\equiv_{RS}$ with an unbounded action set, $\equiv_{TS}$  with $|\act|\geq 2$, and $\equiv_{2S}$  with $|\act|\geq 2$ is \conp-hard.
    \item Deciding whether a formula is characteristic modulo $\equiv_{nS}$, $n\geq 3$, with $|\act|\geq 2$ is \pspace-hard. 
\end{enumerate}
\end{cor}

Combining the results of Subsections~\ref{subsection:primality-nS} and~\ref{subsec:prime-2S} with Corollary~\ref{cor:char-equiv-hard} we show the following two corollaries on the complexity of deciding characteristic formulae modulo $\equiv_{nS}$, $n
\geq 2$.

\begin{cor}\label{cor:n-char-equiv}
    Let $|\act|\geq 2$. Deciding whether a formula $\varphi\in\mathL_{nS}$, where $n\geq 3$, is characteristic for a process modulo $\equiv_{nS}$ is \pspace-complete.
\end{cor}
\begin{proof}
 \pspace-hardness of deciding characteristic formulae modulo $\equiv_{nS}$ for $n\geq 3$ follows from Corollary~\ref{cor:char-equiv-hard}.
 To prove membership of the problem in \pspace, note that satisfiability of formulae in $\mathL_{nS}$ is in \pspace from Corollary~\ref{prop:n-s-sat}. Moreover, deciding whether every two processes that satisfy $\varphi$ are $n$-nested-simulation equivalent can be done in polynomial space from Proposition~\ref{prop:twosim-win-algo}. Thus, the problem lies in \pspace, because of Proposition~\ref{prop:char-mod-equiv}.
\end{proof}

\begin{cor}\label{cor:char-2s-equiv}
    Let $|\act|\geq 2$.  Deciding whether a formula in $\mathL_{2S}$ is characteristic for a process modulo $\equiv_{2S}$ is in \dpc.
\end{cor}
\begin{proof}
    Similarly to the previous case, this is immediate from the definition of the class \dpc, Proposition~\ref{prop:char-mod-equiv}, and the fact that satisfiability for $\mathL_{2S}$ is in \NP (by Proposition~\ref{prop:sat-rs-ts-2s-np-complete}), and the problem of checking whether all processes satisfying a formula in $\mathL_{2S}$ are 2-nested-simulation equivalent is in \conp (by Proposition~\ref{prop:2s-equivalent-processes-conp}).
\end{proof}

Finally, we can adjust the FPT algorithm for  $\mathL_{TS}$-primality for the case of $\equiv_{TS}$.

\begin{cor}\label{cor:char-ts-equiv-bounded-depth}
    Let $|\act|\geq 2$.  Deciding whether a formula in $\mathL_{TS}$ with modal depth $d$ is characteristic for a process modulo $\equiv_{TS}$ can be done in time $\mathcal{O}(\operatorname{tow}(k,d)^3\cdot |\varphi|)$.
\end{cor}
\begin{proof}
    The algorithm is similar to the one in the proof of Theorem~\ref{thm:ts-bounded-depth}.
    For this case, the algorithm first checks whether $p \curle_{TS} q$ and $q \curle_{TS} p$ for all processes $p,q \in P^d$.
    Then, if $P^d_{sat} = \emptyset$, then the algorithm rejects the input.
    Finally, the algorithm then iterates through all processes in $P^d_{sat}$ once, looking for a pair $p,q \models \varphi$, such that $p \mathrel{{\centernot\equiv}_{TS}} q$.
    The analysis of the algorithm is similar.
\end{proof}

The results of this section are depicted in Table~\ref{table:characteristic-equiv}.

\section{Conclusions}\label{section:conclusions}

In this paper, we studied the complexity of determining whether a formula is characteristic for some finite, loop-free process in each of the logics providing modal characterizations of the simulation-based semantics in van Glabbeek's branching-time spectrum~\cite{Glabbeek01}. Since, as shown in~\cite{AcetoMFI19}, characteristic formulae in each of those logics are exactly the satisfiable and prime ones, we gave complexity results for the satisfiability and primality problems, and investigated the boundary between logics for which those problems can be solved in polynomial time and those for which they become computationally hard. Our results show that computational hardness already manifests itself in ready simulation semantics~\cite{BloomIM95,LarsenS91} when the size of the action set is not a constant. Indeed, in that setting, the mere addition of formulae of the form $[a]\ff$ to the logic that characterizes the simulation preorder yields a logic whose satisfiability and primality problems are \NP-hard and \conp-hard respectively. Moreover, we show that deciding primality in the logics characterizing the $n$-nested simulation preorders, $n\geq 3$, is \pspace-complete in the presence of at least two actions. 

The work we present in this article opens several avenues for future research that we are currently pursuing. First of all, even though we succeeded in providing matching (or closely matching) lower and upper bounds on the complexity of the problems we studied in this paper, we have only a \us-hardness result for the problem of deciding whether a formula is characteristic within $\mathL_{TS}$, in the presence of at least two actions (Proposition~\ref{cor:char-ts-complexity}). It would be interesting to prove that the problem belongs to \dpc, as that of deciding characteristic formulae within $\mathL_{RS}$  when $\act$ is unbounded (Corollary~\ref{cor:char-rs-unbounded-complexity}), and deciding whether a formula in $\mathL_{2S}$ is characteristic for a process within $\mathL_{2S}$ when $|\act|\geq 2$ (Corollary~\ref{cor:2S-char-dp}).
Moreover, several of our complexity results depend on having at least two actions. It is natural to wonder whether they also hold  over a singleton set of actions.   
Moreover, we plan to study the complexity of deciding whether formulae are characteristic in the extensions of the modal logics we have considered in this article with greatest fixed points. Indeed, in those extended languages, one can define characteristic formulae for finite processes. It is known that deciding whether a formula is characteristic is \pspace-complete for \hml, but becomes \expc-complete for its extension with fixed-point operators---see reference~\cite{AcetoAFI20}. It would be interesting to see whether similar results hold for the other logics. Finally, building on the work presented in~\cite{AcetoMFI19}, we intend to study the complexity of the algorithmic questions considered in this article for (some of) the linear-time semantics in van Glabbeek's spectrum.

In~\cite{AcetoACI25}, we also studied the complexity of constructing characteristic formulae in each of the logics we consider in this paper, both when such formulae are presented in explicit form and in declarative form. In particular, one of our results in the aforementioned reference identifies a sharp difference between trace simulation and the other semantics when it comes to constructing characteristic formulae. For all the semantics apart from trace simulation, there are characteristic formulae that have  declaration size and equational length that are polynomial in the size of the processes they characterize and they can be efficiently computed. In contrast, for trace simulation, even if characteristic formulae are always of polynomial declaration size and polynomial equational length, they \emph{cannot} be efficiently computed, unless $\cP=\NP$. We will provide a full account of our results on the complexity of constructing characteristic formulae for processes in a subsequent companion paper. 

\section*{Acknowledgment}
  \noindent This paper is dedicated to the memory of our colleague and friend Rance Cleaveland (1961--2024), who used characteristic formulae to compute behavioural relations, logically and efficiently.

\iffalse
  %% the following bibliography is gererated manually for the sake of brevity
  %% only; please use a separate .bib file in your submission

\appendix
\section{}
  Here is a check-list to be completed before submitting the paper to
  LMCS:
\begin{itemize}[label=$\triangleright$]
\item your submission includes the latest version of lmcs.cls, that is, it does
  not rely on the version of the class file provided by arXiv
\item the text of your submission is contained in a single file,
  except for macros and graphics
\item your graphics use only one format
\item you have employed the Journal's original proclamation environments,
  or suitable extensions thereof
\item you have loaded the hyperref package
\item you have \emph{not} loaded the times package
\item you have not routinely adjusted vertical spacing manually by issuing
  \texttt{\textbackslash vspace} or \texttt{\textbackslash vskip} commands
\item you have used the command \texttt{\textbackslash sloppy} only
  locally and in emergency cases
\item your displayed equations use the
  \texttt{\textbackslash[\dots\textbackslash]} construct
\item your abstract only contains as few math-expressions as possible and no
  references
\item your references are supplied in bibtex format in a separate \verb|.bib| file
\end{itemize}

  This listing also shows how to override the default bullet $\bullet$
  of the \texttt{itemize}-envronment by a different symbol, in this
  case \texttt{\textbackslash triangleright}.

\fi

  \bibliographystyle{alphaurl}
  \bibliography{bibliography}

\begin{appendix}
  
\section{Satisfiability for \texorpdfstring{$\mathL_{CS}$}{LCS}}\label{subsection:cs-satisfiability}

We provide detailed proofs for the complexity of the satisfiability problem for $\mathL_{CS}$. To decide satisfiability in the case of complete simulation, we show that  we need much less information than $I(\varphi)$ gives for a formula $\varphi\in\mathL_{CS}$. Alternatively, we associate $\varphi$ to a set $J(\varphi)$, which is one of $\emptyset, \{\emptyset\}, \{\alpha\},\{\emptyset,\alpha\}$. Note that the main difference here is that we let $\alpha$ symbolize every possible set of actions.

\begin{defi}\label{def:cs-J(phi)}
Let $\varphi\in\mathL_{CS}$. We define $J(\varphi)$ inductively as follows:
    \begin{enumerate}[(a)]
        \item $J(\true)=\{\emptyset,\alpha\}$,
        \item $J(\ff)=\emptyset$,
        \item $J(\zero)= \{\emptyset\}$,
        \item $J(\langle a\rangle \varphi)=\begin{cases}
           \emptyset, &\text{if } J(\varphi)=\emptyset,\\
           \{\alpha\}, &\text{otherwise}
        \end{cases}$
        \item $J(\varphi_1\vee\varphi_2)=J(\varphi_1)\cup J(\varphi_2)$,
        \item $J(\varphi_1\wedge\varphi_2)=J(\varphi_1)\cap J(\varphi_2)$.
    \end{enumerate}
\end{defi}

\begin{lem}\label{lem:J(phi)-property}
    For every $\varphi\in\mathL_{CS}$, $\varphi$ is unsatisfiable iff $J(\varphi)=\emptyset$.
\end{lem}
\begin{proof}
    We prove the lemma by proving the following claims simultaneously by induction on the structure of $\varphi$.
    
        \begin{clm}\label{cl-one}$\varphi\equiv\zero$ iff $J(\varphi)=\{\emptyset\}$.
        \end{clm}
        \begin{proof}
 Note that $\varphi\equiv\zero$ iff $\varphi$ is satisfied exactly by $\mathtt{0}$.\\
$(\Rightarrow)$ Let $\varphi$ be satisfied exactly by $\mathtt{0}$. Then $\varphi$ can have one of the following forms.
\begin{description}
    \item[$\varphi=\zero$] Then $J(\varphi)=\{\emptyset\}$.
    \item[$\varphi=\varphi_1\vee\varphi_2$] Then either both $\varphi_i$, $i=1,2$, are satisfied exactly by $\mathtt{0}$, in which case, from inductive hypothesis, $J(\varphi_1)= J(\varphi_2) =\{\emptyset\}$, and so $J(\varphi)=J(\varphi_1)\cup J(\varphi_2)=\{\emptyset\}$, or one of them, w.l.o.g.\ assume that this is $\varphi_1$, is satisfied exactly by $\mathtt{0}$ and $\varphi_2$ is unsatisfiable, in which case $J(\varphi_2)=\emptyset$ from inductive hypothesis of Claim A.4, and $J(\varphi)=J(\varphi_1)\cup J(\varphi_2)=\{\emptyset\}\cup\emptyset=\{\emptyset\}$.
    \item[$\varphi=\varphi_1\wedge\varphi_2$] Then, one of three cases is true. 
    \begin{itemize}
        \item  Both $\varphi_i$, $i=1,2$, are satisfied exactly by $\mathtt{0}$. Then, $J(\varphi)=J(\varphi_1)\cap J(\varphi_2)=\{\emptyset\}\cap\{\emptyset\}=\{\emptyset\}$.
        \item W.l.o.g.\ $\varphi_1$ is satisfied exactly by $\mathtt{0}$ and $\varphi_2\equiv\true$. Then, $J(\varphi)=J(\varphi_1)\cap J(\varphi_2)=\{\emptyset\}\cap \{\emptyset,\alpha\}=\{\emptyset\}$.
        \item W.l.o.g.\ $\varphi_1$ is satisfied exactly by $\mathtt{0}$ and $\varphi_2$ is satisfied in $\mathtt{0}$ and some other processes. From inductive hypothesis of Claim A.6, $J(\varphi_2)=\{\emptyset,\alpha\}$. Then, $J(\varphi)=J(\varphi_1)\cap J(\varphi_2)=\{\emptyset\}\cap \{\emptyset,\alpha\}=\{\emptyset\}$.
    \end{itemize} 
\end{description}
$(\Leftarrow)$ Let $J(\varphi)=\{\emptyset\}$. Then,  one of the following holds.
\begin{description}
    \item[$\varphi=\zero$] Trivial.
    \item[$\varphi_1\vee\varphi_2$] Then, since $J(\varphi)=J(\varphi_1)\cup J(\varphi_2)$, it must be the case that w.l.o.g.\ $J(\varphi_1)=\{\emptyset\}$ and either $J(\varphi_2)=\{\emptyset\}$ or $J(\varphi_2\}=\emptyset$. Hence, either both $\varphi_i$, $i=1,2$, are satisfied exactly by $\mathtt{0}$, or $\varphi_1$ is satisfied exactly by $\mathtt{0}$ and $\varphi_2$ is unsatisfiable. In both cases, $\varphi$ is satisfied exactly by $\mathtt{0}$.
    \item[$\varphi_1\wedge\varphi_2$] 
    Since $J(\varphi)=J(\varphi_1)\cap J(\varphi_2)$, it must be the case that either $J(\varphi_1)=J(\varphi_2)=\{\emptyset\}$, or w.l.o.g.\ $J(\varphi_1)=\{\emptyset\}$ and $J(\varphi_2)=\{\emptyset,\alpha\}$. In the former case, $\varphi_1\equiv\varphi_2\equiv\zero$, and so $\varphi_1\wedge\varphi_2\equiv\zero$. In the latter case,  $\varphi_1\equiv\zero$, $\zero\models\varphi_2$ and $\varphi_2\not\models\zero$, and so $\varphi_1\wedge\varphi_2$ is satisfied exactly by $\mathtt{0}$.\qedhere
\end{description}
        \end{proof}
        
        \begin{clm}\label{cl-two} $\varphi\equiv\ff$ iff $J(\varphi)=\emptyset$.
        \end{clm}
        \begin{proof}
   Note that $\varphi\equiv\ff$ iff $\varphi$ is unsatisfiable.\\
$(\Rightarrow)$ Let $\varphi$ be unsatisfiable. Then, $\varphi$ can have one of the following forms.
\begin{description}
    \item[$\varphi=\ff$] Trivial.
    \item[$\varphi=\langle a\rangle \varphi'$] Then, $\varphi'$ is unsatisfiable, so from inductive hypothesis $J(\varphi')=\emptyset$ and $J(\varphi)=\emptyset$.
    \item[$\varphi=\varphi_1\vee\varphi_2$] Then, both $\varphi_1$ and $\varphi_2$ are unsatisfiable, and $J(\varphi)=J(\varphi_1)\cup J(\varphi_2)=\emptyset\cup\emptyset=\emptyset$.
    \item[$\varphi=\varphi_1\wedge\varphi_2$] Then, one of the following cases holds.
        \begin{itemize}
        \item W.l.o.g.\ $\varphi_1$ is unsatisfiable. Then, $J(\varphi)=J(\varphi_1)\cap J(\varphi_2)=\emptyset\cap J(\varphi_2)=\emptyset$.
        \item W.l.o.g.\ $\varphi_1$ is satisfied exactly by $\mathtt{0}$ and $\varphi_2$ is satisfied only by processes that are not $\mathtt{0}$. From inductive hypothesis of Claims 1 and 3, $J(\varphi_1)=\{\emptyset\}$, and $J(\varphi_2)=\{\alpha\}$, respectively. Thus, $J(\varphi)=J(\varphi_1)\cap J(\varphi_2)=\emptyset$.
         \end{itemize}  
\end{description}
$(\Leftarrow)$ Assume that $J(\varphi)=\emptyset$. Then, one of the following holds.
\begin{description}
    \item[$\varphi=\ff$] Trivial.
    \item[$\varphi=\langle a\rangle \varphi'$] Then, $J(\varphi')=\emptyset$, which implies that $\varphi'$ is unsatisfiable, and so $\varphi$ is unsatisfiable.
    \item[$\varphi=\varphi_1\vee\varphi_2$] Then for both $i=1,2$, $J(\varphi_i)=\emptyset$, and from inductive hypothesis, both $\varphi_i$ are unsatisfiable, which implies that $\varphi$ is unsatisfiable as well.
    \item[$\varphi=\varphi_1\wedge\varphi_2$] We distinguish between the following cases.
    \begin{itemize}
        \item W.l.o.g.\ $J(\varphi_1)=\emptyset$. From inductive hypothesis, $\varphi_1$ is unsatisfiable, and so $\varphi$ is also unsatisfiable.
        \item W.l.o.g.\ $J(\varphi_1)=\{\emptyset\}$ and $J(\varphi_2)=\{\alpha\}$. Then, from inductive hypothesis of Claims 1 and 3, $\varphi_1$ is satisfied exactly by $\mathtt{0}$  and $\varphi_2$ is not satisfied in $\mathtt{0}$, respectively, which implies that $\varphi$ is unsatisfiable.\qedhere
    \end{itemize}
\end{description}
        \end{proof}
        \begin{clm}\label{cl-three} $\varphi$ is satisfiable and $\zero\not\models\varphi$ iff $J(\varphi)=\{\alpha\}$.
        \end{clm}
        \begin{proof}
$(\Rightarrow)$ Let $\varphi$ be satisfiable and $\zero\not\models\varphi$. Then, $\varphi$ can have one of the following forms.
\begin{description}
    \item[$\varphi=\langle a\rangle \varphi'$] Then, $\varphi'$ is satisfiable, and so $J(\varphi)=\{\alpha\}$.
    \item[$\varphi_1\wedge\varphi_2$] One of the following is true.
    \begin{itemize}
        \item For both $i=1,2$, $\varphi_i$ is satisfiable and $\zero\not\models\varphi_i$. From inductive hypothesis, for both $i=1,2$, $J(\varphi_i)=\{\alpha\}$, and $J(\varphi)=J(\varphi_1)\cap J(\varphi_2)=\{\alpha\}$ as well.
        \item For both $i=1,2$, $\varphi_i$ is satisfiable and w.l.o.g.\ $\zero\models\varphi_1$ and $\zero\not\models\varphi_2$. Suppose that $\varphi_1\equiv \zero$. Then, $\varphi_1\wedge\varphi_2$ is not satisfiable, contradiction. So, $\zero\models\varphi_1$ and $\varphi_1\not\models \zero$. From inductive hypothesis of Claims 3 and 4, $J(\varphi_2)=\{\alpha\}$ and $J(\varphi_1)=\{\emptyset,\alpha\}$, respectively. Thus, $J(\varphi)=\{\alpha\}$.
    \end{itemize}
    \item[$\varphi_1\vee\varphi_2$]  Then, for both $i=1,2$, $\varphi_i$ is  satisfiable and $\zero\not\models\varphi_i$, or w.l.o.g.\ $\varphi_1$ is  satisfiable, $\zero\not\models\varphi_1$ and $\varphi_2$ is unsatisfiable. In both cases, $J(\varphi)=\{\alpha\}$.
\end{description}
$(\Leftarrow)$ Let $J(\varphi)=\{\alpha\}$. Then,
\begin{description}
    \item[$\varphi=\langle a\rangle \varphi'$] Since $J(\langle a\rangle \varphi')\neq\emptyset$, $J(\varphi')\neq\emptyset$, which implies that $\varphi'$ is satisfiable. So, $\langle a\rangle \varphi'$ is satisfiable and $\zero\not\models\langle a\rangle \varphi'$.
    \item[$\varphi_1\wedge\varphi_2$] One of the following is true.
    \begin{itemize}
        \item $J(\varphi_1)=J(\varphi_2)=\{\alpha\}$, and so $\varphi_i$ is satisfiable and $\zero\not\models\varphi_i$ for both $i=1,2$. In this case, there are processes $p_1,p_2\neq\mathtt{0}$ such that $p_1\models\varphi_1$ and $p_2\models\varphi_2$. Since $p_1,p_2\neq\mathtt{0}$, it holds that $p_i\curle_{CS} p_1+p_2$ for both $i=1,2$. From Proposition~\ref{logical_characterizations}, $p_1+p_2\models\varphi_1\wedge\varphi_2$, and so $\varphi$ is satisfiable. It is immediate from  $\zero\not\models\varphi_i$ that $\zero\not\models\varphi_1\wedge\varphi_2$.
        \item W.l.o.g.\ $J(\varphi_1)=\{\emptyset,\alpha\}$ and $J(\varphi_2)=\{\alpha\}$.  Then, $\varphi_2$ is satisfiable, $\zero\not\models\varphi_2$, $\zero\models\varphi_1$, and $\varphi_1\not \models\zero$. So, there are processes $p_1,p_2\neq\mathtt{0}$ such that $p_1\models\varphi_1$ and $p_2\models\varphi_2$. As in the previous case, $p_1+p_2\models\varphi_1\wedge\varphi_2$, which implies that $\varphi$ is satisfiable, and $\zero\not\models\varphi_1\wedge\varphi_2$ because of $\zero\not\models\varphi_2$. 
    \end{itemize}
    \item[$\varphi_1\vee\varphi_2$]  We distinguish between two cases. $J(\varphi_i)=\{\alpha\}$ for both $i=1,2$. Then, from inductive hypothesis, for both $i=1,2$, $\varphi_i$ is satisfiable and $\zero\not\models\varphi_i$, which also holds for $\varphi_1\vee\varphi_2$. Otherwise, w.l.o.g.\ $J(\varphi_1)=\{\alpha\}$ and $J(\varphi_2)=\emptyset$, which implies that $\varphi_1$ is satisfiable, $\zero\not\models\varphi_1$, and $\varphi_2$ is unsatisfiable. So, $\varphi_1\vee\varphi_2$ is satisfiable and $\zero\not\models\varphi_1\vee\varphi_2$.\qedhere
\end{description}
        \end{proof}
        \begin{clm}\label{cl-four} $\zero\models\varphi$ and $\varphi\not\models\zero$ iff $J(\varphi)=\{\emptyset,\alpha\}$.
        \end{clm}
        \begin{proof}
     $(\Rightarrow)$ Let  $\zero\models\varphi$ and $\varphi\not\models\zero$ be both true.
\begin{description}
    \item[$\varphi=\true$] Trivial.
    \item[$\varphi=\varphi_1\wedge\varphi_2$] Then, $\zero\models\varphi$ implies that for both $i=1,2$, $\zero\models\varphi_i$, and  $\varphi\not\models\zero$  means that there is $p\neq\mathtt{0}$ such that $p\models\varphi_1\wedge\varphi_2$, or equivalently, $p\models\varphi_1$ and $p\models\varphi_2$. Thus, $J(\varphi)=J(\varphi_1)\cap J(\varphi_2)=\{\emptyset,\alpha\}\cap\{\emptyset,\alpha\}=\{\emptyset,\alpha\}$. 
    \item[$\varphi=\varphi_1\vee\varphi_2$] $\zero\models\varphi_1\vee\varphi_2$ and $\varphi_1\vee\varphi_2\not\models\zero$ implies one of the following cases.
    \begin{itemize}
        \item $\zero\models\varphi_i$ and $\varphi_i\not\models\zero$ for some $i=1,2$. From inductive hypothesis and the fact that $J(\varphi)=J(\varphi_1)\cup J(\varphi_2)$, we have that $J(\varphi)=\{\emptyset,\alpha\}$.
        \item $\varphi_i\equiv\zero$ for some $i=1,2$, assume w.l.o.g that $\varphi_1\equiv\zero$, $\zero\not\models\varphi_2$, and $\varphi_2\not\models\zero$. From $\varphi_2\not\models\zero$, we have that $\varphi_2$ is satisfiable. From inductive hypothesis of Claims 2 and 3, $J(\varphi_1)=\{\emptyset\}$, and $J(\varphi_2)=\{\alpha\}$, respectively. Thus, $J(\varphi_1\vee\varphi_2)=\{\emptyset,\alpha\}$.
    \end{itemize}
\end{description}
$(\Leftarrow)$ Let $J(\varphi)=\{\emptyset,\alpha\}$.
\begin{description}
    \item[$\varphi=\true$] Trivial.
    \item[$\varphi=\varphi_1\wedge\varphi_2$] For both $i=1,2$, $J(\varphi_i)=\{\emptyset,\alpha\}$. So, for both $i=1,2$, $\zero\models\varphi_i$ and $\varphi_i\not\models\zero$, which implies that $\zero\models\varphi_1\wedge\varphi_2$ and for both $i=1,2$, there is $p_i\neq\mathtt{0}$ such that $p_i\models\varphi_i$. As shown above, $p_1+p_2\models\varphi_1\wedge\varphi_2$, and so $\varphi\not\models\zero$.
     \item[$\varphi=\varphi_1\vee\varphi_2$] One of the following cases is true.
     \begin{itemize}
         \item W.l.o.g.\ $J(\varphi_1)=\emptyset$ and $J(\varphi_2)=\{\emptyset,\alpha\}$. Then, $\varphi_1$ is unsatisfiable, $\zero\models\varphi_2$, and $\varphi_2\not\models\zero$. Then, $\zero\models\varphi$ and $\varphi\not\models\zero$.
         \item W.l.o.g.\ $J(\varphi_1)=\{\emptyset\}$ and $J(\varphi_2)=\{\alpha\}$. Then $\varphi_1\equiv\zero$, $\varphi_2$ is satisfiable, and $\zero\not\models\varphi_2$. As a result, $\zero\models\varphi_1\vee\varphi_2$. Suppose that $\varphi_1\vee\varphi_2\models\zero$. Then, from Lemma~\ref{lem:disjunction_lemma}, $\varphi_1\models\zero$ and $\varphi_2\models\zero$. Since $\varphi_2$ is satisfiable,  $\varphi_2\models\zero$ implies that $\varphi_2\equiv\zero$, which contradicts with $\zero\not\models\varphi_2$. So, $\varphi_1\vee\varphi_2\not\models\zero$.
         \item W.l.o.g.\ $J(\varphi_1)=\{\emptyset\}$ and $J(\varphi_2)=\{\emptyset,\alpha\}$. This can be proven similarly to the previous case.
         \item W.l.o.g.\ $J(\varphi_1)=\{\alpha\}$ and $J(\varphi_2)=\{\emptyset,\alpha\}$. This can be proven similarly to the previous two cases.
         \item For both $i=1,2$, $J(\varphi_i)=\{\emptyset,\alpha\}$. Then, $\zero\models\varphi_i$ and $\varphi_i\not\models\zero$ for both $i=1,2$. Consequently, $\zero\models\varphi_1\vee\varphi_2$ and from Lemma~\ref{lem:disjunction_lemma}, $\varphi_1\vee\varphi_2\not\models\zero$.\qedhere
     \end{itemize}
\end{description}
\end{proof}
The lemma is immediate from Claims A.3--A.6.
\end{proof}

\begin{prop}\label{cor:cs-satisfiability}
    Satisfiability of formulae in $\mathL_{CS}$ is linear-time decidable.
\end{prop}
\begin{proof}
    Let $\varphi\in\mathL_{CS}$. Consider algorithm \conscs that recursively computes $J(\varphi)$. We can easily prove by induction on the structure of $\varphi$, that \conscs requires linear time in $|\varphi|$. \conscs accepts $\varphi$ iff $J(\varphi)\neq\emptyset$. The correctness of the algorithm is immediate from Lemma~\ref{lem:J(phi)-property}.
\end{proof}

We prove the following lemma which will be used to prove Proposition A.9 below.

\begin{lem}\label{lem:substitution_bs}
    Let $\varphi,\psi,\chi\in \mathL_{BS}$.  If $\varphi\models\psi$, then $\chi\models\chi[\varphi/\psi]$.
\end{lem}
\begin{proof}
  We prove the lemma by induction on the structure of $\chi$.
  \begin{itemize}
      \item If $\varphi\not\in\sub(\chi)$ or $\chi=\varphi$, then the lemma is trivially true. 
       \item If $\chi=\phi\vee\phi'$, by inductive hypothesis, $\phi\models\phi[\varphi/\psi]$ and $\phi'\models\phi'[\varphi/\psi]$. So, $\phi\models\phi[\varphi/\psi]\vee\phi'[\varphi/\psi]$ and $\phi'\models\phi[\varphi/\psi]\vee\phi'[\varphi/\psi]$. From Lemma~\ref{lem:disjunction_lemma}, $\phi\vee\phi'\models\phi[\varphi/\psi]\vee\phi'[\varphi/\psi]$.
      \item If $\chi=\phi\wedge\phi'$, by inductive hypothesis $\phi\models\phi[\varphi/\psi]$ and $\phi'\models\phi'[\varphi/\psi]$. Then, $\phi\wedge\phi'\models\phi[\varphi/\psi]$ and $\phi\wedge\phi'\models\phi'[\varphi/\psi]$. It holds that for every $p$ such that $p\models\phi\wedge\phi'$, $p\models\phi[\varphi/\psi]$ and $p\models\phi'[\varphi/\psi]$. So, $p\models\phi[\varphi/\psi]\wedge \phi'[\varphi/\psi]$. As a result, $\phi\wedge\phi'\models\phi[\varphi/\psi]\wedge \phi'[\varphi/\psi]$. 
      \item If $\chi=\langle a \rangle\phi$, by inductive hypothesis, $\phi\models\phi[\varphi/\psi]$. Then, $\langle a \rangle\phi\models\langle a \rangle\phi[\varphi/\psi]$ from Lemma~\ref{lem:disjunction_lemma}.
      \item If $\chi=[a]\phi$, by inductive hypothesis, $\phi\models\phi[\varphi/\psi]$. It suffices to show that for every $\varphi,\psi\in\mathL_{BS}$, $\varphi\models\psi$ implies $[a]\varphi\models [a]\psi$. Assume $\varphi\models\psi$ is true and let $p\models[a]\varphi$. This means that for every $p\myarrowa p'$, $p'\models\varphi$. By hypothesis, $p'\models\psi$, and so $p\models [a]\psi$.\qedhere
  \end{itemize}
\end{proof}

\begin{prop}\label{prop:cs-satisfiability-satisfiable}
There is a polynomial-time algorithm that on input a satisfiable $\varphi\in\mathL_{CS}$, it returns $\varphi'$ such that (a) $\varphi\equiv\varphi'$, and (b) every $\psi\in\sub(\varphi')$ is satisfiable.
\end{prop}
\begin{proof}
   Let $\varphi\in\mathL_{CS}$ be a satisfiable formula. Consider algorithm \subcs that computes $J(\psi)$ for every $\psi\in\sub(\varphi)$ and stores  $J(\psi)$ in memory. For every $\psi\in\sub(\varphi)$ such that $J(\psi)=\emptyset$, \subcs substitutes $\psi$ with $\ff$ in $\varphi$. We denote by $\varphi^{\mathrm{ff}}$ the obtained formula. Then, \subcs repeatedly applies the rules $\langle a\rangle\ff\ruleff \ff$, $\ff\vee\psi\ruleff\psi$, $\psi\vee\ff\ruleff\psi$, $\ff\wedge\psi\ruleff \ff$, and $\psi\wedge\ff\ruleff \ff$ on $\varphi^{\mathrm{ff}}$ until no rule can be applied, and returns the resulting formula, which we denote by $\varphi'$. Since every substitution has replaced a formula $\psi$ with some $\psi'\equiv\psi$, from Lemma~\ref{lem:substitution_bs}, $\varphi\equiv\varphi'$. Suppose that there is some $\psi\in\sub(\varphi')$ such that $\psi$ is unsatisfiable. Suppose that $\ff\not\in\sub(\psi)$. Then, either $\psi\in\sub(\varphi)$ or $\psi$ is the result of some substitution. In the latter case, since $\ff\not\in\sub(\psi)$, $\psi$ can only be the result of rules of the form $\ff\vee\psi\ruleff\psi$ (and $\psi\vee\ff\ruleff\psi$), and so again $\psi\in\sub(\varphi)$. However, if $\psi\in\sub(\varphi)$, then since $J(\psi)=\emptyset$, \subcs would have substituted $\psi$ with $\ff$ in $\varphi$ and $\psi\not\in\sub(\varphi')$, contradiction. So, $\ff\in\sub(\psi)$, which implies that some rule can be applied on $\varphi'$, contradiction. So, $\psi$ cannot be unsatisfiable.
\end{proof}
  
\section{The formula primality problem for \texorpdfstring{$\mathL_{CS}$}{LCS}}\label{subsection:cs-primality-appendix}

In this section we consider formulae in $\mathL_{CS}$ that contain only satisfiable subformulae and examine the complexity of deciding whether such a formula is prime. We describe a preprocessing phase during which appropriate rules are applied on $\varphi$, so that the primality of the resulting formula $\varphi'$ can give information on the primality of $\varphi$. Moreover, primality of $\varphi'$ can be efficiently decided by  an appropriate variant of algorithm \algos.

First, we make some observations about formulae in $\mathL_{CS}$ that will be used throughout this section. Let  $\varphi\in\mathL_{CS}$ be a satisfiable formula that does not contain disjunctions and $\true$. We associate a process $p_\varphi$ to $\varphi$ and prove that $\varphi$ is characteristic within $\mathL_{CS}$ for $p_\varphi$.

\begin{defi}\label{def:cs-associated-process}
    Let $\varphi\in\mathL_{CS}$ be a %%satisfiable 
    formula given by the grammar $\varphi::=\zero ~|~ \langle a \rangle \varphi ~|~ \varphi\wedge\varphi$. We define process $p_\varphi$ inductively as follows.
\begin{itemize}
    \item If $\varphi=\zero$, then $p_\varphi=\mathtt{0}$.
     \item If $\varphi=\langle a\rangle \varphi'$, then $p_\varphi=a. p_{\varphi'}$.
     \item If $\varphi=\varphi_1\wedge \varphi_2$, then $p_\varphi=p_{\varphi_1}+p_{\varphi_2}$.
\end{itemize}
\end{defi}

\begin{lem}\label{lem:cs-associated-process}
    Let $\varphi\in\mathL_{CS}$ be a satisfiable formula given by the grammar $\varphi::=\zero ~|~ \langle a \rangle \varphi ~|~ \varphi\wedge\varphi$. Then, $\varphi$ is prime.
    In particular, $\varphi$ is characteristic within $\mathL_{CS}$ for $p_\varphi$.
\end{lem}
 \begin{proof} 
   The proof is analogous to the proof of Lemma~\ref{lem:prime-no-ff-disj}, where we also use the fact that $\varphi$ is satisfiable.
\end{proof}

As a corollary, in the case of complete simulation, all formulae that do not contain disjunctions and $\true$ are prime.

\begin{cor}\label{cor:grammar_prime_cs}
    Let $\varphi\in\mathL_{CS}$ be given by the grammar $\varphi::= \ff ~|~ \zero ~|~ \varphi\wedge\varphi~|~ \langle a\rangle \varphi$. Then, $\varphi$ is prime.
\end{cor}

Let $\varphi$ be given by the grammar $\varphi::= \zero ~|~ \langle a\rangle \varphi ~|~ \varphi\wedge\varphi  ~|~ \varphi\vee\varphi$ and $\bigvee_{i=1}^k \varphi_i$ be $\varphi$ in DNF. Propositions~\ref{prop:primality-LCS} and~\ref{prop:primality-LCS-2}, Lemma~\ref{lem:cs-common-divisor-pairs} and Corollaries~\ref{cor:cs-common-divisor-pairs} and~\ref{cor:primality-LCS-3} are variants of Propositions~\ref{prop:primality-LS} and~\ref{prop:primality-LS-2}, Lemma~\ref{lem:common-divisor-pairs} and Corollaries~\ref{cor:common-divisor-pairs} and~\ref{cor:primality-LS-3}, respectively, that hold in the case of complete simulation. Their proofs are completely analogous to those of their respective statements in Subsection~\ref{subsection:primality-simulation}, where we also need that every $\varphi_i$ is prime from Corollary~\ref{cor:grammar_prime_cs}, and every satisfiable $\varphi_i$ is characteristic within $\mathL_{CS}$ for $p_{\varphi_i}$ from Lemma~\ref{lem:cs-associated-process}.

\begin{prop}\label{prop:primality-LCS}
     Let $\varphi$ be given by the grammar $\varphi::= \zero ~|~ \langle a\rangle \varphi ~|~ \varphi\wedge\varphi  ~|~ \varphi\vee\varphi$; let also $\bigvee_{i=1}^k \varphi_i$ be $\varphi$ in DNF. Then, 
     $\varphi$ is prime iff $\varphi\models \varphi_j$ for some $1\leq j \leq k$.
\end{prop}

\begin{lem}\label{lem:cs-common-divisor-pairs}
    Let $\varphi$ be given by the grammar $\varphi::= \zero ~|~ \langle a\rangle \varphi ~|~ \varphi\wedge\varphi  ~|~ \varphi\vee\varphi$ such that every $\psi\in\sub(\varphi)$ is satisfiable;
    let also $\bigvee_{i=1}^k \varphi_i$ be $\varphi$ in DNF. If for every pair $p_{\varphi_i},p_{\varphi_j}$, $1\leq i,j\leq k$, there is some process $q$ such that $q\curle_{CS} p_{\varphi_i}$, $q\curle_{CS} p_{\varphi_j}$, and $q\models\varphi$, then there is some process $q$ such that $q\curle_{CS} p_{\varphi_i}$ for every $1\leq i\leq k$, and $q\models\varphi$. 
\end{lem}

\begin{cor}\label{cor:cs-common-divisor-pairs}
    Let $\varphi$ be given by the grammar $\varphi::= \zero ~|~ \langle a\rangle \varphi ~|~ \varphi\wedge\varphi  ~|~ \varphi\vee\varphi$ such that every $\psi\in\sub(\varphi)$ is satisfiable;
    let also $\bigvee_{i=1}^k \varphi_i$ be $\varphi$ in DNF. If for every pair $p_{\varphi_i},p_{\varphi_j}$, $1\leq i,j\leq k$, there is some process $q$ such that $q\curle_{CS} p_{\varphi_i}$, $q\curle_{CS} p_{\varphi_j}$, and $q\models\varphi$, then there is some $1\leq m\leq k$, such that $p_{\varphi_m}\curle_{CS} p_{\varphi_i}$ for every $1\leq i\leq k$. 
\end{cor}

\begin{prop}\label{prop:primality-LCS-2}
    Let $\varphi$ be given by the grammar $\varphi::= \zero ~|~ \langle a\rangle \varphi ~|~ \varphi\wedge\varphi  ~|~ \varphi\vee\varphi$ such that every $\psi\in\sub(\varphi)$ is satisfiable;
    let also $\bigvee_{i=1}^k \varphi_i$ be $\varphi$ in DNF. Then, $\varphi$ is prime iff for every pair $p_{\varphi_i},p_{\varphi_j}$, $1\leq i,j\leq k$, there is some process $q$ such that $q\curle_{CS} p_{\varphi_i}$, $q\curle_{CS} p_{\varphi_j}$, and $q\models \varphi$.
\end{prop}

\begin{cor}\label{cor:primality-LCS-3}
    Let $\varphi$ be given by the grammar $\varphi::= \zero ~|~ \langle a\rangle \varphi ~|~ \varphi\wedge\varphi  ~|~ \varphi\vee\varphi$ such that every $\psi\in\sub(\varphi)$ is satisfiable;
    let also $\bigvee_{i=1}^k \varphi_i$ be $\varphi$ in DNF. Then, $\varphi$ is prime iff for every pair $\varphi_i$, $\varphi_j$ there is some $1\leq m \leq k$ such that $\varphi_i\models \varphi_m$ and $\varphi_j\models\varphi_m$.
\end{cor}

Let $\varphi\in\mathL_{CS}$ be a formula such that every $\psi\in\sub(\varphi)$ is satisfiable. We transform  $\varphi$ into a formula that we denote by $\varphi^\diamond$, such that in case $\varphi$ is prime, $\varphi^\diamond\equiv\varphi$, $\varphi^\diamond$ is prime, and primality of $\varphi^\diamond$ can be efficiently checked. To this end, we apply a set of rules on $\varphi$. First, we consider the following rewriting rules: $\true\wedge\psi\rulett \psi$ and $\true\vee\psi\rulett\true$ modulo commutativity---i.e.\ we also consider the rules $\psi\wedge\true\rulett \psi$ and $\psi\vee\true\rulett\true$.  We write  $\varphi\rulettsub\varphi'$ if $\varphi'=\varphi[\psi/\psi']$, where $\psi\rulett\psi'$ for some $\psi\in\sub(\varphi)$, and $\varphi\rulettstar\varphi'$ to denote that there is a sequence $\varphi\rulettsub\varphi_1\cdots\rulettsub\varphi'$ and there is no $\varphi''$ such that $\varphi'\rulettsub\varphi''$. 

\begin{lem}\label{lem:phitt-poly-time-complexity}
    Let $\varphi\in\mathL_{CS}$ and $\varphi\rulettstar\varphi^{tt}$. Then, $\varphi^{tt}$ is unique, can be computed in polynomial time, and $\varphi^{tt}\equiv\varphi$.
\end{lem}

\begin{lem}\label{lem:phitt-diamond-true}
     Let $\varphi\in\mathL_{CS}$ and $\varphi\rulettstar\varphi^{tt}$. If $\varphi$ is not a tautology and $\true\in\sub(\varphi^{tt})$, then every occurrence of $\true$ is in the scope of some $\langle a\rangle$.
\end{lem}

Assume that we have a formula $\varphi\in\mathL_{CS}$, such that for every $\psi\in\sub(\varphi)$, $\psi$ is satisfiable and if $\psi=\true$, then $\psi$ occurs only in the scope of some $\langle a\rangle$. We consider the following rewriting rules modulo commutativity and associativity: 
\begin{enumerate}
    \item $\zero\vee\zero\rulezero \zero$,
    \item $\zero\wedge\varphi\rulezero\zero$, 
    \item $(\zero\vee\varphi_1)\wedge\varphi_2\rulezero \varphi_1\wedge\varphi_2$, where $\varphi_2\neq\zero$ and $\varphi_2\neq\zero\vee\varphi_2'$, 
    \item $(\zero\vee\varphi_1)\wedge(\zero\vee\varphi_2)\rulezero \zero\vee (\varphi_1\wedge\varphi_2)$.
\end{enumerate}  
Note that the following rules can be derived:
\begin{enumerate}\setcounter{enumi}{4}
 \item $\zero\vee (\zero\vee\varphi)\rulezero \zero\vee\varphi$ from rule 1 and associativity, 
    \item $(\zero\vee\varphi_1)\vee\varphi_2\rulezero \zero\vee (\varphi_1\vee\varphi_2)$ from associativity, and  
    \item $(\zero\vee\varphi_1)\vee(\zero\vee\varphi_2)\rulezero \zero\vee (\varphi_1\vee\varphi_2)$ from rule 1, commutativity, and associativity.
\end{enumerate}
We apply these rules on a formula $\varphi$ from the innermost to the outermost subformulae. Formally, we write  $\varphi\rulezerosub\varphi'$ if $\varphi'=\varphi[\psi/\psi']$, where $\psi\rulezero\psi'$ for some $\psi\in\sub(\varphi)$ and there is no $\psi''\in\sub(\psi)$ on which a rule can be applied. For every $\varphi\in\mathL_{CS}$, if there is no $\varphi'$ such that $\varphi\rulezerosub\varphi'$,  we say that $\varphi$ is in \emph{zero normal form}.
We write $\varphi\rulezerostar\varphi'$  if there is a (possibly empty) sequence $\varphi\rulezerosub\varphi_1\cdots\rulezerosub\varphi'$,   
and $\varphi'$ is in zero normal form. 

\begin{lem}\label{lem:phizero-poly-time-complexity}
    Let $\varphi\in\mathL_{CS}$ and $\varphi\rulezerostar\varphi^{0}$. Then, $\varphi^{0}$ is unique and can be computed in polynomial time.
\end{lem}

\begin{lem}\label{lem:guarded-zeros-property}
    Let $\varphi\in\mathL_{CS}$ such that for every $\psi\in\sub(\varphi)$, $\psi$ is satisfiable and if $\psi=\true$, then $\psi$ occurs in the scope of some $\langle a\rangle$; let also  $\varphi\rulezerostar\varphi^0$. Then, (a) $\varphi\equiv\varphi^0$ and (b) $\zero\models\varphi$ iff either $\varphi^0=\zero$ or $\varphi^0=\zero\vee\varphi'$, where $\zero\not\models\varphi'$.
\end{lem}
\begin{proof} Let $\varphi\rulezerosub\varphi'$, where $\varphi'=\varphi[\psi/\psi']$ and every $\psi''\in\sub(\psi)$ is in zero normal form. It suffices to show that $\psi\equiv\psi'$.  Then, from Lemma~\ref{lem:substitution_bs}, $\varphi\equiv\varphi'$. We prove by mutual induction that (i) $\psi\equiv\psi'$ and (ii) for every $\phi\in\varphi$ such that $\zero\models\phi$, either $\phi\rulezerostar\zero$ or $\phi\rulezerostar\zero\vee\phi'$, where $\zero\not\models\phi'$. For part (i), the only interesting cases are the following two.
\begin{description}
    \item[$\psi=\zero\wedge\psi''$.] In this case $\psi\rulezero \zero$. Note that since $\zero\wedge\psi''$ is satisfiable, there is $p$ such that $p\models \zero\wedge\psi''$, which is equivalent to $p\models\zero$ and $p\models\psi''$. Since $\mathtt{0}$ is the only process satisfying $\zero$, $\mathtt{0}$ is also the only process satisfying both $\zero$ and $\psi''$, which means that  $\psi\equiv\zero$ and (i) holds. 
    \item[$\psi=(\zero\vee\psi_1)\wedge\psi_2$,] where $\psi_2\neq \zero$ and $\psi_2\neq\zero\vee\psi_2'$. In this case, $\psi\rulezero\psi_1\wedge\psi_2$. Since $\psi_2$ is in zero normal form, from inductive hypothesis of (ii), $\zero\not\models\psi_2$.   
    Let $p$ be a process such that $p\models\psi$. It holds that $p\models(\zero\vee\psi_1)\wedge\psi_2$ iff ($p\models\zero$ or $p\models\psi_1$) and $p\models\psi_2$ iff ($p\models \zero$ and $p\models \psi_2$) or ($p\models\psi_1$ and $p\models\psi_2$). Since $\zero\not\models\psi_2$, ($p\models \zero$ and $p\models \psi_2$) is not true. Thus, we have that ($p\models\psi_1$ and $p\models\psi_2$), and $\psi\models\psi_1\wedge\psi_2$. Since $\psi_1\wedge\psi_2\models (\zero\vee\psi_1)\wedge\psi_2$ is also true,  $\psi\equiv\psi_1\wedge\psi_2$ holds. 
    \end{description} 
To prove part (ii), let $\phi\in\sub(\varphi)$ such that $\zero\models\phi$. Note that $\phi$ cannot be $\true$. Therefore, $\phi$ can have one of the following forms.
\begin{description}
    \item[$\phi=\zero$.] Trivial.
    \item[$\phi=\phi_1\vee\phi_2$,] where $\zero\models\phi_i$ for some $i=1,2$. Assume that w.l.o.g.\ $\zero\models\phi_1$ and $\zero\not\models\phi_2$.  Let $\phi_2\rulezerostar\phi_2'$. From inductive hypothesis of (i) and Lemma~\ref{lem:substitution_bs}, $\phi_2\equiv\phi_2'$ and so $\zero\not\models\phi_2'$. From inductive hypothesis, either $\phi_1\rulezerostar\zero$  or $\phi_1\rulezerostar\zero\vee \phi_1'$, where $\zero\not\models\phi_1'$. Thus, there is either a sequence $\phi_1\vee\phi_2\rulezerosub\dots\rulezerosub \zero\vee\phi_2'$, where $\zero\not\models\phi_2'$, or $\phi_1\vee\phi_2\rulezerosub\dots\rulezerosub (\zero\vee\phi_1')\vee\phi_2'\rulezero \zero\vee (\phi_1'\vee\phi_2')$, where $\zero\not\models\phi_1'\vee\phi_2'$, respectively. If $\zero\models\phi_1$ and $\zero\models\phi_2$, the proof is analogous.
    \item[$\phi=\phi_1\wedge\phi_2$,] where $\zero\models\phi_i$ for both $i=1,2$. From inductive hypothesis, for both $i=1,2$, either $\phi_i\rulezerostar\zero$ or $\phi_i\rulezerostar\zero\vee \phi_i'$, where $\zero\not\models\phi_i'$. Assume that for both $i=1,2$, $\phi_i\rulezerostar\zero\vee \phi_i'$, where $\zero\not\models\phi_i'$. Then, there is a sequence $\phi_1\wedge\phi_2\rulezerosub \dots\rulezerosub (\zero\vee \phi_1')\wedge(\zero\vee \phi_2')\rulezero \zero\vee(\phi_1'\wedge\phi_2')$, where $\zero\not\models\phi_1'\wedge\phi_2'$. The other cases can be similarly proven.
\end{description}
 Let $\varphi=\varphi_1\rulezerosub\dots\rulezerosub\varphi_n=\varphi^0$. From (i), for every $1\leq i\leq n-1$, $\varphi_i\equiv\varphi_{i+1}$, and so it holds that $\varphi\equiv\varphi^0$. Part (ii) proven above implies that if $\zero\models\varphi$, then either $\varphi^0=\zero$ or $\varphi^0=\zero\vee\varphi'$, where $\zero\not\models\varphi'$. Conversely, if $\varphi^0=\zero$ or $\varphi^0=\zero\vee\varphi'$, then $\zero\models\varphi^0$ is immediate. From (a), $\zero\models\varphi$ holds as well. Consequently, $\zero\models\varphi$ iff $\varphi^0=\zero$ or $\varphi^0=\zero\vee\varphi'$, where $\zero\not\models\varphi'$.    
\end{proof}

\begin{rem}\label{rem:cs-rules}
    Note that since we follow an innermost reduction strategy, when we apply a rule on $(\zero\vee\varphi_1)\wedge \varphi_2$, $\varphi_2$ is already in zero normal form. If $\zero\models\varphi_2$, Lemma~\ref{lem:guarded-zeros-property} guarantees that either $\varphi_2=\zero$ or $\varphi_2=\zero\vee\varphi_2'$, where $\zero\not\models \varphi_2'$, and so either rule 2 or rule 4 is applied, respectively. Otherwise, if $\zero\not\models\varphi_2$, rule 3 is applied. An alternative way to check whether rule 3 is to be applied on $(\zero\vee\varphi_1)\wedge \varphi_2$ is to compute $J(\varphi_2)$ and check that $\emptyset\not\in J(\varphi_2)$, which from Claims 1 and 4 in the proof of Lemma~\ref{lem:J(phi)-property},  is equivalent to $\zero\not\models\varphi_2$.
\end{rem}

Lemma~\ref{lem:conjunction_lemma_simulation} takes the following form in the case of complete simulation.

\begin{lem}\label{lem:conjunction_lemma_cs}
Let $\varphi_1\wedge\varphi_2\in\mathL_{CS}$ such that for every $\psi\in\sub(\varphi_1\wedge\varphi_2)$, $\psi$ is satisfiable, if $\psi=\true$, then $\true$ occurs in the scope of some $\langle a\rangle$, and $\psi$ is in zero normal form. Then, 
$\varphi_1\wedge\varphi_2\models\langle a\rangle\psi \text{ iff } \varphi_1\models\langle a\rangle\psi \text{ or } \varphi_2\models\langle a\rangle\psi.$
\end{lem}
\begin{proof}
    $(\Leftarrow)$ If $\varphi_1\models\langle a\rangle\psi$ or $\varphi_2\models\langle a\rangle\psi$, then $\varphi_1\wedge\varphi_2\models\langle a\rangle\psi$ immediately holds.\\
    $(\Rightarrow)$ Let $\varphi_1\wedge\varphi_2\models\langle a\rangle\psi$. We distinguish between the following cases.
    \begin{description}
        \item[$\zero\not\models\varphi_1$ and $\zero\not\models\varphi_2$] Assume that $\varphi_1\not\models\langle a\rangle\psi$. Then, there is $p_1\neq\mathtt{0}$ such that $p_1\models\varphi_1$ and $p_1\not\models\langle a\rangle \psi$. Let $p_2\models\varphi_2$. Since $\varphi_2$ is satisfiable there is such a process $p_2$ and $p_2\neq\mathtt{0}$ because of $\zero\not\models\varphi_2$. The fact that $p_1,p_2\neq\mathtt{0}$ implies that $p_1\curle_{CS} p_1+p_2$ and $p_2\curle_{CS} p_1+p_2$. From Proposition~\ref{logical_characterizations}, $p_1+p_2\models\varphi_1\wedge\varphi_2$, and so $p_1+p_2\myarrowa p'$ such that $p'\models\psi$. Thus, either $p_1\myarrowa p'$ or $p_2\myarrowa p'$. Since $p_1\not\models\langle a\rangle \psi$, it holds that $p_2\myarrowa p'$. As a result, $\varphi_2\models\langle a\rangle\psi$.
         \item[$\zero\models\varphi_1$ and $\zero\models\varphi_2$] Then, $\zero\models\varphi_1\wedge\varphi_2$ and $\zero\not\models\langle a\rangle\psi$, which  contradicts our assumption that $\varphi_1\wedge\varphi_2\models\langle a\rangle\psi$. This case is therefore not possible.
        \item[W.l.o.g.\ $\zero\models\varphi_1$ and $\zero\not\models\varphi_2$] Since $\varphi_1$ is in zero normal form, from Lemma~\ref{lem:guarded-zeros-property}, $\varphi_1=\zero$ or $\varphi_1=\zero\vee\varphi_1'$, where $\zero\not\models\varphi_1'$. Then, either $\varphi_1\wedge\varphi_2=\zero\wedge\varphi_2$ or $\varphi_1\wedge\varphi_2=(\zero\vee\varphi_1')\wedge\varphi_2$, respectively, which implies that either $\varphi_1\wedge\varphi_2$ is unsatisfiable, since $\zero\not\models\varphi_2$, or $\varphi_1\wedge\varphi_2\rulezero \varphi_1'\wedge\varphi_2$, since $\varphi_2\neq\zero$ and $\varphi_2\neq\zero\vee\varphi_2'$. The former case contradicts the satisfiability of $\varphi_1\wedge\varphi_2$ and the latter case contradicts the fact that  $\varphi_1\wedge\varphi_2$ is in zero normal form. So this case is not possible either.\qedhere
    \end{description}
\end{proof}

\begin{lem}\label{lem:cs-satisfiable-dnf}
    Let $\varphi\in \mathL_{CS}$ be in zero normal form and for every $\psi\in\sub(\varphi)$, $\psi$ is satisfiable, and if $\psi=\true$, then $\true$ occurs in the scope of some $\langle a\rangle$; let also $\bigvee_{i=1}^k\varphi_i$ be $\varphi$ in DNF. Then, $\varphi_i$ is satisfiable for every $1\leq i\leq k$.
\end{lem}
\begin{proof}
    Consider an algorithm that takes $\varphi$ and returns $\varphi$ in DNF. We prove the lemma by induction on the structure of $\varphi$. 
    \begin{description}
        \item[$\varphi$ does not contain disjunctions.] Trivial.
        \item[$\varphi=\varphi_1\vee\varphi_2$.] The DNF of $\varphi$ is $\varphi_1'\vee\varphi_2'$, where $\varphi_i'$ is the DNF of $\varphi_i$ for both $i=1,2$. By the inductive hyppothesis, the claim is true for $\varphi_i$, $i=1,2$, and so the lemma immediately holds for $\varphi$.
        \item[$\varphi=\langle a\rangle\varphi'$.] The DNF of $\varphi$ is $\bigvee_{i=1}^k\langle a\rangle\varphi'_i$, where $\bigvee_{i=1}^k\varphi'_i$ is the DNF of $\varphi'$. Formula $\varphi'$ satisfies the hypothesis of the lemma, and so from inductive hypothesis, every $\varphi_i'$ is satisfiable, which implies that every $\langle a\rangle\varphi'_i$ is also satisfiable.
        \item[$\varphi=\varphi_1\wedge(\varphi_2\vee\varphi_3)$.] The DNF of $\varphi$ is $\bigvee_{i=1}^{k_{1}}\varphi_{12}^i\vee\bigvee_{i=1}^{k_{2}}\varphi_{13}^i$, where $\bigvee_{i=1}^{k_{1}}\varphi_{12}^i,\bigvee_{i=1}^{k_{2}}\varphi_{13}^i$ are the DNFs of $\varphi_1\wedge\varphi_2$ and $\varphi_1\wedge\varphi_3$, respectively. We show that, for both $j=2,3$, the formula $\varphi_1\wedge\varphi_j$   satisfies the hypothesis of the lemma. Suppose w.l.o.g.\ that $\varphi_1\wedge\varphi_2$ is not satisfiable. Since $\varphi_1,\varphi_2$ are satisfiable, we have that w.l.o.g.\ $\mathtt{0}\models\varphi_1$ and $\mathtt{0}\not\models\varphi_2$. Then, $\mathtt{0}\models\varphi_3$ holds, since otherwise, $\varphi$ is unsatisfiable. Then, either $\varphi_1=\zero$ or $\varphi_1=\zero\vee\varphi_1'$, where $\zero\not\models\varphi_1'$, $\varphi_2\neq\zero$ and $\varphi_2\neq\zero\vee\varphi_j'$, and either $\varphi_3=\zero$ or $\varphi_3=\zero\vee\varphi_3'$,  where $\zero\not\models\varphi_3'$, because of Lemma~\ref{lem:guarded-zeros-property} and the fact that $\varphi_1,\varphi_2$, and $\varphi_3$ are in zero normal form. Any combination of these forms leads to contradiction. For example, assume that $\varphi_1=\zero\vee\varphi_1'$ and $\varphi_3=\zero\vee\varphi_3'$. Then, $\varphi=(\zero\vee\varphi_1')\wedge(\varphi_2\vee(\zero\vee\varphi_3'))\rulezerostar \zero \vee (\varphi'\wedge\varphi_2\wedge\varphi_3')$, which contradicts the fact that $\varphi$ is in zero normal form. Every other case can be addressed in a similar way and proven to lead to contradiction. Consequently, every $\psi\in\sub(\varphi_1\wedge\varphi_j)$ is either a subformula of some $\varphi_i$, $i\in\{1,2,3\}$, or $\varphi_1\wedge\varphi_j$, and so $\psi$ is satisfiable. The other parts of the hypothesis of the lemma immediately hold for both $\varphi_1\wedge\varphi_j$, where $j=2,3$. From inductive hypothesis, for every $1\leq n\leq k_{1}$ and $1\leq m\leq k_{2}$, $\varphi_{12}^m$ and $\varphi_{13}^n$ are satisfiable. This implies that every $\varphi_i$, $1\leq i\leq k$, is satisfiable.\qedhere
    \end{description}
\end{proof}

Finally, we consider the rule $\langle a\rangle \true\rulediam\true$ and rules $\true\vee\psi\rulett\true$, $\true\wedge\psi\rulett\psi$ modulo commutativity. As before, we write  $\varphi\rulediamsub\varphi'$ if $\varphi'=\varphi[\psi/\psi']$, where $\psi\rulediam\psi'$ or $\psi\rulett\psi'$ for some $\psi\in\sub(\varphi)$, and $\varphi\rulediamstar\varphi'$ to denote that there is a sequence $\varphi\rulediamsub\varphi_1\cdots\rulediamsub\varphi'$ and there is no $\varphi''$ such that $\varphi'\rulediamsub\varphi''$. 

\begin{lem}\label{lem:cs_phi_diamond_poly_time}
    Let $\varphi\in\mathL_{CS}$ be satisfiable and $\varphi\rulediamstar\varphi^{\diamond}$. Then, $\varphi^{\diamond}$ is unique and can be computed in polynomial time.
\end{lem}

\begin{lem}\label{lem:cs_true_substitutions}
    Let $\varphi\in\mathL_{CS}$ be satisfiable and $\varphi\rulediamstar\varphi^{\diamond}$. Then, either $\true\not\in\sub(\varphi^{\diamond})$ or $\varphi^{\diamond}=\true$.
\end{lem}
\begin{proof}
    Immediate from the definition of $\varphi\rulediamstar\varphi^{\diamond}$.
\end{proof}

We prove Lemma~\ref{lem:phitt-vs-phi}, which is one of the main results of this subsection. We first provide some definitions and statements needed in its proof.

\begin{defi}\label{def:cs-associated-process-extended}
    Let $\varphi\in\mathL_{CS}$ be a  
    formula given by the grammar $\varphi::=\zero ~|~ \true ~|~ \varphi\wedge\varphi~\mid ~ \langle a \rangle \varphi $. We define process $p_\varphi$ inductively as follows.
\begin{itemize}
    \item If either $\varphi=\zero$ or $\varphi=\true$ , then $p_\varphi=\mathtt{0}$.
     \item If $\varphi=\langle a\rangle \varphi'$, then $p_\varphi=a. p_{\varphi'}$.
     \item If $\varphi=\varphi_1\wedge \varphi_2$, then $p_\varphi=p_{\varphi_1}+p_{\varphi_2}$.
\end{itemize}
\end{defi}

\begin{lem}\label{lem:cs-associated-process-extended}
     Let $\varphi\in\mathL_{CS}$ be a  satisfiable
    formula given by the grammar $\varphi::=\zero ~|~ \true ~\mid ~ \varphi\wedge\varphi ~|~\langle a \rangle \varphi  $. Then, $p_\varphi\models\varphi$.
\end{lem}
\begin{proof}
    We prove the lemma by structural induction on $\varphi$ and limit ourselves to presenting the case when $\varphi = \varphi_1\wedge\varphi_2$.   In the remainder of our argument, we will use the following claim, which can be easily shown by induction on the structure of formulae: 
    \begin{quote}
       Let $\varphi\in\mathL_{CS}$ be a  satisfiable formula given by the grammar $\varphi::=\zero ~|~ \true ~\mid ~ \varphi\wedge\varphi ~|~\langle a \rangle \varphi  $. Then,  $\mathtt{0}\models\varphi$ iff $\varphi\equiv\zero$ or $\varphi \equiv\true$.
    \end{quote}
    Our goal is to show that 
    $p_\varphi=p_{\varphi_1}+p_{\varphi_2} \models \varphi_1\wedge\varphi_2 = \varphi$. 
    
    By the inductive hypothesis, we have that $p_{\varphi_1}\models\varphi_1$ and $p_{\varphi_2}\models\varphi_2$.
    We now proceed by considering the following cases: 
    \begin{enumerate}
        \item Neither $p_{\varphi_1}$ nor $p_{\varphi_2}$ is equivalent to $\mathtt{0}$, 
        \item Both $p_{\varphi_1}$ and $p_{\varphi_2}$ are equivalent to $\mathtt{0}$, and 
        \item W.l.o.g.  $p_{\varphi_1}$ is equivalent to $\mathtt{0}$ and $p_{\varphi_2}$ is not. 
    \end{enumerate}
    In the first case, $p_{\varphi_i} \curle_{CS} p_{\varphi_1}+p_{\varphi_2}$, for $i=1,2$. Therefore,  by Proposition~\ref{logical_characterizations}, $p_{\varphi_1}+p_{\varphi_2}\models \varphi_1\wedge\varphi_2$ and we are done. 

    In the second case, observe, first of all, that $p_{\varphi_1}+p_{\varphi_2}$ is equivalent to $\mathtt{0}$. By the aforementioned claim, we infer that $\varphi_i\equiv \zero$ or $\varphi_i \equiv\true$, for $i=1,2$. Thus, $\varphi_1\wedge\varphi_2 \equiv \zero$ or $\varphi_1\wedge\varphi_2 \equiv \true$.  In both cases, 
    $p_{\varphi_1}+p_{\varphi_2}\models \varphi_1\wedge\varphi_2$ and we are done. 

    In the third case, we use the aforementioned claim to infer that $\varphi_1\equiv \zero$ or $\varphi_1 \equiv\true$. Moreover, $p_{\varphi_1}+p_{\varphi_2}$ is equivalent to $p_{\varphi_2}$. Since $p_{\varphi_2}$  is not equivalent to $\mathtt{0}$, the formula $\varphi_1\wedge\varphi_2$ is satisfiable and $p_{\varphi_2}\models\varphi_2$, it follows that $\varphi_1 \equiv\true$. Thus, $\varphi_1\wedge\varphi_2 \equiv \varphi_2$. Since $p_{\varphi_2} \curle_{CS} p_{\varphi_1}+p_{\varphi_2}$, by Proposition~\ref{logical_characterizations}, $p_{\varphi_1}+p_{\varphi_2}\models \varphi_2 \equiv \varphi_1\wedge\varphi_2$ and we are done. 
\qedhere
\end{proof}

\begin{lem}\label{lem:phitt-vs-phi}
    Let $\varphi\in \mathL_{CS}$ be in zero normal form and for every $\psi\in\sub(\varphi)$, $\psi$ is satisfiable, and if $\psi=\true$, then $\true$ occurs in the scope of some $\langle a\rangle$; let also $\varphi\rulediamstar\varphi^{\diamond}$. Then, 
     $\varphi$ is prime iff $\varphi^{\diamond}\models \varphi$ and $\varphi^{\diamond}$ is prime.
\end{lem}
\begin{proof}
   ($\Leftarrow$)   Let $\varphi^{\diamond}\models \varphi$ and $\varphi^{\diamond}$ be prime. It holds that  $\langle a\rangle \true \models\true$, $\true\wedge\psi\models\psi$, and $\true\vee\psi \models\true$, for every $\psi\in\mathL_{CS}$. Thus, from Lemma~\ref{lem:substitution_bs} and the definition of $\varphi^{\diamond}$, $\varphi\models\varphi^{\diamond}$. As a result $\varphi\equiv\varphi^{\diamond}$ and so $\varphi$ is prime.\\
   ($\Rightarrow$) Assume that $\varphi$ is prime. As we just showed,  $\varphi\models \varphi^{\diamond}$. Thus, it suffices to show that $\varphi^{\diamond}\models \varphi$. Then, we have that $\varphi\equiv\varphi^{\diamond}$ and $\varphi^{\diamond}$ is prime.  The proof of $\varphi^{\diamond}\models \varphi$ is by induction on the type of the rules $\psi\rulediamsub\psi'$.
        \begin{itemize}
            \item Let $\varphi^{\diamond}$ be the result of substituting only one occurrence of $\langle a\rangle\true$  with $\true$ in $\varphi$, and $p\models\varphi^{\diamond}$; let also $\bigvee_{i=1}^{k}\varphi^{\diamond}_i$ and $\bigvee_{i=1}^{k'}\varphi_i$ be the DNFs  of $\varphi^{\diamond}$ and $\varphi$, respectively. 
            It holds that $p\models \varphi^{\diamond}_i$ for some $1\leq i\leq k$. Formula $\varphi$ is satisfiable and prime, so from Proposition~\ref{prop:charact-via-primality} there is a process $p_{min}$ for which $\varphi$ is characteristic.
            We prove that $p_{min}\curle_{CS} p$, which combined with  Corollary~\ref{cor:characteristic} implies that $p\models\varphi$. If $\varphi^{\diamond}_i=\varphi_j$ for some $1\leq j\leq k'$, then $p\models \varphi$. Otherwise, $\varphi^{\diamond}_i$ coincides with $\varphi_j$ for some $1\leq j\leq k'$, where an occurrence of $\langle a\rangle \true$ has been substituted with $\true$. 
            Consider the process $p_{\varphi^{\diamond}_i}$ constructed from $\varphi^{\diamond}_i$ according to Definition~\ref{def:cs-associated-process-extended}, so that $p_{\varphi^{\diamond}_i}\models\varphi^{\diamond}_i$ as stated in Lemma~\ref{lem:cs-associated-process-extended}. The construction of $p_{\varphi^{\diamond}_i}$ implies that there is some $p_{tt}$ such that $p_{\varphi^{\diamond}_i}\myarrowtau p_{tt}$, $p_{tt}=\mathtt{0}$ and $t\in \act^*$, and $p_{tt}$ corresponds to subformula $\true$ that substituted $\langle a\rangle\true$ in $\varphi$. Define process $p_{\varphi^{\diamond}_i}^1$ to be a copy of $p_{\varphi^{\diamond}_i}$ extended with $p_{tt}^1\myarrowa p_1=\mathtt{0}$, and $p_{\varphi^{\diamond}_i}^2$  to be a copy of $p_{\varphi^{\diamond}_i}$ extended with $p_{tt}^2\myarrowa p_2\myarrowa p_3=\mathtt{0}$. From Lemma~\ref{lem:cs-satisfiable-dnf}, $\varphi_j$ is satisfiable and note that $p_{\varphi^{\diamond}_i}^1$ is $p_{\varphi_j}$, which implies that $p_{\varphi^{\diamond}_i}^1\models\varphi_j$ because of Lemma~\ref{lem:cs-associated-process-extended}. Moreover, it immediately holds that $p_{\varphi^{\diamond}_i}^2\models\varphi_j$ as well. Therefore, both $p_{\varphi^{\diamond}_i}^1$ and $p_{\varphi^{\diamond}_i}^2$ satisfy $\varphi$. This means that $p_{min}\curle_{CS} p_{\varphi^{\diamond}_i}^j$ for both $j=1,2$. Let  $p_{min}\myarrowtau q$ for some $t\in \act^*$. Then, there are some $q_j$, $j=1,2$, such that $p_{\varphi^{\diamond}_i}^j\myarrowtau q_j$ and $q\curle_{CS} q_j$.  
            \begin{itemize}
                \item Assume that there is some $q$ such that $p_{min}\myarrowtau q$ and $q\curle_{CS} p_1$. Thus, $q=p_1=\mathtt{0}$. On the other hand, $q\curle_{CS} p_2$ does not hold, since $p_2\neq \mathtt{0}$. So there is $p'$ such that $p_{\varphi^{\diamond}_i}^2\myarrowtau p'$, $p'\neq p_2$ and $q\curle_{CS} p'$. Moreover, $p'\neq p_3$, since $p_{\varphi^{\diamond}_i}^2\notmyarrowtau p_3$. But then, $p'$ is a copy of a process $p''$ such that $p_{\varphi^{\diamond}_i}\myarrowtau p''$ and $q\curle_{CS} p''$.
                \item Assume that there is some $q$ such that $p_{min}\myarrowtau q$ and $q\curle_{CS} p'$, where $p' \neq p_1$. Similar arguments can show that there is some $p''$ such that  $p_{\varphi^{\diamond}_i}\myarrowtau p''$ and $q\curle_{CS} p''$.
            \end{itemize} 
            As a result, $p_{min}\curle_{CS} p_{\varphi^{\diamond}_i}$. In a similar way, we can prove that $p_{min}$ is complete-simulated by any process that satisfies $\varphi^{\diamond}_i$, and so $p_{min}\curle_{CS} p$.
            \item Let $\varphi^{\diamond}=\varphi[\true\wedge\psi/\psi]$. From Lemma~\ref{lem:substitution_bs} and the fact that $\varphi^{\diamond}\models\varphi$, as $\true\wedge\psi\equiv\psi$.
            \item Let $\varphi^{\diamond}=\varphi[\true\vee\psi/\true]$. Similarly to the previous case, from Lemma~\ref{lem:substitution_bs} and the fact that  $\varphi^{\diamond}\models\varphi$, since $\true\vee\psi\equiv\true$.\qedhere
        \end{itemize}
\end{proof}

\begin{cor}\label{cor:phitt-vs-phi}
    Let $\varphi\in\mathL_{CS}$ be a formula such that  every $\psi\in\sub(\varphi)$ is satisfiable; let also $\varphi\rulettstar\varphi^{tt}\rulezerostar\varphi^0\rulediamstar\varphi^{\diamond}$. Then, every $\psi\in\sub(\varphi^\diamond)$ is satisfiable and $\varphi$ is prime iff $\varphi^\diamond\models\varphi$ and $\varphi^\diamond$ is prime.
\end{cor}
\begin{proof}
    Let $\psi\in \sub(\varphi^{tt})$. Then, there is some $\psi'\in\sub(\varphi)$, such that $\psi'\rulettstar\psi$, and so $\psi'\models\psi$. This implies that $\psi$ is satisfiable. Analogously, we can show that every $\psi\in\sub(\varphi^\diamond)$ is satisfiable.  It holds that $\varphi\equiv\varphi^{tt}\equiv\varphi^0$ from Lemmas~\ref{lem:phitt-poly-time-complexity} and~\ref{lem:guarded-zeros-property}(a). Formula $\varphi^0$ satisfies the hypothesis of Lemma~\ref{lem:phitt-vs-phi} and so $\varphi^0$ is prime iff $\varphi^\diamond\models\varphi^0$ and $\varphi^\diamond$ is prime. Combining the aforementioned facts, we have that $\varphi$ is prime iff $\varphi^\diamond\models\varphi$ and $\varphi^\diamond$ is prime.
\end{proof}

\begin{cor}\label{cor:primality-phitt}
     Let $\varphi\in\mathL_{CS}$ be a formula such that  every $\psi\in\sub(\varphi)$ is satisfiable; let also $\varphi\rulettstar\varphi^{tt}\rulezerostar\varphi^0\rulediamstar\varphi^{\diamond}$ and $\bigvee_{i=1}^k \varphi^{\diamond}_i$ be $\varphi^{\diamond}$ in DNF. Then, 
     $\varphi^{\diamond}$ is prime iff $\varphi^{\diamond}\models \varphi^{\diamond}_j$ for some $1\leq j \leq k$, such that $\varphi^{\diamond}_j\neq \true$.
\end{cor}
\begin{proof} Let $\varphi^{\diamond}$ be prime. By the definition of primality and the fact that $\varphi^{\diamond}\models \bigvee_{i=1}^k \varphi^{\diamond}_i$, we have that $\varphi^{\diamond}\models \varphi^{\diamond}_j$ for some $1\leq j \leq k$. Suppose that $\varphi^{\diamond}_j=\true$. Since $\true \vee \psi$,  $\psi\in\mathL_{CS}$,  is not prime, $\bigvee_{i=1}^k \varphi^{\diamond}_i$ is also not prime. From Lemma~\ref{lem:DNF-equiv}, $\varphi^{\diamond}$ is not prime, which contradicts our assumption. So $\varphi^{\diamond}_j\neq\true$. Conversely, let $\varphi^{\diamond} \models \varphi^{\diamond}_j$ for some $1\leq j \leq k$, such that $\varphi^{\diamond}_j\neq\true$. From Lemma~\ref{lem:cs_true_substitutions}, $\varphi^{\diamond}$ and $\varphi_j^{\diamond}$ do not contain $\true$. To prove that $\varphi^{\diamond}$ is prime, let $\varphi^{\diamond}\models \bigvee_{l=1}^m \phi_l$. From Lemmas~\ref{lem:DNF-equiv} and~\ref{lem:disjunction_lemma}, $\varphi^{\diamond}_i\models \bigvee_{l=1}^m \phi_l$, for every $1\leq i\leq k$. In particular, $\varphi^{\diamond}_j\models \bigvee_{l=1}^m \phi_l$. Since $\varphi^{\diamond}_j$ does not contain disjunctions and $\true$, from Corollary~\ref{cor:grammar_prime_cs}, $\varphi^{\diamond}_j$ is prime. Consequently, $\varphi^{\diamond}_j\models \phi_s$ for some $1\leq s \leq m$. Finally, since $\varphi^{\diamond}\models\varphi^{\diamond}_j$, it holds that $\varphi^{\diamond}\models\phi_s$. 
\end{proof}

\begin{exa}  
Formula $\varphi=\langle a\rangle\langle a\rangle\true\wedge\langle a\rangle \zero$ is not prime and $\varphi^\diamond=\langle a\rangle \zero \not\models\varphi$, whereas the prime formula $\psi=(\langle a\rangle\true\wedge\langle a\rangle\zero)\vee\langle a\rangle\true$ has $\psi^\diamond=\langle a\rangle\zero$, which logically implies $\psi$.
\end{exa}

We can prove now the following main proposition.

\begin{prop}\label{prop:phitt-primality}
  Let $\varphi\in\mathL_{CS}$ be a formula such that every $\psi\in\sub(\varphi)$ is satisfiable; let also  $\varphi\rulettstar\varphi^{tt}\rulezerostar\varphi^0\rulediamstar\varphi^{\diamond}$. There is a polynomial-time algorithm that decides whether $\varphi^{\diamond}$ is prime.
\end{prop}
\begin{proof}
    We describe algorithm \algott which takes $\varphi^\diamond$ and decides whether $\varphi^\diamond$ is prime. If $\varphi^\diamond=\true$, \algott rejects. Otherwise, from Lemma~\ref{lem:cs_true_substitutions}, $\true\not\in\varphi^\diamond$. Then, \algott constructs the alternating graph $G_{\varphi^{\diamond}}=(V,E,A,s,t)$ by starting with vertex $(\varphi^\diamond,\varphi^\diamond\Rightarrow\varphi^\diamond)$ and repeatedly applying the rules for complete simulation, i.e.\ the rules from Table~\ref{tab:S-rules}, where rule (tt) is replaced by the following one:
    \begin{prooftree}
    \AxiomC{$\zero,\zero\Rightarrow\zero$ }
    \RightLabel{\scriptsize(0)}
    \UnaryInfC{\textsc{True}}
\end{prooftree}
Then, the algorithm solves \reacha on $G_{\varphi^\diamond}$, where $s$ is $(\varphi^\diamond,\varphi^\diamond\Rightarrow\varphi^\diamond)$ and $t=\textsc{True}$, and accepts $\varphi^\diamond$ iff there is an alternating path from $s$ to $t$.  From Corollary~\ref{cor:phitt-vs-phi}, every $\psi\in\sub(\varphi^\diamond)$ is satisfiable. Correctness of \algott  is  immediate  from the following claim.\\
\noindent\textbf{\textcolor{darkgray}{Claim A.}} $\varphi^\diamond$ is prime iff there is an alternating path in $G_{\varphi^\diamond}$ from $s$ to $t$.\\
\noindent\textbf{\textcolor{darkgray}{Proof of Claim A.}} ($\Leftarrow$) The proof of this implication is similar to the proof of Lemma~\ref{lem:sim-algo-correct-1}. If $(\varphi_1,\varphi_2\Rightarrow\psi)$ is a vertex in the alternating path from $s$ to $t$, then property $P_1$ is true for $\varphi_1,\varphi_2,\psi$ and this can be proven by induction on the type of the rules. We only include two cases that are different here. 
\begin{description}
    \item[Case (0).] $P_1$ trivially holds for $\zero,\zero,\zero$.
    \item[Case (L$\wedge_1$).] Let $\bigvee_{i=1}^{k_{12}}\varphi_{12}^i$ be $\varphi_1\wedge\varphi_2$ in DNF. The argument is the same as in the proof of Lemma~\ref{lem:sim-algo-correct-1}. In particular, if $P_1$ is true either for $\varphi_1,\varphi,\langle a\rangle\psi$ or for $\varphi_2,\varphi,\langle a\rangle\psi$, then either $\varphi_1^{i_1},\varphi^{j}\models\langle a\rangle \psi^{k}$ for some $i_1,j,k$, or $\varphi_2^{i_2},\varphi^j\models\langle a\rangle \psi^k$ for some $i_2,j,k$. From the easy direction of Lemma~\ref{lem:conjunction_lemma_cs}, $\varphi_1^{i_1}\wedge\varphi_2^{i_2},\varphi^j\models\langle a\rangle \psi^k$, where $\varphi_1^{i_1}\wedge\varphi_2^{i_2}=\varphi_{12}^i$ for some $1\leq i\leq k_{12}$.
\end{description}
As a result, $P_1$ is true for $\varphi^\diamond,\varphi^\diamond,\varphi^\diamond$, and from Corollary~\ref{cor:primality-LCS-3}, $\varphi^\diamond$ is prime.\\
($\Rightarrow$) This implication can be proven similarly to Lemma~\ref{lem:sim-algo-correct-2}. We prove the parts that exhibit some differences from those in Lemma~\ref{lem:sim-algo-correct-2}.\\
\noindent\textbf{\textcolor{darkgray}{Claim 1(a).}} For every vertex $x=(\varphi_1,\varphi_2\Rightarrow \psi)$  in $G_{\varphi^{\diamond}}$ such that $\varphi_1,\varphi_2,\psi\neq\zero$ and $\varphi_1,\varphi_2,\psi$ satisfy $P_2$, one of the rules for complete simulation can be applied on $x$.\\
\noindent\textbf{\textcolor{darkgray}{Proof of Claim 1(a).}} If a rule cannot be applied on $x$, then it must be the case that:
\begin{itemize}
    \item either $\varphi_1=\langle a\rangle \varphi_1'$, $\varphi_2=\langle b\rangle \varphi_2'$, and $\psi=\langle c\rangle \psi'$, where $a=b=c$ is not true, which leads to contradiction as we have already proven in the proof of Lemma~\ref{lem:sim-algo-correct-2},
    \item or all of $\varphi_1,\varphi_2,\psi$ are $\langle a\rangle \varphi$ or $\zero$, and there is at least one of each kind. For example, if $\varphi_1=\langle a\rangle \varphi_1'$, $\varphi_2=\langle a\rangle \varphi_2'$, and $\psi=\zero$, then $P_2$ does not hold for $\varphi_1,\varphi_2,\psi$, contradiction. All other cases can be proven similarly.
\end{itemize}
\noindent\textbf{\textcolor{darkgray}{Claim 1(b).}} If an existential rule is applied on $x=(\varphi_1,\varphi_2\Rightarrow \psi)\in V$, where $\varphi_1,\varphi_2,\psi\neq\zero$ and $\varphi_1,\varphi_2,\psi$ satisfy $P_2$, then there is some $z=(\varphi_1',\varphi_2'\Rightarrow \psi')\in V$ such that $(x,z)\in E$ and $\varphi_1',\varphi_2',\psi'$ satisfy $P_2$. \\
\noindent\textbf{\textcolor{darkgray}{Proof of Claim 1(b).}} All cases of the induction proof can be proven in the same manner as in Lemma~\ref{lem:sim-algo-correct-2}. In particular, consider (L$\wedge_i$). The hypothesis of Lemma~\ref{lem:conjunction_lemma_cs} holds for $\varphi_1\wedge\varphi_2$. So, Lemma~\ref{lem:conjunction_lemma_cs} can be used in place of Lemma~\ref{lem:conjunction_lemma_simulation}.\\
\noindent\textbf{\textcolor{darkgray}{Claim 2.}} If $x$ is a vertex $(\varphi_1,\varphi_2\Rightarrow\psi)$ in $G_{\varphi^{\diamond}}$ such that $\varphi_1,\varphi_2,\psi$ satisfy $P_2$, then there is an alternating path from $x$ to \textsc{True}.\\
\noindent\textbf{\textcolor{darkgray}{Proof of Claim 2.}} All cases of the induction proof are the same except for the case $x=(\zero,\zero\Rightarrow\zero)$. Then, $P^G(x,\textsc{True})$ trivially holds.\\
Claim 1(c) needs no adjustment. As $\phi^{\diamond}$ is prime, from Corollary~\ref{cor:primality-phitt}, $\phi^{\diamond},\phi^{\diamond},\phi^{\diamond}$ satisfy $P_2$ and there is an alternating path from $s$ to $t$. This completes the proof of Claim A. 

The polynomial-time complexity of \algott relies on the polynomial size of $G_{\varphi^{\diamond}}$ and linear-time solvability of \reacha.
\end{proof}

\begin{rem}\label{rem:type-ordering-rules}
  At this point, we comment on the type of the rules and the ordering in which they are applied on a given formula $\varphi$ in this subsection. Note that formulae which are satisfied by $\zero$ have a simple zero normal form, i.e.\ their zero normal form is either $\zero$ or $\zero\vee\varphi'$, where $\zero\not\models\varphi'$. This is possible since we initially applied rules $\true\vee\psi\rulett \true$ and $\true\wedge\psi\rulett \psi$ and we obtained $\varphi^{tt}$, such that the zero normal form of every tautology in $\varphi^{tt}$ is also $\zero\vee\varphi'$, where $\zero\not\models\varphi'$. Next, we apply rules that result in the equivalent formula $\varphi^0$, which is in zero normal form. Formula $\varphi^0$ has a DNF $\bigvee_{i=1}^{k}\varphi_i^0$, where every $\varphi_i^0$ is satisfiable as shown in Lemma~\ref{lem:cs-satisfiable-dnf}, which is a crucial property in the proof of Lemma~\ref{lem:phitt-vs-phi}. A formula that is not in zero normal form can have a DNF where some disjuncts are unsatisfiable. For example, the DNF of $\psi=(\langle a\rangle\true\vee\langle b\rangle \true)\wedge(\zero\vee\langle a\rangle \zero)$ is $(\langle a\rangle\true\wedge\zero)\vee(\langle a\rangle\true\wedge\langle a\rangle\zero)\vee(\langle b\rangle \true\wedge\zero)\vee(\langle b\rangle \true\wedge\langle a\rangle\zero)$, where $\langle a\rangle\true\wedge\zero$ and $\langle b\rangle\true\wedge\zero$ are unsatisfiable. However, the zero normal form of $\psi$ is $\psi^0=\zero\vee(\langle a\rangle \zero\wedge\langle a\rangle\true\wedge\langle b\rangle\true)$ which is in DNF and every disjunct is satisfiable. Moreover, Lemma~\ref{lem:conjunction_lemma_cs}, which is necessary for proving our main result, does not hold for formulae that are not in zero normal form. For instance, $(\zero\vee\langle a\rangle\zero)\wedge(\langle a\rangle\true\vee\langle b\rangle\true)\models\langle a\rangle\zero$, but $\zero\vee\langle a\rangle\zero\not\models\langle a\rangle\zero$ and $\langle a\rangle\true\vee\langle b\rangle\true\not \models\langle a\rangle\zero$. Finally, we apply rules to obtain $\varphi^\diamond$, which, in the case that is not $\true$, does not contain $\true$, so it has a DNF the disjuncts of which are satisfiable and prime. As a result, $\varphi^\diamond$ satisfies various desired properties that allow us to use a variant of \algos that checks primality of $\varphi^\diamond$.
\end{rem}

\begin{prop}\label{prop:find-p-cs-algo-phitt}
    Let $\varphi\in\mathL_{CS}$ be a formula such that every $\psi\in\sub(\varphi)$ is satisfiable; let also  $\varphi\rulettstar\varphi^{tt}\rulezerostar\varphi^0\rulediamstar\varphi^\diamond$. If $\varphi^{\diamond}$ is prime, there is a polynomial-time algorithm that constructs a process for which  $\varphi^{\diamond}$ is characteristic within $\mathL_{CS}$.
\end{prop}
\begin{proof}
    As in the case of simulation and the proof of Corollary~\ref{cor:find-p-algo}, there is an algorithm that finds an alternating path $\mathP_a$ in $G_{\varphi^\diamond}$ from $s$ to $t$ and associates a process to every vertex of $\mathP_a$ so that the process associated to $s$ is a process for which $\varphi$ is characteristic within $\mathL_{CS}$. 
\end{proof}

\csprime*
\begin{proof}
    We describe algorithm \algocs that decides whether $\varphi$ is prime in polynomial time. On input $\varphi$, \algocs computes $\varphi^{tt},\varphi^0,$ and $\varphi^\diamond$ such that $\varphi\rulettstar\varphi^{tt}\rulettstar\varphi^{0}\rulediamstar\varphi^\diamond$. Then, it checks whether $\varphi^{\diamond}$ is prime by calling \algott($\varphi^{\diamond}$). If $\varphi^{\diamond}$ is not prime, \algocs rejects. Otherwise, \algocs computes $p$ for which $\varphi^{\diamond}$ is characteristic within $\mathL_{CS}$. Finally, it checks whether $p\models\varphi$. \algocs accepts iff $p\models\varphi$.

     If $p\models\varphi$, from Lemma~\ref{def:characteristic}, for every $q$ such that $q\models\varphi^{\diamond}$, it holds that $q\models\varphi$. Thus, $\varphi^{\diamond}\models\varphi$. Correctness of \algocs now follows immediately from Corollary~\ref{cor:phitt-vs-phi}. The polynomial-time complexity of the algorithm is a corollary of Lemmas~\ref{lem:phitt-poly-time-complexity}, \ref{lem:phizero-poly-time-complexity}, and~\ref{lem:cs_phi_diamond_poly_time}, which state that $\varphi^{tt}$, $\varphi^0$, and $\varphi^\diamond$, respectively, can be computed in polynomial time, Proposition~\ref{prop:phitt-primality} that shows polynomial-time complexity of \algott, and Propositions~\ref{prop:find-p-cs-algo-phitt} and~\ref{model-checking-complexity}, which demonstrate that $p$ can be efficiently computed and $p\models\varphi$ can be also solved in polynomial time, respectively.
\end{proof}

\cschar*
\begin{proof}
Let $\varphi\in\mathL_{CS}$. Consider the algorithm $\mathrm{Char}_{CS}$ that on input $\varphi$ proceeds as follows: it checks whether $\varphi$ is satisfiable by calling $\conscs(\varphi)$. If $\varphi$ is unsatisfiable, $\mathrm{Char}_{CS}$  rejects $\varphi$. Otherwise, it calls $\subcs(\varphi)$ to compute $\varphi'$ which is logically equivalent to $\varphi$ and contains no unsatisfiable subformulae. Then, it calls \algocs on input $\varphi'$ to decide whether $\varphi'$ is prime. It accepts iff $\algocs(\varphi')$ accepts.
The correctness and the polynomial-time complexity of $\mathrm{Char}_{CS}$ is immediate from Propositions~\ref{prop:charact-via-primality}, \ref{cor:cs-satisfiability}, and~\ref{prop:cs-satisfiability-satisfiable}, and~\ref{prop:cs-primality}.
\end{proof}

\begin{cor}\label{cor:find-p-cs-algo}
     Let $\varphi\in\mathL_{CS}$. If $\varphi$ is satisfiable and prime, then there is a polynomial-time algorithm that constructs a process for which  $\varphi$ is characteristic within $\mathL_{CS}$.
\end{cor}

\section{The formula primality problem for \texorpdfstring{$\mathL_{RS}$}{LRS} with a bounded action set}\label{subsection:rs-primality-bounded-appendix}

The following proposition allows us to consider specific formulae in $\mathL_{RS}$.

\begin{prop}\label{prop:rs-satisfiability-satisfiable}
Let $|\act|=k$, where $k\geq 1$ is a constant. There is a polynomial-time algorithm that on input a satisfiable $\varphi\in\mathL_{RS}$, it returns $\varphi'$ such that (a) $\varphi\equiv\varphi'$, and (b) if  $\psi\in\sub(\varphi')$ is unsatisfiable, then $\psi=\ff$ and occurs in $\varphi'$ in the scope of some $[a]$.
\end{prop}
\begin{proof}
    Let $\varphi\in\mathL_{RS}$ be a satisfiable formula. Consider algorithm \subrs that computes $I(\psi)$ for every $\psi\in\sub(\varphi)$ and stores  $I(\psi)$ in memory. For every $\psi\in\sub(\varphi)$ such that $I(\psi)=\emptyset$, \subrs substitutes $\psi$ with $\ff$ in $\varphi$. We denote by $\varphi^{\mathrm{ff}}$ the obtained formula. Then, \subrs repeatedly applies the rules $\langle a\rangle\ff\ruleff \ff$, $\ff\vee\psi\ruleff\psi$, $\psi\vee\ff\ruleff\psi$, $\ff\wedge\psi\ruleff \ff$, and $\psi\wedge\ff\ruleff \ff$ on $\varphi^{\mathrm{ff}}$ until no rule can be applied, and returns the resulting formula, which we denote by $\varphi'$. Since every substitution has replaced a formula $\psi$ with some $\psi'\equiv\psi$, from Lemma~\ref{lem:substitution_bs}, $\varphi\equiv\varphi'$. From the type of the substitutions made on $\varphi$, every unsatisfiable formula has been substituted with $\ff$, and all occurrences of $\ff$ have been eliminated except for the ones that are in the scope of some $[a]$. Moreover, it is not hard to see that the algorithm requires polynomial time.
\end{proof}

We first introduce the notion of \emph{saturated formulae}, which intuitively captures the following  property: if a saturated formula $\varphi$ is satisfied in $p$, then $\varphi$ describes exactly which actions label the outgoing edges of $p$.

\begin{defi}\label{def:saturated-fromula}
    Let $\varphi\in\mathL_{RS}$ such that if $\psi\in\sub(\varphi)$ is unsatisfiable, then $\psi=\ff$ and occurs in the scope of some $[a]$.
    We say that $\varphi$ is saturated if $I(\varphi)$ is a singleton.
\end{defi}

\begin{rem}\label{rem:saturated}
    From now on, when we say that $\varphi$ is saturated, we imply that if $\psi\in\sub(\varphi)$ is unsatisfiable, then $\psi=\ff$ and occurs in the scope of some $[a]$.
\end{rem}

\begin{lem}\label{lem:rs-not-saturated-property}
    Let $\varphi\in\mathL_{RS}$ be such that if $\psi\in\sub(\varphi)$ is unsatisfiable, then $\psi=\ff$ and it occurs within the scope of some $[a]$. If $\varphi$ is not saturated, there are two processes $p_1$ and $p_2$, such that $p_i\models\varphi$ for both $i=1,2$ and $I(p_1)\neq I(p_2)$.
\end{lem}
\begin{proof}
    From the assumptions of the lemma, Definition~\ref{def:saturated-fromula}, and Lemma~\ref{lem:I(phi)-property}, $|I(\varphi)|>1$. Let $S_1,S_2\in I(\varphi)$, such that $S_1\neq S_2$. From Lemma~\ref{lem:I(phi)-property}, there are $p_1$, $p_2$, such that $I(p_i)=S_i$ and $p_i\models\varphi$, where $i\in\{1,2\}$.
\end{proof}

\begin{cor}\label{cor:rs-not-saturated-property}
    Let $\varphi\in\mathL_{RS}$ be such that if $\psi\in\sub(\varphi)$ is unsatisfiable, then $\psi=\ff$ and it occurs within the scope of some $[a]$. If $\varphi$ is prime, then $\varphi$ is saturated.
\end{cor}
\begin{proof}
    Suppose that $\varphi$ is satisfiable, prime, and not saturated. Let $p$ be a process for which $\varphi$ is characteristic within $\mathL_{RS}$. Consider two processes $p_1$, $p_2$ that satisfy $\varphi$ and $I(p_1)\neq I(p_2)$, whose existence is guaranteed by Lemma~\ref{lem:rs-not-saturated-property}. Then, $p\curle_{RS} p_1$ and $p\curle_{RS} p_2$, which contradicts $I(p_1)\neq I(p_2)$.
\end{proof}

Note that a saturated formula $\varphi$ might not describe exactly the labels of the outgoing edges of processes reachable from $p$, where $p\models\varphi$. To focus on this first level of edges that start from $p$, we construct a propositional formula corresponding to $\varphi$, where any formula of the form $\langle a\rangle \varphi'$ that requires an edge labelled with $a$ leaving from $p$ corresponds to a propositional variable $x_a$.

\begin{defi}\label{def:rs-propositional-formula}
    Let $\varphi\in\mathL_{RS}$ be such that if $\psi\in\sub(\varphi)$ is unsatisfiable, then $\psi=\ff$ and it occurs within the scope of some $[a]$. The mapping $\sat(\varphi):\mathL_{RS}\rightarrow\mathL_{prop}$, where $\mathL_{prop}$ is the set of propositional formulae, is inductively defined as follows:
    \begin{itemize}
        \item $\sat(\true)=\mathtt{TRUE}$,
        \item $\sat([a]\ff)=\neg x_a$, 
        \item $\sat(\langle a\rangle\varphi')= x_a$,
        \item $\sat(\varphi_1\wedge\varphi_2)= \sat(\varphi_1)\wedge\sat(\varphi_2)$,
        \item $\sat(\varphi_1\vee\varphi_2)= \sat(\varphi_1)\vee\sat(\varphi_2)$,
    \end{itemize}
    where $\mathtt{TRUE}$ denotes a propositional tautology.
\end{defi}

\begin{rem}\label{rem:rs-propositional-formula}
When we construct $\sat(\varphi)$, if $\varphi=\langle a\rangle \varphi'$, we can attach  $\varphi'$ as a label to $x_a$ by setting $\sat(\langle a\rangle\varphi')=x_a^{\varphi'}$, where $\varphi'$ acts as a label for this occurrence of $x_a$. Then, given a propositional formula $\psi$ over the set of variables $\mathrm{VAR}_k=\{x_{a_1},\dots,x_{a_k}\}$, together with labels for each positive occurrence of the variables, we can construct the formula in $\mathL_{RS}$ that corresponds to $\psi$. 
\end{rem}

We prove below that, in the case of a saturated formula $\varphi$, $\sat(\varphi)$ has a unique satisfying assignment $s$ and $I(\varphi)$ determines $s$.

\begin{lem}\label{lem:rs-propositional-formula-sat-assignment}
   Let $\varphi\in\mathL_{RS}$ be such that if $\psi\in\sub(\varphi)$ is unsatisfiable, then $\psi=\ff$ and it occurs within the scope of some $[a]$. If  $p$ is a process such that $p\models\varphi$, then the truth assignment $s:\mathrm{VAR}_k\rightarrow\{\mathrm{true,false}\}$ such that $s(x_a)=\mathrm{true}$ iff $p\myarrowa p'$ for some $p'$ satisfies 
   %is satisfying for 
   $\sat(\varphi)$. Conversely, if $t$ is a satisfying truth assignment for $\sat(\varphi)$, then there is a process $p$ such that $p\models\varphi$ and, for each action $a$, there is some $p'$ such that $p\myarrowa p'$  iff $t(x_a)=\mathrm{true}$.
\end{lem}
\begin{proof}
    The proof is by induction on the structure of $\varphi$.
 \begin{itemize}
     \item If either $\varphi=\true$  or $\varphi=[a]\ff$, then the lemma follows easily. %can be easily proven.
     \item Let $\varphi=\langle a\rangle\varphi'$ and assume that $p\models\varphi$. This means that $p\myarrowa p'$, for some $p'\models\varphi'$. 
     %, and perhaps there is  $b\neq a$ such that $p\myarrowb p''$. 
     The assignment $s$ defined in the proviso of the first claim in the lemma is such that $\sat(\varphi)=x_a$ and therefore %is satisfiable for 
     satisfies $\sat(\varphi)$. Conversely, if there is a satisfying truth assignment $t$ for $\sat(\varphi)=x_a$, then %$\sat(\varphi)=x_a$, and so 
     $t(x_a)=\mathrm{true}$. Since $\varphi'$  is satisfiable by the proviso of the lemma, there is a process $p'$ such that $p'\models\varphi'$. Consider the process $p$ with transitions $p\myarrowa p'$ and $p\myarrowb \mathtt{0}$ for every $x_b\neq x_a$ such that $t(x_b)=\mathrm{true}$. By construction,  $p\models\varphi$.
     \item Let  $\varphi=\varphi_1\wedge\varphi_2$ and assume that $p\models\varphi$. Since $p\models\varphi_1$ and $p\models\varphi_2$, the inductive hypothesis yields that the assignment $s$ defined in proviso of this claim in the lemma satisfies both $\sat(\varphi_1)$ and $\sat(\varphi_2)$. So, $s$ also satisfies $\sat(\varphi_1)\wedge\sat(\varphi_2)=\sat(\varphi_1\wedge\varphi_2)$. Conversely, assume that we have a satisfying assignment $t$ for $\sat(\varphi_1\wedge\varphi_2)=\sat(\varphi_1)\wedge\sat(\varphi_2)$. So, $t$ satisfies both $\sat(\varphi_1)$ and $\sat(\varphi_2)$. By the inductive hypothesis, there are $p_1$ and $p_2$ such that $p_1\models\varphi_1$ and $p_2\models \varphi_2$, respectively, and $p_1\myarrowa p_1'$ iff $p_2\myarrowa p_2'$ iff $t(x_a)=\mathrm{true}$, which imply that $I(p_1)=I(p_2)$. Consider now the process $p_1+p_2$. It holds that $p_i\curle_{RS}p_1+p_2$ for both $i=1,2$, and so we have that $p_1+p_2\models \varphi_1\wedge\varphi_2$.
     \item Assume that  $\varphi=\varphi_1\vee\varphi_2$ and let $p\models\varphi$. Since $p\models\varphi_1$ or $p\models\varphi_2$, the assignment $s$ defined in the proviso of this claim in the lemma is satisfying for  $\sat(\varphi_1)$ or $\sat(\varphi_2)$ from the inductive hypothesis. So, $s$ is also satisfying for $\sat(\varphi_1)\vee\sat(\varphi_2)=\sat(\varphi_1\vee\varphi_2)$. Conversely, let $t$ be a satisfying assignment for $\sat(\varphi_1\vee\varphi_2)=\sat(\varphi_1)\vee\sat(\varphi_2)$. Assume w.l.o.g.\ that $t$ is satisfying for $\sat(\varphi_1)$. By the inductive hypothesis, there is some process $p_1$ such that $p_1\models \varphi_1$ and, for every action $a$, there is some $p'$ such that $p_1\myarrowa p'$ iff $t(x_a)=\mathrm{true}$. Since  $p_1\models\varphi$ also holds, we are done.\qedhere
 \end{itemize}
\end{proof}

\begin{cor}\label{cor:rs-propositional-formula-saturated-sat-assignment}
   Let $\varphi\in\mathL_{RS}$ be saturated and $I(\varphi)=\{S\}$. Then, $s:\mathrm{VAR}_k\rightarrow\{\mathrm{true,false}\}$ is a satisfying truth assignment for $\sat(\varphi)$ iff $s(x_a)=\mathrm{true}\Longleftrightarrow a\in S$.
\end{cor}
\begin{proof}
$(\Rightarrow)$ Let $s$ be a satisfying assignment for $\sat(\varphi)$. Then, from Lemma~\ref{lem:rs-propositional-formula-sat-assignment}, there is $p$ such that $p\models\varphi$ and $a\in I(p)\Longleftrightarrow s(x_a)=\mathrm{true}$. From Lemma~\ref{lem:I(phi)-property}, $I(p)=S$ and so $a\in S\Longleftrightarrow s(x_a)=\mathrm{true}$.\\
$(\Leftarrow)$ Let $s$ be a truth assignment such that $s(x_a)=\mathrm{true}\Longleftrightarrow a\in S$. Then, there is  $p$ such that $p\models\varphi$, which means that $I(p)=S$ from Lemma~\ref{lem:I(phi)-property}.  Thus, $s(x_a)=\mathrm{true}\Longleftrightarrow a\in I(p)$. From Lemma~\ref{lem:rs-propositional-formula-sat-assignment}, $s$ is satisfying for $\sat(\varphi)$.
\end{proof}

Next, if $\varphi$ is saturated, we can simplify $\sat(\varphi)$---and consequently, $\varphi$ as well---so that the resulting propositional formula has a DNF with only satisfiable disjuncts. 

\begin{defi}\label{def:rs-simplification}
    Let $\varphi\in\mathL_{RS}$ be saturated, $I(\varphi)=\{S\}$, and $\psi=\sat(\varphi)$. We denote by $\simpl(\psi)$ the formula we obtain by making the following substitutions in $\psi$:
    \begin{itemize}
        \item for every $a\in S$, substitute $\neg x_a$ with $\mathtt{FALSE}$ in $\psi$,
        \item for every $a\not\in S$, substitute $x_a$ with $\mathtt{FALSE}$ in $\psi$, and
        \item apply rules $\mathtt{FALSE}\vee\psi\rightarrow \psi$ and $\mathtt{FALSE}\wedge\psi\rightarrow\mathtt{FALSE}$ modulo commutativity,
    \end{itemize}
    where $\mathtt{FALSE}$ denotes a propositional contradiction. 
    %We denote by $\simpl(\psi)$ the result of simplifying $\psi$ and by $\simpl(\varphi)$ the formula in $\mathL_{RS}$ that corresponds to $\simpl(\psi)$. We also say that a formula in $\varphi\in\mathL_{RS}$ is simplified if $\varphi=\simpl(\varphi)$.
\end{defi}

\begin{lem}\label{lem:rs-simplification}
    Let $\varphi\in\mathL_{RS}$ be saturated, $I(\varphi)=\{S\}$, and $\psi=\sat(\varphi)$. Then, $\simpl(\psi)$ is logically equivalent to $\psi$. Moreover, $\mathtt{FASLE}$ does not occur in $\simpl(\psi)$, if $a\in S$, then $\neg x_a$ does not occur in $\simpl(\psi)$, and if $a\not\in S$, then $x_a$ does not occur in $\simpl(\psi)$.
\end{lem}
\begin{proof}
    Let $I(\varphi)=\{S\}$. From Corollary~\ref{cor:rs-propositional-formula-saturated-sat-assignment}, $\psi$ has a unique assignment $s$ and it holds that $s(x_a)=\mathrm{true}\Longleftrightarrow a\in S$. So, if any of the substitutions presented in Definition~\ref{def:rs-simplification}, is made on $\psi$, we obtain a formula that is only satisfied by  truth assignment $s$. It is easy to see that the second part of the lemma is also true.
\end{proof}

\begin{lem}\label{lem:rs-simplification-dnf}
    Let $\varphi\in\mathL_{RS}$ be saturated, $I(\varphi)=\{S\}$, and $\psi=\sat(\varphi)$. If $\bigvee_{i=1}^k\psi_i$ denotes the DNF of $\simpl(\psi)$, then $\psi_i$ is satisfiable for every $1\leq i\leq k$.
\end{lem}
\begin{proof}
Let $s:\mathrm{VAR}_k\rightarrow\{\mathrm{true,false}\}$ be such that $s(x_a)=\mathrm{true}$ iff $a\in S$. From Corollary~\ref{cor:rs-propositional-formula-saturated-sat-assignment} and Lemma~\ref{lem:rs-simplification}, $s$ is the only satisfying assignment for $\simpl(\psi)$. The lemma is immediate from the following two claims, which we prove below.
\begin{description}
    \item[Claim 1.] Let $\psi'\in\sub(\psi)$. Then, $s$ is satisfying for $\psi'$.
    \item[Claim 2.] Let $\phi$ be a propositional formula and $\bigvee_{i=1}^n\phi_i$ denote the DNF of $\phi$. If there is a truth assignment $t$ such that is satisfying for every subformula of $\phi$, then $t$ is satisfying for $\phi_i$, for every $1\leq i\leq n$.
\end{description}
\noindent\textbf{\textcolor{darkgray}{Proof of Claim 1.}}  If $s(x_a)=\mathrm{false}$, then $a\not\in S$, and from Lemma~\ref{lem:rs-simplification}, $x_a$ does not appear in $\psi'$. Analogously, if $s(x_a)=\mathrm{true}$, $\neg x_a$ does not appear in $\psi'$. Hence, $s$ assigns the false value only to literals that do not appear in $\psi'$, and so $s$ is satisfying for $\psi'$.\\
\noindent\textbf{\textcolor{darkgray}{Proof of Claim 2.}} We prove the claim by structural induction on $\phi$.
\begin{itemize}
    \item If $\phi$ does not contain disjunctions, then  $\phi$ is already in DNF, and $t$ is satisfying for $\phi$.
    \item If $\phi=\phi_1\vee\phi_2$, then  $t$ is satisfying for  every subformula of $\phi_i$, where $i=1,2$. Let $\bigvee_{i=1}^{k_1}\phi_1^i$ and $\bigvee_{i=1}^{k_2}\phi_2^i$ denote the DNFs of $\phi_1$ and $\phi_2$, respectively. From inductive hypothesis, $t$ is satisfying for $\phi_j^i$, for every $j=1,2$ and $1\leq i\leq k_j$. This is sufficient since $\bigvee_{i=1}^n\phi_i=\bigvee_{i=1}^{k_1}\phi_1^i\vee\bigvee_{i=1}^{k_2}\phi_2^i$.
    \item If $\phi=\phi_1\wedge\phi_2$, then  $t$ is satisfying for  every subformula of $\phi_i$, where $i=1,2$. Let $\bigvee_{i=1}^{k_1}\phi_1^i$ and $\bigvee_{i=1}^{k_2}\phi_2^i$ denote the DNFs of $\phi_1$ and $\phi_2$, respectively.  It holds that for every $1\leq i\leq k$, $\phi_i=\phi_1^{i_1}\wedge \phi_2^{i_2}$ for some $1\leq i_1\leq k_1$ and $1\leq i_2\leq k_2$. Since from inductive hypothesis, $t$ is satisfying for $\phi_j^i$, for every $j=1,2$ and $1\leq i\leq k_j$, $t$ is satisfying for $\phi_i$, for every $1\leq i\leq k$.\qedhere
\end{itemize}
\end{proof}

Properties of the simplified version of $\sat(\varphi)$ can be transferred to $\varphi$.

\begin{defi}\label{def:rs-saturated-simplified}
    Let $\varphi\in\mathL_{RS}$ be saturated and $\psi=\sat(\varphi)$. We denote by $\simpl(\varphi)$ the formula in $\mathL_{RS}$ that corresponds to $\simpl(\psi)$ as described in Remark~\ref{rem:rs-propositional-formula}. We say that $\varphi$ is simplified if $\varphi=\simpl(\varphi)$.
\end{defi}

\begin{cor}\label{cor:rs-simplification}
    Let $\varphi\in\mathL_{RS}$ be saturated and $I(\varphi)=\{S\}$. Then, $\simpl(\varphi)\equiv\varphi$; if $a\in S$, then any $[a]\ff$ occurs in $\simpl(\varphi)$ only in the scope of some $\langle b\rangle$, $b\in \act$; and if $a\not\in S$, then any $\langle a\rangle\varphi'$, where $\varphi'\in\mathL_{RS}$, occurs in $\simpl(\varphi)$ only in the scope of some $\langle b\rangle$, $b\in \act$.
\end{cor}

\begin{cor}\label{cor:rs-simplification-dnf}
    Let $\varphi\in\mathL_{RS}$ be saturated, $I(\varphi)=\{S\}$, and $\psi=\sat(\varphi)$; let also  $\bigvee_{i=1}^k\psi_i$ denote the DNF of $\simpl(\psi)$ and $\bigvee_{i=1}^k\varphi_i$ denote the formula in $\mathL_{RS}$ that corresponds to $\bigvee_{i=1}^k\psi_i$  as described in Remark~\ref{rem:rs-propositional-formula}. Then, every $\varphi_i$ is satisfiable and disjunctions occur in $\varphi_i$ only in the scope of some $\langle a\rangle$, where $a\in \act$.
\end{cor}

Given a formula $\varphi\in\mathL_{RS}$ such that if $\psi\in\sub(\varphi)$ is unsatisfiable, then $\psi=\ff$ and occurs in the scope of some $[a]$, we process the formula by running Algorithm~\ref{algo:satur} on input $\varphi$. We show that the resulting formula, denoted by $\varphi^s$, has properties that allow us to use a variant of \algos to check its primality. In the case that $\varphi^s$ is prime and logically implies $\varphi$, then $\varphi$ is also prime. As the reader may have already notice, the strategy is similar to the case of complete simulation.

\begin{algorithm}
\caption{Saturation of a formula in $\mathL_{RS}$}
\begin{algorithmic}[1]
\DontPrintSemicolon
\Procedure{Satur}{$\varphi$}
  %  \State {$\varphi'\gets\varphi$\;} 
     \State \Repeat{$\varphi'=\varphi$}{
     {$\varphi'\gets\varphi$}\;
    {compute $\varphi^{tt}$ such that $\varphi\rulettstar\varphi^{tt}$\;
      $\varphi\gets\varphi^{tt}$}\;
       \lIf{$|I(\varphi)|\neq 1$}{$\varphi\gets\true$} % \Comment{Call Recursion again}
     \lElse{$\varphi\gets\simpl(\varphi)$}
      \ForAll{occurrences of $\langle a\rangle\psi$ in $\varphi$ not in the scope of some $\langle b\rangle$, $b\in \act$}
    {\lIf{$|I(\psi)|\neq 1$}{substitute $\langle a\rangle\psi$ with $\true$ in $\varphi$}
     \Else{$\psi\gets\simpl(\psi)$\;
           substitute $\langle a\rangle\psi$ with $\langle a \rangle\Call{Satur}{\psi}$ in $\varphi$}}
    }
     \State {$\varphi\gets\simpl(\varphi)$}
     \State {return $\varphi$\;}
\EndProcedure
\end{algorithmic}
\label{algo:satur}
\end{algorithm}

\begin{lem}\label{lem:satur-poly-time}
    Let $\varphi\in\mathL_{RS}$ be such that if $\psi\in\sub(\varphi)$ is unsatisfiable, then $\psi=\ff$ and occurs in the scope of some $[a]$. Then, $\Call{Satur}{\varphi}$ runs in polynomial time.
\end{lem}
\begin{proof}
  It is not hard to see that all steps of Algorithm~\ref{algo:satur} run in polynomial time. Specifically, $I(\varphi)$ can be computed in polynomial time since $|\act|=k$, where $k$ is a constant. 
\end{proof}

\begin{lem}\label{lem:saturated-algo}
    Let $\varphi\in\mathL_{RS}$ be such that if $\psi\in\sub(\varphi)$ is unsatisfiable, then $\psi=\ff$ and occurs in the scope of some $[a]$. Then, $\Call{Satur}{\varphi}$ returns either $\true$ or a saturated and simplified formula $\varphi^s$ such that $\true\not\in\sub(\varphi^s)$ and for every $\langle a\rangle \psi\in\sub(\varphi^s)$, $\psi$ is saturated and simplified. 
\end{lem}
\begin{proof}
    Immediate from the substitutions made by Algorithm~\ref{algo:satur}.
\end{proof}

\begin{lem}\label{lem:rs-non-saturated-substitution}
Let $\varphi\in\mathL_{RS}$ be prime, saturated, and simplified; let also $\langle a\rangle \varphi'\in\sub(\varphi)$ such that there is an occurrence of $\langle a\rangle \varphi'$ in $\varphi$, which is not in the scope of any $\langle b\rangle$, $b\in \act$, and $\varphi'$ is not saturated. If $\varphi^s$ denotes $\varphi$ where this occurrence of $\langle a\rangle \varphi'$ has been substituted with $\true$, then  $\varphi^s\models\varphi$.
\end{lem}
\begin{proof}
Consider the propositional formulae $\psi=\sat(\varphi)$ and $\psi^s=\sat(\varphi^s)$. Let $\bigvee_{i=1}^k\psi_i$ and $\bigvee_{i=1}^{k'} \psi^s_i$ denote the DNFs of $\psi$ and $\psi^s$, respectively. Formula $\psi$ satisfies the assumptions of Lemma~\ref{lem:rs-simplification-dnf} and so every $\psi_i$ is satisfiable. 
 %Moreover, $\psi$ has one satisfying assignment $s$ that makes every literal appearing in $\psi$ true. 
 %Note that for every $1\leq i\leq k'$, $\psi^s_i$ is either some $\psi_j$ for some $1\leq j\leq k$ or some $\psi_j$, where a variable $x_a$ has been substituted with $\mathtt{TRUE}$. So, $s$ is also satisfying for $\psi^s_i$
 Let $\bigvee_{i=1}^k\varphi_i$ and $\bigvee_{i=1}^{k'} \varphi^s_i$ denote the formulae in $\mathL_{RS}$ that correspond to $\bigvee_{i=1}^k\psi_i$ and $\bigvee_{i=1}^{k'} \psi^s_i$,  respectively, as described in Remark~\ref{rem:rs-propositional-formula}. From Corollary~\ref{cor:rs-simplification-dnf}, every $\varphi_i$, $1\leq i\leq k$, is satisfiable and disjunctions occur in $\varphi_i$ and $\varphi^s_j$, $1\leq j\leq k'$, only in the scope of some $\langle b\rangle$, $b\in \act$. Moreover, $\bigvee_{i=1}^k\varphi_i\equiv\varphi$ and $\bigvee_{i=1}^{k'} \varphi^s_i\equiv\varphi^s$. 

 Let $p\models \varphi^s$ and $p_{min}$ be a process for which $\varphi$ is characteristic within $\mathL_{RS}$. Such a $p_{min}$ exists since $\varphi$ is satisfiable and prime. We show that $p_{min}\curle_{RS} p$, which combined with  Corollary~\ref{cor:characteristic} implies that $p\models\varphi$. Note that $p\models \varphi^s_i$ for some $1\leq i\leq k'$. If $\varphi^s_i=\varphi_j$ for some $1\leq j\leq k$, then $p\models\varphi$. Otherwise, $\varphi^s_i$ coincides with some $\varphi_j$, where an occurrence of $\langle a\rangle \varphi'$ has been substituted with $\true$. We describe now how to construct a process $p_{\varphi^s_i}$ that satisfies $\varphi^s_i$. %W.l.o.g.\ $\varphi^s_i=\bigwedge_{j=1}^m\psi_j$, where $\psi_j$ is either $[a]\ff$, $\true$, or $\langle a\rangle\psi$.
 We define $p_{\varphi^s_i}$ to be such that for every $\langle a_i\rangle\varphi_{a_i}\in\sub(\varphi^s_i)$, such that $\langle a_i\rangle\varphi_{a_i}$ is not in the scope of some $\langle b\rangle$, $b\in \act$, it holds that $p\myarrowa p_{\varphi_{a_i}}$, where $p_{\varphi_{a_i}}$ is some process that satisfies $\varphi_{a_i}$, and $p_{\varphi^s_i}$ has no other outgoing edges. We also consider two copies of $p_{\varphi^s_i}$, namely $p_{\varphi^s_i}^1$ and $p_{\varphi^s_i}^2$, that are as follows:  $p_{\varphi^s_i}^1=p_{\varphi^s_i}+a.p_1$ is  and $p_{\varphi^s_i}^2=p_{\varphi^s_i}+a.p_2$, where $p_1\models\varphi'$ and $p_2\models\varphi'$ and $I(p_1)\neq I(p_2)$. Such processes exist because $\varphi'$ is not saturated and Lemma~\ref{lem:rs-not-saturated-property} holds. Note that $p_{\varphi^s_i}^j\models\varphi_j$ for both $j=1,2$, which means that $p_{\varphi^s_i}^j\models\varphi$ for both $j=1,2$, and $p_{min}\curle_{RS} p_{\varphi^s_i}^j$ for both $j=1,2$.
 Let  $p_{min}\myarrowa q$. Then, there are some $q_j$, $j=1,2$, such that $p_{\varphi^s_i}^j\myarrowa q_j$ and $q\curle_{RS} q_j$.  
            \begin{itemize}
                \item Assume that there is some $q$ such that $p_{min}\myarrowa q$ and $q\curle_{RS} p_1$. Thus, $I(q)=I(p_1)$. On the other hand, $q\curle_{RS} p_2$ does not hold, since $I(p_2)\neq I(p_1)=I(q)$. So there is $p'$ such that $p_{\varphi^{\diamond}_i}^2\myarrowa p'$, $p'\neq p_2$, and $q\curle_{RS} p'$. This means that $p'$ is a copy of a process $p''$ such that $p_{\varphi^s_i}\myarrowa p''$ and $q\curle_{RS} p''$.
                \item Assume that there is some $q$ such that $p_{min}\myarrowtau q$ and $q\curle_{RS} p'$, where $p' \neq p_1$. Simpler arguments can show that there is some $p''$ such that  $p_{\varphi^s_i}\myarrowtau p''$ and $q\curle_{RS} p''$.
            \end{itemize} 
            As a result, $p_{min}\curle_{RS} p_{\varphi^s_i}$. Any process that satisfies $\varphi^s_i$ has the form of $p_{\varphi^s_i}$, so this completes the proof of the lemma.
\end{proof}

\begin{lem}\label{lem:rs-non-saturated-substitution-2}
Let $\varphi\in\mathL_{RS}$ be prime, saturated, and simplified; let also $\langle a\rangle \varphi'\in\sub(\varphi)$ such that there is an occurrence of $\langle a\rangle \varphi'$ in $\varphi$ in the scope of some $\langle b\rangle$, $b\in \act$, and $\varphi'$ is not saturated. If $\varphi^s$ denotes $\varphi$ where this occurrence of $\langle a\rangle \varphi'$ has been substituted with $\true$, then  $\varphi^s\models\varphi$.
\end{lem}
\begin{proof}
 We provide a  proof sketch. Formally, the proof is by induction on the number of $\langle b\rangle$ in the scope of which $\langle a\rangle\varphi'$ occurs in $\varphi$. Let $\langle b\rangle \psi\in\sub(\varphi)$ be such that $\langle a\rangle\varphi'$ occurs in $\psi$ not in the scope of any $\langle c\rangle$, $c\in \act$.   From Corollary~\ref{cor:rs-simplification-dnf}, every saturated formula $\varphi$ corresponds to an equivalent $\bigvee_{i=1}^k\varphi_i$ with the properties outlined in the corollary. By combining all these formulae corresponding to  the saturated formulae examined before this occurrence of $\langle a\rangle\varphi'$ by procedure \textsc{Satur}, we can prove that there is $\bigvee_{i=1}^m\varphi_i\equiv\varphi$, every $\varphi_i$ is satisfiable, $\langle a\rangle \varphi'$ occurs in some $\varphi_i$'s and the structure of $\varphi_i$'s is such that a similar argument to the one in the proof of Lemma~\ref{lem:rs-non-saturated-substitution} works. 
\end{proof}

Next, we show one of the main results of this subsection. 

\begin{prop}\label{prop:primality-LRS-algo}
    Let $\varphi\in\mathL_{RS}$ be a formula such that  if $\psi\in\sub(\varphi)$ is unsatisfiable, then $\psi=\ff$ and occurs in the scope of some $[a]$; let also $\varphi^s$ denote the output of $\Call{Satur}{\varphi}$. Then, 
    $\varphi$ is prime iff $\varphi^s\models \varphi$ and $\varphi^s$ is prime.
\end{prop}
\begin{proof}
($\Leftarrow$)   Let $\varphi^s\models \varphi$ and $\varphi^s$ be prime. It holds that  $\langle a\rangle \psi \models\true$, $\true\wedge\psi\models\psi$, and $\true\vee\psi \models\true$, for every $\psi\in\mathL_{RS}$. Thus, from Lemma~\ref{lem:substitution_bs} and the type of substitutions made to compute $\varphi^s$, $\varphi\models\varphi^s$. As a result $\varphi\equiv\varphi^s$ and so $\varphi$ is prime.\\
($\Rightarrow$) Assume that $\varphi$ is prime. As we just showed,  $\varphi\models \varphi^s$. Thus, it suffices to show that $\varphi^s\models \varphi$. Then, we have that $\varphi\equiv\varphi^s$ and $\varphi^s$ is prime.  The proof of $\varphi^s\models \varphi$ is by induction on the type of substitutions made to derive $\varphi^s$ from $\varphi$. 
If either $\varphi^s=\varphi[\true\vee\psi/\true]$ or $\varphi^s=\varphi[\true\wedge\psi/\psi]$, then $\varphi^s\equiv\varphi$ as already shown in the proof of Lemma~\ref{lem:phitt-vs-phi}. If \textsc{Satur} is called on input a prime formula $\varphi$, then $\varphi$ is saturated from Corollary~\ref{cor:rs-not-saturated-property} and substitution in line 4 is not made. Moreover, every recursive call is on saturated formulae and again this type of substitution is not made.
The only interesting case is when $\langle a\rangle \psi$ occurs in a saturated subformula of $\varphi$ not in the scope of some $\langle b\rangle$ and $\psi$ is not saturated. Then, this occurrence of $\langle a\rangle \psi$ is substituted with $\true$ in $\varphi$. Then, Lemmas~\ref{lem:rs-non-saturated-substitution} and~\ref{lem:rs-non-saturated-substitution-2} show that $\varphi^s\models\varphi$.
\end{proof}

 If a formula $\varphi^s\neq\true$ is the output of $\Call{satur}{\varphi}$, where the only unsatisfiable subformulae that $\varphi\in\mathL_{RS}$ contains are occurrences of $\ff$ in the scope of some $[a]$, then $\varphi^s$ has properties shown in the following statements. To start with, a variant of Lemma~\ref{lem:conjunction_lemma_cs} holds for saturated formulae.

\begin{lem}\label{lem:conjunction_lemma_rs}
Let $\varphi\in\mathL_{RS}$ such that if $\psi\in\sub(\varphi)$ is unsatisfiable, then $\psi=\ff$ and occurs in the scope of some $[a]$; let also $\varphi^s$ be the output of $\Call{satur}{\varphi}$. For every $\varphi_1\wedge\varphi_2,\langle a\rangle\psi\in\sub(\varphi^s)$, the following are true: 
\begin{enumerate}[(a)]
    \item $\varphi_1\wedge\varphi_2\models\langle a\rangle\psi$ iff $\varphi_1\models\langle a\rangle\psi$ or $\varphi_2\models\langle a\rangle\psi$, and
    \item $\varphi_1\wedge\varphi_2\models[a]\ff$ iff $\varphi_1\models[a]\ff$ or $\varphi_2\models[a]\ff$.
\end{enumerate}
\end{lem}
\begin{proof} Note that in the case of $\varphi^s=\true$, the lemma is trivial. Assume that $\true\not\in\sub(\varphi^s)$. The direction from left to right is easy for both cases. We prove the converse direction for (a) and (b). Let $\langle a\rangle \varphi'\in\sub(\varphi)$ such that $\varphi_1\wedge\varphi_2$ occurs in $\varphi'$ not in the scope of some $\langle b\rangle$, where $b\in \act$. Since $\varphi'$ is saturated, it holds that $I(\varphi')=\{S\}$ for some $S\subseteq A$. \\
(a) ($\Rightarrow$) Since $\true\not\in\sub(\varphi^s)$, then $\psi$ is not a tautology. Let $\varphi_1\wedge\varphi_2\models\langle a\rangle\psi$. Assume that $\varphi_1\not\models\langle a\rangle\psi$, and let $p_1$ such that $p_1\models\varphi_1$ and $p_1\not\models\langle a\rangle \psi$ and $p_2$  be such that $p_2\models\varphi_2$. Then, we construct process $p_i'$ by modifying $p_i$,  so that $I(p_i')=S$ and $p_i'\models\varphi_i$, for $i=1,2$. First, we set $p_i'=p_i+\sum_{a\in S\setminus I(p_i)} a.q$, where $q$ is any process that does not satisfy $\psi$. Second, for every $a\in I(p_i)\setminus S$, we remove every transition $p_i'\myarrowa p'$ from $p_i'$. We argue that $p_i'\models\varphi_i$. If  $a\in S\setminus I(p_i)$, from Corollary~\ref{cor:rs-simplification} and the fact that $\varphi_1\wedge\varphi_2$ occurs in $\varphi'$ not in the scope of some $\langle b\rangle$, if $[a]\ff$  occurs in $\varphi_i$ then it occurs in the scope of some $\langle b\rangle$, $b\in \act$. So, $p_i+a.q\models\varphi_i$. Similarly, if $a\in I(p_i)\setminus S$, then any $\langle a\rangle\psi$ can  occur in $\varphi_i$ only in the scope of some $\langle b\rangle$, $b\in \act$. Therefore, if $p_i'\models \varphi_i$ and we remove all transitions $p_i'\myarrowa p'$, the resulting process still satisfies $\varphi_i$. We now consider process $p_1'+p_2'$. As $I(p_1'+p_2')=I(p_1')=I(p_2')$, we have that $p_i'\curle_{RS} p_1'+p_2'$, and so $p_1'+p_2'\models\varphi_1\wedge\varphi_2$, which in turn implies that $p_1'+p_2'\models\langle a\rangle\psi$. This means that for some $i=1,2$, $p_i'\myarrowa p'$ such that $p'\models\psi$. Since $p_1\not\models\langle a\rangle \psi$, if $p_1'\myarrowa p''$, then $p_1\myarrowa p''$ and $p''\not\models \psi$, or $p''=q$ and $q\not\models\psi$. So, $p_2\models\langle a\rangle\psi$.\\
(b) ($\Rightarrow$) The proof is along the lines of the previous proof. Assume that $\varphi_1\not\models[a]\psi$, and let $p_1$ such that $p_1\models\varphi_1$ and $p_1\not\models[a] \psi$. Suppose that the same holds for $p_2$ and let $p_2$  such that $p_2\models\varphi_2$ and $p_2\not\models[a] \psi$. So, both $p_1$ and $p_2$ have an $a$-transition. As in the case of (a), we can construct $p_i'$ by removing and adding transitions. In this case,  for every $b\in S\setminus I(p_i)$, we add transitions $b.q$ as above, where $q$ can be any process, and we remove all transitions $p_i\myarrowb p'$ for every $b\in I(p_i)\setminus S$ except for the $a$-transitions. At the end, $p_i'\models\varphi_i$ and $I(p_i)=S\cup\{a\}$. As in the proof of (a), $p_1'\wedge p_2'\models\varphi_1\wedge\varphi_2$ and so $p_1'\wedge p_2'\models [a]\ff$. But  $I(p_1)=I(p_2)=S\cup\{a\}$, contradiction. As a result, $\varphi_2\models [a]\ff$.
\end{proof}

If  $\varphi^s\neq\true$ and it does not contain disjunctions, then $\varphi^s$ is characteristic within $\mathL_{RS}$.

\begin{lem}\label{lem:rs-associated-process}
    Let $\varphi\in\mathL_{RS}$ be a formula given by the grammar $\varphi::=\varphi\wedge\varphi~\mid ~\langle a\rangle\varphi~\mid ~[a]\ff$. If $\varphi$ is saturated and for every $\langle a\rangle \psi\in\sub(\varphi)$, $\psi$ is saturated, then $\varphi$ is prime and a process $p_\varphi$ for which $\varphi$ is characteristic within $\mathL_{RS}$ can be constructed in polynomial time.
\end{lem}
\begin{proof}
   Since a saturated formula is satisfiable, $\varphi$ is prime iff $\varphi$ is characteristic within $\mathL_{RS}$. We describe a polynomial-time recursive algorithm that constructs $p_\varphi$ and we prove that $\varphi$ is characteristic for $p_\varphi$. Since $\varphi$ is saturated, $I(\varphi)=\{S\}$, for some $S\subseteq A$. Moreover, w.l.o.g.\ $\varphi=\bigwedge_{i=1}^k\varphi_i$, where $k\geq 1$, every $\varphi_i$ is of the form $\langle a\rangle\varphi'$ or $[a]\ff$, and $\varphi'$ is given by the same grammar as $\varphi$. We construct $p_\varphi$ such that for every $\varphi_i=\langle a\rangle\varphi'$, $p_\varphi\myarrowa p_{\varphi'}$, where  $p_{\varphi'}$ is constructed recursively, and $p_{\varphi}$ has no other outgoing edge. First, we show that $p_\varphi\models\varphi$.  Let $\varphi_i=\langle a\rangle\varphi'$. Then, $p_\varphi\myarrowa p_{\varphi'}$ and from inductive hypothesis, $p_{\varphi'}\models\varphi'$. So, $p_\varphi\models\varphi_i$. If $\varphi_i=[a]\ff$, then there is no $\varphi_i$ such that $\varphi_i=\langle a\rangle\varphi'$, since $\varphi$ is satisfiable. So $p_\varphi\notmyarrowa$ and $p_\varphi\models \varphi_i$. As a result, $p_\varphi\models \varphi_i$ for every $1\leq i\leq k$. To prove that $\varphi$ is characteristic for $p_\varphi$, we show that for every $p$, $p\models\varphi$ iff $p_\varphi\curle_{RS} p$. Let $p\models\varphi$. Since $p_\varphi\models\varphi$ is also true, from Lemma~\ref{lem:I(phi)-property}, $I(p)=I(p_\varphi)=S$. If $p_\varphi\myarrowa p'$, then from construction, $p'=p_{\varphi'}$, where $\varphi_i=\langle a\rangle\varphi'$ for some $1\leq i\leq k$, and as we showed above, $p_{\varphi'}\models\varphi'$. So,  $p\models\varphi_i$ and $p\myarrowa p''$ such that $p''\models\varphi'$. From inductive hypothesis, $p_{\varphi'}\curle_{RS} p''$. So $p_\varphi\curle_{RS} p$. Conversely, assume that $p_\varphi\curle_{RS} p$. Let $\varphi_i=\langle a\rangle \varphi'$. We have that $p_\varphi\myarrowa p_{\varphi'}$ and $p_{\varphi'}\models\varphi'$. Hence, $p\myarrowa p''$ such that $p_{\varphi'}\curle_{RS} p''$, and so $p''\models\varphi'$. So, $p\models\langle a\rangle \varphi'$. Let $\varphi_i=[a]\ff$. Then, $p_\varphi\notmyarrowa$, and so $p\notmyarrowa$, which implies that $p\models [a]\ff$. So, $p\models\varphi_i$, for every $1\leq i\leq k$.
\end{proof}

\begin{lem}\label{lem:rs-dnf-disjuncts}
    Let $\varphi\in\mathL_{RS}$ be a saturated formula given by the grammar $\varphi::=\varphi\wedge\varphi~\mid~\varphi\vee\varphi~\mid~\langle a\rangle\varphi~\mid~[a]\ff$ such that for every $\langle a\rangle\varphi'\in\sub(\varphi)$, $\varphi'$ is saturated; let also $\bigvee_{i=1}^k\varphi_i$ be the DNF of $\varphi$. Then, $\varphi_i$ is saturated and for every $\langle a\rangle\varphi'\in\sub(\varphi_i)$, $\varphi'$ is saturated.
\end{lem}
\begin{proof} 
From Lemma~\ref{lem:DNF-equiv}, $\varphi\equiv\bigvee_{i=1}^k\varphi_i$, and so $I(\varphi)=\bigcup_{i=1}^k I(\varphi_i)$. Consequently, $I(\varphi_i)=I(\varphi)$ for every $1\leq i\leq k$, which implies that $I(\varphi_i)$ is a singeton and $\varphi_i$ is saturated. If $\langle a\rangle \varphi'\in\sub(\varphi_i)$ for some $1\leq i\leq k$, then $\langle a\rangle \varphi'\in\sub(\varphi)$ and so $\varphi'$ is saturated.
\end{proof}

 Let $\varphi\in\mathL_{RS}$ be a saturated formula given by the grammar $\varphi::=\varphi\wedge\varphi~\mid~\varphi\vee\varphi~\mid~\langle a\rangle\varphi~\mid~[a]\ff$ such that for every $\langle a\rangle\varphi'\in\sub(\varphi)$, $\varphi'$ is saturated; let also $\bigvee_{i=1}^k\varphi_i$ be the DNF of $\varphi$. The proofs of Proposition~\ref{prop:primality-LRS}--Corollary~\ref{cor:primality-LRS-3} are analogous to the proofs of Proposition~\ref{prop:primality-LS}--Corollary~\ref{cor:primality-LS-3}, where here we also make use of the fact that every $\varphi_i$ is characteristic for $p_{\varphi_i}$ from Lemmas~\ref{lem:rs-associated-process} and~\ref{lem:rs-dnf-disjuncts}.

\begin{prop}\label{prop:primality-LRS}
    Let $\varphi\in\mathL_{RS}$ be a saturated formula given by the grammar $\varphi::=\varphi\wedge\varphi~\mid~\varphi\vee\varphi~\mid~\langle a\rangle\varphi~\mid~[a]\ff$ such that for every $\langle a\rangle\varphi'\in\sub(\varphi)$, $\varphi'$ is saturated; let also $\bigvee_{i=1}^k\varphi_i$ be the DNF of $\varphi$. Then, 
     $\varphi$ is prime iff $\varphi\models \varphi_j$ for some $1\leq j \leq k$.
\end{prop}

\begin{lem}\label{lem:rs-common-divisor-pairs}
    Let $\varphi\in\mathL_{RS}$ be a saturated formula given by the grammar $\varphi::=\varphi\wedge\varphi~\mid~\varphi\vee\varphi~\mid~\langle a\rangle\varphi~\mid~[a]\ff$ such that for every $\langle a\rangle\varphi'\in\sub(\varphi)$, $\varphi'$ is saturated; let also $\bigvee_{i=1}^k\varphi_i$ be the DNF of $\varphi$. If for every pair $p_{\varphi_i},p_{\varphi_j}$, $1\leq i,j\leq k$, there is some process $q$ such that $q\curle_{RS} p_{\varphi_i}$, $q\curle_{RS} p_{\varphi_j}$, and $q\models\varphi$, then there is some process $q$ such that $q\curle_{RS} p_{\varphi_i}$ for every $1\leq i\leq k$, and $q\models\varphi$. 
\end{lem}

\begin{cor}\label{cor:rs-common-divisor-pairs}
   Let $\varphi\in\mathL_{RS}$ be a saturated formula given by the grammar $\varphi::=\varphi\wedge\varphi~\mid~\varphi\vee\varphi~\mid~\langle a\rangle\varphi~\mid~[a]\ff$ such that for every $\langle a\rangle\varphi'\in\sub(\varphi)$, $\varphi'$ is saturated; let also $\bigvee_{i=1}^k\varphi_i$ be the DNF of $\varphi$. If for every pair $p_{\varphi_i},p_{\varphi_j}$, $1\leq i,j\leq k$, there is some process $q$ such that $q\curle_{RS} p_{\varphi_i}$, $q\curle_{RS} p_{\varphi_j}$, and $q\models\varphi$, then there is some $1\leq m\leq k$, such that $p_{\varphi_m}\curle_{RS} p_{\varphi_i}$ for every $1\leq i\leq k$. 
\end{cor}

\begin{prop}\label{prop:primality-LRS-2}
    Let $\varphi\in\mathL_{RS}$ be a saturated formula given by the grammar $\varphi::=\varphi\wedge\varphi~\mid~\varphi\vee\varphi~\mid~\langle a\rangle\varphi~\mid~[a]\ff$ such that for every $\langle a\rangle\varphi'\in\sub(\varphi)$, $\varphi'$ is saturated; let also $\bigvee_{i=1}^k\varphi_i$ be the DNF of $\varphi$. Then, $\varphi$ is prime iff for every pair $p_{\varphi_i},p_{\varphi_j}$, $1\leq i,j\leq k$, there is some process $q$ such that $q\curle_{RS} p_{\varphi_i}$, $q\curle_{RS} p_{\varphi_j}$, and $q\models \varphi$.
\end{prop} 
%  \begin{proof} Analogous to the proof of Proposition~\ref{prop:primality-LS-2}, where we also need Corollary~\ref{cor:rs-common-divisor-pairs}.
% \end{proof}

\begin{cor}\label{cor:primality-LRS-3}
    Let $\varphi\in\mathL_{RS}$ be a saturated formula given by the grammar $\varphi::=\varphi\wedge\varphi~\mid~\varphi\vee\varphi~\mid~\langle a\rangle\varphi~\mid~[a]\ff$ such that for every $\langle a\rangle\varphi'\in\sub(\varphi)$, $\varphi'$ is saturated; let also $\bigvee_{i=1}^k\varphi_i$ be the DNF of $\varphi$. Then, $\varphi$ is prime iff for every pair $\varphi_i$, $\varphi_j$ there is some $1\leq m \leq k$ such that $\varphi_i\models \varphi_m$ and $\varphi_j\models\varphi_m$.
\end{cor}

\begin{cor}\label{cor:primality-phis}
     Let $\varphi\in\mathL_{RS}$ be a formula such that if $\psi\in\sub(\varphi)$ is unsatisfiable, then $\psi=\ff$ and occurs in the scope of some $[a]$; let also $\varphi^s$ be the output of $\Call{satur}{\varphi}$  and $\bigvee_{i=1}^k \varphi^s_i$ be $\varphi^s$ in DNF. Then, 
     $\varphi^s$ is prime iff $\varphi^s\models \varphi^s_j$ for some $1\leq j \leq k$, such that $\varphi^s_j\neq \true$.
\end{cor}
\begin{proof} The proof is analogous to the proof of Corollary~\ref{cor:primality-phitt}, where here we need that from Lemmas~\ref{lem:saturated-algo}, \ref{lem:rs-associated-process}, and~\ref{lem:rs-dnf-disjuncts}, $\varphi^s_j$ is prime, for every $1\leq i\leq k$.
\end{proof}

\begin{prop}\label{prop:phis-primality}
  Let $\varphi\in\mathL_{RS}$ be a formula such that if $\psi\in\sub(\varphi)$ is unsatisfiable, then $\psi=\ff$ and occurs in the scope of some $[a]$; let also  $\varphi^s$ be the output of $\Call{satur}{\varphi}$. There is a polynomial-time algorithm that decides whether $\varphi^s$ is prime.
\end{prop}
\begin{proof}
    We describe algorithm \algosat, which takes $\varphi^s$ and decides whether $\varphi^s$ is prime. If $\varphi^s=\true$, \algosat rejects. Otherwise, from Lemma~\ref{lem:saturated-algo}, $\true\not\in\varphi^s$. Then, \algosat constructs the alternating graph $G_{\varphi^s}=(V,E,A,s,t)$ by starting with vertex $(\varphi^s,\varphi^s\Rightarrow\varphi^s)$ and repeatedly applying the rules for ready simulation, i.e.\ the rules from Table~\ref{tab:S-rules}, where rule (L$\wedge_i$), where $i=1,2$, is replaced by the following two rules:\\
    \begin{minipage}{0.45 \textwidth}
        \begin{prooftree}
       \AxiomC{$\varphi_1\wedge\varphi_2,\varphi\Rightarrow \langle a\rangle\psi$}
       \RightLabel{\scriptsize(L$\wedge_i^\diamond$)}
        \UnaryInfC{$\varphi_1,\varphi\Rightarrow \langle a\rangle \psi~|_{\exists}~ \varphi_2,\varphi\Rightarrow \langle a\rangle \psi$}
        \end{prooftree}
    \end{minipage}
    \begin{minipage}{0.45 \textwidth}
        \begin{prooftree}
       \AxiomC{$\varphi_1\wedge\varphi_2,\varphi\Rightarrow [a]\ff$}
       \RightLabel{\scriptsize(L$\wedge_i^\Box$)}
        \UnaryInfC{$\varphi_1,\varphi\Rightarrow [a]\ff~|_{\exists}~ \varphi_2,\varphi\Rightarrow [a]\ff$}
        \end{prooftree}
    \end{minipage}\\
    
    \noindent and (tt) is replaced by the rule ($\Box$) as given below:
     \begin{prooftree}
        \AxiomC{$[a]\ff,[a]\ff\Rightarrow[a]\ff$ }
        \RightLabel{\scriptsize($\Box$)}
         \UnaryInfC{\textsc{True}}
        \end{prooftree}
Then, \algosat calls $\reacha(G_{\varphi^s})$, where $s$ is $(\varphi^s,\varphi^s\Rightarrow\varphi^s)$ and $t=\textsc{True}$, and accepts $\varphi^s$ iff there is an alternating path from $s$ to $t$.  The correctness of \algosat can be proven similarly to the correctness of \algott, by using analogous results proven in this subsection for ready simulation, i.e.\ Lemma~\ref{lem:conjunction_lemma_rs} and Corollaries~\ref{prop:primality-LRS-2} and~\ref{cor:primality-phis}.
\end{proof}

Similarly to the cases of the simulation and complete simulation preorders, \algosat can be modified to return a process for which the input formula is characteristic.

\begin{prop}\label{prop:find-p-rs-algo-phitt}
   Let $\varphi\in\mathL_{RS}$ be a formula such that if $\psi\in\sub(\varphi)$ is unsatisfiable, then $\psi=\ff$ and occurs in the scope of some $[a]$; let also  $\varphi^s$ be the output of $\Call{satur}{\varphi}$. If $\varphi^s$ is prime, there is a polynomial-time algorithm that constructs a process for which  $\varphi^s$ is characteristic within $\mathL_{RS}$.
\end{prop}
\begin{proof}
   A process for which $\varphi^s$ is characteristic within $\mathL_{RS}$ can be constructed by calling algorithm \algosat on $\varphi^s$ and following steps analogous to the ones described for the simulation preorder in Corollary~\ref{cor:find-p-algo}.
\end{proof}

We can now prove that prime formulae can be decided in polynomial time.

\rsprimehard*
\begin{proof}
  It suffices to compute $\varphi^s:=\Call{satur}{\varphi}$ and check whether $\varphi^s$ is prime and $\varphi^s\models\varphi$. Checking whether $\varphi^s\models\varphi$ holds is equivalent to constructing a process $p$ for which $\varphi^s$ is characteristic within $\mathL_{RS}$ and checking whether $p\models\varphi$. The correctness and polynomial-time complexity of this procedure is an immediate corollary of Lemma~\ref{lem:satur-poly-time} and Propositions~\ref{prop:primality-LRS-algo}, \ref{prop:phis-primality}, and~\ref{prop:find-p-rs-algo-phitt}.
\end{proof}

Finally, deciding characteristic formulae in $\mathL_{RS}$ can be done in polynomial time.

\begin{cor}\label{cor:rs-decide-characteristic}
Let $|\act|=k$, $k\geq 1$.  Deciding characteristic formulae within $\mathL_{RS}$ is polynomial-time solvable.
\end{cor}
\begin{proof}
    The corollary follows from Corollary~\ref{cor:sat-s-cs-rs-poly} and Propositions~\ref{prop:rs-satisfiability-satisfiable} and~\ref{prop:rs-primality}.
\end{proof}

\begin{cor}\label{cor:find-p-rs-algo}
Let $|\act|=k$, $k\geq 1$ and $\varphi\in\mathL_{RS}$. If $\varphi$ is satisfiable and prime, then there is a polynomial-time algorithm that constructs a process for which  $\varphi$ is characteristic within $\mathL_{RS}$.
\end{cor}

\section{The \conp-membership of the formula primality problem for \texorpdfstring{$\mathL_{RS}$}{LRS} with an unbounded action set}\label{subsection:rs-primality-unbounded-appendix}

The \conp-hardness of deciding pimality in $\mathL_{RS}$ was demonstrated in the proof of Proposition~\ref{prop:decide-prime-rs-infinite-actions-hard} in the main body of the paper. To prove that the problem belongs to \conp, we first provide some properties of prime formulae. 

\begin{defi}\label{def:rs-associated-process-extended}
    Let $\varphi\in\mathL_{RS}$ be a  
    formula given by the grammar $\varphi::=\true ~\mid~ \varphi\wedge\varphi~\mid ~ \langle a \rangle \varphi ~\mid~ [a]\ff$. We define process $p_\varphi$ inductively as follows.
\begin{itemize}
    \item If either $\varphi=[a]\ff$ or $\varphi=\true$ , then $p_\varphi=\mathtt{0}$.
     \item If $\varphi=\langle a\rangle \varphi'$, then $p_\varphi=a. p_{\varphi'}$.
     \item If $\varphi=\varphi_1\wedge \varphi_2$, then $p_\varphi=p_{\varphi_1}+p_{\varphi_2}$.
\end{itemize}
\end{defi}

\begin{lem}\label{lem:rs-associated-process-extended}
    Let $\varphi\in\mathL_{RS}$ be a satisfiable 
    formula given by the grammar $\varphi::=\true ~\mid~ \varphi\wedge\varphi~\mid ~ \langle a \rangle \varphi ~\mid~ [a]\ff$. Then,
    \begin{enumerate}[(a)]
        \item $p_\varphi\models\varphi$, and
        \item if $\true\not\in\sub(\varphi)$, $\varphi$ is saturated and for every $\langle a\rangle\varphi'\in\sub(\varphi)$, $\varphi'$ is saturated, then $\varphi$ is characteristic within $\mathL_{RS}$ for $p_\varphi$.
    \end{enumerate} 
\end{lem}
\begin{proof}
    The proof of (a) is similar to the proof of Lemma~\ref{lem:cs-associated-process-extended}, whereas the proof of (b) is analogous to the proof of Lemma~\ref{lem:rs-associated-process}.
\end{proof}

\begin{lem}\label{lem:rs-prime-property-prelemma-1}
    Let $\varphi$ be a satisfiable and saturated formula given by the grammar $\varphi::=\true ~\mid~ \varphi\wedge\varphi~\mid ~ \langle a \rangle \varphi ~\mid~ [a]\ff$. If $\varphi\models\psi$ for some prime $\psi$ and $\Call{Satur}{\varphi}\neq\true$, then $\Call{Satur}{\varphi}\models\psi$.
\end{lem}
\begin{proof}
    We denote $\Call{Satur}{\varphi}$ by $\varphi^s$. Note that $\psi$ is characteristic within $\mathL_{RS}$ for some process, which we denote by $p_\psi$. For every process $p$, $p\models\varphi\implies p\models\psi$ by assumption, and from Remark~\ref{Remark:charforms}, $p_\psi\curle_{RS} p$. Since $\varphi^s\neq \true$ and $\varphi^s$ does not contain disjunctions, $\varphi^s$ is characteristic within $\mathL_{RS}$ from Lemmas~\ref{lem:saturated-algo} and~\ref{lem:rs-associated-process}. Thus, $\varphi^s$ is characteristic within $\mathL_{RS}$ for $p_{\varphi^s}$, where $p_{\varphi^s}$ is the process that corresponds to $\varphi^s$ and is constructed as described in Definition~\ref{def:rs-associated-process-extended}. We prove that $p_\psi\curle_{RS} p_{\varphi^s}$ by induction on the type of substitutions made by procedure \textsc{Satur} in $\varphi$.
    \begin{itemize}
        \item Let $\varphi^s=\varphi[\true\wedge\varphi'/\varphi']$. Then $\varphi^s\equiv\varphi$ and as was shown above $p_\psi\curle_{RS} p$ for every $p$ that satisfies $\varphi$, or equivalently, for every $p$ that satisfies $\varphi^s$. In particular, $p_{\varphi^s}$ described in Definition~\ref{def:rs-associated-process-extended}, satisfies $\varphi^s$ from Lemma~\ref{lem:rs-associated-process-extended}.
        \item  Let $\varphi^s$ be derived from $\varphi$ by substituting an occurrence of $\langle a\rangle\varphi'$ with $\true$, where $\varphi'$ is a non-saturated formula. Then, there is a process $p_{tt}=\mathtt{0}$ such that $p_{\varphi^s}\myarrowtau p_{tt}$, $t\in \act^*$, and $p_{tt}$ corresponds to this occurrence of $\true$ that substituted $\langle a\rangle\varphi'$ in $\varphi$. We consider two copies of $p_{\varphi^s}$, namely $p_{\varphi^s}^1$ and $p_{\varphi^s}^2$, that are as follows: for $i=1,2$, $p_{\varphi^s}^i$ is $p_{\varphi^s}$ where $p_{tt}$ is substituted with $p_{tt}^i=a.p_i$, where $p_1,p_2$ are two processes that satisfy $\varphi'$ and $I(p_1)\neq I(p_2)$. The existence of $p_1,p_2$ is guaranteed by Lemma~\ref{lem:rs-not-saturated-property}. Then, $p_{\varphi^s}^i\models\varphi$, and by assumption, $p_{\varphi^s}^i\models\psi$ and $p_\psi\curle_{RS}p_{\varphi^s}^i$ from Remark~\ref{Remark:charforms}, for both $i=1,2$. If there is $p_\psi\myarrowtau p'$, such that $p_{\varphi^s}^1\myarrowtau p_1$ and $p'\curle_{RS} p_1$, then $p'\not\curle_{RS} p_2$, since $I(p_1)\neq I(p_2)$, and so there is $p_{\varphi^s}^2\myarrowtau q_2$ such that $p'\curle_{RS} q_2$ and $q_2\neq p_2$. Note that $q_2$ is the copy of some $q$ such that $p_{\varphi^s}\myarrowtau q$ and $p'\curle_{RS} q$ by the definition of $p_{\varphi^s}^2$. As a result, for every $p_\psi\myarrowa p'$, there is $p_{\varphi^s}\myarrowa p''$ such that $p'\curle_{RS} p''$. So, $p_\psi\curle_{RS}p_{\varphi^s}$. 
    \end{itemize}
    Since $\psi,\varphi^s$ are characteristic within $\mathL_{RS}$ for $p_\psi$ and $p_{\varphi^s}$, respectively, we have that $\varphi^s\models\psi$ from Remark~\ref{Remark:charforms}.
\end{proof}

\begin{lem}\label{lem:rs-prime-property-prelemma-2}
    Let $\varphi$ be a satisfiable and saturated formula given by the grammar $\varphi::=\true ~\mid~ \varphi\wedge\varphi~\mid ~ \langle a \rangle \varphi ~\mid~ [a]\ff$. If $\varphi\models\psi$ for some prime $\psi$, then $\Call{Satur}{\varphi}\neq\true$.
\end{lem}
\begin{proof}
   Let $\varphi^s$ denote $\Call{Satur}{\varphi}$ and suppose that  $\varphi^s=\true$. Let also $p_\psi$ and $p_{\varphi^s}$ be as in the proof of Lemma~\ref{lem:rs-prime-property-prelemma-1}. In the proof of Lemma~\ref{lem:rs-prime-property-prelemma-1}, we showed that $p_\psi\curle_{RS} p_{\varphi^s}$. In this case, from Definition~\ref{def:rs-associated-process-extended}, $p_{\varphi^s}=\mathtt{0}$, which means that $p_\psi=\mathtt{0}$ and $\psi\equiv\zero$. Since $\varphi\models\psi$, $\varphi\equiv\zero$ as well. Then, it is not possible that $\varphi^s=\true$ because of the type of the substitutions made by \textsc{Satur}, contradiction.
\end{proof}

\begin{lem}\label{lem:rs-prime-property}
    Let $\varphi\in\mathL_{RS}$ and $\bigvee_{i=1}^k\varphi_i$ be the DNF of $\varphi$. Then, $\varphi$ is prime iff there is some prime $\varphi_i$, where $1\leq i\leq k$, such that for every satisfiable $\varphi_j$, where $1\leq j\leq k$, %$\Call{satur}{\varphi_j}\neq\true$ and 
    $\Call{satur}{\varphi_j}\models\varphi_i$.
\end{lem}
\begin{proof} For every $\varphi\in\mathL_{RS}$, let $\varphi^s$ denote $\Call{satur}{\varphi}$. \\
    ($\Leftarrow$) Assume that for every satisfiable $\varphi_j$, $\varphi^s_j\models\varphi_i$, for some prime $\varphi_i$. For every $\varphi\in\mathL_{RS}$, it holds that $\varphi\models\varphi^s$ as was shown in the proof of Proposition~\ref{prop:primality-LRS-algo}. Hence, $\varphi_j\models\varphi^s_j$ and so $\varphi_j\models\varphi_i$. Consequently, for every $\varphi_j$, $\varphi_j\models\varphi_i$. From Lemmas~\ref{lem:DNF-equiv} and~\ref{lem:disjunction_lemma} and the fact that $\varphi_i\models\varphi$, $\varphi\equiv\varphi_i$, which implies that $\varphi$ is prime.\\
    ($\Rightarrow$) Let $\varphi$ be prime. Then, from Lemma~\ref{lem:DNF-equiv} and Definition~\ref{def:prime-formula}, there is some $\varphi_i$ such that $\varphi\models\varphi_i$. This means that $\varphi\equiv\varphi_i$ and so $\varphi_i$ is prime. From Lemma~\ref{lem:disjunction_lemma}, $\varphi_j\models\varphi_i$ for every $1\leq j\leq k$. Assume that there is some $1\leq j\leq k$ such that $\varphi_j$ is satisfiable. Then, from Lemmas~\ref{lem:rs-prime-property-prelemma-1} and~\ref{lem:rs-prime-property-prelemma-2}, $\varphi_j^s\neq\true$ and $\varphi^s_j\models\varphi_i$.
\end{proof}

\begin{lem}\label{lem:rs-prime-property-2}
    Let $\varphi\in\mathL_{RS}$ and $\bigvee_{i=1}^k\varphi_i$ be the DNF of $\varphi$.  Then, $\varphi$ is prime iff (a) for every satisfiable  $\varphi_i$, $1\leq i\leq k$, $\Call{Satur}{\varphi_i}\neq\true$, and (b) for every satisfiable $\varphi_i$, $\varphi_j$, $1\leq i,j\leq k$, there is some $\varphi_m$, $1\leq m\leq k$, such that  $\Call{Satur}{\varphi_i}\models\varphi_m$ and $\Call{Satur}{\varphi_j}\models\varphi_m$. 
\end{lem}
\begin{proof}  For every $\varphi\in\mathL_{RS}$, let $\varphi^s$ denote $\Call{satur}{\varphi}$. \\
    ($\Rightarrow$)  This direction is immediate from Lemmas~\ref{lem:rs-prime-property-prelemma-2} and~\ref{lem:rs-prime-property}.\\
    ($\Leftarrow$) If $\varphi$ is unsatisfiable, then $\varphi$ is prime and we are done. Assume that $\varphi$ is satisfiable and (a) and (b) are true. It suffices to show that there is some $\varphi_m$, $1\leq m\leq k$, such that for every $1\leq i\leq k$, $\varphi_i^s\models\varphi_m$. Then, $\varphi_m^s\models\varphi_m$ and consequently, $\varphi_m\equiv\varphi_m^s\neq\true$, which implies that $\varphi_m$ is prime. From Lemma~\ref{lem:rs-prime-property}, $\varphi$ is prime. Let $\varphi_1^s,\dots,\varphi_k^s$ be $k$ satisfiable formulae such that for every pair $\varphi_i^s$, $\varphi_j^s$ there is some $\varphi_m$ such that  $\varphi_i^s\models\varphi_m$ and $\varphi_j^s\models\varphi_m$. We prove by strong induction on $k$ that there is $\varphi_m$ such that for every $\varphi_i^s$, $\varphi_i^s\models\varphi_m$. For $k=2$, the argument is trivial. Let the argument hold for $k\leq n-1$ and assume we have $n$ satisfiable formulae $\varphi_1^s,\dots,\varphi_n^s$. From assumption, we have  that for every pair $\varphi_i^s,\varphi_n^s$, $1\leq i\leq n-1$, there is $\varphi_{in}$ such that $\varphi_i^s\models\varphi_{in}$ and $\varphi_n^s\models\varphi_{in}$.  Then, since $\varphi_{1n},\dots,\varphi_{n-1,n}$ are at most $n-1$ formulae, from inductive hypothesis there is some $\varphi_m$, $1\leq m\leq n$, such that $\varphi_{in}^s\models\varphi_m$, for every $1\leq i\leq n-1$. As a result,  $\varphi_i^s,\varphi_n^s\models\varphi_{in}\models\varphi_{in}^s\models\varphi_m$, which means that for every $1\leq i\leq n$, $\varphi_i^s\models\varphi_m$, which was to be shown.
\end{proof}

\begin{lem}\label{lem:check-saturated-without-disjunctions}
    Let $\varphi\in\mathL_{RS}$ be a satisfiable formula given by the grammar $\varphi::=\varphi\wedge\varphi~\mid~\langle a\rangle\varphi~\mid~ [a]\ff$. Then, deciding whether $\varphi$ is saturated can be done in polynomial time.  
\end{lem}
\begin{proof}
     W.l.o.g.\ $\varphi=\bigwedge_{i=1}^m\varphi_i$, where $\varphi_i$ is either $[a]\ff$ or $\langle a\rangle\varphi'$, where $\varphi'$ is given by the same grammar as $\varphi$. Let $\psi=\sat(\varphi)$. Then, $\psi=\bigwedge_{i=1}^m\psi_i$, where $\psi_i$ is either $x_a$ or $\neg x_a$.
      It is not hard to see that $\varphi$ is saturated iff for every $x_{a}$, exactly one of $x_{a}$, $\neg x_{a}$ occurs in $\psi$.
\end{proof}

\begin{lem}\label{lem:RS-poly-sat-no-disj}
 Let $\act$ be unbounded and $\varphi$ be a formula in $\mathL_{RS}$ that does not contain disjunctions. Then, the satisfiability of $\varphi$ can be decided in polynomial time.
\end{lem}
\begin{proof}
Consider the algorithm that first repeatedly applies the rules $\langle a\rangle\ff\ruleff \ff$, $\ff\wedge\psi\ruleff \ff$, and $\psi\wedge\ff\ruleff \ff$ on $\varphi$ until no rule can be applied. If the resulting formula $\varphi'$ is $\ff$, it rejects. Otherwise, it checks that for every $\varphi_1\wedge\varphi_2\in \sub(\varphi')$, it is not the case that $\varphi_1=\langle a\rangle \varphi_1'$ and $\varphi_2=[a]\ff$, for some $a\in \act$. This is a necessary and sufficient condition for acceptance.
\end{proof}

\begin{prop}\label{prop:conp-algo-prime-rs-unbounded}
Let $\act$ be unbounded. The formula primality problem for $\mathL_{RS}$ is in \conp.
\end{prop}
\begin{proof}
 We now describe algorithm \algorsu that decides primality of $\varphi\in \mathL_{RS}$. Let $\bigvee_{i=1}^k\varphi_i$ be the DNF of $\varphi$. Given $\varphi\in\mathL_{RS}$, \algorsu calls the \conp algorithm for unsatisfiability of $\varphi$. If a universal guess of that algorithm accepts, \algorsu accepts. Otherwise,  \algorsu universally guesses a pair $\varphi_i$, $\varphi_j$. Since these formulae do not contain disjunctions, \algorsu decides in polynomial time whether $\varphi_i$, $\varphi_j$ are satisfiable as explained in the proof of Lemma~\ref{lem:RS-poly-sat-no-disj}. If at least one of them is unsatisfiable, \algorsu accepts. Otherwise, it computes $\varphi_i^s=\Call{Satur}{\varphi_i}$ and $\varphi_j^s=\Call{Satur}{\varphi_j}$, which can be done in polynomial time from Lemma~\ref{lem:check-saturated-without-disjunctions}. If at least one of $\varphi_i^s,\varphi_j^s$ is $\true$, then it rejects. Otherwise, \algorsu needs to check whether there is some $\varphi_m$ such that  $\varphi_i^s\models\varphi_m$ and $\varphi_j^s\models\varphi_m$. It does that by constructing a DAG $G_\varphi^{ij}$ similarly to the proof of  Proposition~\ref{prop:phis-primality}. It starts with vertex $s=(\varphi_i,\varphi_j\Rightarrow\varphi)$ and applies the rules that were introduced in the proof of  Proposition~\ref{prop:phis-primality} for ready simulation. Next, it solves \reacha on $G_\varphi^{ij}$, where $t=\textsc{True}$. \algorsu accepts iff there is an alternating path from $s$ to $t$. The correctness of these last steps can be proven similarly to the case of $\mathL_{RS}$ with a bounded action set. In particular, a variant of Lemma~\ref{lem:conjunction_lemma_rs} holds here since $\varphi_i$, $\varphi_j$ are satisfiable and prime.  \algorsu  is correct due to Lemma~\ref{lem:rs-prime-property-2}.
 \end{proof}

  \end{appendix}
\end{document}